\documentclass[prb,aps,showpacs,twocolumn,preprintnumbers,
amsmath,amssymb,superscriptaddress,longbibliography]{revtex4-2}
\usepackage[english]{babel}
\usepackage{amsmath,amssymb,amsfonts}
\usepackage{graphicx}
\usepackage[colorlinks=True,linkcolor=red,citecolor=blue,urlcolor=blue]{hyperref}

\usepackage{booktabs}
\usepackage[dvipsnames]{xcolor}
\usepackage{braket}
\usepackage{bm}
\usepackage{bbm}
\usepackage{enumitem}
\usepackage{makecell}
\usepackage{float}
\usepackage{multirow}
\usepackage{longtable}
\usepackage[normalem]{ulem}
\usepackage{array}
\usepackage{makecell}
\usepackage{subfigure}
\usepackage{graphicx}
\graphicspath{{NewFigures/}}
\usepackage{dcolumn}
\usepackage{xcolor}

\usepackage{verbatim}

\newcolumntype{C}{>{$}c<{$}}

\renewcommand{\Re}{\mathop{\mathop{Re}}}
\renewcommand{\Im}{\mathop{\mathrm{Im}}}

\allowdisplaybreaks

\def\CH{\textcolor{magenta}}

\allowdisplaybreaks

\begin{document}

\title{Nonlocality-induced critical-length hierarchy from non-Hermitian competition}

\author{Mengjie Yang}
\affiliation{Department of Physics, National University of Singapore, Singapore 117551, Singapore}

\author{Alexander N. Poddubny}
\email{poddubny@weizmann.ac.il}
\affiliation{Department of Physics of Complex Systems, Weizmann Institute of Science, Rehovot 7610001, Israel}

\author{Ching Hua Lee}
\email{phylch@nus.edu.sg}
\affiliation{Department of Physics, National University of Singapore, Singapore 117551, Singapore}

\date{\today}

\begin{abstract}
Spectral transitions in non-Hermitian lattices often arise from the competition between non-reciprocal skin accumulation and inter-component hybridization. In short-range systems formed by two coupled chains, this competition conventionally leads to the logarithmic critical-length law $N_c\sim\ln D$, where $D$ is the transverse separation between the chains. Here we show that long-range hoppings fundamentally reorganizes this critical behavior, producing a hierarchy of distinct scaling laws. When only the hybridization couplings are power-law decaying with exponent $\alpha$, the onset becomes algebraic, $N_c\sim D^{\alpha/3}$. When the hoppings within each chain are themselves also power-law decaying, in addition to the hybridization couplings, the system enters a scale-covariant regime for $\alpha<2$, in which the criticality threshold equation depends only on the system aspect ratio $N_c/D$. 
At $\alpha=2$ and beyond, this regime is followed by a marginal logarithmically corrected and algebraically corrected regimes, respectively.  
We identify two new non-local mechanisms that enable this unconventional critical hierarchy: a nonanalytic band-edge dispersion from long-range intra-chain hoppings, and parity-mixing hybridization induced by non-reciprocity. Our results show that nonlocality systematically removes the physical length scales i.e. skin depth underlying conventional critical non-Hermitian skin behavior, offering a platform-independent framework testable in programmable topoelectrical circuits, photonic lattices and digital quantum simulators.
\end{abstract}

\maketitle

\section{Introduction}

Non-Hermitian lattices exhibit intrinsically nonlocal responses when reciprocity is broken: spectra and eigenmodes can depend sensitively on boundary conditions and impurities, with the non-Hermitian skin effect (NHSE)~\cite{lee2016anomalous,alvarez2018non,yao2018edge,kunst2018biorthogonal,lee2019anatomy,okuma2020topological,lin2023topological,okuma2023non,longhi2019topological,kawabata2019symmetry,lee2019hybrid,song2019non,borgnia2020non,helbig2020generalized,li2020critical,longhi2020non,lee2020unraveling,zou2021observation,zhang2021observation,yang2022concentrated,zhang2021tidal,li2022non,yang2022designing,xue2022non,longhi2022self,gu2022transient,jiang2023dimensional,manna2023inner,xiong2024non,xue2024topologically,tai2023zoology,lei2024activating,shimomura2024general,yang2024percolation,gliozzi2024many,yang2025non,li2025phase,yoshida2024non,hamanaka2025multifractality,yan2024transport,das2024quantized,yang2025beyond,cheng2026non,shi2025chiral,yang2026reversing} as a paradigmatic example. For short-ranged models, this boundary sensitivity can often be organized by generalized Brillouin zone~\cite{yao2018edge,lee2019anatomy,song2019non,yang2020non,longhi2020non,helbig2020generalized,li2025phase,meng2025generalized,yuan2026non} and transfer-matrix approaches~\cite{kunst2019non,luo2021transfer,li2026non}, which reduce boundary phenomena to local recursion data and yield closed-form scaling laws for the standard finite-size crossovers.

This picture is qualitatively reshaped by long-range couplings. For power-law-decaying hoppings, the Bloch symbol can acquire singular or nonanalytic structures, so boundary modes and finite-size splittings are no longer governed by the same finite-range local recursion logic as in short-ranged chains~\cite{lepori2017long,gu2016holographic,jones2023bulk,perezgonzalez2019ssh,vodola2014kitaev,viyuela2016topological,alecce2017extended,jager2020edge,patrick2017topological,qi2021topological,chavez2021disorder,singh2017effect,gong2016topological,kim2024longrange}. Such nonlocal couplings can further modify the NHSE itself, producing scale-free skin modes, altered spectral transitions, and localization properties absent in finite-range models~\cite{wang2023scaling,xu2021non,guo2024scale,liu2025non,qin2026anyon}. Thus, even at the simplest single-chain level in 1D, nonlocality does not merely mean ``more connectivity'': it creates interference pathways across multiple length scales, reshapes the relevant gaps, and renormalizes finite-size crossovers. As we show below, these nonlocality features in a single 1D non-Hermitian chain propagates into, and ultimately governs, a new fundamental hierarchy of complex spectral transitions when such chains are coupled together.

Coupling non-Hermitian chains together, however weakly, allows their individual NHSE amplification channels to concatenate and form amplifying feedback loops. Such amplification can cause the energy spectrum to transition abruptly from real to complex. When all hoppings are \emph{local}, this occurs only when the resultant feedback loops are long enough to allow for sufficient NHSE growth -- this scenario is famously known as the critical NHSE, where a real-to-complex transition occurs by tuning the system size, leading to unconventional critical scaling~\cite{li2020critical,lee2020unraveling,yang2024percolation}. 

Long-range couplings naturally arise in realistic platforms when interactions between otherwise localized degrees of freedom are mediated by collective phonon modes, shared optical cavities, waveguides, or circuit buses~\cite{britton2012engineered,landig2016quantum,evans2018photon,zhang2023superconducting,pellerin2024wavefunction}.  
This raises the general question of whether realistic coupled systems with effective non-local descriptions possess 
finite-size instability thresholds that are correspondingly reshaped, or merely rescaled. We provide rigorous support of the former, that it is indeed qualitatively reshaped: long-range couplings fundamentally alters the spectral instability, with the precise nature of the reorganization depending sensitively on how nonlocality is distributed across the lattice. By tracking a minimal representative ladder model with fully local, nonlocal-rungs, and fully nonlocal geometries, we identify a sophisticated hierarchy of three qualitatively distinct critical behaviors: the conventional logarithmic growth characterizing purely short-range competitive NHSE systems~\cite{li2020critical}; a distinct spectral onset once long-range coupling is introduced on the inter-chain rungs; and a novel scale-covariant regime in the fully nonlocal limit, where the critical threshold no longer fixes an absolute chain length but instead fixes a geometric aspect ratio, independent of overall system size. 

\section{Results}

\subsection{Background: nonlocality signatures in a single non-Hermitian chain.}

\begin{figure}
    \centering
    \includegraphics[width=0.99\linewidth]{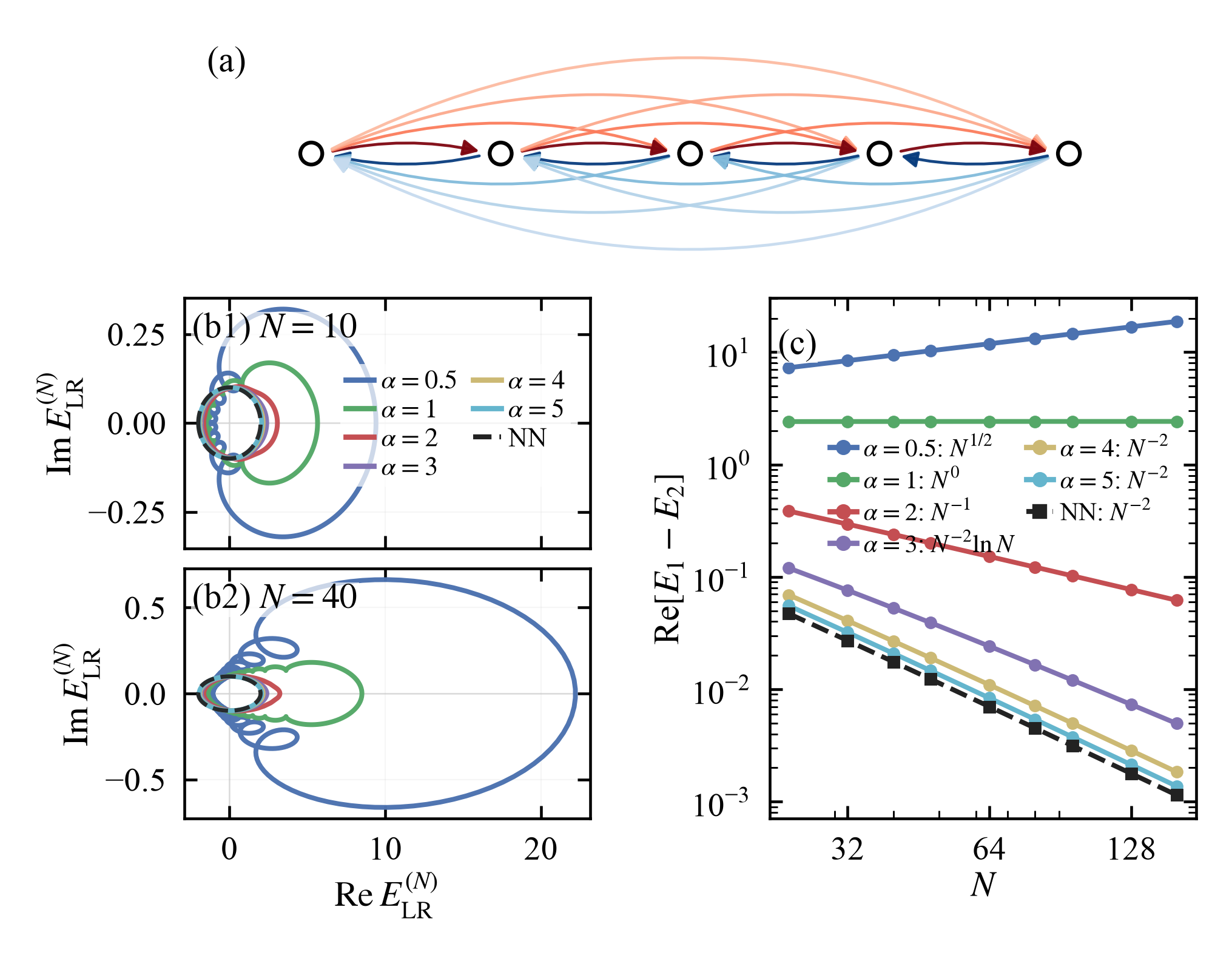}
   \caption{Single-chain signatures of nonlocality. (a) Schematic of a non-reciprocal long-range chain with algebraically decaying asymmetric hoppings [Eq.~\eqref{eq:hamLR}]. (b1,b2) Finite-cutoff Bloch spectra of Eq.~\eqref{eq:LR_Bloch} for $N=10$ and $40$, $\gamma=0.05$, and $\alpha=0.5,1,2,3,4,5$, with the nearest-neighbor Hatano--Nelson chain shown as the black dashed reference. Reducing $\alpha$ strongly enlarges the spectral excursion and exposes the nonanalytic long-range band edge. (c) Open-boundary band-edge spacing $\Delta\operatorname{Re}E=\operatorname{Re}(E_1-E_2)$, demonstrating the crossover from the long-range scaling $\Delta E\sim N^{1-\alpha}$ to the quadratic nearest-neighbor scaling $\Delta E\sim N^{-2}$.}
    \label{fig:sgchain}
\end{figure}

Before discussing how coupled nonlocal chains can compete to produce a hierarchy of critical non-Hermitian transitions, we first review the relevant key features of an individual nonlocal chain. Working under open boundary conditions (OBCs) throughout this entire work, we first consider the 1D length-$N$ chain with asymmetric power law-decaying long-range  (LR) hoppings [Fig.~\ref{fig:sgchain}(a)]:
\begin{equation}
    H^{\mathrm{LR}}(+\gamma)=\sum_{n\neq m}\frac{1+\gamma\operatorname{sgn}(n-m)}{|n-m|^\alpha}c^\dagger_n c_m, \label{eq:hamLR}
\end{equation}
where $\alpha$ controls the power-law decay and $\gamma$ introduces the hopping asymmetry (non-reciprocity). The lattice nodes correspond to dimensionless positions $n=1,2,...,N$ in units of lattice spacing. 
Small $\alpha$ correspond to strongly nonlocal hoppings. The large-$\alpha$ limit suppresses all hoppings with $|n-m|>1$ according to $|n-m|^{-\alpha}$; for $\alpha \geq 4$, it is in practice well-approximated by the nearest-neighbor (NN) Hatano--Nelson limit~\cite{hatano1996localization,hatano1997vortex}
\begin{equation}
    H^{\mathrm{NN}}(\gamma)=\sum_{n=1}^{N-1}\left[(1+\gamma)c_{n+1}^\dagger c_n+(1-\gamma)c_n^\dagger c_{n+1}\right], \label{eq:hamNN}
\end{equation}
with only nearest-neighbor ($|n-m|=1$) hoppings. In that limit, the OBC right eigenstates are exactly skin-deformed standing waves~\cite{yao2018edge,okuma2020topological}, $\psi_\ell(x)\propto e^{\kappa x}\sin(k_\ell x)$, where $\kappa=\ln\sqrt{(1+\gamma)/(1-\gamma)}$ is the inverse skin depth and $k_\ell=\ell\pi/(N+1)$, $\ell=1,\dots,N$, is the mode index labeling the $N$ discrete standing-wave momenta. 

Long-range hopping does not merely add more bonds to the chain---it reshapes the entire single-chain problem, at the levels of both its band structure (Fig.~\ref{fig:sgchain}) and its eigenstates, consistent with related studies of non-Hermitian chains with power-law or extended-range couplings~\cite{lepori2017long,jones2023bulk,perezgonzalez2019ssh,wang2023scaling,xu2021non,li2021extended,wu2022nonhermiticity,shi2024entanglement,rafiulislam2024knots,lai2025skin,zhao2025tentacle}, as detailed in Appendix~\ref{sec:single_chain_properties}. As will be explained later, of key importance to the real-to-complex phase transition are the two band-edge eigenenergies $E_1$ and $E_2$. To estimate them analytically, we use the finite-size Bloch symbol, namely the Fourier transform of the translation-invariant hopping amplitudes truncated to the hopping distances available in an $N$-site chain
\begin{eqnarray}
E_{\rm LR}^{(N)}(k)
&=&\sum_{r=1}^{N-1}
\frac{(1+\gamma)e^{\mathrm{i}kr}+(1-\gamma)e^{-\mathrm{i}kr}}{r^\alpha}\notag\\
&=& 2\Re\,\operatorname{Li}_{\alpha}^{[N-1]}(e^{\mathrm{i}k})+2\mathrm{i}\gamma\,\Im\,\operatorname{Li}_{\alpha}^{[N-1]}(e^{\mathrm{i}k})\notag\\
\label{eq:LR_Bloch}
\end{eqnarray}
evaluated at $k=\frac{\pi}{N}$ and $\frac{2\pi}{N}$, where $\operatorname{Li}*{\alpha}^{[N-1]}(z)\equiv \sum*{r=1}^{N-1}z^r/r^\alpha$ is the truncated polylogarithm function. Conventionally, a Bloch symbol is evaluated at a real crystal momentum $k$, or equivalently at $\beta=e^{\mathrm{i}k}$ on the unit circle. A non-Bloch symbol instead analytically continues $e^{\mathrm{i}k}$ to a generally complex factor $\beta$ with $|\beta|\neq1$, thereby incorporating the exponential spatial deformation associated with the NHSE under OBCs. Equation~\eqref{eq:LR_Bloch} is the former: its finite cutoff retains the hopping distances available in the finite chain, while its values at the lowest standing-wave momenta $k\simeq\pi/N$ and $2\pi/N$ provide analytical approximations to the two OBC band-edge energies.

From Fig.~\ref{fig:sgchain}(b1,b2), $E_{\rm LR}^{(N)}(k)$ exhibits more and more intricately small loops i.e. ``epicycles" with increasing $N$ and decreasing $\alpha$. These complex spectral loops represent effective feedback channels in the translation invariant region, with amplification rate proportional to their Im$\,E$ radius. Evidently, further or stronger nonlocal hoppings increase the number and strength of the feedback amplification channels~\footnote{In the $\alpha\rightarrow \infty$ local limit, $E_{\rm LR}^{(N)}(k)$ is just a simple ellipse, corresponding to the one and only amplification channel around the PBC loop}.

We next discuss the lowest eigenenergy gap $\Delta E_{\rm LR} =\text{Re}(E_1-E_2)$ of a nonlocal chain, which will play an important role in controlling the effects of coupling such chains. In the large $N$ limit, $E^{(N)}_\text{LR}$ [Eq.~\eqref{eq:LR_Bloch}] is non-analytic, and can be shown to give 
\begin{equation}
    \Delta E_{\rm LR}=\text{Re}\lim_{N\to\infty}\left[E^{N}_\text{LR}\left(\frac{\pi}{N}\right)-E^{N}_\text{LR}\left(\frac{2\pi}{N}\right)\right]  \sim N^{1-\alpha}
    \label{eq:gap_LR}
\end{equation}
for sufficiently small $\alpha$. Two contrasting regimes are apparent: in the very nonlocal hopping regime of $\alpha<1$, the lowest gap increases with chain length $N$, while for less nonlocal  $\alpha>1$, it decreases. However, this lowest gap $\text{Re}(E_1-E_2)$ does not decrease indefinitely as the chain becomes even more local -- in the nearest neighbor hopping limit of $\alpha\rightarrow \infty$, the gap instead tends to 
\begin{equation}
    \Delta E_{\rm NN}=2\text{Re}\lim_{N\to\infty}\left[\cos\frac{\pi}{N}-\cos\frac{2\pi}{N}\right]  \simeq \frac{3\pi^2}{N^2}\sim N^{-2}.
    \label{eq:gap_NN}
\end{equation}
In other words, as the chain becomes more local beyond $\alpha=1$, the lowest gap decreases with $N$ faster and faster, till it saturates at the inverse square law $\sim N^{-2}$ [Fig.~\ref{fig:sgchain}(c)].  Incidentally at the marginal power $\alpha=3$, $\Delta E_{\rm LR}\sim N^{-2}\ln N$ [see Appendix~\ref{sec:single_chain_properties}], with the subleading $\ln N$ factor disappearing as $\alpha$ is further increased.

\begin{figure*}
\centering
\includegraphics[width=0.95\textwidth]{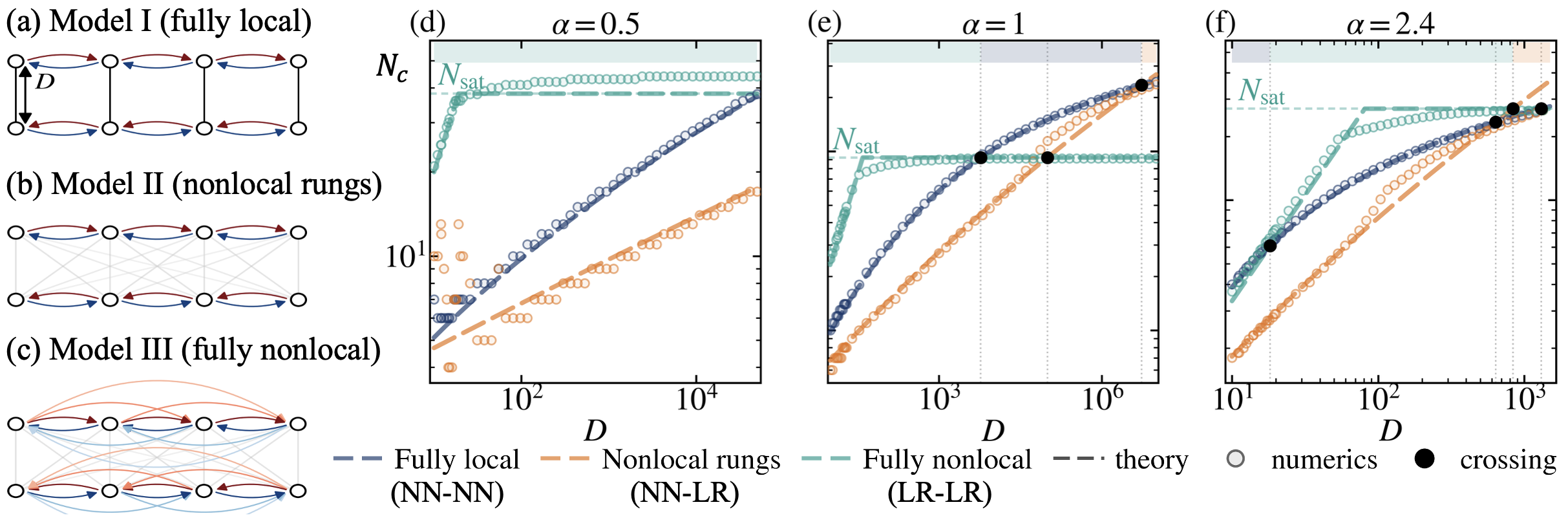}
\caption{Hierarchy of critical-length scalings in three competitive non-Hermitian skin lattices. (a)--(c) Schematics of the fully local (model I, NN--NN, Eq.~\eqref{eq:H_model_I_main}), nonlocal-rungs (model II, NN--LR, Eq.~\eqref{eq:H_model_II_main}), and fully nonlocal (model III, LR--LR, Eq.~\eqref{eq:H_model_III_main}) geometries, with opposite non-Hermitian pumping on the two legs and interchain separation $D$. The dominant scaling laws: $N_c\sim \ln D$, $N_c\sim D^{\alpha/3}$, and a scale-covariant fully nonlocal onset with finite-window saturation/crossover. (d)--(f) Critical length $N_c$ versus $D$ for $\alpha=0.5,1,2.4$ at fixed non-reciprocity. Open circles are obtained by direct diagonalization of the finite open-boundary single-particle lattice Hamiltonians, dashed curves are the analytical estimates [Eq.~\eqref{eq:NNNN_active_threshold_main},~\eqref{eq:NNLR_Nc_main}, and~\eqref{eq:NcIII_full_main} for model I, II, and III, respectively], and black markers indicate crossings between different onset mechanisms. The horizontal dashed line marks the saturation scale $N_{\rm sat}$ [Eq.~\eqref{eq:LRLR_Lsat_main}] of the fully nonlocal lattice. Calculation has been performed for $\gamma=0.05$. }
\label{fig:scalings}
\end{figure*}

\subsection{Coupled antagonistic nonlocal NHSE chain}

\subsubsection{Three contrasting nonlocal ladder configurations}

Here, we proceed to couple two NHSE chains, and investigate how nonlocality interplays with the coupling to induce real-complex transitions. It turns out that the amplification behavior is far more interesting than what is possible for a single nonlocal chain (which is already quite intricate). Without loss of generality, we set the coupled chains to have equal and opposite NHSE pumping strengths $\pm\gamma$~\footnote{This purely anti-symmetrical NHSE configuration can be easily gauge-transformed into a pair with any combination of NHSE strengths. For a pair of chains with inverse skin depths $\pm\kappa=\pm\ln\sqrt\frac{1+\gamma}{1-\gamma}$, threading an imaginary flux $h$ transforms their inverse skin depths viz. $\kappa \rightarrow \kappa +h$ and $-\kappa \rightarrow -\kappa +h$, which can take on any desired pair of values with suitable $\kappa$ and $h$.}. Each chain can have the nonlocality turned on or off, as given by $H^{\rm LR}(\pm\gamma)$ or $H^{\rm NN}(\pm\gamma)$ [Eqs.~\eqref{eq:hamLR} and \eqref{eq:hamNN}] respectively.

To systematically investigate the rich variety of possible nonlocal phenomena, we consider the following 3 variations in how the nonlocality enters the antagonistic chains and their couplings:
\begin{align}
\text{Model I:}&\qquad\mathcal{H}^{\rm I} = \begin{pmatrix} H^{\rm NN}(+\gamma) & H_{\perp}^{\rm NN} \\ (H_{\perp}^{\rm NN})^{\dagger} & H^{\rm NN}(-\gamma) \end{pmatrix}, \label{eq:H_model_I_main} \\
\text{Model II:}&\qquad\mathcal{H}^{\rm II} = \begin{pmatrix} H^{\rm NN}(+\gamma) & H_{\perp}^{\rm LR} \\ (H_{\perp}^{\rm LR})^{\dagger} & H^{\rm NN}(-\gamma) \end{pmatrix}, \label{eq:H_model_II_main} \\
\text{Model III:}&\qquad\mathcal{H}^{\rm III} = \begin{pmatrix} H^{\rm LR}(+\gamma) & H_{\perp}^{\rm LR} \\ (H_{\perp}^{\rm LR})^{\dagger} & H^{\rm LR}(-\gamma) \end{pmatrix}. \label{eq:H_model_III_main}
\end{align}
In all of these 3 configurations, illustrated in Figure~\ref{fig:scalings}(a-c), the two NHSE chains form the ``legs" of an OBC ladder separated by a spacing $D$, and the couplings between them form the ``rungs". The legs occupy the diagonal blocks of  $\mathcal{H}^{\rm I}$, $\mathcal{H}^{\rm II}$, and $\mathcal{H}^{\rm III}$, and their rungs occupy the off-diagonal blocks. We also consider two variations of the rung hoppings: The local version with rung couplings existing only between nearest-neighbor (NN) leg nodes:
\begin{equation}
H_{\perp}^{\rm NN} = D^{-\alpha} \sum_n c^\dagger_{+,n}c_{-,n}, \label{eq:H_int_NN}
\end{equation}
and the long-range (LR) version with all pairs of leg nodes connected:
\begin{equation}
H_{\perp}^{\rm LR} = \sum_{n,m} \frac{c^\dagger_{+,n}c_{-,m}}{\left(D^2+|n-m|^2\right)^{\alpha/2}}. \label{eq:H_int_LR}
\end{equation}
In all but $H^\text{NN}$, the coupling strength between any two nodes decay with power-law exponent $\alpha$, be they on the same leg or not.

The three ladder configuration models described by Eqs.~\ref{eq:H_model_I_main} to \ref{eq:H_model_III_main} encode increasing extents of nonlocality: $\mathcal{H}^{\rm I}$ [Figure~\ref{fig:scalings}(a)] is the usual critical NHSE ladder~\cite{li2020critical} with both legs and rungs being local, and is used as a comparison baseline. Introducing nonlocality first in the rungs (interchain couplings), we obtain $\mathcal{H}^{\rm II}$ [Figure~\ref{fig:scalings}(b)] with local (NN) legs and long-range  (LR) rungs. Further introducing nonlocality also to the legs, we obtain $\mathcal{H}^{\rm III}$ [Figure~\ref{fig:scalings}(c)] with long-range (LR) legs and long-range  (LR) rungs. 
Equivalently, we also refer to these 3 configurations as NN--NN, NN--LR, and LR--LR, where the first entry describe the intra-chain (leg) hoppings and the second specifies the inter-chain (rung) hoppings~\footnote{There exists another possible configuration LR--NN, but it is mathematically more subtle without illustrating additional new physics, and is discussed separately in Appendices~\ref{sec:app_model_definitions} and \ref{sec:app:Nc_modelIV}.}.

\subsubsection{Overview of key results}

To understand nonlocal NHSE amplification, the main question to answer is: Given a common ladder length $N$, which of the three configurations $\mathcal{H}^{\rm I},\mathcal{H}^{\rm II}$  or $\mathcal{H}^{\rm III}$ is the most stable? It turns out that the answer is immensely subtle, often out of the reach of intuitive explanations. 

A stable system necessarily has a real energy spectrum (Im$E=0$), since Im$E>0$ indicates unlimited exponential growth. We characterize system stability through $N_c$, \textbf{the critical system length where a real-to-complex spectral transition occurs.} In other words, it is the smallest $N$ supporting an eigenenergy pair $E,E^*$ with $\operatorname{Im}E\neq0$. A system of length $N$ is stable if $N<N_c$; otherwise, the smallest $N_c$ is, the more unstable it likely is. 

Table~\ref{table} summarizes our key analytic results on $N_c$, which will be rigorously derived in the following subsections. We observe that, because of the markedly different ways $N_c$ scale among our 3 ladder configurations, at a given ($N$, $\alpha$), \emph{any} of the 3 configurations can be the most stable candidate. The behavior of $N_c$ against ladder spacing $D$ is plotted in Figure~\ref{fig:scalings}(d-f) for three representative values of $\alpha$: 0.5 (very long-ranged), 1 (long-ranged) and 2.4 (somewhat short-ranged), with very good agreement between analytic predictions (dashed) and numerical results (small circles).  

\begin{table}[t]
\caption{Summary of the key contrasting behavior in our three competitive NHSE model configurations. For each model, the critical length $N_c$ follows from the pair-resolved EP-colliding-pair criterion, Eqs.~\eqref{eq:active_instability_main}--\eqref{eq:min_threshold_main}, evaluated with the model-specific projected coupling matrix elements. The third column gives our key new results: the closed-form critical onset conditions, with their asymptotic limiting scaling behavior given in the last column.}
\centering
\begin{tabular}{cccc}
\toprule
Model & Legs--Rungs & Key result & Limiting $N_c$ behavior \\
\midrule
I
& NN--NN
&  Eq.~\eqref{eq:NNNN_active_threshold_main}
& $(\alpha/\gamma)\ln D$ \\
[0.4em]

II
& NN--LR
&  Eq.~\eqref{eq:NNLR_Nc_main}
& $D^{\alpha/3}$ \\
[0.4em]

III
& LR--LR
&  Eq.~\eqref{eq:NcIII_full_main}
& \makecell[l]{
$\alpha{<}2$: $\left[2.4-0.026\alpha+0.33\alpha^2\right.$\\
\hspace{1.3em}$\left.-\gamma(11-12\alpha+5.7\alpha^2)\right]D$\\
\hspace{1.3em}[Eqs.~\eqref{eq:LRLR_AB_compact_main} and \eqref{eq:AB_lt2_empirical_main}]\\
$\alpha{=}2$: log-corrected [Eq.~\eqref{eq:LRLR_alpha2_main}]\\
$\alpha{>}2$: algebraic [Eq.~\eqref{eq:LRLR_ratio_23_main}]\\
capped at $N_{\rm sat}$ [Eq.~\eqref{eq:LRLR_Lsat_main}]
} \\
\bottomrule
\end{tabular}
\label{table}
\end{table}

Below, we discuss general trends in the hierarchy of stability of our 3 ladder configurations (models), as suggested by Figure~\ref{fig:scalings}(d-f). At small ladder spacings $D$, the strong inter-chain coupling makes all three models I to III susceptible to amplification (due to the small $N_c$), but the relative extents depend on the detailed balance between local skin overlap and aggregate feedback contributions from the nonlocal rungs. 
As $D$ increases, the $N_c$ curves for the 3 models diverge in qualitatively different ways: the fully local ladder's (model I) grows only logarithmically, the nonlocal rungs ladder's (model II) grows algebraically with exponent $\alpha/3$, while the fully nonlocal ladder's (model III) has an approximately linear leading growth in the strongly nonlocal regime, with finite-size corrections at and above $\alpha=2$, before reaching the cutoff scale $N_{\rm sat}$ [see Appendix~\ref{sec:app:Nc_LRLR} for the derivation]. 

Consequently, the most stable configuration can change as $D$ varies, in a manner reminiscent of a collection of first-order phase transitions: a configuration that first exhibits complex $E$ at small ladder spacing $D$ may be overtaken at larger spacings. For instance, at $\alpha=1$, the fully nonlocal (LR-LR) configuration (green) starts off as the most stable configuration (largest $N_c$) at small $D$.  But, as $D$ increases and the ladder becomes wider,  its $N_c$ eventually gets eclipsed by that of the fully local (NN-NN) and then the nonlocal rungs (NN-LR) configurations;
black dots in Fig.~\ref{fig:scalings} highlight these first-order crossings. 
Saliently, there is an apparent ``non-monotonicity" in the stability vs. nonlocality trend: in the window of $D$ demarcated by the central two black dots, the fully nonlocal configuration is neither the most nor the least stable. This rather counterintuitive observation will be discussed further in Sec.~\ref{sec:hierarchy}. In all, it is evident that nonlocality does not simply delay or advance the onset of amplification; rather, its precise effect is qualitatively distinct in different ladder configurations, depending intimately on $D$ and $\alpha$.  

\subsection{Key mechanisms for real-to-complex transitions in coupled nonlocal NHSE ladders}

Having surveyed the critical length $N_c$ behavior in our three ladder configurations, we now derive its underlying mechanism. Consider an initially real energy spectrum. As a parameter i.e. $N$, $D$ or $\alpha$ is tuned, the eigenenergies evolve along the real line. A real-to-complex transition occurs at when a pair of real eigenvalues first collide to form an exceptional point (EP) before escaping into the complex plane [see Appendix~\ref{sec:app_effective_hamiltonian}]. 

\subsubsection{Theoretical framework}

To obtain the critical length $N_c$ for this to occur, it suffices to project the Hamiltonian onto the subspace of the pair of single-chain modes, denoted by $|a\rangle$ and $|b\rangle$, that participates in the first EP collision~\footnote{This is generally valid very close to the real-to-complex transition, where this pair is much closer to each other than any other eigenmode.}. 
Since each of the two single-chain modes appears on both legs of the ladder, projecting onto this two-mode EP sector gives the $4\times4$ basis
${|a,+\rangle,|b,+\rangle,|a,-\rangle,|b,-\rangle}$
(see Appendix~\ref{sec:app_effective_projection}),
\begin{equation}
H_{\mathrm{EP}}=
\begin{pmatrix}
E_a & 0 & u_{aa} & u_{ab} \\
0 & E_b & u_{ba} & u_{bb} \\
v_{aa} & v_{ab} & E_a & 0 \\
v_{ba} & v_{bb} & 0 & E_b
\end{pmatrix}.
\label{eq:Hact_main}
\end{equation}
The off-diagonal blocks contain the projected rung hopping matrix elements
$
u_{ij}=\langle i,+|H_{\rm int}|j,-\rangle,
v_{ij}=\langle i,-|H_{\rm int}^{\dagger}|j,+\rangle
, i,j\in{a,b}$. 
Diagonalizing Eq.~\eqref{eq:Hact_main}, the first real-to-complex transition occurs when the relevant EP discriminant becomes negative,
\begin{equation}
\Lambda_{\eta}^2+4u_{ab}v_{ba}<0,
\label{eq:active_instability_main}
\end{equation}
where $\Lambda_{\eta} = \Delta E_{ab} +\eta( u_{aa}-u_{bb})$, $\Delta E_{ab}\equiv E_a-E_b$ is the single-chain spacing, and $\eta=\pm$ selects the branch that first undergoes the EP collision. 
Thus, the transition is governed by the competition between the single-chain spacing $\Delta E_{ab}$,
the projected rung hopping spacing $u_{aa}-u_{bb}$, and the product of the projected off-diagonal hybridizations $u_{ab}v_{ba}$.
Since the eigenmode overlaps and energies generically depend on the chain length $N$~\footnote{This is because oppositely localized skin modes overlap less with increasing system length. Being a single-chain property, it is \emph{not} a consequence of the critical NHSE.}, for any given candidate EP collision pair $(a,b)$, Eq.~\eqref{eq:active_instability_main} defines a pair-resolved threshold length
\begin{equation}
N_{ab} = \min\left\{ N: \min_{\eta=\pm} \Lambda_{\eta}^2 + 4u_{ab}v_{ba} < 0 \right\}\CH{.}
\label{eq:pair_threshold_main}
\end{equation}
The observed critical transition length is then selected by the earliest such collision,
\begin{equation}
N_c=\min_{a,b}N_{ab}. 
\label{eq:min_threshold_main}
\end{equation}
The formulation above is completely general, and is not a perturbative approximation. Importantly, it explains why different ladder configurations (Models I to III) may involve distinct sectors of the single-chain spectrum, based on energetic considerations on their different state overlap profiles. In this precise sense, the EP-colliding pair is not assumed in advance: it is selected by the minimum-threshold condition Eq.~\eqref{eq:min_threshold_main}. Local and nonlocal rungs weigh the spatial profiles of the single-chain modes differently, thereby controlling which pair of levels collides first.

\subsubsection{Scaling laws for critical length $N_c$}
We now apply Eqs.~\eqref{eq:active_instability_main}--\eqref{eq:min_threshold_main} to our 3 models in Fig.~\ref{fig:scalings}. From this common framework, we will see how the scaling of the relevant projected quantities -- the colliding-level spacing $\Delta E_{ab}$ and the rung-induced hybridizations entering Eq.~\eqref{eq:active_instability_main} -- interplay to determine the critical scaling of the entire ladder.

\vspace{\baselineskip}
\noindent\textbf{Model I. Fully local NN--NN ladder.---}
As a baseline example, we employ our framework to the simplest case with no nonlocal hopping: the fully local NN--NN ladder $\mathcal{H}^{\rm I}$ [Eq.~\ref{eq:H_model_I_main}]. Even though this model is not new~\cite{li2020critical}, 
our study gives the first complete finite-size projected expression for its critical length [Eq.~\eqref{eq:NNNN_active_threshold_main}], beyond the known logarithmic estimate [Eq.~\eqref{eq:local_Nc_main}] that often turns out to be significantly inaccurate.

For this Model I, each leg is just an OBC Hatano-Nelson chain with eigenenergies $2\cos k_\ell$ and biorthogonal eigenmodes $\psi_{\ell,\pm}^R(n)=\sqrt{2/(N+1)}e^{\pm\kappa n}\sin(k_\ell n)$ and $\psi_{\ell,\pm}^L(n)=\sqrt{2/(N+1)}e^{\mp\kappa n}\sin(k_\ell n)$, where $\pm$ labels the legs, $\kappa=\ln\sqrt{(1+\gamma)/(1-\gamma)}$ and $k_\ell=\ell\pi/(N+1)$ labels the modes. 
It can be shown (see Appendix~\ref{sec:app:Nc_modelI}) that the first EP collision is given by the $l=1,2$ subspace with maximal energy.
Their energy spacing is 
\begin{equation}
\Delta E_{\rm NN} = E_1 - E_2 = 2(\cos k_1 - \cos k_2)\simeq 3\pi^2N^{-2}. 
\label{eq:NNgap}
\end{equation}
Evaluating the EP subspace projections of the local inter-chain couplings yields the complete set of projected matrix elements for the rung hoppings:
\begin{equation}
\begin{aligned}
u_{11} &= \frac{2D^{-\alpha}}{N+1} e^{\kappa(N+1)} \sum_{n=1}^N A_n^2, \\
u_{22} &= \frac{2D^{-\alpha}}{N+1} e^{\kappa(N+1)} \sum_{n=1}^N B_n^2, \\
u_{12} &= \frac{2D^{-\alpha}}{N+1} e^{\kappa(N+1)} \sum_{n=1}^N A_n B_n, \\
v_{21} &= -u_{12}.
\end{aligned}
\label{eq:u_and_v_coefficients}
\end{equation}
where $A_n = e^{-\kappa n}\sin(k_1 n)$ and $B_n = e^{-\kappa n}\sin(k_2 n)$. Substituting the projected rung matrix elements in Eq.~\eqref{eq:u_and_v_coefficients} into the threshold criterion Eq.~\eqref{eq:active_instability_main}, including the rung-induced diagonal level shifts $u_{11}$ and $u_{22}$ through the combination $\eta(u_{11}-u_{22})$, yields the complete finite-size projected threshold condition determining the critical length $N_c^{\rm I}$ for Model I:
\begin{equation}
D^{-\alpha} = (N_c^{\rm I}+1)e^{-\kappa(N_c^{\rm I}+1)}(\cos k_1 - \cos k_2)/\mathcal{D}_{12},
\label{eq:NNNN_active_threshold_main}
\end{equation}
where $\mathcal{D}_{12} = \sum_{n=1}^N (A_n^2 + B_n^2) + 2\left|\sum_{n=1}^N A_n B_n\right|$ (see Appendix~\ref{sec:app:Nc_modelI} for details). This closed-form expression, which is new in the literature, is plotted as the dark blue dashed curves in  
Fig.~\ref{fig:scalings}(d-f), with excellent agreement with numerics. Importantly, due to the implicit form of Eq.~\eqref{eq:NNNN_active_threshold_main}, the critical length threshold $N_c^{\rm I}$ does not vary with the model parameters in any simple manner across all scales, despite the apparent simplicity of Model I. 

In the asymptotic skin-overlap regime $\kappa N_c\gg1$, the extensive lattice sums in the denominator simplify significantly, 
reducing Eq.~\eqref{eq:NNNN_active_threshold_main} to $e^{\kappa N_c^{\rm I}}D^{-\alpha}\sim 1$ i.e.  the familiar critical-NHSE logarithmic scaling law 
\begin{equation}
N_c^{\rm I}\simeq-\frac{1}{\kappa}\ln D^{-\alpha}\simeq\frac{\alpha}{\gamma}\ln D.
\label{eq:local_Nc_main}
\end{equation}
Physically, the logarithmic dependence originates from balancing an exponentially small skin-overlap scale against the algebraically weak inter-chain coupling $D^{-\alpha}$. Increasing the chain length enhances the opposite-skin overlap exponentially, so only a logarithmic increase of $N_c$ is needed to compensate for a power-law increase of $D$. 
Note that the above arguments assumed very strongly localized NHSE modes satisfying $\kappa N_c\gg1$: In the more experimentally accessible regime plotted in Fig.~\ref{fig:scalings}(d-f), excellent agreement with numerics is possible only with our full new expression Eq.~\eqref{eq:NNNN_active_threshold_main}, not its approximation Eq.~\eqref{eq:local_Nc_main}.

This logarithmic balance between an exponentially small overlap and a parametrically weak competing scale is not unique to the NHSE. Analogous structures arise when the Wentzel-Kramers-Brillouin (WKB) tunneling splitting competes with a weak bias in an asymmetric double well~\cite{Song2008DoubleWell}, or in the Mott--Berezinskii resonant-pair mechanism for localized disordered systems, where matching an exponentially decaying hybridization to a small frequency scale produces a logarithmic resonant length~\cite{Mott1968Localized,Falco2017MottBerezinskii}.

\vspace{\baselineskip}
\noindent
\textbf{Model II. Long-range rungs: NN--LR ladder.---}
We next proceed to introduce nonlocality in the antagonistic NHSE ladder. Here we shall only make the interchain couplings (rungs) power-law decaying, and leave the intrachain (leg) couplings local. The case with nonlocality in both the rungs and the legs (Model III) will be studied later.

Since the ladder legs are still local like before, the relevant EP-colliding modes in the parameter regime considered here remain the NN band-edge pair $(\psi_1,\psi_2)$, with the same energy spacing $\Delta E_{\rm NN}=E_1-E_2\simeq 3\pi^2N^{-2}$ [Eq.~\eqref{eq:NNgap}] as before. The new distinction lies entirely in the projection of the rungs to the colliding EP subspace. To facilitate analytic progress, we use the large-separation, or constant-kernel, approximation
\begin{equation}
\left[D^2+(n-m)^2\right]^{-\alpha/2}
\simeq D^{-\alpha},
\qquad D\gg |n-m|,
\label{eq:constant_kernel_main}
\end{equation}
under which the projected long-range rung contributions factorize into independent wavefunction sums,
\begin{equation}
u_{ij}\simeq D^{-\alpha}\Sigma_{i,+}\Sigma_{j,-},
\qquad
\Sigma_{i,\pm}\equiv\sum_n\psi_{i,\pm}(n),
\label{eq:NNLR_factorized_rungs_main}
\end{equation}
for $i,j\in{1,2}$. This approximation retains the collective contribution of all inter-chain site pairs while neglecting the longitudinal variation of their distances relative to the transverse separation $D$. Its derivation and numerical range of validity are discussed in Appendices~\ref{sec:app:Nc_NNLR} and \ref{sec:app_checks}. 
For $i,j\in\{1,2\}$, the constant-kernel projection must be understood biorthogonally: the factors entering $u_{ij}$ are the spatial sums of the left state on the $(+)$ leg and the right state on the $(-)$ leg. With the convention $\psi_{i,\pm}^{R}(n)\propto e^{\pm\kappa n}\sin(in\pi/N)$ and $\psi_{i,\pm}^{L}(n)\propto e^{\mp\kappa n}\sin(in\pi/N)$, both sums entering $u_{ij}$ carry $e^{-\kappa n}$. Expanding this factor for weak NHSE gives $S_i-\kappa X_i$, where $S_i\equiv\sum_n \sqrt{2/N}\sin(i\pi n/N)$ and $X_i\equiv\sum_n n\sqrt{2/N}\sin(i\pi n/N)$. Evaluating these sums gives $S_1\simeq 2\sqrt{2N}/\pi$, while $S_2=0$ by parity and $X_2\simeq -N^{3/2}/(\sqrt{2}\pi)$. Thus the mode-1--mode-2 channel is parity-forbidden at $\kappa=0$ and is activated linearly by the skin deformation. Further details are provided in Appendix~\ref{sec:app:Nc_NNLR}.

These preceding arguments above show that, to leading nonvanishing order in the non-reciprocity $\kappa$, the effects of the long-range rungs can be encapsulated in the following rung matrix elements in the projected EP-collision subspace:
\begin{equation}
\begin{aligned}
u_{11} &\simeq \frac{8N}{\pi^2 D^\alpha}, \\
u_{12} &\simeq \frac{2\kappa N^2}{\pi^2 D^\alpha}, \\
v_{21} &\simeq -\frac{2\kappa N^2}{\pi^2 D^\alpha}, \\
u_{22} &\simeq -\frac{\kappa^2N^3}{2\pi^2 D^\alpha}.
\end{aligned}
\label{eq:u_and_v_approximations}
\end{equation}
Essentially, each occurrence of $\psi_2$ contributes an additional factor proportional to $\kappa N$: since $S_2=0$, mode 2's leading spatial overlap is set entirely by its non-reciprocity-induced first moment $\kappa X_2\propto \kappa N^{3/2}$, which exceeds mode 1's zeroth moment $S_1\propto N^{1/2}$ by a relative factor proportional to $\kappa N$. The relative sign between $u_{12}$ and $v_{21}$ follows from the opposite skin deformations on the two legs; the individual sign of either matrix element depends on the arbitrary phase convention chosen for mode 2, whereas the negative product $u_{12}v_{21}$ entering Eq.~\eqref{eq:active_instability_main} is convention independent.

\begin{figure}[!htp]
    \centering
    \includegraphics[width=.99\linewidth]{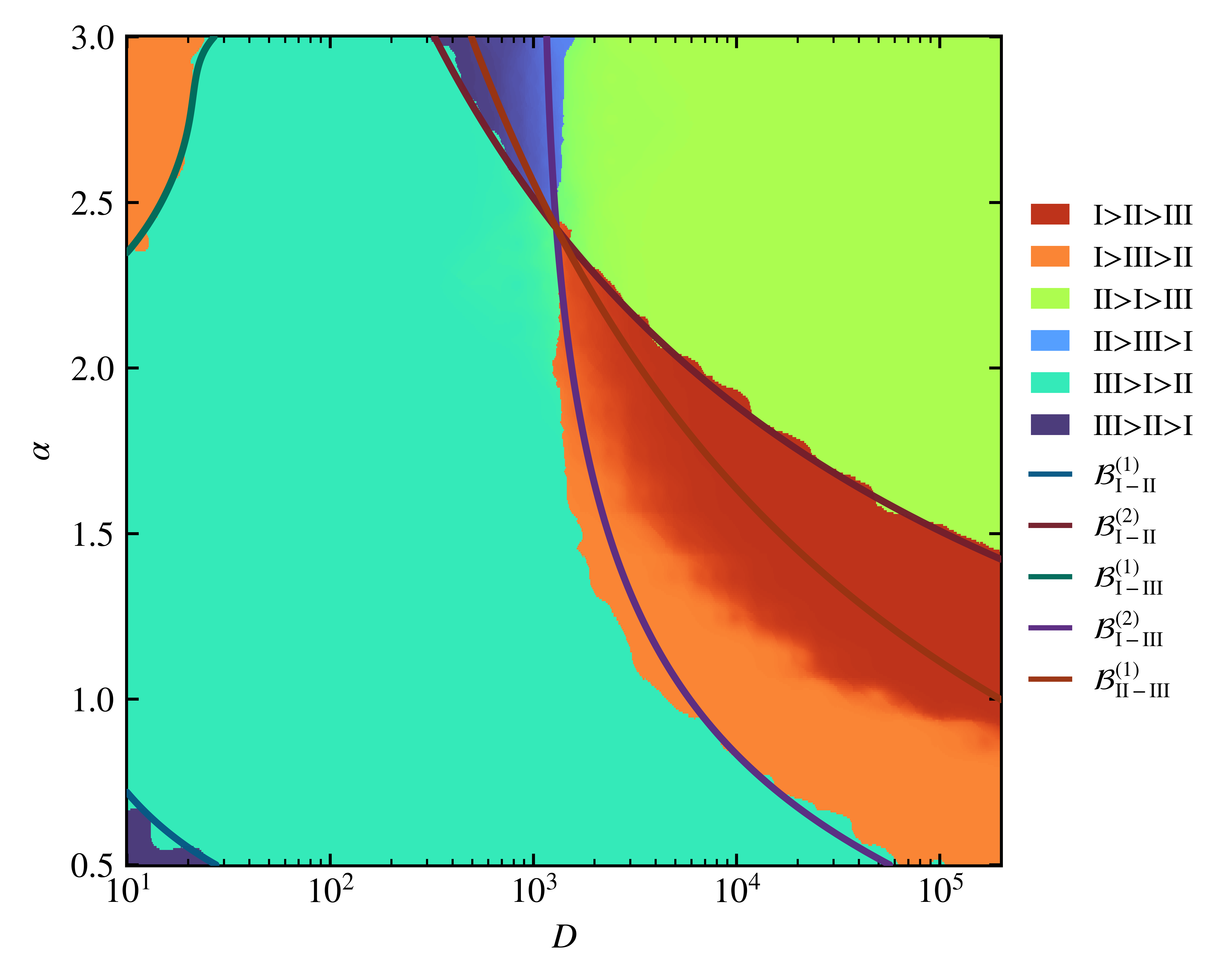}
    \caption{ Global critical-length hierarchy in the $(D,\alpha)$ plane at $\gamma=0.05$. The background colors denote the six possible orderings of the critical lengths $N_c^{\rm I}$, $N_c^{\rm II}$, and $N_{c,\mathrm{full}}^{\rm III}$ for the fully local, nonlocal-rung, and fully nonlocal lattices, respectively. The five colored curves are the distinct connected analytical hierarchy-boundary solutions: the deep-blue and wine-red curves denote $\mathcal B_{\rm I-II}^{(1)}$ and $\mathcal B_{\rm I-II}^{(2)}$ [Eqs.~\eqref{eq:BI-II_branch1} and \eqref{eq:BI-II_branch2}], respectively; the dark-teal and deep-violet curves denote $\mathcal B_{\rm I-III}^{(1)}$ and $\mathcal B_{\rm I-III}^{(2)}$ [Eqs.~\eqref{eq:BI-III_branch1} and \eqref{eq:BI-III_branch2}], respectively; and the burnt-red curve denotes $\mathcal B_{\rm II-III}^{(1)}$ [Eq.~\eqref{eq:BII-III_branch1}]. All five curves are obtained by equating the corresponding critical-length conditions and are not independent fits. }
    \label{fig:phase_diagram}
\end{figure}

Substituting the NN band-edge spacing $\Delta E_{\rm NN}$ from Eq.~\eqref{eq:NNgap} and the projected rung matrix elements from Eq.~\eqref{eq:u_and_v_approximations} into the EP criterion Eq.~\eqref{eq:active_instability_main} gives the critical-length threshold equation
\begin{equation}
\left[N_c^{\rm II}\right]^3
\left(1+\eta_1\kappa N_c^{\rm II}/4\right)^2
\simeq
\eta_0(3\pi^4/8)D^\alpha.
\label{eq:NNLR_Nc_main}
\end{equation}
Here $\eta_0$ and $\eta_1$ are two dimensionless prefactors arising from the full distance-dependent rung kernel, finite-size wavefunction sum and energy spacing corrections, and the  dominant projected subspace truncation, all which can renormalize the effective parameters without altering the key functional dependence of the $N_c^{\rm II}$ threshold. For the Model-II analytical curves in Fig.~\ref{fig:scalings} and the corresponding hierarchy boundaries in Fig.~\ref{fig:phase_diagram}, we use the common effective values $\eta_0=3.0$ and $\eta_1=3.7$.

Although Eq.~\eqref{eq:NNLR_Nc_main} seems like a rather complicated scaling rule, in the weak NHSE limit $\eta_1\kappa N_c^{\rm II}\ll1$, it reduces to the algebraic scaling law
\begin{equation}
N_c^{\rm II}
\sim
\left[\eta_0(3\pi^4/8)\right]^{1/3}D^{\alpha/3}.
\end{equation}
The origin of the exponent $\alpha/3$ is the balance between the quadratic NN band-edge spacing, $\Delta E_{\rm NN}\sim N^{-2}$, and the collectively enhanced long-range rung projection, $u_{11}\sim ND^{-\alpha}$. Since the spatial sum of each lowest band-edge mode scales as $N^{1/2}$, the factorized rung matrix element contains their product and therefore grows linearly with $N$. The threshold condition thus gives $N^{-2}\sim ND^{-\alpha}$, or $N^3\sim D^\alpha$, yielding $N_c^{\rm II}\sim D^{\alpha/3}$. The higher-moment contributions in $u_{12}$ and $u_{22}$ produce the additional $\kappa N_c^{\rm II}$ correction in Eq.~\eqref{eq:NNLR_Nc_main}, but do not alter the leading $\alpha/3$ exponent in the weak-NHSE regime. 
This scaling law for Model II, being algebraic, has a distinct origin from the logarithmic scaling of the previous fully local ladder (Model I). The legs, being local as before, still provide the NN quadratic band-edge spacing $\Delta E_{\rm NN}\sim N^{-2}$.  But importantly, the long-range rungs no longer acts as local point-to-point tunneling couplings. Instead, its projection aggregates the contributions from many nodes of the two constituent chains, producing a hybridization that \emph{grows}  with the spatial extent of the band-edge modes. The transition threshold is therefore governed by a power-law balance between the finite-size level spacing and distributed long-range hybridization. In a sense, this mechanism is reminiscent of the collective light--matter coupling mechanism in the Tavis--Cummings model, where many individually weak couplings combine into a size-enhanced collective matrix element $\propto\sqrt{N}$, thereby producing a system-size-dependent strong-coupling threshold~\cite{Tavis1968Exact,Breeze2017Dicke}.

\vspace{\baselineskip}
\noindent
\textbf{Model III. Fully nonlocal LR--LR ladder.---}
Finally, we also introduce the power-law nonlocality in the legs too, such that all ladder couplings (rungs and legs) are long-ranged (LR--LR).

In the parameter regime used for the analytical Model-III branches below, the real-to-complex transition is effected by the EP collision of the band-edge pair of eigenstates $(\psi_1,\psi_2)$, whose real energy gap scales like $\Delta E_{\rm LR}\simeq C(\alpha)N^{1-\alpha}$ for $0<\alpha<3$ [Eq.~\eqref{eq:gap_LR}]~\footnote{The physical threshold is always determined by the minimum condition in Eq.~\eqref{eq:min_threshold_main}; the band-edge formulas below apply to the parameter window in which $(1,2)$ is the first-colliding pair.}. But here, we have explicitly introduced an $O(1)$ prefactor function $C(\alpha)$ to capture the nontrivial $\alpha$-dependence -- see Appendix~\ref{sec:single_chain_properties} for details.

The effects of the two sources of nonlocality can be separated conceptually. First, the long-range rungs generate the same collective inter-chain projection as in Model II. Using the constant-kernel approximation introduced in Eq.~\eqref{eq:constant_kernel_main}, their projected matrix elements again factorize as
\begin{equation}
u_{ij}\simeq D^{-\alpha}\Sigma_{i,+}\Sigma_{j,-},
\qquad
\Sigma_{i,\pm}=\sum_n\psi_{i,\pm}(n),
\label{eq:LRLR_factorized_rungs_main}
\end{equation}
for $i,j\in{1,2}$. Thus, rung nonlocality still produces a collective hybridization proportional to the product of the spatial sums of the participating modes.

The long-range legs modify different ingredients of the EP balance. In contrast to Model II, they replace the quadratic NN band-edge spacing $\Delta E_{\rm NN}\sim N^{-2}$ by the nonanalytic long-range spacing $\Delta E_{\rm LR}\simeq C(\alpha)N^{1-\alpha}$. They also change how non-reciprocity deforms the two band-edge modes. Within the band-edge approximation used here, the reciprocal eigenstates are represented by the same sine standing waves $\psi_1^{(0)}$ and $\psi_2^{(0)}$, with
$S_1\equiv\sum_n\psi_1^{(0)}(n)\neq0$ and
$S_2\equiv\sum_n\psi_2^{(0)}(n)=0$.
However, instead of the simple Hatano--Nelson skin deformation responsible for the $\kappa N$ dependence in Model II, the long-range intra-chain non-reciprocity mixes these two modes through an $\alpha$-dependent amplitude $\chi_N(\alpha)$. Hence, the rung nonlocality determines the collective factorized form of the inter-chain coupling, whereas the leg nonlocality determines both the anomalous band-edge spacing and the size dependence of the non-reciprocity-induced mode mixing.

This is captured perturbatively in $\gamma$ through the eigenstates $\psi_{i,\pm}(n)$ 
of $H^{\rm LR}(\pm\gamma)=H_0\pm\gamma V$ [Eq.~\eqref{eq:hamLR}], where $H_0$ is the Hermitian ($\gamma=0$) long-range chain and 
\begin{equation} V=\sum_{n\neq m}\frac{\mathrm{sgn}(n-m)}{|n-m|^\alpha}c_n^\dagger c_m \label{eq:V_def} \end{equation} 
is the antisymmetric non-reciprocity kernel. These perturbed eigenstates take the forms
\begin{equation}
\begin{aligned}
\psi_{1,\pm}(n) &\approx \psi_1^{(0)}(n) \pm \gamma\chi_N(\alpha)\,\psi_2^{(0)}(n), \\
\psi_{2,\pm}(n) &\approx \psi_2^{(0)}(n) \mp \gamma\chi_N(\alpha)\,\psi_1^{(0)}(n),
\end{aligned}
\label{eq:LR_perturbed_states}    
\end{equation}
where
\begin{equation}
\chi_N(\alpha) \equiv \frac{\langle\psi_2^{(0)}|V|\psi_1^{(0)}\rangle}{\Delta E_{\rm LR}}
\label{eq:chiN_def_main}
\end{equation}
is the non-reciprocity-induced mixing amplitude, with sign fixed by the antisymmetry $\langle\psi_1^{(0)}|V|\psi_2^{(0)}\rangle=-\langle\psi_2^{(0)}|V|\psi_1^{(0)}\rangle$. 
This mixing amplitude governs the EP collision dynamics of our LR--LR ladder, and turns out to possess a special form of $N$-dependence (unlike Model II's linear $\kappa N/4$ dependence):
\begin{equation}
\chi_N(\alpha)
\sim
\begin{cases}
\chi(\alpha), & 0<\alpha<2,\\[1mm]
\bar\chi_2\ln N, & \alpha=2,\\[1mm]
\bar\chi(\alpha)N^{\alpha-2}, & 2<\alpha<3,\\[1mm]
\bar\chi_3 N/\ln N, & \alpha=3,\\[1mm]
\bar\chi_{>}(\alpha)N, & \alpha>3,
\end{cases}
\label{eq:chiN_piecewise_main}
\end{equation}
where $\chi(\alpha)$, $\bar\chi_2$, $\bar\chi(\alpha)$,
$\bar\chi_3$, and $\bar\chi_{>}(\alpha)$ are independent of $N$. 

To derive Eq.~\eqref{eq:chiN_piecewise_main}, we note that $\chi_N(\alpha)$ is the ratio of two separately $N$-dependent quantities, as defined in Eq.~\eqref{eq:chiN_def_main}. The numerator is the non-reciprocity-induced matrix element between the two Hermitian band-edge standing waves. Its large-$N$ behavior is controlled by the antisymmetric combination
$\psi_2^{(0)}(n)\psi_1^{(0)}(m)-\psi_2^{(0)}(m)\psi_1^{(0)}(n)$,
weighted by the long-range kernel $|n-m|^{-\alpha}$, giving (see Appendix~\ref{sec:single_chain_properties})
\begin{equation}
  \langle\psi_2^{(0)}|V|\psi_1^{(0)}\rangle
  \sim
  \begin{cases}
    N^{1-\alpha}, & 0<\alpha<2,\\[1mm]
    N^{-1}\ln N, & \alpha=2,\\[1mm]
    N^{-1}, & \alpha>2,
  \end{cases}
  \label{eq:V21_piecewise_main}
\end{equation}
while the denominator retains the single scaling law $\Delta E_{\rm LR}\sim N^{1-\alpha}$ up to $\alpha=3$. For $\alpha<2$, the numerator and gap are therefore both controlled by the \emph{same} long-distance tail of the non-reciprocity kernel which cancel in their ratio, leaving an $N$-independent ratio $\chi(\alpha)$ which we elaborate on in Appendix~\ref{sec:single_chain_properties}. For $\alpha>2$, however, the numerator is instead dominated by short hopping distances and decouples from the gap, which continues to fall as $N^{1-\alpha}$ up to $\alpha=3$.

Substituting the perturbed states in Eq.~\eqref{eq:LR_perturbed_states} into the factorized rung matrix elements in Eq.~\eqref{eq:LRLR_factorized_rungs_main} gives, to leading order,
\begin{equation}
\begin{aligned}
u_{11} &\simeq \frac{8N}{\pi^2D^\alpha}, \\
u_{12} &\simeq \gamma\chi_N(\alpha)\frac{8N}{\pi^2D^\alpha}, \\
v_{21} &\simeq -\gamma\chi_N(\alpha)\frac{8N}{\pi^2D^\alpha}, \\
u_{22} &\simeq -\gamma^2|\chi_N(\alpha)|^2\frac{8N}{\pi^2D^\alpha}.
\end{aligned}
\label{eq:u_and_v_gamma_chi_main}
\end{equation}
Equations~\eqref{eq:u_and_v_approximations} and \eqref{eq:u_and_v_gamma_chi_main} have the same underlying structure. In both Models II and III, the long-range rungs produce the collective diagonal element
$u_{11}\simeq 8N/(\pi^2D^\alpha)$, while non-reciprocity activates an antisymmetric off-diagonal channel with $v_{21}\simeq-u_{12}$ and generates a negative diagonal correction $u_{22}$ that is quadratic in the corresponding mode-mixing amplitude. The essential difference lies in how this mixing amplitude is generated by the legs. For the NN legs of Model II, the exact Hatano--Nelson skin deformation gives the dimensionless factor $\kappa N/4$, so that
$|u_{12}|/u_{11}\sim\kappa N/4$, with $v_{21}=-u_{12}$, and
$u_{22}/u_{11}\sim-(\kappa N/4)^2$.
For the long-range legs of Model III, the corresponding factor is instead
$\gamma\chi_N(\alpha)$, arising from the non-reciprocity-induced mixing of the two long-range band-edge modes. Unlike the linear $\kappa N$ dependence in Model II, $\chi_N(\alpha)$ has the piecewise size dependence given in Eq.~\eqref{eq:chiN_piecewise_main}. Moreover, the two models differ in their intrinsic band-edge protection: Model II retains the NN spacing $\Delta E_{\rm NN}\sim N^{-2}$, whereas Model III has the nonanalytic long-range spacing $\Delta E_{\rm LR}\sim N^{1-\alpha}$ for $0<\alpha<3$. These two differences are ultimately responsible for their distinct critical-length scaling laws.

Substituting these expressions into the projected EP criterion Eq.~\eqref{eq:active_instability_main}, gives the window
\begin{equation}
u_{11}\left(1-\gamma|\chi_N(\alpha)|\right)^2 < \Delta E_{\rm LR} < u_{11}\left(1+\gamma|\chi_N(\alpha)|\right)^2 ,
\label{eq:LRLR_EP_window_main}
\end{equation}
Since $u_{11}\sim ND^{-\alpha}$ grows linearly with $N$ while $\Delta E_{\rm LR}\sim C(\alpha)N^{1-\alpha}$ grows at most sublinearly, their ratio $\Delta E_{\rm LR}/u_{11}\sim N^{-\alpha}$ is monotonically decreasing in $N$ for every $\alpha>0$ -- regardless of whether $\Delta E_{\rm LR}$ itself grows ($\alpha<1$) or falls ($\alpha>1$) with $N$. The window in Eq.~\eqref{eq:LRLR_EP_window_main} is therefore always entered from above at a single, well-defined crossing, which uniquely determines the critical threshold length $N_c^{\rm III}$ implicitly as
\begin{equation}
C(\alpha)\left[N_c^{\rm III}\right]^{1-\alpha}
\simeq
\frac{8N_c^{\rm III}}{\pi^2D^\alpha}
\left[1+\gamma\left|\chi_{N_c^{\rm III}}(\alpha)\right|\right]^2 .
\label{eq:LRLR_Nc_balance_main}
\end{equation}
Because $\chi_N(\alpha)$ in Eq.~\eqref{eq:LRLR_Nc_balance_main} is itself evaluated at $N=N_c^{\rm III}$, the piecewise scaling Eq.~\eqref{eq:chiN_piecewise_main} turns this single threshold equation into not one but \emph{three}  qualitatively different threshold laws. 

\noindent\textbf{Strongly nonlocal regime (i) $0<\alpha<2$:}

\noindent Here $\chi_N(\alpha)\to\chi(\alpha)$ is asymptotically $N$-independent, and Eq.~\eqref{eq:LRLR_Nc_balance_main} simply gives
\begin{equation}
\frac{N_c^{\rm III}}{D}
=
\left[\frac{\pi^2C(\alpha)}{8}\right]^{1/\alpha}
\left(1+\gamma|\chi(\alpha)|\right)^{-2/\alpha}.
\label{eq:LRLR_ratio_lt2_main}
\end{equation}
Here the two lengths $N_c^{\rm III}$ and $D$ only enter in the form of their ratio on the left hand side i.e. the solution is \emph{scale-covariant} i.e. invariant under a simultaneous dilation $N\to\lambda N$, $D\to\lambda D$. 
While this scale-covariance may be naively dismissed as a simple consequence of the ladder being fully non-local i.e. with all couplings exhibiting power-law decay, the fact is that the this scale-covariance will not survive at larger values of $\alpha$, as we shall show soon after~\footnote{ 
In this case, this is because both $\Delta E_{\rm LR}$ and $u_{11}$ acquire the same factor $\lambda^{1-\alpha}$, but that does not hold for $\alpha\geq 2$. }

So far, we have left our critical threshold equation Eq.~\eqref{eq:LRLR_ratio_lt2_main} in terms of relatively complicated $\alpha$-dependent factors $C(\alpha)$ and $\chi(\alpha)$. Referring the reader to Appendix~\ref{sec:app:Nc_LRLR} for their detailed definitions and forms, we mention that, in the usual cases of weak mixing  $\gamma|\chi(\alpha)|\ll1$, Eq.~\eqref{eq:LRLR_ratio_lt2_main} can be written as
\begin{equation}
\frac{N_c^{\rm III}}{D} \approx A(\alpha)-\gamma B(\alpha),
\label{eq:LRLR_AB_compact_main}
\end{equation}
with $A(\alpha)=\left[\frac{\pi^2C(\alpha)}{8}\right]^{1/\alpha}$ and $B(\alpha)=\tfrac{2}{\alpha}A(\alpha)|\chi(\alpha)|$ well-approximated by 
\begin{equation}
\begin{aligned}
A(\alpha)&\simeq 2.4-0.026\alpha+0.33\alpha^2,\\
B(\alpha)&\simeq 11-12\alpha+5.7\alpha^2
\end{aligned}
\label{eq:AB_lt2_empirical_main}
\end{equation}
over the explored range $0<\alpha<2$.

The agreement between this scale-covariant prediction and exact diagonalization is demonstrated in Figs.~\ref{fig:scalings}(d) and \ref{fig:scalings}(e) for $\alpha=0.5$ and $1$, respectively, where the analytical Model-III curves closely follow the numerical critical lengths. The broader dependence on $D$, $\gamma$, and $\alpha$ within the unsaturated $\alpha<2$ regime is shown in Appendix Fig.~\ref{fig:modelIII_Nc_vs_D_gamma}.

\noindent\textbf{Marginal case (ii) $\alpha=2$:} At this marginal power exponent, $\chi_N(2)\simeq\bar\chi_2\ln N$, and Eq.~\eqref{eq:LRLR_Nc_balance_main} becomes
\begin{equation}
\left[N_c^{\rm III}\right]^2
\left(1+\gamma|\bar\chi_2|\ln N_c^{\rm III}\right)^2
\simeq
\frac{\pi^2C(2)}{8}D^2.
\label{eq:LRLR_alpha2_main}
\end{equation}
The leading $N_c^{\rm III}\propto D$ growth survives, but is now dressed by a logarithmic finite-size correction; in the weak-mixing limit $\gamma|\bar\chi_2|\ln N_c^{\rm III}\ll1$,
\begin{equation}
\frac{N_c^{\rm III}}{D}
\simeq
\left[\frac{\pi^2C(2)}{8}\right]^{1/2}
\left[1-\gamma|\bar\chi_2|
\ln\!\left(\left[\frac{\pi^2C(2)}{8}\right]^{1/2}D\right)\right],
\label{eq:LRLR_alpha2_weak_main}
\end{equation}
with the effective linear $\gamma$ coefficient depending  logarithmically on $D$ itself. 

\noindent\textbf{Almost local regime (iii) $2<\alpha<3$:} 

\noindent Here $\chi_N(\alpha)\simeq\bar\chi(\alpha)N^{\alpha-2}$ grows algebraically, and Eq.~\eqref{eq:LRLR_Nc_balance_main} becomes the following implicit equation for $N_c^{\rm III}$,
\begin{equation}
\frac{N_c^{\rm III}}{D}
=
A_{>}(\alpha)
\left[1+\gamma B_{>}(\alpha)\left(N_c^{\rm III}\right)^{\alpha-2}\right]^{-2/\alpha},
\label{eq:LRLR_ratio_23_main}
\end{equation}
with $A_{>}(\alpha)\equiv[\pi^2C(\alpha)/8]^{1/\alpha}$, $B_{>}(\alpha)\equiv|\bar\chi(\alpha)|$. Because $N_c^{\rm III}$ now also appears inside the mixing correction, the aspect ratio $N_c^{\rm III}/D$ is no longer exactly dilation-fixed as in regime (i): the correction is controlled by $\gamma D^{\alpha-2}$, which grows with $D$ rather than saturating, implying that $\alpha<2$ form Eq.~\eqref{eq:LRLR_AB_compact_main} cannot simply be continued across $\alpha=2$. 

Here, the $\alpha$-dependent coefficients $A_{>}(\alpha)$ and $B_{>}(\alpha)$ are well approximated over the fitted interval $2<\alpha<3$ by 
\begin{equation} 
\begin{aligned} 
A_{>}(\alpha) &\simeq 3.63+2.18(\alpha-2)-0.098(\alpha-2)^2,\\ 
B_{>}(\alpha) &\simeq 2.63-5.02(\alpha-2)+2.48(\alpha-2)^2. 
\end{aligned} \label{eq:AgtBgt_empirical_main} \end{equation} 
Their extraction from the unsaturated $(1,2)$-channel numerical thresholds is detailed in Appendix~\ref{sec:app:Nc_LRLR}, particularly Eqs.~\eqref{eq:Agt_alpha23_empirical} and \eqref{eq:Bgt_alpha23_empirical}. These finite-window parametrizations apply only for $2<\alpha<3$ and should not be extrapolated beyond $\alpha=3$.

At $\alpha=3$, the band-edge spacing crosses to the marginal form $\Delta E_{\rm LR}\sim N^{-2}\ln N$, while $\chi_N(3)\sim N/\ln N$. Consequently, the leading weak-mixing balance becomes $\frac{\left(N_c^{\rm III}\right)^3} {\ln N_c^{\rm III}} \sim D^3, \alpha=3, $ up to the corresponding non-reciprocity-induced mixing correction. For $\alpha>3$, the analytic quadratic band-edge scaling $\Delta E_{\rm LR}\sim N^{-2}$ is restored and $\chi_N(\alpha)\sim N$, so the weak-mixing balance crosses to $ \left(N_c^{\rm III}\right)^3\sim D^\alpha, \alpha>3, $or $N_c^{\rm III}\sim D^{\alpha/3}$ before the mixing and finite-window cutoff corrections are included.

The agreement of this implicit $2<\alpha<3$ threshold with exact diagonalization is illustrated in Fig.~\ref{fig:scalings}(f) for $\alpha=2.4$, where the analytical Model-III curve reproduces the numerical critical lengths, including the crossover to the finite-window cutoff.

In a nutshell, the above discussions has revealed the special significance of $\alpha=2$ as the threshold demarcating different types of $N_c^\text{III}$ scaling behavior. This is because for $\alpha<2$, the long-distance part of the antisymmetric mixing kernel sets both the mixing element and the band-edge gap, leading to their ratio being size-independent; for $\alpha>2$ the mixing element is instead set by short hops and scales as $N^{-1}$, decoupled from the gap's continued $N^{1-\alpha}$ falloff, so their ratio grows as $N^{\alpha-2}$.

Finally, in all three regimes, the candidate inter-chain EP threshold is bounded by the intrinsic finite-size crossover scale of each long-range leg. This single-chain crossover, previously identified for long-range non-Hermitian chains~\cite{wang2023scaling}, occurs at the length $N_{\rm sat}$ determined by
\begin{equation}
e^{\kappa(N_{\rm sat}-2)}
=
(N_{\rm sat}-1)^{\alpha}.
\label{eq:LRLR_Lsat_main}
\end{equation}
The imaginary-gauge derivation of Eq.~\eqref{eq:LRLR_Lsat_main} and its application to Model III are presented in Appendix~\ref{sec:app:Nc_LRLR}.
The physical critical length of Model III is therefore
\begin{equation}
N_{c,\mathrm{full}}^{\rm III}
\simeq
\min\!\left\{
N_{c}^{\rm III},
N_{\rm sat}
\right\},
\label{eq:NcIII_full_main}
\end{equation}
where $N_{c}^{\rm III}$ is the candidate inter-chain EP threshold
given by Eq.~\eqref{eq:LRLR_ratio_lt2_main} for $0<\alpha<2$,
Eq.~\eqref{eq:LRLR_alpha2_main} for $\alpha=2$, and
Eq.~\eqref{eq:LRLR_ratio_23_main} for $2<\alpha<3$.

This threshold is plotted as the dashed curve for the fully nonlocal lattice in Fig.~\ref{fig:scalings}(d--f); in particular, the $\alpha=2.4$ panel uses the implicit $2<\alpha<3$ branch.

\subsection{Hierarchy of critical lengths $N_c$ } 
\label{sec:hierarchy}

Having derived the scaling laws $N_c^{\rm I}$ [Eq.~\eqref{eq:local_Nc_main}], $N_c^{\rm II}$ [Eq.~\eqref{eq:NNLR_Nc_main}], and $N_{c,\mathrm{full}}^{\rm III}$ [Eq.~\eqref{eq:NcIII_full_main}] for our 3 ladder configurations, we now put them together and see how they define a rich phase diagram in an extended $(D,\alpha)$ parameter space, from very nonlocal ($\alpha=0.5$) to almost local ($\alpha=3.0$), across over 5 orders of magnitude of ladder width $D$. 
In Fig.~\ref{fig:phase_diagram}, each of the $3!=6$ colors denotes a different ordering hierarchy of the three critical lengths $N_c^{\rm I}$, $N_c^{\rm II}$, and $N_{c,\mathrm{full}}^{\rm III}$ obtained by numerical diagonalization.  
We also overlay their phase boundaries with our analytically obtained thresholds (colored thick curves), each obtained by equating two of the three threshold expressions $N_c^{\rm I}$, $N_c^{\rm II}$, and $N_{c,\mathrm{full}}^{\rm III}$.

Remarkably, all $3!=6$ possible orderings of the three critical lengths are realized within the parameter window of Fig.~\ref{fig:phase_diagram}. A larger critical length indicates that the corresponding ladder retains a real spectrum up to a larger system size and is therefore more stable against amplification. For example, the ordering III$>$I$>$II means that Model III is the most stable and Model II is the first to undergo the real-to-complex transition.

The analytical hierarchy boundaries are obtained by equating the corresponding critical-length conditions derived above for Models I--III [Eqs.~\eqref{eq:NNNN_active_threshold_main}, \eqref{eq:NNLR_Nc_main}, and \eqref{eq:NcIII_full_main}]. Solving these pairwise equalities over the parameter window of Fig.~\ref{fig:phase_diagram} yields the following five distinct connected boundary branches:
\begin{align}
\mathcal B_{\rm I-II}^{(1)}:\quad
&N_c^{\rm I}(D,\alpha,\gamma)
=
N_c^{\rm II}(D,\alpha,\gamma),
\label{eq:BI-II_branch1}
\\
\mathcal B_{\rm I-II}^{(2)}:\quad
&N_c^{\rm I}(D,\alpha,\gamma)
=
N_c^{\rm II}(D,\alpha,\gamma),
\label{eq:BI-II_branch2}
\\
\mathcal B_{\rm I-III}^{(1)}:\quad
&N_c^{\rm I}(D,\alpha,\gamma)
=
N_{c,\,2<\alpha<3}^{\rm III}(D,\alpha,\gamma),
\label{eq:BI-III_branch1}
\\
\mathcal B_{\rm I-III}^{(2)}:\quad
&N_c^{\rm I}(D,\alpha,\gamma)
=
N_{\rm sat}(\alpha,\gamma),
\label{eq:BI-III_branch2}
\\
\mathcal B_{\rm II-III}^{(1)}:\quad
&N_c^{\rm II}(D,\alpha,\gamma)
=
N_{\rm sat}(\alpha,\gamma).
\label{eq:BII-III_branch1}
\end{align}
The superscripts distinguish disconnected solutions of the same pairwise equality, rather than different fitting forms. In particular, $\mathcal B_{\rm I-II}^{(1)}$ and $\mathcal B_{\rm I-II}^{(2)}$ are two distinct roots of the Model-I--Model-II equality; $\mathcal B_{\rm I-III}^{(1)}$ follows from the unsaturated $2<\alpha<3$ branch of Model III; and $\mathcal B_{\rm I-III}^{(2)}$ and $\mathcal B_{\rm II-III}^{(1)}$ arise after the Model-III threshold reaches $N_{\rm sat}$. These boundaries are therefore direct consequences of the previously derived critical-length laws, rather than independent fits, and they reproduce the overall structure and most of the numerical hierarchy boundaries in Fig.~\ref{fig:phase_diagram}.

The structure of the diagram can be understood from the three competing length laws. Model I grows only logarithmically, $N_c^{\rm I}\sim(\alpha/\gamma)\ln D$, because it is controlled by exponential skin overlap. Model II grows algebraically, $N_c^{\rm II}\sim D^{\alpha/3}$, because long-range rungs coherently enhance the inter-chain hybridization while the legs retain the ordinary $N^{-2}$ band-edge protection. In the strongly nonlocal regime, Model III has the leading scale-covariant growth $N_c^{\rm III}\propto D$, with the corrections derived above at and beyond $\alpha=2$, before the physical branch is cut off by $N_{\rm sat}$.

These different functional dependences explain the main regions in Fig.~\ref{fig:phase_diagram}. For small $\alpha$, the rung kernel is very long-ranged, so Model II is strongly hybridized and often has the smallest $N_c$; it is therefore the earliest to complexify in regions whose ordering ends with ``$>$II'', such as I$>$III$>$II or III$>$I$>$II. For larger $\alpha$, the long-range rung becomes less effective and the algebraic threshold of Model II grows rapidly with $D$, allowing it to become the most stable branch in regions beginning with ``II$>$''. The fully nonlocal lattice behaves differently: its scale-covariant growth can make it stable at intermediate separations, but once the candidate $N_c^{\rm III}$ reaches $N_{\rm sat}$, the physical threshold $N_{c,\mathrm{full}}^{\rm III}$ stops increasing while $N_c^{\rm I}$ and $N_c^{\rm II}$ continue to grow. This produces the regions where Model III becomes the earliest-onset branch at large $D$ or large $\alpha$.

We remind the reader that the colored regions in Fig.~\ref{fig:phase_diagram} should not be read as thermodynamic phases. They  represent different ordering hierarchies of real-to-complex transitions, separated by pairwise crossings of finite-size thresholds belonging to different ladder configurations (models). Some color transitions correspond to a change in the earliest instability mechanism, for example when the smallest threshold switches from Model II to Model III. Other color changes only exchange the order of the two later-onset branches while the first complexifying mechanism remains the same. The physically most important information is therefore both the full ordering and, in particular, which model appears at the end of the inequality (becomes unstable first).

\section{Discussion}

Our results identify a nonlocal route by which competitive non-Hermitian lattices depart from the conventional critical-NHSE paradigm: from logarithmic scaling, to algebraic onset, to a scale-covariant threshold controlled by the ladder aspect ratio $N_c/D$. Long-range hopping does not merely renormalize the coupling strength entering this transition; it changes both sides of the EP-discriminant condition, Eq.~\eqref{eq:active_instability_main}: the single-chain level spacing $\Delta E_{ab}$, dressed by the diagonal rung shift $\eta(u_{aa}-u_{bb})$, on one side, and the off-diagonal rung hybridization product $u_{ab}v_{ba}$ that must overcome it, on the other. The identity of the colliding EP pair $(a,b)$ is likewise not fixed in advance, but must be selected self-consistently from the finite open-boundary spectrum via Eq.~\eqref{eq:min_threshold_main}, with the projected rung matrix elements $u_{ij},v_{ij}$ requiring a model-specific evaluation of the skin-deformed wavefunction overlaps. It is this combination of nonanalytic band-edge scaling, biorthogonal skin deformation, and architecture-dependent rung projection that produces three qualitatively distinct scaling regimes, rather than three different prefactors sharing a common scaling law.

More broadly, this shows that finite-size criticality in non-Hermitian lattices is not an intrinsic property of a given amplification mechanism, but depends sensitively on how far the underlying couplings reach. Any competitive non-Hermitian system whose boundary sensitivity is conventionally organized by a local recursion relation---generalized Brillouin-zone~\cite{yao2018edge,lee2019anatomy,song2019non,yang2020non,longhi2020non,helbig2020generalized,li2025phase,meng2025generalized,yuan2026non} and transfer-matrix~\cite{kunst2019non,luo2021transfer,li2026non} constructions being the standard examples---is susceptible to this new kind of reorganization even if only one channel of its coupling is made long-ranged. The resulting logarithmic-to-algebraic-to-scale-covariant hierarchy is therefore unlikely to be specific to the ladder geometry studied here, and offers a general diagnostic for whether a non-Hermitian platform's boundary sensitivity is genuinely local in origin.

Experimentally, the real-to-complex threshold can be mapped out in platforms with programmable nonlocal couplings, including photonics~\cite{pan2018photonic,xiao2020non,zhu2020photonic,song2020two,ao2020topological,lin2024observation,liu2024observation}, mechanical materials~\cite{brandenbourger2019non,ghatak2020observation,wen2022unidirectional,xiu2023synthetically,li2024observation}, non-Hermitian topolectrical circuits~\cite{sahin2025topolectrical,hofmann2019chiral,ezawa2019electric,helbig2020generalized,hofmann2020reciprocal,liu2020gain,liu2021non,stegmaier2021topological,zhang2020non,zhang2022observation,yuan2023non,zhu2023higher,hohmann2023observation,zhang2024observation,zou2024experimental,zhang2023electrical,stegmaier2024realizing,stegmaier2024topological,shang2024observation,zhang2024observation2,guo2024scale,halder2024circuit,sahin2025protected,li2025realization}, and programmable quantum simulators~\cite{shen2026simulating,smith2019simulating,gou2020tunable,koh2022simulation,kirmani2022probing,frey2022realization,chertkov2023characterizing,chen2023high,liu2024simulating,yang2023simulating,iqbal2023creation,shen2025observation,koukoutsis2024quantum,koh2024realization,koh2025interacting,shen2025robust,zhang2025observation}. Operationally, one scans the chain length at fixed $(D,\alpha,\gamma)$ and detects the first resolvable imaginary splitting of the spectrum: in topoelectrical circuits this can be extracted by reconstructing the admittance spectrum from node-resolved voltage responses, and in amplification-based measurements more generally, the relevant observable is the largest relative modal growth rate $\Gamma_{\max}$ after subtracting uniform background loss, rather than voltage growth alone; $N_c$ is then the first length for which $\Gamma_{\max}$ exceeds the experimental resolution floor.

Because $\alpha$, $D$, and the choice of architecture (Models I--III) each set a different coupling matrix rather than a single tunable knob, most platforms cannot sweep them within one fabricated sample: mapping Fig.~\ref{fig:scalings} over $D$, or comparing architectures across Fig.~\ref{fig:phase_diagram}, generally requires assembling a family of samples with different rung and leg coupling profiles rather than continuously tuning a single device. This, however, does not present any practical difficulty in platforms with fully programmable coupling matrices -- electric circuits with reconfigurable admittance networks, or programmable quantum simulators -- where $D$, $\alpha$, and the model architecture can be reprogrammed electronically without physical refabrication. Detailed measurement protocols for both scenarios are given in Appendix~\ref{sec:app_experimental_extraction}.

The same sensitivity of $N_c$ to the aspect ratio geometry suggests a new mode of sensing applications in the spirit of exceptional-point and non-Hermitian circuit sensors: a small structural change can switch the system across an exceptional-point threshold, producing a measurable spectral splitting or modal-amplification onset~\cite{chen2017exceptional,hodaei2017enhanced,yuan2023non}.


%

\onecolumngrid
\appendix 

\section{Detailed definitions and introduction for the models studied}
\label{sec:app_model_definitions}

This Appendix fixes the notation and model map used in the main-text Hamiltonians [Eqs.~\eqref{eq:H_model_I_main}--\eqref{eq:H_model_III_main}], the three schematics in Fig.~\ref{fig:scalings}(a--c), and the hierarchy comparison in Fig.~\ref{fig:phase_diagram}. Its purpose is to make explicit which hopping channel is local or nonlocal in each model, so that the origin of the different critical-length laws can be followed without repeatedly reconstructing the geometry from the block matrices. The same conventions are used in all later Appendices.

\subsection{Global overview and model map}
\label{sec:app_model_map}

This subsection is the navigation key for the model labels used in main-text Fig.~\ref{fig:scalings}, Table~\ref{table}, and Fig.~\ref{fig:phase_diagram}. In particular, it separates the three models entering the main hierarchy from Model IV, which is retained only as a control for placing nonlocality on the legs alone.

The main text focuses on three ladders, Models I--III. We include the LR--NN geometry only as Model IV, a control included in the Appendix. All four systems considered in this Appendix are competitive NHSE ladders composed of two oppositely pumped chains~\cite{hatano1996localization,lee2019anatomy,ashida2020non,bergholtz2021exceptional}. What changes from one model to another is the \emph{distribution of nonlocality}: whether long-range hopping is placed on neither channel, only rungs, both channels, or only legs. Because this distribution controls projected two-mode hybridization and finite-size level spacings, it also changes the resulting onset mechanism and $N_c$ scaling law.

Throughout this Appendix the counter-pumped legs follow the main-text notation: nearest-neighbor legs are denoted $H^{\rm NN}(+\gamma)$ and $H^{\rm NN}(-\gamma)$, while long-range legs are denoted $H^{\rm LR}(+\gamma)$ and $H^{\rm LR}(-\gamma)$.

We fix the global naming-and-numbering convention used throughout this Appendix as follows: fully local NN--NN (Model I), nonlocal rungs NN--LR (Model II), fully nonlocal LR--LR (Model III), and nonlocal-legs LR--NN control geometry (Model IV). We keep repeating ``name + Model number'' in model-specific titles to make navigation unambiguous in long derivations.

\begin{table}[H]
  \centering
  \caption{Model map used throughout the Appendix. Models I--III are the three ladders analyzed in the main text. Model IV is retained only as the LR--NN control in the Appendix.}
  \begin{tabular}{llll}
    \toprule
    Model number & Legs & Rungs & Expected scaling character \\
    \midrule
    Model I & NN & NN & logarithmic critical scaling \\
    Model II  / nonlocal  & NN legs & LR  & algebraic onset \\
    Model III  & LR & LR  & scale-covariant onset \\
    Model IV & LR & NN  & approximation-sensitive \\
    \bottomrule
  \end{tabular}
\end{table}

This table is intended as a navigation map rather than a replacement for the full derivations below. We now proceed in the fixed order Model I $\to$ Model II $\to$ Model III, and then discuss Model IV as an Appendix control case.

\subsection{Hamiltonian building blocks}
\label{sec:ham_building_block}

The building blocks below are the detailed versions of the intra-chain and inter-chain terms quoted in the main text. They are introduced here so that the projected matrix elements in Appendices~\ref{sec:app_effective_hamiltonian}--\ref{sec:app_Nc_derivations} can be traced back to a definite microscopic coupling rather than treated as abstract parameters.

The main text writes the isolated leg Hamiltonians as
$H^{\rm NN}(\pm\gamma)$ and $H^{\rm LR}(\pm\gamma)$, and writes the coupled
ladders as block Hamiltonians $\mathcal H$. We use the same notation here. The
main text studies and compares the three ladders Models I--III; Model IV is
kept here only as an LR--NN control. Schematically, each ladder has the
block form
\begin{equation}
  \mathcal H
  =
  \begin{pmatrix}
    H^{\rm leg}(+\gamma) & H_{\perp} \\
    H_{\perp}^{\dagger} & H^{\rm leg}(-\gamma)
  \end{pmatrix}\:,
\end{equation}
where $H^{\rm leg}$ is chosen as either $H^{\rm NN}$ or $H^{\rm LR}$, and
$H_{\perp}$ denotes the upper-right inter-chain block.

To construct the four distinct geometries, we define two types of single-leg
Hamiltonians and two types of inter-chain blocks.

\noindent \underline{\textbf{Intra-chain hopping.}}

We consider two possibilities for each leg.
\begin{enumerate}
  \item Long-range (LR): Hopping between any two sites on the same leg, with power-law decay:
    \begin{equation}
      H^{\rm LR}(\pm\gamma)
      =
      \sum_{n\neq m}
      \frac{1\pm \gamma \operatorname{sgn}(n-m)}
      {|x_n-x_m|^{\alpha}}c_n^\dagger c_m
      \label{eq:H-intra-lr}
    \end{equation}
    Here $c_n^{\dagger}$ ($c_n$) create (annihilate) a particle at site $n$ on the given leg, $\gamma$ controls the non-reciprocity, and the $\pm$ sign corresponds to rightward ($+$) or leftward ($-$) non-Hermitian pumping.

  \item Nearest-neighbor (NN): the $|n-m|=1$ limit, retaining only adjacent-site hoppings:
    \begin{equation}
      H^{\rm NN}(\pm\gamma)
      =
      \sum_{n=1}^{N-1}
      \Bigl[(1\pm\gamma)c_{n+1}^\dagger c_n
      +(1\mp\gamma)c_n^\dagger c_{n+1}\Bigr]
      \label{eq:H-intra-nn}
    \end{equation}
\end{enumerate}

\noindent \underline{\textbf{Inter-chain hopping.}}

We consider two forms for the coupling block $H_{\perp}$ between the legs.
Chain indices $\pm$ appear here because two distinct legs are involved. The
Hermitian-conjugate process appears as the lower-left block
$H_{\perp}^{\dagger}$ in the full Hamiltonian, so the upper-right block itself
is written without an additional Hermitian-conjugate term.
\begin{itemize}
  \item Long-range rung block:
    \begin{equation}
      H_{\perp}^{\rm LR}
      =
      \sum_{n,m}
      \frac{c_{+,n}^{\dagger}c_{-,m}}
      {[D^2+(x_n-x_m)^2]^{\alpha/2}}
      \label{eq:H-inter-lr}
    \end{equation}

  \item Local/NN rung block:
    \begin{equation}
      H_{\perp}^{\rm NN}
      =
      D^{-\alpha}\sum_{n}c_{+,n}^{\dagger}c_{-,n}
      \label{eq:H-inter-local}
    \end{equation}
\end{itemize}
For consistency with the main-text numerics, the LR rung in
Eq.~\eqref{eq:H-inter-lr} keeps the full distance dependence. For the local
coupling in Eq.~\eqref{eq:H-inter-local}, only vertical ($n=m$) couplings of
strength $D^{-\alpha}$ are retained.

\subsection{The four models: definitions and overview}
\label{sec:app_four_models}

These block Hamiltonians make precise the three architectures compared in main-text Fig.~\ref{fig:scalings} and the additional LR--NN control. Their significance is that changing only the location of the long-range couplings changes which term in the EP balance is size enhanced, and therefore changes the functional form of $N_c(D)$.

By combining the components defined above in Sec.~\ref{sec:ham_building_block},
we construct the four geometries discussed in the text and in this Appendix.
The main-text ladders are denoted by $\mathcal H^{\rm I}$,
$\mathcal H^{\rm II}$, and $\mathcal H^{\rm III}$; the LR--NN control geometry included only in the Appendix is denoted by $\mathcal H^{\rm IV}$.

The four geometries are constructed in the following order:

\begin{itemize}
  \item \textbf{fully local NN--NN ladder (Model I)}\\
    All hopping is short-range: nearest-neighbor within each leg and local between legs. This model uses the nearest-neighbor legs $H^{\rm NN}(\pm\gamma)$ [Eq.~\eqref{eq:H-intra-nn}] and the local/NN rung block $H_{\perp}^{\rm NN}$ [Eq.~\eqref{eq:H-inter-local}].
    \begin{equation}
      \mathcal H^{\rm I}
      =
      \begin{pmatrix}
        H^{\rm NN}(+\gamma) & H_{\perp}^{\rm NN} \\
        (H_{\perp}^{\rm NN})^{\dagger} & H^{\rm NN}(-\gamma)
      \end{pmatrix}
      \label{eq:H_model_I}
    \end{equation}

  \item \textbf{nonlocal rungs NN--LR ladder (Model II)}\\
    Nearest-neighbor hopping within each leg, but long-range coupling between legs. This model uses $H^{\rm NN}(\pm\gamma)$ [Eq.~\eqref{eq:H-intra-nn}] and $H_{\perp}^{\rm LR}$ [Eq.~\eqref{eq:H-inter-lr}].
    \begin{equation}
      \mathcal H^{\rm II}
      =
      \begin{pmatrix}
        H^{\rm NN}(+\gamma) & H_{\perp}^{\rm LR} \\
        (H_{\perp}^{\rm LR})^{\dagger} & H^{\rm NN}(-\gamma)
      \end{pmatrix}
      \label{eq:H_model_II}
    \end{equation}

  \item \textbf{fully nonlocal LR--LR ladder (Model III)}\\
    Long-range hopping both within and between legs. This model uses $H^{\rm LR}(\pm\gamma)$ [Eq.~\eqref{eq:H-intra-lr}] and the long-range rung block $H_{\perp}^{\rm LR}$ [Eq.~\eqref{eq:H-inter-lr}].
    \begin{equation}
      \mathcal H^{\rm III}
      =
      \begin{pmatrix}
        H^{\rm LR}(+\gamma) & H_{\perp}^{\rm LR} \\
        (H_{\perp}^{\rm LR})^{\dagger} & H^{\rm LR}(-\gamma)
      \end{pmatrix}
      \label{eq:H_model_III}
    \end{equation}

  \item \textbf{nonlocal-legs LR--NN control geometry (Model IV)}\\
    Long-range hopping within each leg, but local/NN coupling between legs. This model uses $H^{\rm LR}(\pm\gamma)$ [Eq.~\eqref{eq:H-intra-lr}] and $H_{\perp}^{\rm NN}$ [Eq.~\eqref{eq:H-inter-local}].
    \begin{equation}
      \mathcal H^{\rm IV}
      =
      \begin{pmatrix}
        H^{\rm LR}(+\gamma) & H_{\perp}^{\rm NN} \\
        (H_{\perp}^{\rm NN})^{\dagger} & H^{\rm LR}(-\gamma)
      \end{pmatrix}
      \label{eq:H_model_IV}
    \end{equation}
\end{itemize}

Throughout this Appendix, the only valid mapping is fully local NN--NN (Model I), nonlocal rungs NN--LR (Model II), fully nonlocal LR--LR (Model III), and nonlocal-legs LR--NN control geometry (Model IV). The derivations are organized in this order because it gives the clearest progression from a local model to different distributions of nonlocality, while keeping Model IV clearly separated as a control outside the main hierarchy.

With the four models fixed, we next develop the projected EP criterion and the single-chain ingredients used in the model-specific derivations. The constant-kernel approximation for long-range rungs is introduced only where it enters the Model II and Model III estimates, and it is benchmarked separately in the final numerical-check section.

\section{Effective Hamiltonian for the Real-to-Complex Transition}
\label{sec:app_effective_hamiltonian}

This Appendix provides the numerical motivation and detailed construction behind the projected Hamiltonian in main-text Eq.~\eqref{eq:Hact_main}, the instability criterion Eq.~\eqref{eq:active_instability_main}, and the threshold definitions Eqs.~\eqref{eq:pair_threshold_main} and \eqref{eq:min_threshold_main}. This reduction is the common starting point for the three main-text scaling laws.

For the parameter windows analyzed in the main text, the first onset is controlled by a well-resolved pair of single-chain modes. This motivates projecting the full $2N\times2N$ Hamiltonian onto the corresponding four-dimensional two-leg sector, which is the only reduction needed for the scaling derivations below.

\subsection{Numerical Observation and Model Reduction}
\label{sec:app_numerical_reduction}

The spectra in this subsection justify the two-mode reduction for the representative parameters used below and identify the levels entering the projected Hamiltonian.

\begin{figure}[!htp]
  \centering
  \includegraphics[width=0.5\textwidth]{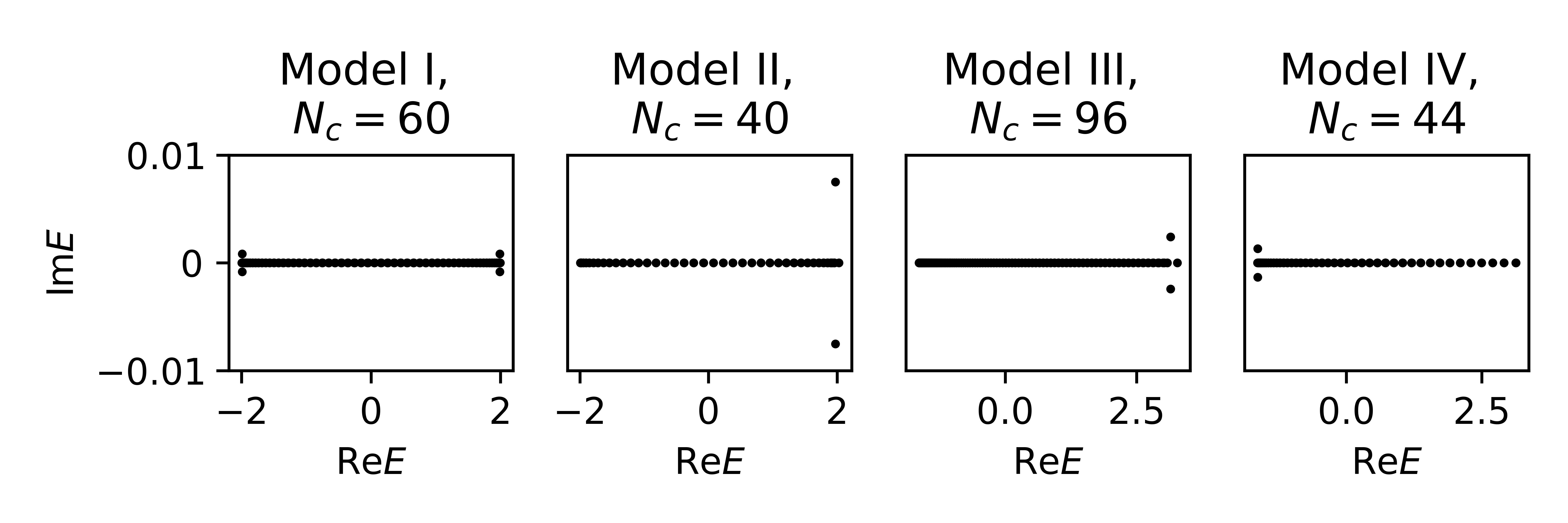}
  \caption{The energy spectrum for the four competitive NHSE ladders at their respective critical lengths $N_c$: fully local NN--NN ladder (Model I), nonlocal rungs NN--LR ladder (Model II), fully nonlocal LR--LR ladder (Model III), and nonlocal-legs LR--NN control geometry (Model IV). For the fully nonlocal LR--LR ladder (Model III), the displayed collision occurs near the right edge of the spectrum. For the nonlocal-legs LR--NN control geometry (Model IV), the corresponding edge-sector onset is analyzed below through the bottom-edge pair. For the fully local NN--NN ladder (Model I) and nonlocal rungs NN--LR ladder (Model II), transitions occur at both the top and bottom of the spectrum. $\alpha=2, \gamma=0.05, D=30$. }
  \label{fig:details_four_models}
\end{figure}

To motivate and validate this reduction, we first plot the numerical spectra of the full Hamiltonian and identify which eigenstates are responsible for the onset of complex eigenvalues.  The parameters are set to $\alpha=2, \gamma=0.05$, and $D=30$. The corresponding critical lengths are $N_c=40$ for fully local NN--NN ladder (Model I), $N_c=60$ for nonlocal rungs NN--LR ladder (Model II), $N_c=96$ for fully nonlocal LR--LR ladder (Model III), and $N_c=44$ for nonlocal-legs LR--NN control geometry (Model IV), as shown in Fig.~\ref{fig:details_four_models}.

For the representative parameters in Fig.~\ref{fig:details_four_models}, the first complex eigenvalues emerge from edge sectors. The Model-III example below tracks the highest-real-part pair, while the LR--NN control later uses the bottom-edge pair.

In the Model-III example below, ``Top-2'' and ``Top-3'' denote the two levels forming the active pair for this parameter set.

To study the spectrum instability, we focus on fully nonlocal LR--LR ladder (Model III) as a representative case. We investigate how the instability develops by introducing a Hamiltonian $\mathcal H^{\rm III}(\lambda)$ that smoothly interpolates from two decoupled chains (at $\lambda=0$) to the fully coupled ladder (at $\lambda=1$). This interpolating Hamiltonian is defined in block matrix form as
\begin{equation}
  \mathcal H^{\rm III}(\lambda) =
  \begin{pmatrix}
    H^{\rm LR}(+\gamma) & \lambda H_{\perp}^{\rm LR} \\
    \lambda (H_{\perp}^{\rm LR})^{\dagger} & H^{\rm LR}(-\gamma)
  \end{pmatrix},
  \label{eq:hamiltonian_interp}
\end{equation}
where $H_{\perp}^{\rm LR}$ is the long-range inter-chain block in Eq.~\eqref{eq:H-inter-lr}, and $\lambda \in [0, 1]$ is the interpolation parameter. The decoupled Hamiltonian discussed previously corresponds to $\mathcal H^{\rm III}(\lambda=0)$.

\begin{figure}[!htp]
  \centering
  \begin{minipage}{0.48\linewidth}
    (a) \\ \vspace{1mm}
    \includegraphics[width=\linewidth]{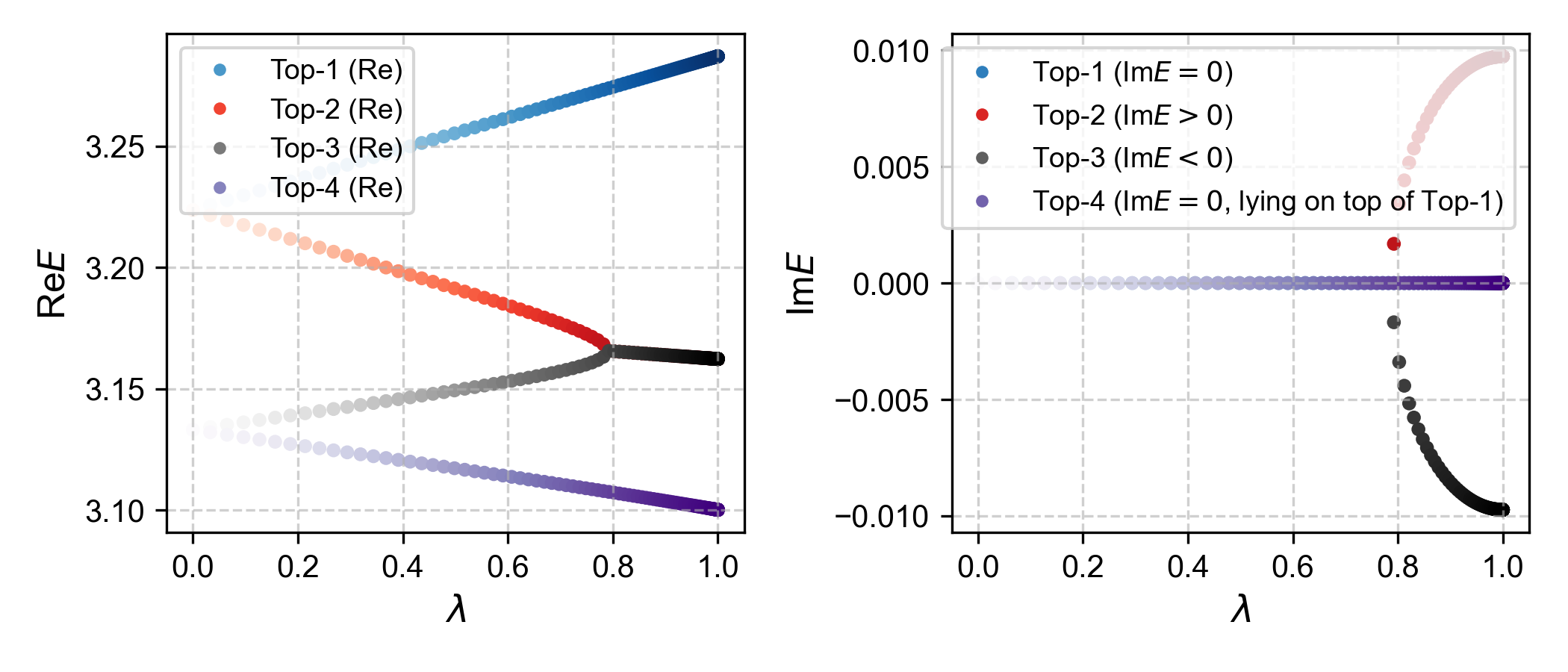}
  \end{minipage}
  \hfill
  \begin{minipage}{0.48\linewidth}
    (b) \\ \vspace{1mm}
    \includegraphics[width=\linewidth]{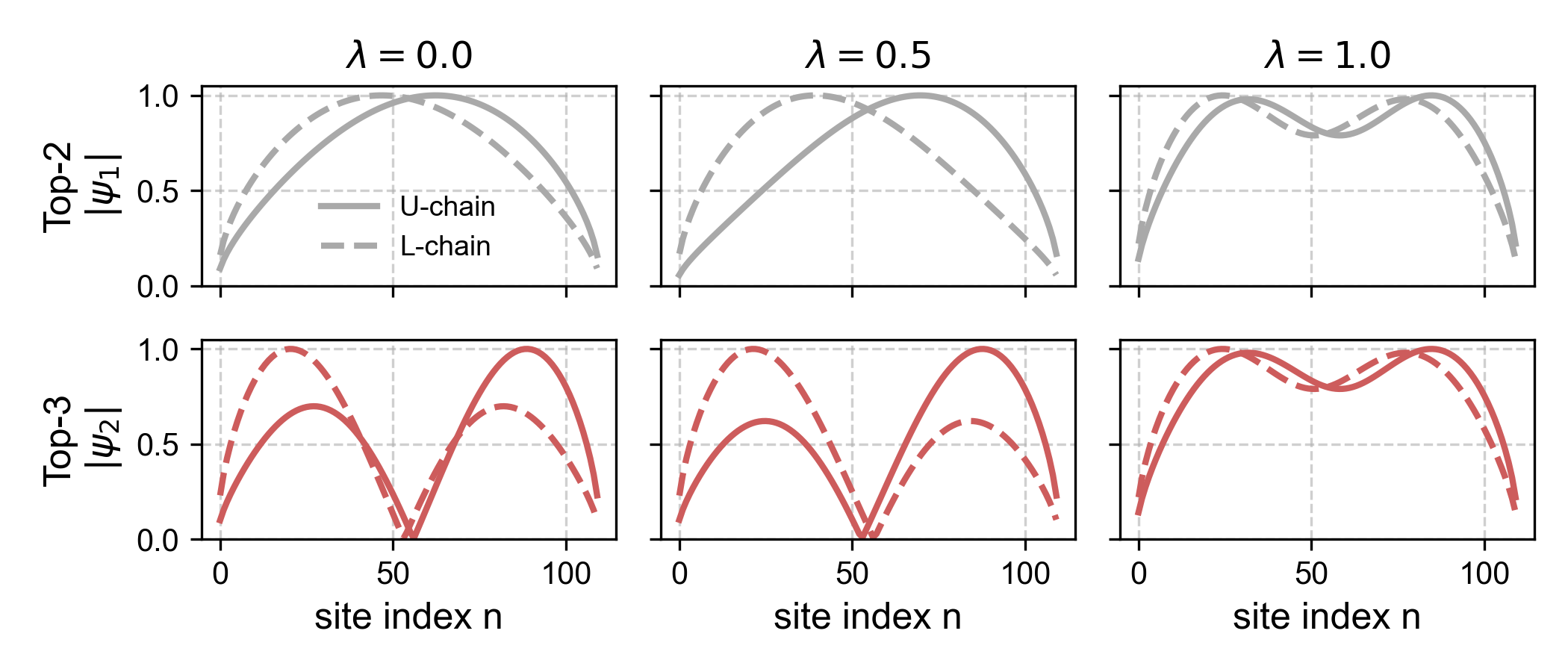}
  \end{minipage}
  \caption{Fully nonlocal LR--LR competitive NHSE ladder (Model III). (a) The complex energy spectrum of the four eigenstates with the largest real parts (Top-1 to Top-4), plotted as a function of the inter-chain coupling strength $\lambda$. Each state is distinguished by a unique color map, where the color intensity corresponds to $\lambda$. The Top-2 and Top-3 eigenvalues merge as $\lambda$ increases.
    (b) Spatial distribution of the eigenstate amplitudes ($|\psi_n|$) for the Top-2 (top row) and Top-3 (bottom row) eigenstates, which originate from the $p=1$ and $q=2$ single-chain modes, respectively.
    The plots illustrate the state evolution at three discrete values of the inter-chain coupling: $\lambda=0.0$ (left column), $\lambda=0.5$ (middle column), and $\lambda=1.0$ (right column).
    In each panel, the amplitude is plotted against the site index $n$. The solid and dashed lines represent the distributions on the $(+)$ and $(-)$ legs, respectively.
    The system parameters are: $N=110$, $\gamma=0.05$, $\alpha=2$, and $D=30$.
  }
  \label{fig:modelIII_interpolation_top_pair}
\end{figure}

By numerically diagonalizing $\mathcal H^{\rm III}(\lambda)$, we trace the evolution of the eigenvalues with the largest real parts, as shown in Fig.~\ref{fig:modelIII_interpolation_top_pair}(a). At the decoupled limit ($\lambda=0$), the spectrum is simply two superimposed copies of the single-chain spectrum. Consequently, the highest energy levels appear as degenerate pairs corresponding to the single-chain eigenenergies: the $E_1$ pair (labeled Top-1 and Top-2) and the $E_2$ pair (labeled Top-3 and Top-4). As $\lambda$ increases, the inter-chain coupling lifts these degeneracies. Notably, one eigenvalue from each pair---specifically, those labeled Top-2 and Top-3---moves towards the other and eventually merges at an exceptional point (EP). The physical nature of this merger becomes clear upon plotting the corresponding eigenstates in Fig.~\ref{fig:modelIII_interpolation_top_pair}(b). At $\lambda=0$, the Top-2 eigenstate is the $p=1$ non-Hermitian skin mode of one of the single chains (e.g., the upper chain), while the Top-3 eigenstate is the $q=2$ skin mode of the same chain. As the coupling $\lambda$ is turned on, these two modes hybridize across the ladder. Their energy eigenvalues coalesce, and this pair is the first one to acquire a non-zero imaginary energy component for the displayed parameter set.

For the representative LR--LR parameters in Fig.~\ref{fig:modelIII_interpolation_top_pair}, the active pair is $(p,q)=(1,2)$. We now construct the corresponding projected Hamiltonian.

We denote the two participating single-chain levels by $(p,q)$. The minimal effective Hamiltonian is constructed within the subspace spanned by these two states on both legs:
$\left\{|p, +\rangle, |q, +\rangle, |p, -\rangle, |q, -\rangle\right\}$.

For notational simplicity, we relabel $(p,q)\to(a,b)$ in the general projected matrix and use $(1,2)$ in the model-specific band-edge estimates.

\subsection{Effective Hamiltonian in the projected two-mode sector}
\label{sec:app_effective_projection}

We now derive the $4\times4$ matrix quoted in main-text Eq.~\eqref{eq:Hact_main}. This step matters because it isolates the three quantities that are compared throughout the paper: the single-chain spacing, the diagonal rung-induced shift, and the off-diagonal hybridization product.

The preceding analysis provides a solid foundation for constructing a minimal effective model based on the $4 \times 4$ sector $\left\{|a, +\rangle, |b, +\rangle, |a, -\rangle, |b, -\rangle\right\}$. Here $|i,\pm\rangle$ denotes the $i$-th right eigenstate of the isolated leg $H^{\rm leg}(\pm\gamma)$, with $H^{\rm leg}=H^{\rm NN}$ or $H^{\rm LR}$ depending on the model, and the corresponding left states are used in the biorthogonal projection.

Projecting the full Hamiltonian onto this basis yields the main-text projected EP Hamiltonian:
\begin{equation}
  H_{\rm EP}^{(a,b)} =
  \begin{pmatrix}
    E_a & 0 & u_{aa} & u_{ab}\\
    0 & E_b & u_{ba} & u_{bb}\\
    v_{aa} & v_{ab} & E_a & 0\\
    v_{ba} & v_{bb} & 0 & E_b
  \end{pmatrix}.
  \label{eq:Heff_4x4}
\end{equation}
The projected upper- and lower-block matrix elements are
\begin{equation}
  u_{ij}=\langle L_{i,+}|H_{\perp}|R_{j,-}\rangle,\qquad
  v_{ij}=\langle L_{i,-}|H_{\perp}^{\dagger}|R_{j,+}\rangle,
  \qquad i,j\in\{a,b\}.
\end{equation}
In a non-Hermitian biorthogonal projection one should not set $v_{ij}=u_{ij}^{*}$ unless this relation is separately justified by an additional symmetry or by a special real-gauge estimate.

As an explicit example, in Fig.~\ref{fig:modelIII_interpolation_top_pair} the coalescing states are labeled ``Top-2'' and ``Top-3'', and in that case the projected pair is $a=1$ and $b=2$.

\subsection{The condition for the real-to-complex transition}
\label{sec:app_transition_condition}

This subsection supplies the discriminant condition used in the main text to define every pair-resolved threshold and, after minimization, the physical critical length. The result is important because it turns the spectral transition into a transparent competition between an intrinsic level spacing and architecture-dependent projected hybridization.

With the reduced Hamiltonian explicitly constructed, we next state the analytical condition under which its eigenvalues first become complex, marking the real-to-complex spectral transition.

Let
\begin{equation}
  \Delta_{ab} \equiv E_a-E_b>0.
\end{equation}
For the pairwise EP collision described by Eq.~\eqref{eq:Heff_4x4}, the relevant discriminant becomes negative when
\begin{equation}
  \left[\Delta_{ab}\pm(u_{aa}-u_{bb})\right]^{2}+4u_{ab}v_{ba}<0.
  \label{eq:central_condition}
\end{equation}
This inequality presents an analytical criterion for determining the system's critical length, $N_c$. The critical length is fundamentally defined as the minimum system size at which the spectrum of the \textit{full} Hamiltonian first exhibits complex eigenvalues. The validity of our effective model hinges on whether the smallest integer $N$ that satisfies Eq.~\eqref{eq:central_condition} accurately predicts the numerically observed value of $N_c$.

Physically, this condition signifies that the transition occurs when the projected inter-chain matrix elements become comparable in magnitude to the intrinsic energy gap $\Delta_{ab}$ of the constituent chains. The problem of finding $N_c$ is thus transformed into a more tractable procedure: calculating the relevant parameters ($\Delta_{ab},u_{ij},v_{ij}$) as functions of $N$ and solving the inequality. In concrete evaluations, we will label this gap by a superscript indicating the intra-chain range whenever possible, e.g., $\Delta_{ab}^{\mathrm{LR}}$ or $\Delta_{ab}^{\mathrm{NN}}$.

To apply Eq.~\eqref{eq:central_condition} to a critical-length estimate, we need three single-chain inputs:

\begin{itemize}
  \item Spectrum: the energies $E_p$ and $E_q$, and hence the spacing $\Delta E_{pq}$ of the pair under consideration.
  \item Eigenstates: the spatial profiles entering the projected matrix elements. Their skin deformation controls the upper- and lower-block overlaps.
  \item Biorthogonal basis: the corresponding left and right eigenvectors, which are both required for a non-Hermitian projection.
\end{itemize}

These ingredients determine the $N$-dependence of $\Delta_{pq}$, $u_{ij}$, and $v_{ij}$ in Eq.~\eqref{eq:central_condition}. Before applying the criterion to specific models, it is useful to state when this low-dimensional reduction is expected to be quantitatively reliable.

For the parameter ranges used in the analytical curves, the participating pair is well isolated and the projected condition accurately locates the onset. The numerical values of $N_c$ shown in the figures are nevertheless obtained from direct diagonalization of the full Hamiltonian.

The resulting discriminant is the criterion used in the model-specific threshold derivations below.

\section{Key properties of a single chain with Long-Range and Nearest-Neighbor couplings}
\label{sec:single_chain_properties}

This Appendix derives the single-chain inputs used in main-text Fig.~\ref{fig:sgchain}(b,c), Eqs.~\eqref{eq:gap_LR} and \eqref{eq:gap_NN}, and the Model-III mixing laws Eqs.~\eqref{eq:chiN_piecewise_main} and \eqref{eq:V21_piecewise_main}. The motivation is that the ladder threshold cannot be understood from the rung coupling alone: its scaling is fixed by how the isolated-chain level spacing and parity structure change with $N$ and $\alpha$. These results are then inserted into the model-specific EP conditions in Appendix~\ref{sec:app_Nc_derivations}.

The previous section reduced the real-to-complex transition to the competition between a single-chain level spacing and projected inter-chain matrix elements. This section collects the single-chain ingredients needed for that reduction: the relevant energy gaps and the eigenstates used to compute the projected matrix elements. More broadly, extended-range hopping in one-dimensional SSH-type chains is known to modify the accessible winding sectors, boundary-state localization, and edge-state dynamics~\cite{hsu2020anderson,dias2022indexing,ghosh2023quench,betancurocampo2024twofold}. 

The critical length is controlled by the gap and wavefunctions of the pair used in the projected sector. The main-text branches require the top-band pair for Models I--III, while the LR--NN control uses the bottom-band pair.

\subsection{Single-chain energy gaps at the band edges}
\label{sec:app_band_edge_gaps}

The gap laws collected here are used directly in the main-text power counting: the NN $N^{-2}$ gap controls Models I and II, whereas the nonanalytic LR $N^{1-\alpha}$ gap controls Model III for $0<\alpha<3$. The contrast between these two gaps is one of the two mechanisms behind the logarithmic--algebraic--scale-covariant hierarchy.

\subsubsection{Top-of-band gap: Long-Range chain, \texorpdfstring{$\Delta \Re E \sim N^{1-\alpha}$}{DeltaE \textasciitilde{} N\textasciicircum{}(1-alpha)}}

This derivation supplies the coefficient $C(\alpha)$ and the anomalous band-edge scaling used in main-text Eqs.~\eqref{eq:gap_LR} and \eqref{eq:LRLR_Nc_balance_main}. Its physical significance is that long-range intra-chain hopping removes the ordinary quadratic band-edge protection that would otherwise force the same $N^{-2}$ input in every architecture.

When the system size is smaller than the critical length, the spectrum is purely real. Therefore, in the following, we only consider the real part of the energy. 
For the Hermitian long-range single chain in Eq.~\eqref{eq:H-intra-lr}, the Bloch dispersion in the thermodynamic (PBC) limit can be written in terms of the polylogarithm as
\begin{equation}
  \Re E^{\mathrm{LR}}(k)=\sum_{n=1}^{\infty}\frac{2\cos(kn)}{n^{\alpha}}
  =\operatorname{Li}_{\alpha}(e^{ik})+\operatorname{Li}_{\alpha}(e^{-ik}).
\end{equation}
The top of the band is located at $k\approx0$. For a finite chain of length $N$, the two highest eigenvalues correspond to the smallest quantized wavevectors, and we may take
$k_1\simeq \pi/N$ and $k_2\simeq 2\pi/N$.

To extract the gap scaling, we use the small-$k$ expansion of the polylogarithm,
\begin{equation}
  \operatorname{Li}_{\alpha}(e^{ik})
  =\Gamma(1-\alpha)(-ik)^{\alpha-1}+\zeta(\alpha)+\cdots,
\end{equation}
which yields, near $k=0$,
\begin{equation}
  \Re E^{\mathrm{LR}}(k)
  =2\Gamma(1-\alpha)\cos\left[\frac{\pi}{2}(\alpha-1)\right]k^{\alpha-1}
  +2\zeta(\alpha)-\zeta(\alpha-2)k^{2}+O(k^{4}).
\end{equation}
The gap between the two highest levels (top-of-band first-colliding pair),
\begin{equation}
  \Delta \Re E^{\mathrm{LR}}_{\mathrm{top}} \equiv \Re E(k_1)- \Re E(k_2),
\end{equation}
is dominated by the nonanalytic term for $0<\alpha<3$ (with the integer cases understood through their limiting forms) as
\begin{equation}
  \Delta \Re E^{\mathrm{LR}}_{\mathrm{top}} \simeq
  2\Gamma(1-\alpha)\cos\left[\frac{\pi}{2}(\alpha-1)\right]
  \left(k_1^{\alpha-1}-k_2^{\alpha-1}\right).
  \label{append_eq:DeltaE}
\end{equation}
Substituting $k_1\simeq \pi/N$ and $k_2\simeq 2\pi/N$ gives the asymptotic system-size scaling
\begin{equation}
  \Delta E^{\mathrm{LR}}_{\mathrm{top}} \simeq C(\alpha)N^{1-\alpha},
  \qquad 0<\alpha<3.
  \label{append_eq:DeltaEScale}
\end{equation}
At $\alpha=2$, this limiting form gives $\Delta E^{\mathrm{LR}}_{\mathrm{top}}\sim N^{-1}$.

To validate the scaling and extract the fitted prefactor $C(\alpha)$ in Eq.~\eqref{append_eq:DeltaEScale}, we performed exact diagonalization of the finite-$N$ Hamiltonian and evaluated $\Delta E^{\mathrm{LR}}_{\mathrm{top}}$ for various $\alpha$.
As shown in Fig.~\ref{app_fig:DeltaE}(a), the numerical data (dots) follow the predicted power law $\Delta E^{\mathrm{LR}}_{\mathrm{top}}\propto N^{1-\alpha}$ (solid lines) across the range of $\alpha$ considered, confirming that the top-of-band gap is controlled by the fractional small-$k$ dispersion induced by the long-range hopping. The $C(\alpha)$ fit is in Fig.~\ref{app_fig:DeltaE}(b).

At the marginal decay exponent $\alpha=3$, the nonanalytic contribution in the polylogarithm merges with the analytic quadratic term. Expanding the truncated long-range lattice sum near $k=0$ gives a leading finite-size correction of the form
\begin{equation}
\Delta E_{\rm LR}(\alpha=3)\sim N^{-2}\ln N,
\end{equation}
before crossing over to the ordinary analytic $N^{-2}$ scaling for $\alpha>3$.

\subsubsection{Top-of-band gap: Nearest-Neighbor chain, \texorpdfstring{$\Delta \Re E \sim N^{-2}$}{DeltaE \textasciitilde{} N\textasciicircum{}(-2)}}

This standard result is retained because it is the quantitative gap inserted into the Model-I and Model-II thresholds in the main text. Keeping it next to the LR result makes clear that the $\alpha/3$ law of Model II comes from changing the rungs while leaving this quadratic leg gap intact.

\begin{figure}
  \centering

  \subfigure[]{
    \includegraphics[width=0.34\linewidth]{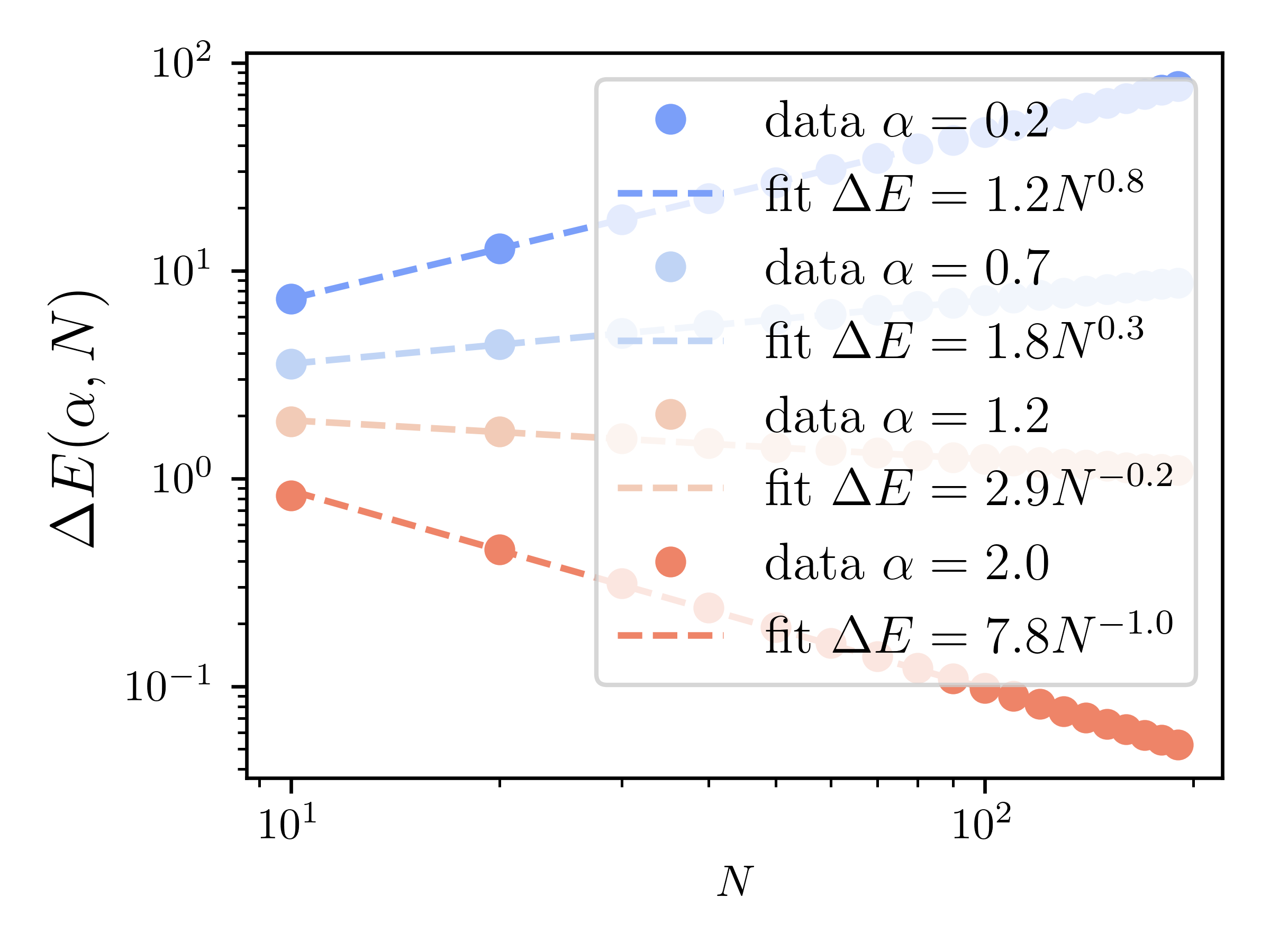}
    \label{app_fig:DeltaE_a}
  }
  \subfigure[]{
    \includegraphics[width=0.34\linewidth]{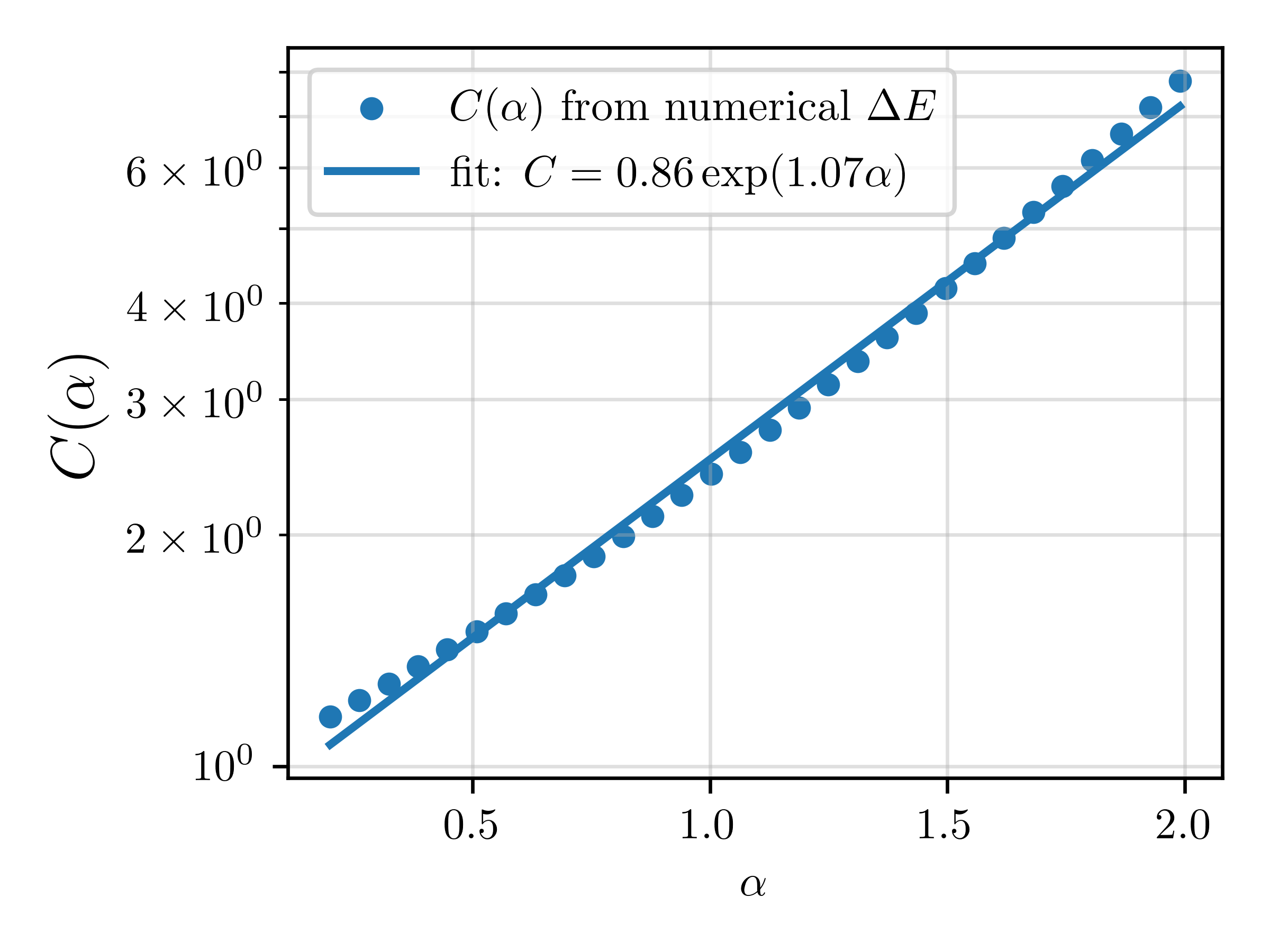}
    \label{app_fig:DeltaE_b}
  }\\
  \subfigure[]{
    \includegraphics[width=0.34\linewidth]{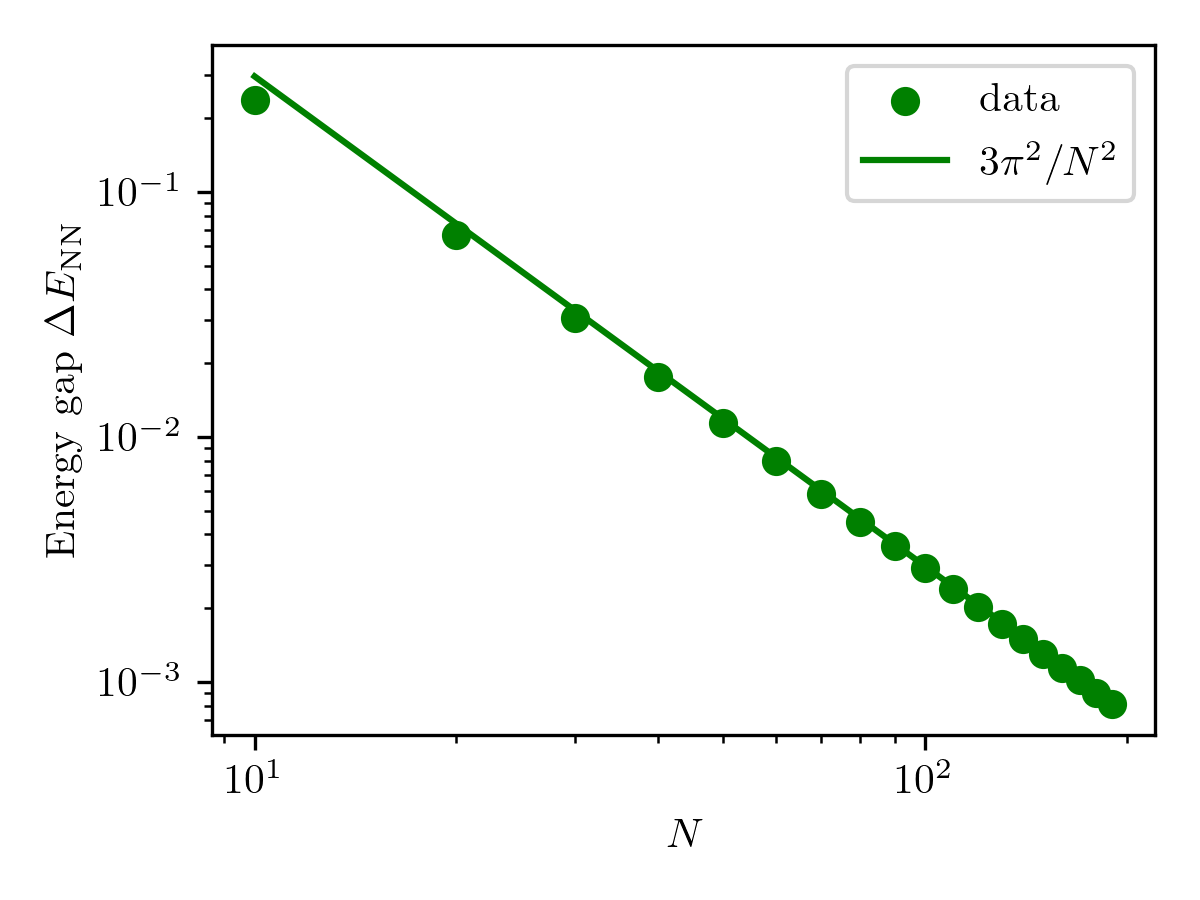}
    \label{app_fig:DeltaE_c}
  }
  \subfigure[]{
    \includegraphics[width=0.34\linewidth]{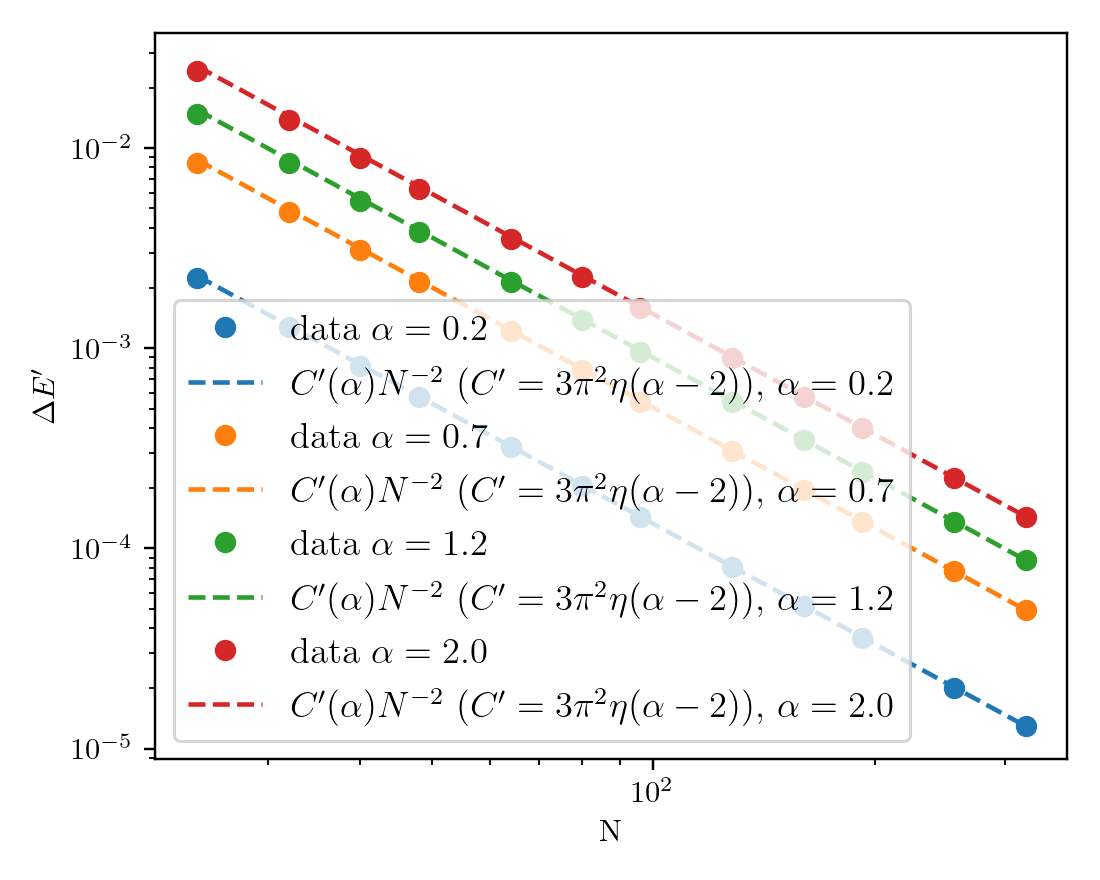}
    \label{app_fig:DeltaE_d}
  }

  \caption{Numerical verification of the scaling of the single-chain energy gap $\Delta\Re E$.
  (a) Top-of-band gap. For the long-range chain, $\Delta E$ vs. $N$ for $\alpha \in \{0.2, 0.7, 1.2, 2.0\}$. Dots are numerical data; solid lines are fits to the $N^{1-\alpha}$ scaling law.
  (b) The prefactor $C(\alpha)$ vs. $\alpha$. Dots are extracted from the fits in (a), while the blue solid line is the fitted form $C(\alpha)\approx 0.86\times \exp(1.07\alpha)$.
  (c) For the nearest-neighbor chain, $\Delta E$ vs. $N$. The numerical data, shown as dots, perfectly match the theoretical $3\pi^2N^{-2}$ scaling, shown as the solid line.
  (d) Bottom-of-band gap. Same long-range-chain parameters as in (a), shown with the theoretical prediction $\Delta \Re E \simeq C'(\alpha)N^{-2}$, where $C'(\alpha)=3\pi^2\eta(\alpha-2)$. Dots are numerical data, and dashed lines are $C'(\alpha)N^{-2}$.}
  \label{app_fig:DeltaE}
\end{figure}

For the nearest-neighbor single chain, the energy spectrum is given by the well-known tight-binding dispersion relation:
\begin{equation}
  E^{\mathrm{NN}}(k) = 2\cos(k).
\end{equation}
Again, the highest energies correspond to the smallest wavevectors. Expanding the cosine for small $k$ gives $E(k) \approx 2(1 - k^2/2) = 2 - k^2$. The top-of-band gap is therefore:
\begin{equation}
  \Delta\Re E_{\mathrm{top}}^{\mathrm{NN}} = \Re E(k_1) -\Re E(k_2) \simeq (2-k_1^2) - (2-k_2^2) = k_2^2 - k_1^2.
  \label{append_eq:DeltaE_NN}
\end{equation}
Substituting the quantized wavevectors $k_1 \approx \pi/N$ and $k_2 \approx 2\pi/N$, we obtain the precise scaling:
\begin{equation}
  \boxed{
    \Delta \Re E_{\mathrm{top}}^{\mathrm{NN}} \simeq \left(\frac{2\pi}{N}\right)^2 - \left(\frac{\pi}{N}\right)^2 = 3\pi^2 N^{-2}
  }\label{append_eq:DeltaEScale_NN}
\end{equation}
This theoretical prediction perfectly matches the numerical data shown in Fig.~\ref{app_fig:DeltaE}(c).

The difference between the scaling laws---$\Delta\Re E \sim N^{-2}$ in Eq.~\eqref{append_eq:DeltaEScale_NN} for nearest-neighbor coupling versus $\Delta\Re E \sim N^{1-\alpha}$ in Eq.~\eqref{append_eq:DeltaE} for long-range interactions---stems fundamentally from the analyticity of the energy dispersion at the band edge ($k=0$).
For local (short-range) Hamiltonians, the energy dispersion $\Re E(k)$ is a smooth, analytic function. Due to inversion symmetry $\Re E(k)=\Re E(-k)$, the expansion around the extremum must be quadratic, $\Re E(k) \propto k^2$, reflecting the standard Laplacian dynamics of a massive particle. Substituting $k \sim 1/N$ yields the diffusive $N^{-2}$ scaling.
In contrast, the power-law hopping $1/r^\alpha$ induces a non-analytic singularity at $k=0$. The Fourier transform of the heavy-tailed coupling leads to a fractional dispersion term $k^{\alpha-1}$, which dominates the analytic quadratic term.

\subsubsection{Bottom-of-band gap for Long-Range chains: \texorpdfstring{$\Delta\Re  E' \sim N^{-2}$}{DeltaE' \textasciitilde{} N\textasciicircum{}(-2)}}
This bottom-edge result is not part of the three main-text hierarchy branches. It is derived to support the LR--NN Model-IV control in Appendix~\ref{sec:app:Nc_modelIV}, where the first collision can arise from the bottom rather than the top of the spectrum.
\label{subsec:bottom_gap_LR}

To analyze the energy spectrum at the bottom of the band, which corresponds to wave vectors near $k=\pi$, we introduce a small momentum deviation $\delta k$ such that $k = \pi - \delta k$. We begin with the general expression for the energy spectrum, $\Re E(k) = \sum_{n=1}^{\infty} 2\cos(kn)/n^\alpha$. The substitution for $k$ transforms the cosine term to $\cos[n(\pi - \delta k)] = (-1)^n \cos(n\delta k)$, which turns the energy expression into an alternating series. By then expanding the cosine for small $\delta k$ as $\cos(n\delta k) \approx 1 - (n\delta k)^2/2$, we can express the full sum in terms of the Dirichlet eta function, $\eta(s) = \sum_{n=1}^{\infty} (-1)^{n-1}/n^s$. This procedure yields the dispersion relation near $k=\pi$,
\begin{equation}
  \Re E^{\mathrm{LR}}(\pi - \delta k) \approx -2\eta(\alpha) + (\delta k)^2 \eta(\alpha-2).
  \label{eq:dispersion_k_pi}
\end{equation}
This shows that the dispersion at the bottom of the band is quadratic in the momentum deviation $\delta k$. The bottom-of-band gap for the LR chain, $\Delta \Re E_{\mathrm{bot}}^{\mathrm{LR}}$, is the difference between the two lowest energy levels, which correspond to the smallest quantized momentum deviations, $\delta k_1 \approx \pi/N$ and $\delta k_2 \approx 2\pi/N$. Using the dispersion relation from Eq.~\eqref{eq:dispersion_k_pi}, the gap is calculated as $\Delta \Re E_{\mathrm{bot}}^{\mathrm{LR}} = \Re E^{\mathrm{LR}}(\pi - \delta k_2) - \Re E^{\mathrm{LR}}(\pi - \delta k_1) \approx \eta(\alpha-2) [ (\delta k_2)^2 - (\delta k_1)^2 ]$. Substituting the quantized values for $\delta k_1$ and $\delta k_2$ gives
\begin{equation}
  \Delta \Re E_{\mathrm{bot}}^{\mathrm{LR}} \approx \eta(\alpha-2) \left( \left(\frac{2\pi}{N}\right)^2 - \left(\frac{\pi}{N}\right)^2 \right) = \eta(\alpha-2) \frac{3\pi^2}{N^2}.
\end{equation}
This result leads directly to the final scaling law for the energy gap at the bottom of the band
\begin{equation}
  \Delta \Re E_{\mathrm{bot}}^{\mathrm{LR}} \simeq C'(\alpha)N^{-2},
  \label{eq:DeltaE_prime_scaling}
\end{equation}
where the prefactor is defined as $C'(\alpha) = 3\pi^2\eta(\alpha-2)$, which is verified in Fig.~\ref{app_fig:DeltaE}(d).

\subsection{Dominant single-chain wavefunctions for Long-Range couplings with finite \texorpdfstring{$\gamma$}{gamma}: top-of-band}
\label{sec:app_LR_top_wavefunctions}

The wavefunction mixing derived here is the origin of $\chi_N(\alpha)$ in the main-text Model-III analysis. Its purpose is to show explicitly how non-reciprocity activates a parity-forbidden off-diagonal rung channel; without this mixing, the negative product $u_{12}v_{21}$ required by the EP criterion would be absent in the two-mode approximation.

The evaluation of the critical length $N_c$ hinges on the projected matrix elements $u_{ij}$ and $v_{ij}$. A crucial observation is that within the unperturbed Hermitian limit ($\gamma=0$), the eigenstates possess definite parity (symmetric $n=1$ and anti-symmetric $n=2$ modes). Consequently, the destabilizing off-diagonal channel $u_{12}v_{21}$ would vanish identically due to parity selection rules.

Therefore, to capture the mechanism of the spectral transition, it is essential to account for the role of non-reciprocity $\gamma$ not merely as a spectral shift, but as a \textit{symmetry-breaking perturbation}. Finite $\gamma$ induces a mixing between the symmetric and anti-symmetric manifolds, skewing the wavefunction towards the boundary (the precursor to the skin effect). This $\mathcal{O}(\gamma)$ correction is the physical origin of the non-zero inter-chain hybridization.

To quantify this effect, we decompose the counter-pumped single-chain Hamiltonians into a Hermitian base and a non-Hermitian perturbation,
\begin{equation}
  H^{\rm leg}(\pm\gamma)=H_{0} \pm \gamma V,
\end{equation}
where $H^{\rm leg}$ is $H^{\rm LR}$ or $H^{\rm NN}$ depending on the model, $H_0$ is the Hermitian base (at $\gamma=0$), and the $+/-$ sign follows the leg label.

\paragraph{Perturbative Framework.}
For small $\gamma$, the true eigenstate $|\pm, n\rangle$ on each leg can be found using first-order perturbation theory. The unperturbed eigenstates of $H_0$, denoted by $|\psi_n^{(0)}\rangle$, are well-approximated by sine functions:
\begin{equation}
  \psi^{(0)}_1(x) = \sqrt{\frac{2}{N}}\sin\left(\frac{\pi x}{N}\right), \quad \psi^{(0)}_2(x) = \sqrt{\frac{2}{N}}\sin\left(\frac{2\pi x}{N}\right).
\end{equation}
Using a two-state approximation, the first-order correction to $|\psi_1^{(0)}\rangle$ is dominated by its coupling to $|\psi_2^{(0)}\rangle$:
\begin{equation}
  |\psi_1^{(1)}\rangle \approx \frac{\langle \psi_2^{(0)} | V | \psi_1^{(0)} \rangle}{\Delta E} |\psi_2^{(0)}\rangle,
\end{equation}
and similarly for $|\psi_2^{(1)}\rangle$. The full perturbed eigenstate on each leg is:
\begin{equation}
  |n,\pm\rangle \approx |\psi^{(0)}_{n}\rangle \pm \gamma |\psi^{(1)}_{n}\rangle, \quad n=1,2,
\end{equation}
where the upper/lower sign applies to the $(+)$/$(-) $ leg.

\paragraph{Application and validation for the long-range model.}
For the long-range model, the perturbation operator is
$V^{(\mathrm{LR})}
= \sum_{n\neq m}
\frac{{\scriptscriptstyle \operatorname{sgn}(n-m)}}{|n-m|^{\alpha}}
c_{n}^\dagger c_{m}$,
so $H^{\rm LR}(\pm\gamma)=H_0 \pm \gamma V^{(\mathrm{LR})}$. The first-order correction to $\psi_1^{(0)}(x)$ is:
\begin{equation}
  \psi_1^{(1)}(x) = \frac{\psi_2^{(0)}(x)}{\Delta E} \left( \sum_{n\neq m} \frac{\text{sgn}(n-m)}{|n-m|^{\alpha}} \psi_2^{(0)}(m) \psi_1^{(0)}(n) \right).\label{appeq:pertWave}
\end{equation}
This mixing of the symmetric $\psi_1^{(0)}$ and antisymmetric $\psi_2^{(0)}$ states breaks the wavefunction's reflection symmetry, causing it to localize towards one boundary.

\begin{figure}[!htp]
  \begin{minipage}{0.23\linewidth}
    (a)\\[1mm]
    \includegraphics[width=\linewidth]{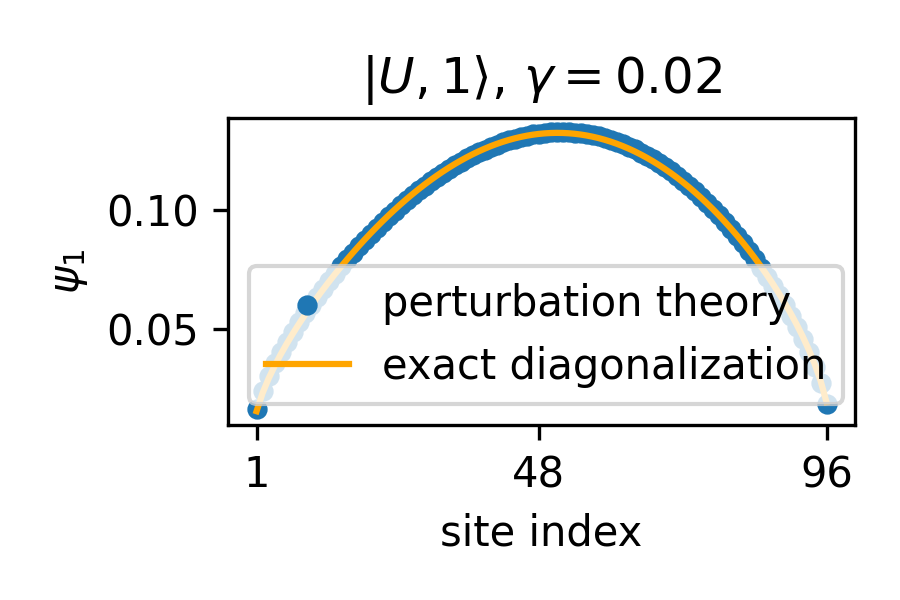}
  \end{minipage}\hfill
  \begin{minipage}{0.23\linewidth}
    (b)\\[1mm]
    \includegraphics[width=\linewidth]{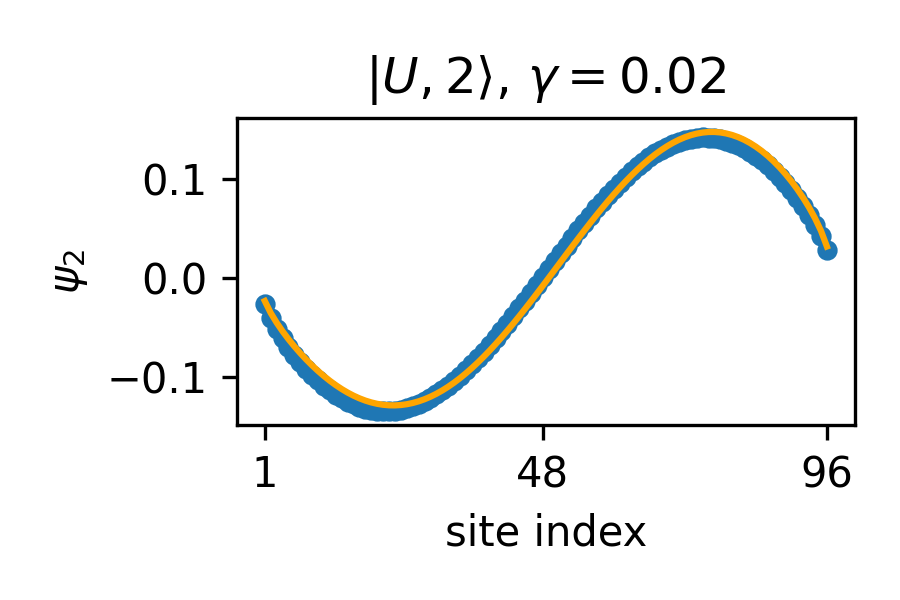}
  \end{minipage}\hfill
  \begin{minipage}{0.23\linewidth}
    (c)\\[1mm]
    \includegraphics[width=\linewidth]{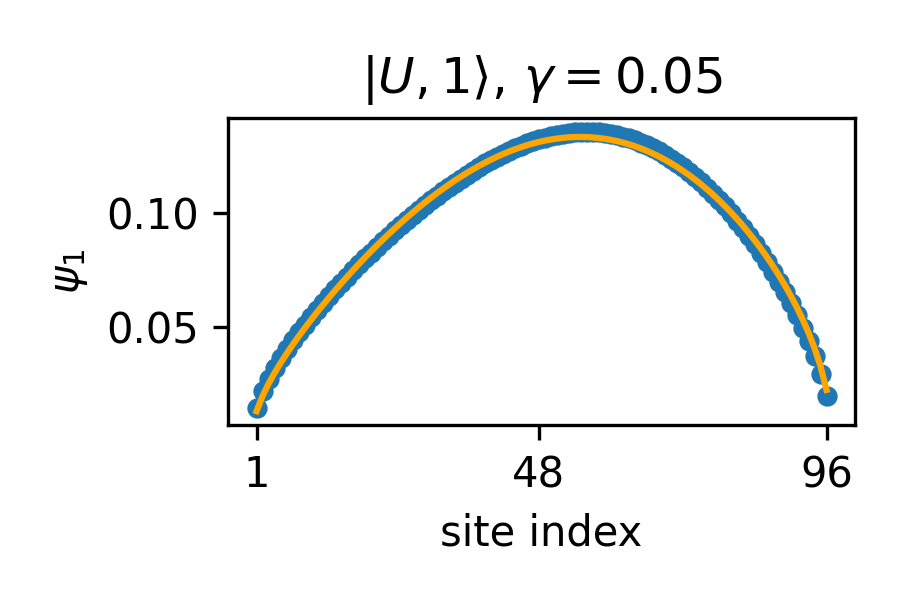}
  \end{minipage}\hfill
  \begin{minipage}{0.23\linewidth}
    (d)\\[1mm]
    \includegraphics[width=\linewidth]{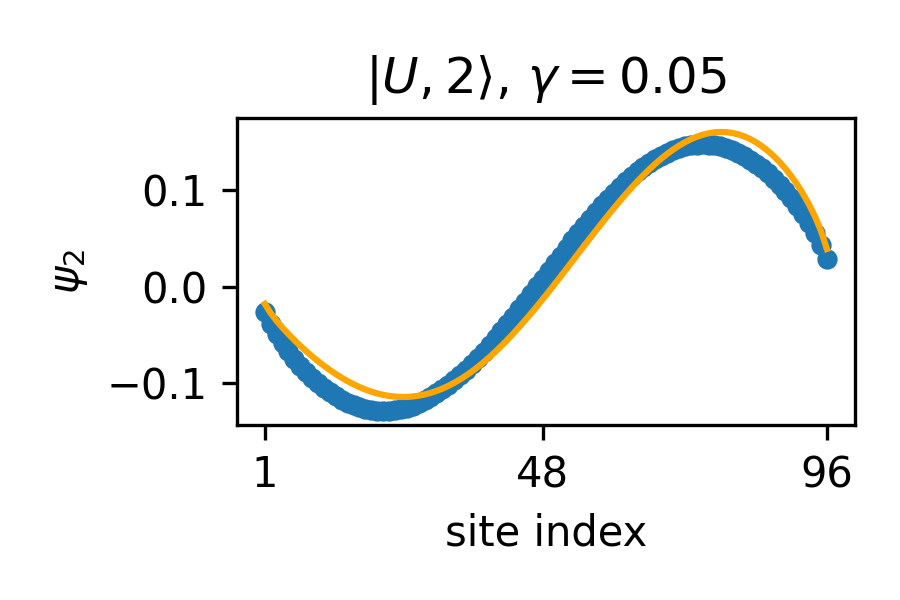}
  \end{minipage}
  \caption{Validation of first-order perturbation theory for the ($N=96, \alpha=2$) single-chain \textbf{LR model}. (a,b) Comparison for $\psi_{1}(x)$ at $\gamma=0.02$ and $0.05$, respectively, and (c,d) for $\psi_{2}(x)$ at the same parameters. Solid lines are from perturbation theory; dots are from exact diagonalization.}
  \label{fig:validation_plots_LR_alpha2}
\end{figure}

\begin{figure}[!htp]
  \begin{minipage}{0.23\linewidth}
    (a)\\[1mm]
    \includegraphics[width=\linewidth]{perturbation_theory_U1_vs_ED_g0p02.png}
  \end{minipage}
  \begin{minipage}{0.23\linewidth}
    (b)\\[1mm]
    \includegraphics[width=\linewidth]{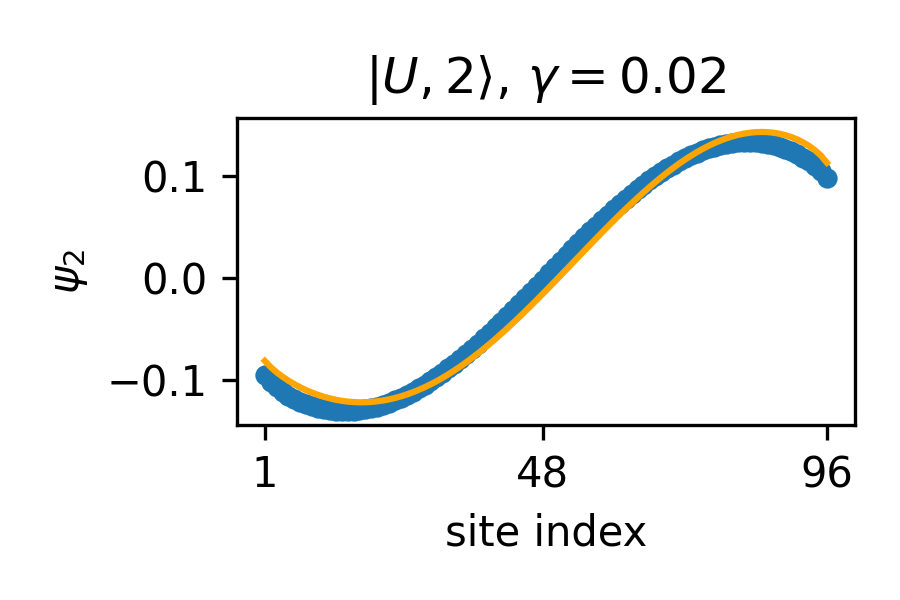}
  \end{minipage}\hfill
  \begin{minipage}{0.23\linewidth}
    (c)\\[1mm]
    \includegraphics[width=\linewidth]{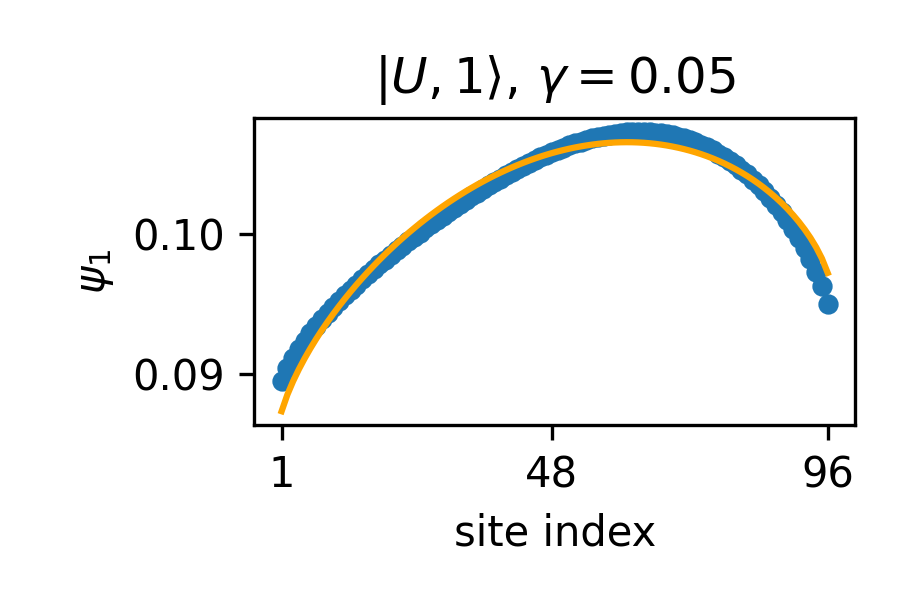}
  \end{minipage}
  \begin{minipage}{0.23\linewidth}
    (d)\\[1mm]
    \includegraphics[width=\linewidth]{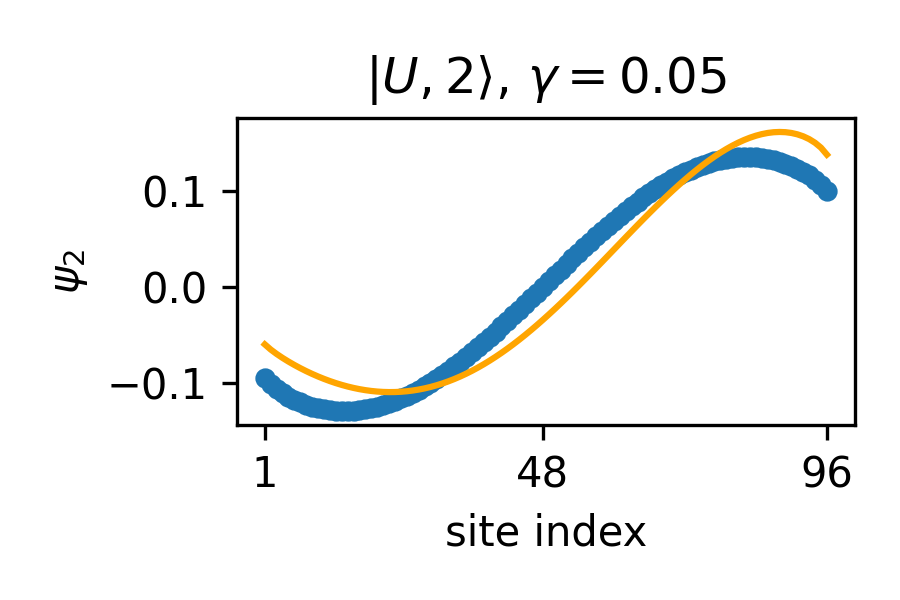}
  \end{minipage}
  \caption{Validation for the ($N=96, \alpha=0.2$) single-chain \textbf{LR model} with a smaller exponent. Panels (a–d) show the comparison between perturbation theory (solid lines) and exact diagonalization (dots) for $\psi_{1}(x)$ and $\psi_{2}(x)$ at $\gamma=0.02$ and $0.05$, demonstrating the robustness of the perturbative approach.}
  \label{fig:validation_plots_LR_alpha0p2}
\end{figure}

We validate this perturbative calculation by comparing it with exact diagonalization results for the LR model. As shown in Fig.~\ref{fig:validation_plots_LR_alpha2} (for $\alpha=2$) and Fig.~\ref{fig:validation_plots_LR_alpha0p2} (for $\alpha=0.2$), the analytical wavefunctions (solid lines) show excellent agreement with the numerical results (dots) for small $\gamma$. This confirms the validity of our approach.

\subsection{Dominant single-chain wavefunctions for Long-Range couplings at the band bottom (finite \texorpdfstring{$\gamma$}{gamma})}
This subsection provides the wavefunctions needed only for the Model-IV control calculation in Appendix~\ref{sec:app:Nc_modelIV}. Including it verifies that the same non-reciprocity-induced parity-mixing mechanism can operate at a different spectral edge, even though that branch is not used in the main hierarchy.
\label{subsec:bottom_wavefunctions_LR}

To characterize the influence of non-reciprocity on the lowest-energy eigenstates (those with the largest momenta), we perform a similar perturbative analysis for the band-bottom states at small but finite $\gamma$, again using
\begin{equation}
  H^{\rm LR}(\pm\gamma)=H_{0} \pm \gamma V.
\end{equation}
Here $H_0$ is the Hermitian long-range chain (at $\gamma=0$) and $\pm\gamma V$ is the non-Hermitian perturbation on the $(+)/(-)$ leg.

For sufficiently small $\gamma$, the true eigenstates can be obtained using first-order perturbation theory. The unperturbed eigenstates of $H_0$ are again well approximated by sine functions. For the two states near the bottom of the band (i.e., with momenta close to $k \simeq \pi$), we take
\begin{equation}
  \psi^{(0)}_{N-1}(x) = \sqrt{\frac{2}{N}}\sin\left(\frac{(N-1)\pi x}{N}\right),
  \qquad
  \psi^{(0)}_{N-2}(x) = \sqrt{\frac{2}{N}}\sin\left(\frac{(N-2)\pi x}{N}\right).
\end{equation}
Within the two-state approximation, the first-order correction to each eigenstate is mainly determined by its coupling to the nearest neighboring state in energy. Defining the energy spacing
\begin{equation}
  \Delta E' \equiv E^{(0)}_{N-1}-E^{(0)}_{N-2},
\end{equation}
we obtain
\begin{equation}
  |\psi^{(1)}_{N-1}\rangle \approx
  \frac{\langle \psi^{(0)}_{N-2} | V | \psi^{(0)}_{N-1} \rangle}{\Delta E'}
  |\psi^{(0)}_{N-2}\rangle,
  \qquad
  |\psi^{(1)}_{N-2}\rangle \approx
  \frac{\langle \psi^{(0)}_{N-1} | V | \psi^{(0)}_{N-2} \rangle}{-\Delta E'}
  |\psi^{(0)}_{N-1}\rangle.
\end{equation}
Thus, the first-order perturbed eigenstates on the $(+)/(-)$ leg are
\begin{equation}
  |\pm,N-1\rangle \approx |\psi^{(0)}_{N-1}\rangle \pm \gamma |\psi^{(1)}_{N-1}\rangle,\qquad
  |\pm,N-2\rangle \approx |\psi^{(0)}_{N-2}\rangle \pm \gamma |\psi^{(1)}_{N-2}\rangle,
\end{equation}
where the upper/lower sign applies to the $(+)/(-)$ leg.

For the long-range (LR) model, $H^{\rm LR}(\pm\gamma)=H_0 \pm \gamma V^{(\mathrm{LR})}$ with
\begin{equation}
  V^{(\mathrm{LR})} = \sum_{n\neq m} \frac{\mathrm{sgn}(n-m)}{|n-m|^{\alpha}} c_{n}^\dagger c_{m}.
\end{equation}
The first-order correction to $\psi^{(0)}_{N-1}(x)$ is then given explicitly by
\begin{equation}
  \psi^{(1)}_{N-1}(x) =
  \frac{\psi^{(0)}_{N-2}(x)}{\Delta E'}
  \left(
    \sum_{n\neq m} \frac{\mathrm{sgn}(n-m)}{|n-m|^{\alpha}}
    \psi^{(0)}_{N-2}(m)\psi^{(0)}_{N-1}(n)
  \right),
  \label{eq:pertWave_bottom_N-1}
\end{equation}
and similarly, the correction to $\psi^{(0)}_{N-2}(x)$ is
\begin{equation}
  \psi^{(1)}_{N-2}(x) =
  - \frac{\psi^{(0)}_{N-1}(x)}{\Delta E'}
  \left(
    \sum_{n\neq m} \frac{\mathrm{sgn}(n-m)}{|n-m|^{\alpha}}
    \psi^{(0)}_{N-1}(m)\psi^{(0)}_{N-2}(n)
  \right).
  \label{eq:pertWave_bottom_N-2}
\end{equation}
Since $V^{(\mathrm{LR})}$ has an antisymmetric kernel in real space [$\mathrm{sgn}(n-m)$], it primarily couples pairs of high-momentum standing waves with opposite parity near $k \simeq \pi$. As a result, the perturbation breaks reflection symmetry and causes the bottom eigenstates to localize toward one boundary. The direction of this boundary accumulation is set by the $\pm$ sign of the leg.

\subsection{Dominant single-chain wavefunctions for nearest-neighbor couplings}
\label{sec:app_NN_wavefunctions}

These exact Hatano--Nelson wavefunctions and their perturbative expansion are used in the Model-I and Model-II projections. The comparison is included to demonstrate that the two-state perturbative language employed for the long-range chain reproduces the known NN skin deformation, thereby placing all three main models within one consistent projection framework.

For the nearest-neighbor model, we can analyze the eigenstates in two instructive ways: first, by finding the exact solution for the corresponding Hatano-Nelson model, and second, by applying the perturbative formalism from the previous section to show its consistency.

\paragraph{Exact Solution: The Hatano-Nelson Model.}
The single-chain NN Hamiltonian on the $(+)$ leg is $H = \sum_{n} \left[ (1 + \gamma) c_{n+1}^\dagger c_n + (1 - \gamma) c_n^\dagger c_{n+1} \right]$. This Hamiltonian can be diagonalized exactly via a non-unitary similarity transformation, yielding the right eigenstates on the $(+)$ leg:
\begin{equation}
  \psi_n^{R}(x) \propto e^{\kappa x} \sin(k_n x),
  \qquad
  \kappa=\frac{1}{2}\ln\left(\frac{1+\gamma}{1-\gamma}\right).
  \label{eq:HN_exact_sol_final}
\end{equation}
This is the classic NHSE, where all eigenstates are exponentially localized at a boundary.

\paragraph{Pedagogical view from perturbation theory and quantitative comparison.}
To demonstrate that the perturbative framework established in the preceding section is fully applicable across different interaction ranges, we apply it here to the nearest-neighbor (NN) limit. This serves as a rigorous consistency check, ensuring that our two-state approximation captures the correct localization physics even in the limit where exact solutions exist. The non-reciprocal NN perturbation operator is $V^{(NN)} = \sum_{n} (c_{n+1}^\dagger c_{n} - c_{n}^\dagger c_{n+1})$, so $H^{\rm NN}(\pm\gamma)=H_0 \pm \gamma V^{(NN)}$. The extent of the non-Hermitian mixing is governed by the off-diagonal matrix element $V_{21} = \langle \psi_2^{(0)}|V^{(NN)}|\psi_1^{(0)} \rangle$ between the dominant symmetric and anti-symmetric modes:
\begin{equation}
  V_{21} = \sum_{n=1}^{N-1} \left[\psi_2^{(0)}(n) \psi_1^{(0)}(n+1) - \psi_2^{(0)}(n+1) \psi_1^{(0)}(n)\right].
\end{equation}
Substituting the sine-wave functions, $\psi_n^{(0)}(x) = \sqrt{2/N}\sin(n\pi x/N)$, into this discrete summation yields an analytical result.
In particular, for the lowest mode $n=1$, we will repeatedly use
\begin{equation}
  \sum_{x=1}^{N} \psi_1^{(0)}(x)
  = \sqrt{\frac{2}{N}}
  \sum_{x=1}^{N}\sin\frac{\pi x}{N}
  = \sqrt{\frac{2}{N}}\cot\frac{\pi}{2N}
  \simeq \frac{2\sqrt{2N}}{\pi},
  \label{eq:sum_psi1_0}
\end{equation}
which immediately implies
$\bigl(\sum_x \psi_1^{(0)}(x)\bigr)^2 \simeq 8N/\pi^2$ for large $N$.

The mixing coefficient is then $J_{\text{pert}} = V_{21}/\Delta E$, which is found to be:
\begin{equation}
  J_{\text{pert}} = \frac{2}{\Delta E\,N}\left[1+\frac{\sin(\frac{5\pi}{2N})}{\sin(\frac{3\pi}{2N})}\right].
\end{equation}
For large $N$, we can use the small-angle approximation $\sin(x) \approx x$, which simplifies the coefficient to:
\begin{equation}
  J_{\text{pert}} \approx \frac{2}{\Delta E N} \left( 1 + \frac{5\pi/2N}{3\pi/2N} \right) = \frac{2}{\Delta E N} \left( 1 + \frac{5}{3} \right) = \frac{16}{3\Delta E N}.
  \label{eq:J_pert}
\end{equation}
The perturbed wavefunction for the $(+)$ leg is thus $\psi_1^+(x) \approx \psi_1^{(0)}(x) + \gamma J_{\text{pert}} \psi_2^{(0)}(x)$, and for the $(-)$ leg $\psi_1^-(x) \approx \psi_1^{(0)}(x) - \gamma J_{\text{pert}} \psi_2^{(0)}(x)$.

Next, we extract an equivalent mixing coefficient from the exact solution. Expanding Eq.~\eqref{eq:HN_exact_sol_final} for small $\gamma$ (so that $\kappa\simeq\gamma$) gives:
\begin{equation}
  \psi_1(x) \propto \sin(k_1 x) + \gamma x \sin(k_1 x).
\end{equation}
The correction term is $\gamma x |\psi_1^{(0)}\rangle$. The component of this correction along the $|\psi_2^{(0)}\rangle$ direction defines the effective mixing coefficient from the exact solution, $J_{\text{exact}}$:
\begin{equation}
  J_{\text{exact}} = \langle \psi_2^{(0)} | x | \psi_1^{(0)} \rangle.
\end{equation}
Using the previously calculated overlap integral, $\langle \psi_2^{(0)} | x | \psi_1^{(0)} \rangle = -16N/(9\pi^2)$, we get:
\begin{equation}
  J_{\text{exact}} = -\frac{16N}{9\pi^2}.
\end{equation}
Now for the final comparison. We substitute the energy gap for the NN model, $\Delta E \approx 3\pi^2/N^2$, into our result for $J_{\text{pert}}$ from Eq.~\eqref{eq:J_pert}:
\begin{equation}
  J_{\text{pert}} \approx \frac{16}{3N} \left( \frac{1}{\Delta E} \right) = \frac{16}{3N} \left(\frac{N^2}{3\pi^2 }\right) = \frac{16N}{9\pi^2}.
\end{equation}
Comparing the two results, we find that their magnitudes are in perfect agreement: $|J_{\text{pert}}| = |J_{\text{exact}}|$. The sign on each leg follows the $\pm\gamma$ convention and correctly captures the direction of localization. This quantitative match provides strong validation of our two-state perturbative approach.

The results of this Appendix feed directly into the main-text hierarchy: Model I uses the NN band-edge gap and exact skin-deformed overlaps, Model II combines the same NN gap with a collectively enhanced LR rung projection, and Model III combines the nonanalytic LR gap with the size-dependent mixing ratio $\chi_N(\alpha)$. Model IV uses the bottom-edge ingredients only as a control. This separation of single-chain inputs makes clear which physical ingredient is responsible for each critical-length law derived next.

\section{Derivation of the Critical Length Scaling $N_c$}
\label{sec:app_Nc_derivations}

This Appendix derives the three threshold formulas plotted as analytical curves in main-text Fig.~\ref{fig:scalings}(d--f) and then compared to construct Fig.~\ref{fig:phase_diagram}. The motivation is not merely to obtain three fits: starting from the same EP criterion, the derivations identify which $N$-dependent quantity changes when nonlocality is placed on the rungs, the legs, or both. The resulting logarithmic, algebraic, and scale-covariant laws are therefore mechanism-resolved predictions.

This section explains how the critical length $N_c(D)$ follows from the instability criterion Eq.~\eqref{eq:central_condition} once the relevant (single-chain) spectral data and wavefunctions are known.

For a candidate first-colliding pair $(p,q)$ of single-chain levels, we define the critical length $N_c^{(p,q)}$ as the smallest $N$ for which the projected effective Hamiltonian in the $(p,q)$ reduced sector satisfies Eq.~\eqref{eq:central_condition}. The physical critical length is then obtained by minimizing over all possible pairs (or, more generally, over all small reduced sectors) as
\begin{equation}
  N_c(D)=\min_{(p,q)} N_c^{(p,q)}(D).
\end{equation}
In practice, for the parameter regimes considered below, the minimizing pair typically lies near a band edge (top or bottom), which allows analytic control.

For a given first-colliding pair $(p,q)$, the effective Hamiltonian is obtained by projecting the full two-leg problem onto
$\{|p,+\rangle,|q,+\rangle,|p,-\rangle,|q,-\rangle\}$.
The instability condition Eq.~\eqref{eq:central_condition} depends on two types
of inputs: (i) the single-chain gap
$\Delta_{pq}\equiv E_p-E_q$, and (ii) the projected matrix elements $u_{ij}$ and $v_{ij}$ (with $i,j\in\{p,q\}$), which quantify how strongly the two counter-pumped sectors hybridize within the chosen reduced sector.

Substituting these quantities into Eq.~\eqref{eq:central_condition} gives the explicit instability criterion for the $(p,q)$ first-colliding pair,
\begin{equation}
  \boxed{\left[\Delta_{pq}\pm(u_{pp}-u_{qq})\right]^2+4u_{pq}v_{qp}<0,}
  \label{eq:central_condition_pq_repeat}
\end{equation}
which states that complexification occurs when the projected inter-chain hybridization overcomes the intrinsic single-chain level spacing. Accordingly, $N_c^{(p,q)}$ is obtained by evaluating the $N$-dependence of $\Delta_{pq}$ and the projected matrix elements for the model under consideration and solving the above inequality for the smallest integer $N$.

In the following real-gauge estimates we sometimes write the upper-block elements $u_{ij}$ as $\delta_{ij}$ to preserve compact intermediate algebra, while the corresponding lower-block element is kept separately whenever its sign matters. The final discriminants are interpreted in the $u/v$ notation of Eq.~\eqref{eq:central_condition}.

In this section, we follow the fixed structural route fully local NN--NN ladder (Model I) $\to$ nonlocal rungs NN--LR ladder (Model II) $\to$ fully nonlocal LR--LR ladder (Model III), and then include nonlocal-legs LR--NN control geometry (Model IV). This order starts from the local baseline and then introduces different placements of nonlocality in a controlled way.

\subsection{Critical-length scaling for fully local NN--NN ladder (Model I)}
\label{sec:app:Nc_modelI}

This subsection derives the finite-size Model-I threshold quoted in main-text Eq.~\eqref{eq:NNNN_active_threshold_main} and its logarithmic limit Eq.~\eqref{eq:local_Nc_main}. It provides the local baseline against which the genuinely nonlocal mechanisms of Models II and III are identified, and supplies the Model-I curve used in both main figures.

For the NN--NN ladder, the first-colliding channel is the band-edge Hatano--Nelson pair $(1,2)$, together with the reflected partner at the opposite edge. We take
\begin{equation}
  k_\ell=\frac{\ell\pi}{N+1},\qquad \ell=1,2,
  \qquad
  \kappa=\frac12\ln\frac{1+\gamma}{1-\gamma}.
\end{equation}
Using the same weak-nonreciprocity energy convention as in the main text, the NN band-edge spacing is
\begin{equation}
  \Delta_{\rm NN}
  =
  E_1-E_2
  \simeq
  2(\cos k_1-\cos k_2).
\end{equation}
The omitted overall $O(\gamma^2)$ bandwidth renormalization does not affect the finite-size mechanism or the logarithmic scaling discussed here.

\begin{figure}
  \centering
  \subfigure[]{\includegraphics[width=0.32\linewidth]{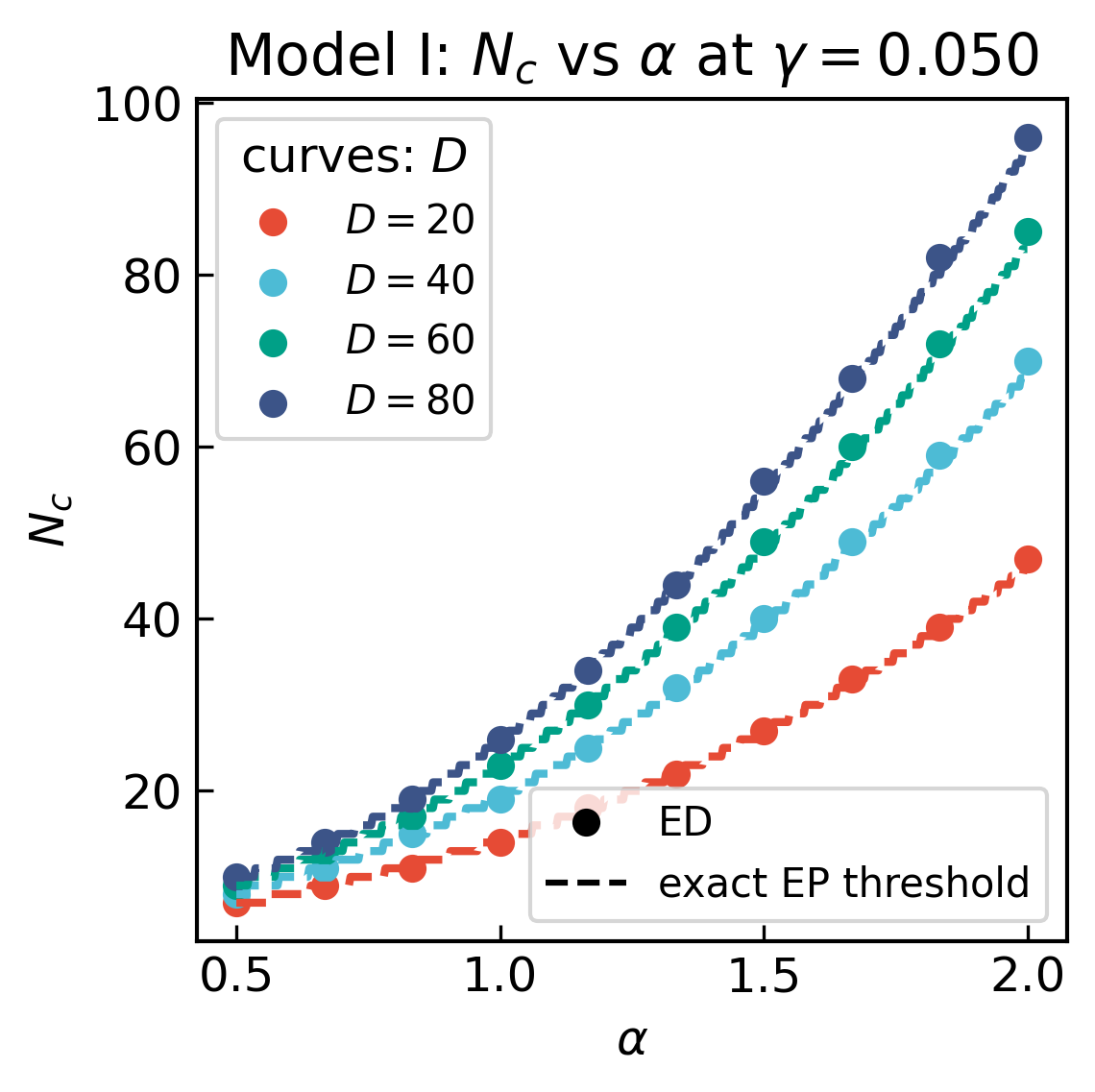}}
  \subfigure[]{\includegraphics[width=0.32\linewidth]{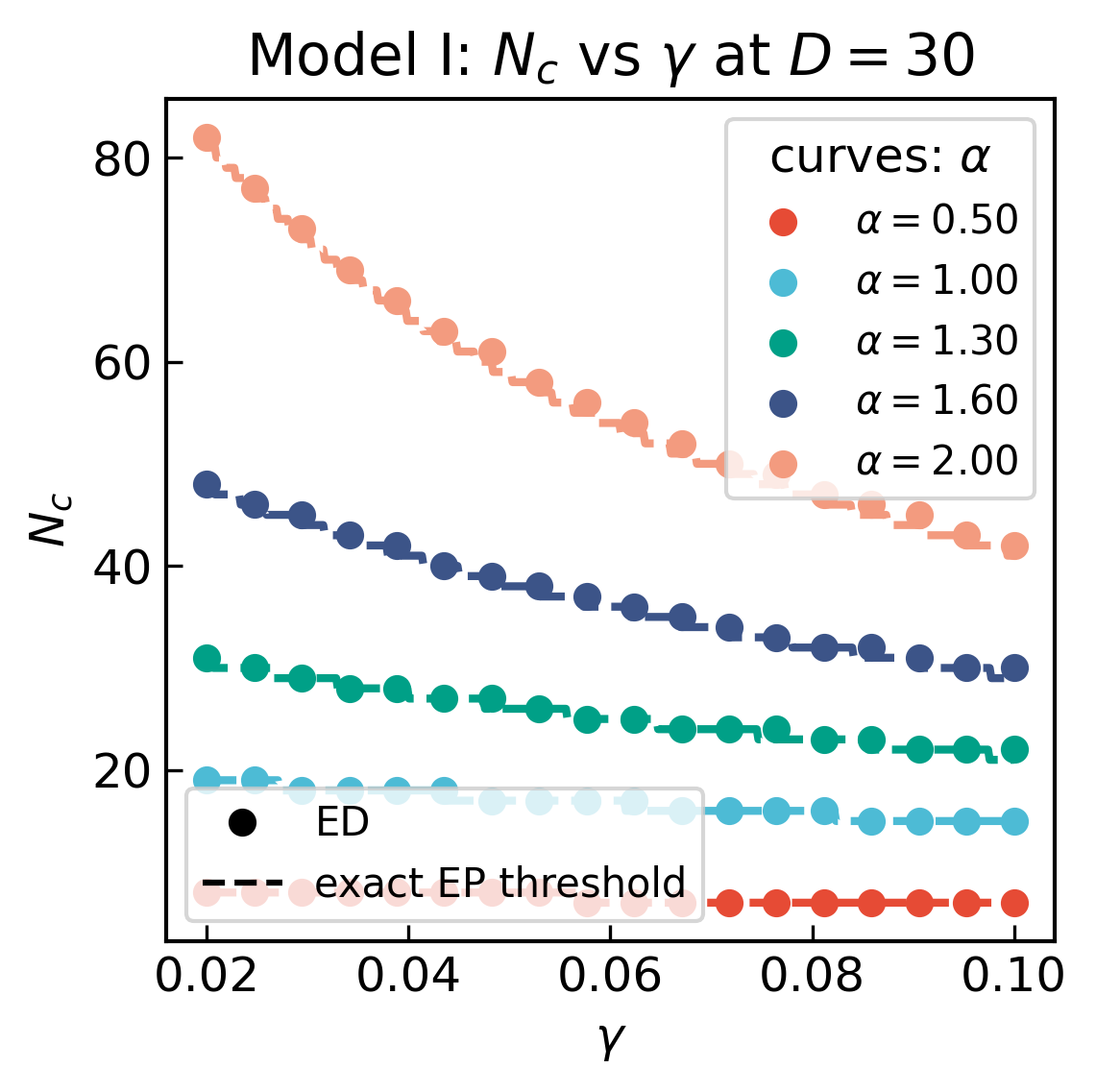}}
  \subfigure[]{\includegraphics[width=0.32\linewidth]{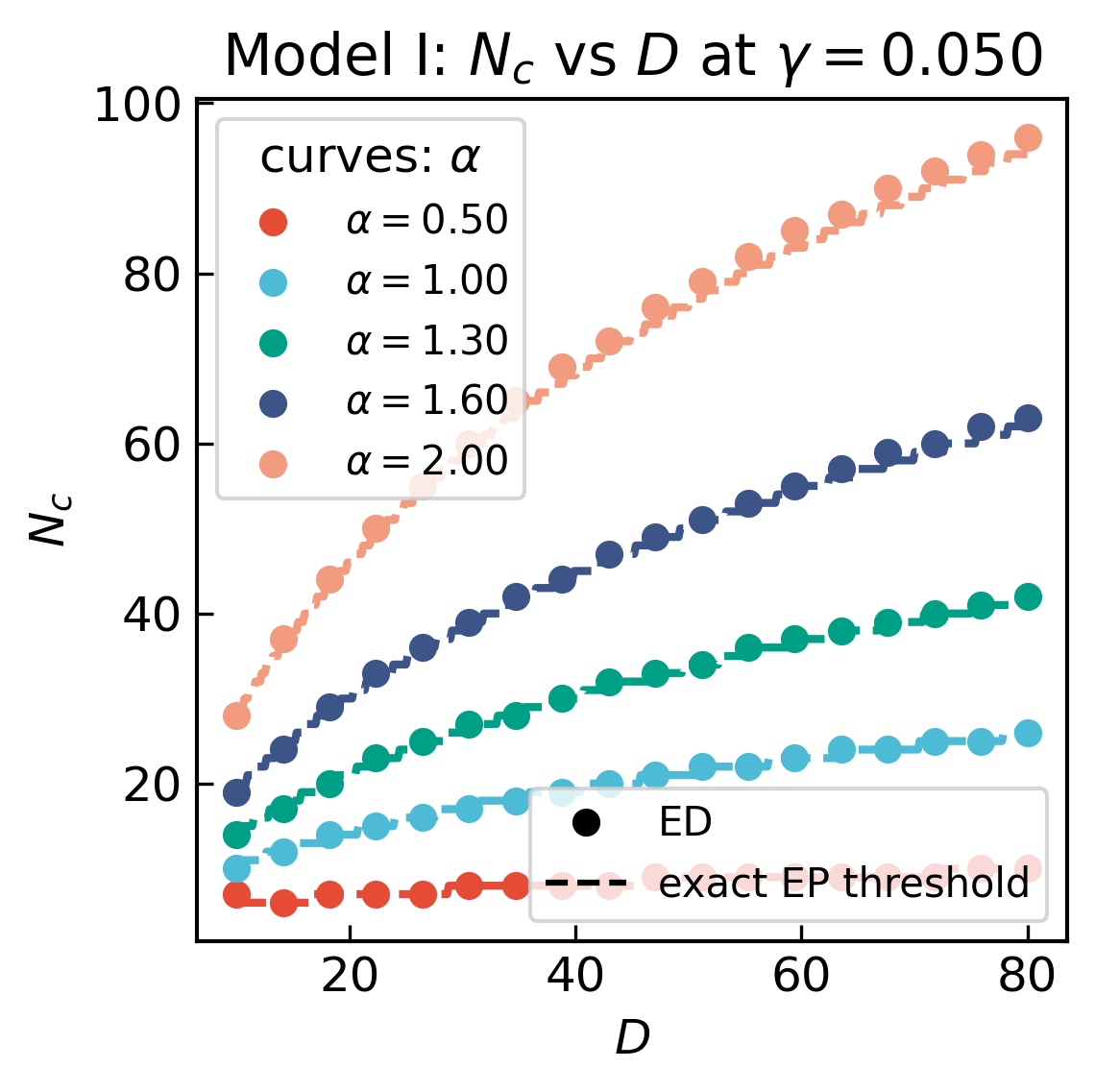}}
  \caption{Critical length $N_c$ scaling for the fully local NN--NN ladder (Model I). Numerical $N_c$ obtained from exact diagonalization is benchmarked against the finite-size exceptional point (EP) threshold condition in Eq.~\eqref{eq:modelI_NNNN_threshold_supp}, obtained by projecting the local rung coupling onto the first-colliding band-edge Hatano--Nelson pair $(1,2)$. In this form, the onset is determined by 
$D^{-\alpha}(N_c+1)e^{-\kappa(N_c+1)}\frac{\cos k_1-\cos k_2}{\mathcal{D}_{12}}$, 
where $\mathcal{D}_{12}$ contains the skin-weighted overlap factors of the two band-edge standing waves. 
(a)~$N_c$ versus $D$ for representative $(\alpha,\gamma)$, showing the logarithmic growth induced by the exponentially small local skin overlap. 
(b)~$N_c$ versus $\gamma$ for representative $(D,\alpha)$, exhibiting the decrease of the critical length as stronger nonreciprocity enhances the exponential skin factor. 
(c)~$N_c$ versus $\alpha$ for representative $(D,\gamma)$, consistent with the larger critical length required when the local rung coupling $D^{-\alpha}$ is more strongly suppressed.}
  \label{fig:modelI_3plots}
\end{figure}

Define the skin-weighted standing-wave factors
\begin{equation}
  A_n=e^{-\kappa n}\sin(k_1 n),
  \qquad
  B_n=e^{-\kappa n}\sin(k_2 n).
\end{equation}
Projecting the local rung block $H_{\perp}^{\rm NN}$ onto this first-colliding pair gives
\begin{align}
  u_{11}
  &=
  \frac{2D^{-\alpha}}{N+1}e^{\kappa(N+1)}
  \sum_{n=1}^{N}A_n^2,\\
  u_{22}
  &=
  \frac{2D^{-\alpha}}{N+1}e^{\kappa(N+1)}
  \sum_{n=1}^{N}B_n^2,\\
  u_{12}
  &=
  \frac{2D^{-\alpha}}{N+1}e^{\kappa(N+1)}
  \sum_{n=1}^{N}A_nB_n,\\
  v_{21}&=-u_{12}.
\end{align}
The EP threshold is therefore
\begin{equation}
  \Delta_{\rm NN}=u_{11}+u_{22}+2|u_{12}|.
\end{equation}
Equivalently, at $N=N_c$,
\begin{equation}
  D^{-\alpha}
  =
  (N_c+1)e^{-\kappa(N_c+1)}
  \frac{\cos k_1-\cos k_2}{\mathcal D_{12}},
  \label{eq:modelI_NNNN_threshold_supp}
\end{equation}
where
\begin{equation}
  \mathcal D_{12}
  =
  \sum_{n=1}^{N_c}(A_n^2+B_n^2)
  +
  2\left|\sum_{n=1}^{N_c}A_nB_n\right|.
\end{equation}
In the large-size skin-overlap regime the sums saturate, and Eq.~\eqref{eq:modelI_NNNN_threshold_supp} reduces to
\begin{equation}
  e^{\kappa N_c}D^{-\alpha}\sim \mathrm{const.},
  \qquad
  N_c^{\rm I}\simeq \frac{\alpha}{\kappa}\ln D
  \simeq \frac{\alpha}{\gamma}\ln D.
\end{equation}

Equation~\eqref{eq:modelI_NNNN_threshold_supp} is the expression used for the quantitative Model-I analytical curves in main-text Fig.~\ref{fig:scalings}; the logarithmic form explains their asymptotic trend and the corresponding Model-I boundaries in Fig.~\ref{fig:phase_diagram}. Its significance is that a power-law change in $D$ is balanced by an exponential skin-overlap factor, which is why the resulting length grows only logarithmically.

\subsection{Critical-length scaling for nonlocal rungs NN--LR ladder (Model II)}
\label{sec:app:Nc_NNLR}

This subsection derives main-text Eq.~\eqref{eq:NNLR_Nc_main}, including the common effective prefactors $\eta_0$ and $\eta_1$ used for all Model-II curves. The calculation is included to show that the algebraic $D^{\alpha/3}$ law is caused by the collective spatial sum of the nonlocal rungs, not by a change in the NN band-edge dispersion.

We next consider the nonlocal-rungs NN--LR ladder (Model II), in which each constituent chain retains nearest-neighbor Hatano--Nelson hopping, while the inter-chain coupling is long-ranged. The transition is again determined by the competition between a finite-size single-chain level spacing and the projected coupling between the two counter-pumped chains. In the parameter regime considered here, the dominant projected sector is formed by the two NN band-edge modes.

For an open Hermitian NN chain of length $N$, it is convenient to define $L=N+1$. The two highest-energy standing waves are
\begin{equation}
  \psi^{(0)}_{1}(x)=\sqrt{\frac{2}{L}}\sin\left(\frac{\pi x}{L}\right),\qquad
  \psi^{(0)}_{2}(x)=\sqrt{\frac{2}{L}}\sin\left(\frac{2\pi x}{L}\right),
  \qquad x=1,\ldots,N.
  \label{appeq:psi0_12_NNLR}
\end{equation}
Their lattice sums satisfy
\begin{equation}
  \sum_{x=1}^{N}\psi^{(0)}_{1}(x)
  =
  \sqrt{\frac{2}{L}}\cot\left(\frac{\pi}{2L}\right),
  \qquad
  \sum_{x=1}^{N}\psi^{(0)}_{2}(x)=0.
  \label{appeq:sums_psi0_12_NNLR}
\end{equation}
The corresponding finite-size band-edge spacing is
\begin{equation}
  \Delta E_{\mathrm{NN}}
  =
  2\left[
  \cos\left(\frac{\pi}{L}\right)
  -
  \cos\left(\frac{2\pi}{L}\right)
  \right]
  \simeq
  \frac{3\pi^2}{L^2}.
  \label{appeq:DeltaE_NN_scaling_NNLR}
\end{equation}

For finite non-reciprocity, the right and left eigenstates of the isolated NN Hatano--Nelson chains are obtained from the Hermitian standing waves through the non-unitary skin transformation
\begin{equation}
  \psi_{p,\pm}^{R}(x)\propto e^{\pm\kappa x}\psi^{(0)}_{p}(x),
  \qquad
  \psi_{p,\pm}^{L}(x)\propto e^{\mp\kappa x}\psi^{(0)}_{p}(x),
  \label{appeq:skin_states_NNLR}
\end{equation}
where
\begin{equation}
  \kappa=\frac{1}{2}\ln\frac{1+\gamma}{1-\gamma}
  =
  \operatorname{arctanh}\gamma.
  \label{appeq:kappa_gamma_relation_NNLR}
\end{equation}
We retain Eq.~\eqref{appeq:DeltaE_NN_scaling_NNLR} consistently with the weak-nonreciprocity convention used in the main-text Model-II threshold; the omitted $O(\gamma^2)$ bandwidth correction is absorbed into the finite-window prefactor $\eta_0$.

The long-range rung block of Model II is
\begin{equation}
  H_{\perp}^{\mathrm{LR}}
  =
  \sum_{n,m=1}^{N}
  \frac{c^{\dagger}_{+,n}c_{-,m}}
  {\left[D^2+(x_n-x_m)^2\right]^{\alpha/2}}.
  \label{appeq:full_rung_kernel_NNLR}
\end{equation}
Projecting this block biorthogonally onto the two-mode sector gives
\begin{equation}
  u_{ij}
  =
  \sum_{n,m=1}^{N}
  \frac{\psi_{i,+}^{L}(x_n)\psi_{j,-}^{R}(x_m)}
  {\left[D^2+(x_n-x_m)^2\right]^{\alpha/2}},
  \qquad i,j\in\{1,2\}.
  \label{appeq:full_projected_elements_NNLR}
\end{equation}
Equation~\eqref{appeq:full_projected_elements_NNLR} is the appropriate projected expression when the full longitudinal dependence of the rung coupling is retained. To extract a compact scaling law, we use the large-separation reduction
\begin{equation}
  \left[D^2+(x_n-x_m)^2\right]^{-\alpha/2}
  \simeq
  D^{-\alpha},
  \qquad D\gg |x_n-x_m|,
  \label{appeq:constant_kernel_condition_NNLR}
\end{equation}
under which
\begin{equation}
  H_{\perp}^{\mathrm{LR}}
  \simeq
  D^{-\alpha}\sum_{n,m=1}^{N}c^{\dagger}_{+,n}c_{-,m}.
  \label{appeq:Hint_const_kernel_NNLR}
\end{equation}
The projected matrix elements then factorize as
\begin{equation}
  u_{ij}
  =
  D^{-\alpha}\Sigma_{i,+}^{L}\Sigma_{j,-}^{R},
  \qquad
  \Sigma_{p,\pm}^{L/R}
  \equiv
  \sum_{x=1}^{N}\psi_{p,\pm}^{L/R}(x).
  \label{appeq:delta_factorized_sums_NNLR}
\end{equation}

In the weak-skin regime, the two sums entering $u_{ij}$ both carry the factor $e^{-\kappa x}$, whereas those entering $v_{ij}$ carry $e^{+\kappa x}$. Thus,
\begin{equation}
  \Sigma_{p,+}^{L}=\Sigma_{p,-}^{R}
  =S_p-\kappa X_p+O\!\left((\kappa L)^2\right),
  \qquad
  \Sigma_{p,-}^{L}=\Sigma_{p,+}^{R}
  =S_p+\kappa X_p+O\!\left((\kappa L)^2\right),
  \label{appeq:Sigma_expand_NNLR}
\end{equation}
with $S_2=0$, and
\begin{equation}
  S_1
  \equiv
  \sum_{x=1}^{N}\psi^{(0)}_1(x)
  \simeq
  \frac{2\sqrt{2L}}{\pi},
  \qquad
  X_2
  \equiv
  \sum_{x=1}^{N}x\psi^{(0)}_2(x)
  \simeq
  -\frac{\sqrt{2}}{2\pi}L^{3/2}.
  \label{appeq:S1_X2_asymptotic_NNLR}
\end{equation}
Consequently, the leading projected elements have the asymptotic structure
\begin{align}
  u_{11}
  &\simeq
  \frac{8L}{\pi^2D^\alpha},
  \label{appeq:u11_asymptotic_NNLR}\\
  u_{12}
  &\simeq
  \frac{\kappa L}{4}u_{11},
  \qquad
  v_{21}\simeq-u_{12},
  \label{appeq:u12_v21_asymptotic_NNLR}\\
  u_{22}
  &\simeq
  -\left(\frac{\kappa L}{4}\right)^2u_{11}.
  \label{appeq:u22_asymptotic_NNLR}
\end{align}

Substitution into the projected instability condition gives
\begin{equation}
  \left|
  \Delta E_{\mathrm{NN}}
  -
  u_{11}
  \left[
  1+\left(\frac{\kappa L}{4}\right)^2
  \right]
  \right|
  <
  2u_{11}\frac{\kappa L}{4},
  \label{appeq:window_pre_square_NNLR}
\end{equation}
or equivalently
\begin{equation}
  u_{11}
  \left(1-\frac{\kappa L}{4}\right)^2
  <
  \Delta E_{\mathrm{NN}}
  <
  u_{11}
  \left(1+\frac{\kappa L}{4}\right)^2.
  \label{appeq:window_square_NNLR}
\end{equation}
Since the NN band-edge spacing decreases as $L^{-2}$, the first onset occurs when it reaches the upper edge of this interval.

\begin{figure}
  \centering
  \subfigure[]{\includegraphics[width=0.32\linewidth]{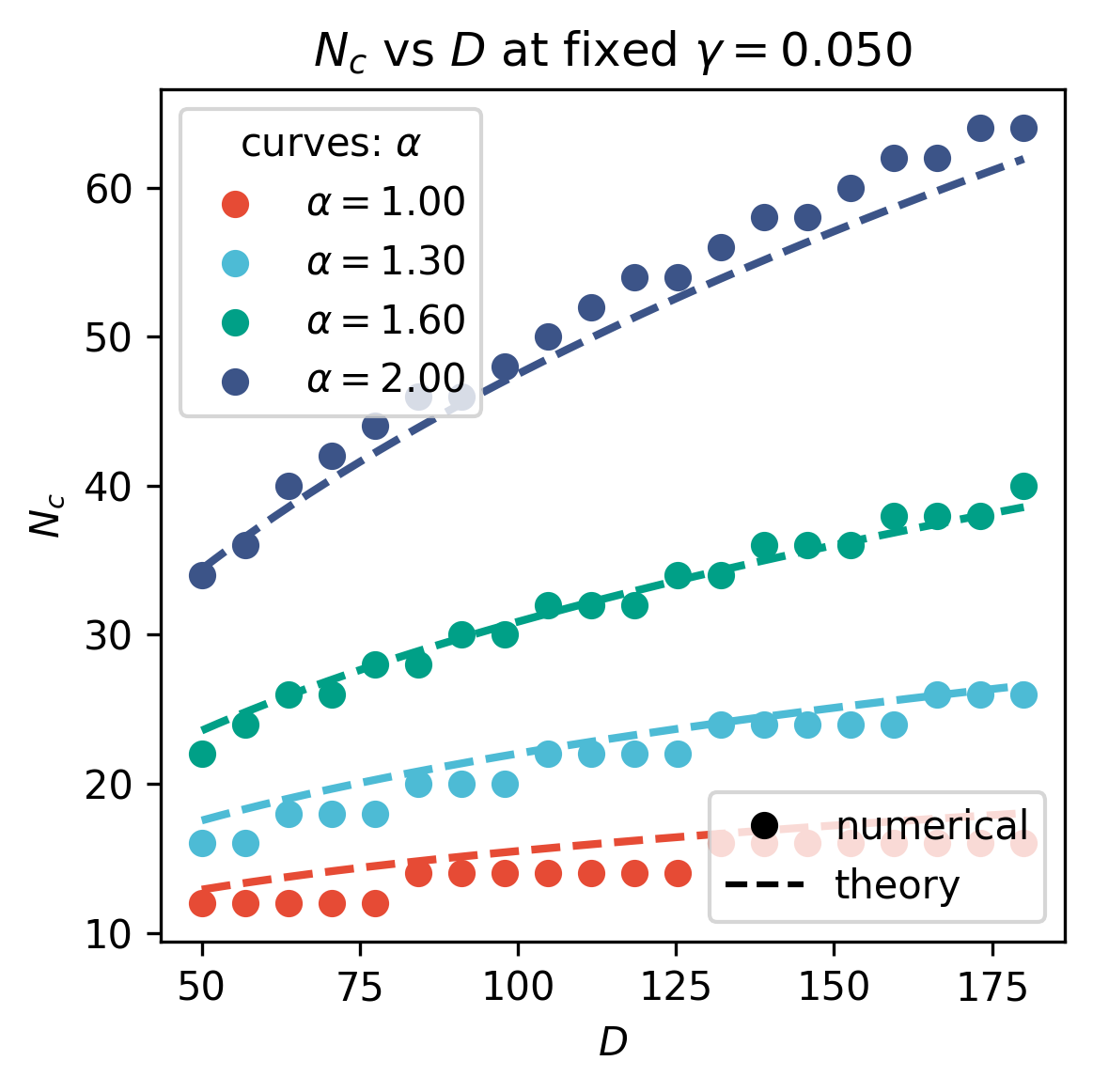}}
  \subfigure[]{\includegraphics[width=0.32\linewidth]{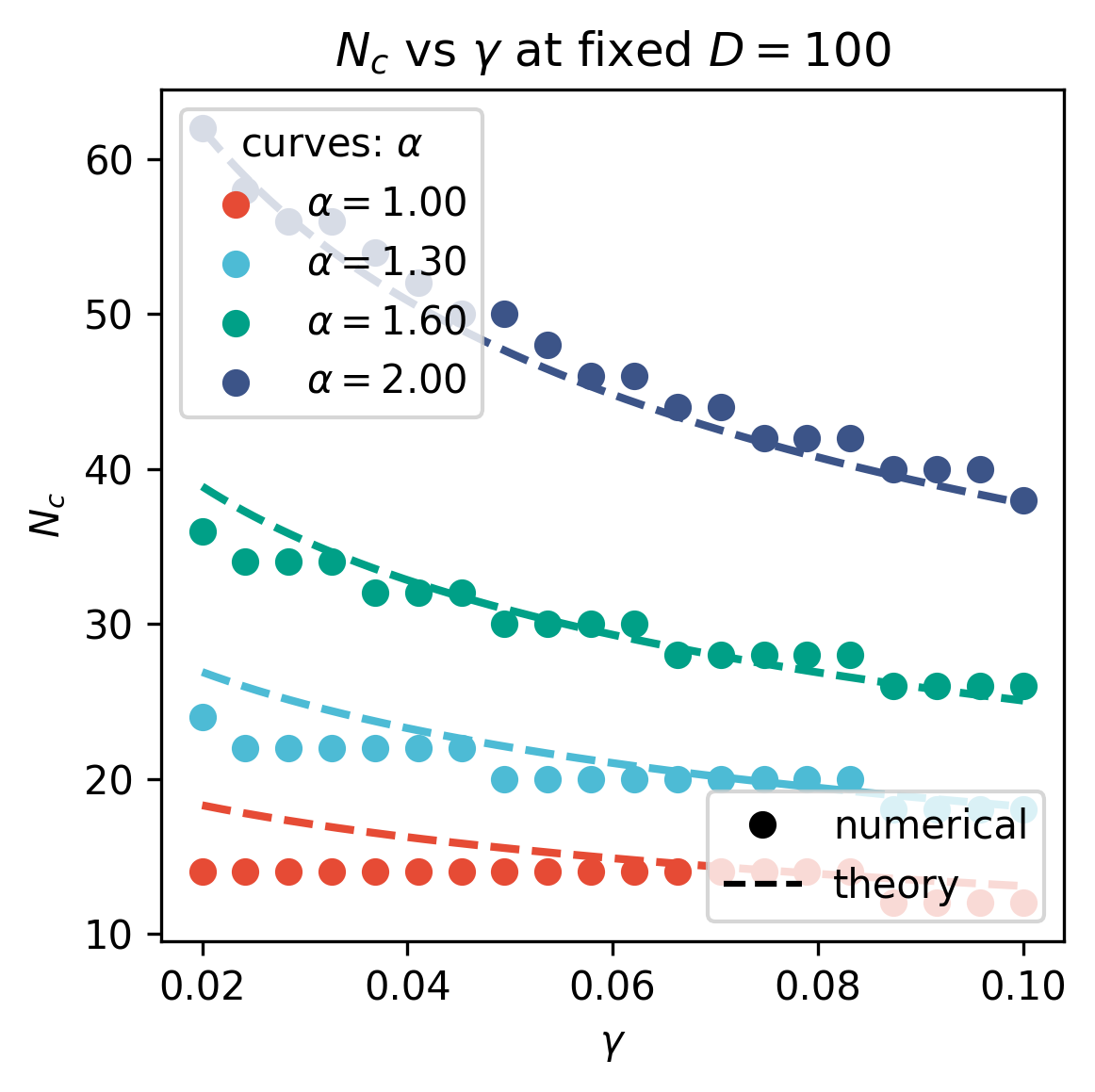}}
  \subfigure[]{\includegraphics[width=0.32\linewidth]{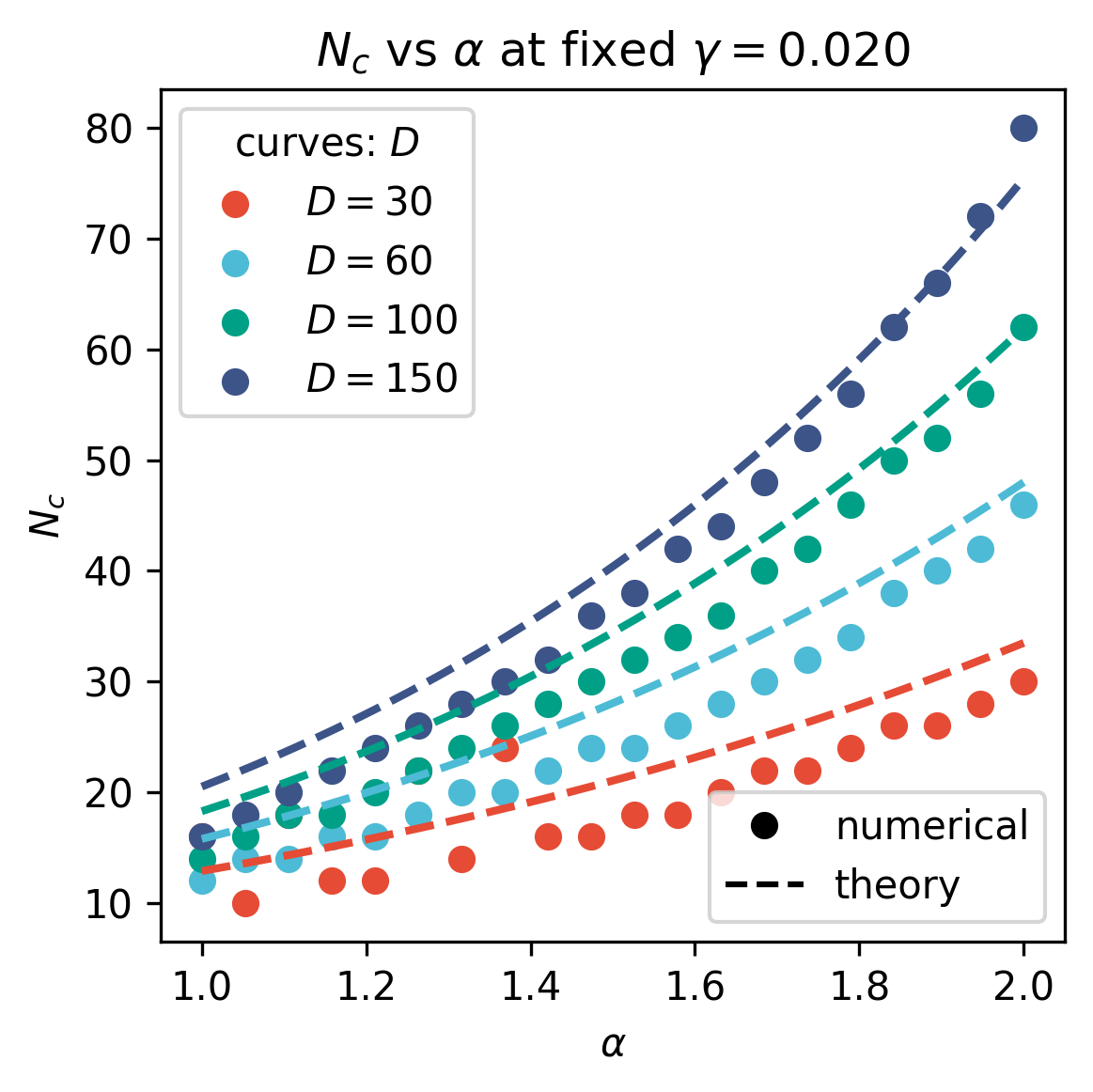}}
  \caption{Critical length $N_c$ for the nonlocal-rungs NN--LR ladder (Model II). Numerical thresholds are compared with the effective onset equation in Eq.~\eqref{appeq:Nc_scaling_equation_kappa_NNLR}, using the common parameters $\eta_0=3.0$ and $\eta_1=3.7$. (a) $N_c$ as a function of the inter-chain separation $D$ for representative values of $(\alpha,\gamma)$. The increase of $N_c$ follows from the algebraic suppression of the long-range rung coupling by $D^{-\alpha}$. (b) $N_c$ as a function of $\gamma$ for representative values of $(D,\alpha)$. Stronger non-reciprocity enhances the projected skin-dependent coupling and reduces the critical length. (c) $N_c$ as a function of $\alpha$ for representative values of $(D,\gamma)$. Increasing $\alpha$ more strongly suppresses the inter-chain coupling and therefore shifts the onset to larger system sizes. The same effective parameters are used in all three panels.}
  \label{fig:modelII_Nc_vs_D_gamma_alpha}
\end{figure}

The preceding reduction fixes the algebraic dependence of the threshold but neglects several dimensionless finite-window corrections, including those associated with the full distance-dependent rung kernel, the finite-$N$ wavefunction sums and level spacing, and the truncation to the dominant two-mode sector. We incorporate their net effect through two dimensionless effective prefactors, $\eta_0$ and $\eta_1$, while leaving the physical Hamiltonian and the numerical definition of $N_c$ unchanged. The resulting threshold equation is
\begin{equation}
  L_c^3
  \left(
  1+\eta_1\frac{\kappa L_c}{4}
  \right)^2
  \simeq
  \eta_0\frac{3\pi^4}{8}D^\alpha,
  \qquad
  L_c=N_c+1.
  \label{appeq:Nc_scaling_effective_L_NNLR}
\end{equation}
For large $N_c$, where $L_c\simeq N_c$, this becomes
\begin{equation}
  \boxed{
  N_c^3
  \left(
  1+\eta_1\frac{\kappa N_c}{4}
  \right)^2
  \simeq
  \eta_0\frac{3\pi^4}{8}D^\alpha
  }.
  \label{appeq:Nc_scaling_equation_kappa_NNLR}
\end{equation}
The original leading asymptotic estimate is recovered for $\eta_0=\eta_1=1$. In the finite parameter window considered in the numerical calculations, we use the common effective values
\begin{equation}
  \eta_0=3.0,
  \qquad
  \eta_1=3.7.
  \label{appeq:eta_values_NNLR}
\end{equation}
These values are held fixed across the Model-II curves and are not adjusted separately for individual values of $D$, $\alpha$, or $\gamma$.

For weak non-reciprocity, $\kappa=\gamma+O(\gamma^3)$, and Eq.~\eqref{appeq:Nc_scaling_equation_kappa_NNLR} reduces to
\begin{equation}
  N_c^3
  \left(
  1+\eta_1\frac{\gamma N_c}{4}
  \right)^2
  \simeq
  \eta_0\frac{3\pi^4}{8}D^\alpha.
  \label{appeq:Nc_scaling_equation_NNLR}
\end{equation}
The effective prefactors modify the quantitative finite-size threshold while preserving the characteristic algebraic dependence generated by the balance between the NN band-edge gap and the coherently projected long-range rung coupling.

Figure~\ref{fig:modelII_Nc_vs_D_gamma_alpha} shows that the effective threshold equation captures the numerical dependence on $D$, $\gamma$, and $\alpha$ over the parameter window considered here. When the first-colliding sector moves away from the band-edge pair or when the large-separation reduction becomes insufficient, the full-kernel projected condition in Eq.~\eqref{appeq:full_projected_elements_NNLR} should be used instead. 

This result is used as the Model-II dashed curve in main-text Fig.~\ref{fig:scalings}(d--f) and as one of the three pairwise-compared thresholds in Fig.~\ref{fig:phase_diagram}. The essential takeaway is the power balance $N^{-2}\sim ND^{-\alpha}$: the NN gap decreases with size while the collectively projected rung coupling increases with size, yielding $N_c^3\sim D^\alpha$.

\subsection{Critical-length scaling for fully nonlocal LR--LR ladder (Model III)}
\label{sec:app:Nc_LRLR}

This subsection supplies the piecewise Model-III threshold summarized in main-text Eqs.~\eqref{eq:LRLR_ratio_lt2_main}, \eqref{eq:LRLR_alpha2_main}, \eqref{eq:LRLR_ratio_23_main}, and \eqref{eq:NcIII_full_main}. It is the central derivation behind the scale-covariant branch, the special role of $\alpha=2$, the $2<\alpha<3$ correction, and the finite-window cutoff shown in Fig.~\ref{fig:scalings}.

We now consider the fully nonlocal LR--LR ladder (Model III), where both the
intra-chain and inter-chain hoppings are long-ranged.  We focus first on the
top-of-band pair $(p,q)=(1,2)$.  The physical threshold is still obtained by
minimizing over all candidate pairs, as in Eq.~\eqref{eq:central_condition_pq_repeat};
the formulas below describe the parameter window in which this band-edge pair is
the first-colliding channel.

The projected EP condition is
\begin{equation}
  \left[\Delta E \pm (u_{11}-u_{22})\right]^2+4u_{12}v_{21}<0,
  \label{appeq:critical_condition_LRLR}
\end{equation}
where $\Delta E=E_1-E_2>0$.  For the analytical estimate we use the
constant-kernel reduction
\begin{equation}
  H_{\perp}^{\rm LR}
  \approx D^{-\alpha}\sum_{n,m}c_{+,n}^{\dagger}c_{-,m},
  \label{appeq:Hint_longrange_approx}
\end{equation}
so that the projected rung matrix elements factorize:
\begin{equation}
  u_{ij}
  =
  D^{-\alpha}
  \left(\sum_{n=1}^{N}\psi_{i,+}^{L}(n)\right)
  \left(\sum_{m=1}^{N}\psi_{j,-}^{R}(m)\right),
  \qquad i,j\in\{1,2\}.
  \label{eq:delta_factor_12}
\end{equation}

For the Hermitian long-range chain, we use
\begin{equation}
  \psi^{(0)}_{1}(x)=\sqrt{\frac{2}{N}}\sin\frac{\pi x}{N},
  \qquad
  \psi^{(0)}_{2}(x)=\sqrt{\frac{2}{N}}\sin\frac{2\pi x}{N},
  \label{eq:psi0_12_again}
\end{equation}
with
\begin{equation}
  \sum_x\psi^{(0)}_1(x)
  =
  \sqrt{\frac{2}{N}}\cot\frac{\pi}{2N},
  \qquad
  \sum_x\psi^{(0)}_2(x)=0.
  \label{eq:sums_psi0_12}
\end{equation}
The non-reciprocal part of an isolated leg is
\begin{equation}
  V_{nm}
  =
  \frac{\operatorname{sgn}(n-m)}{|n-m|^\alpha}
  \quad(n\neq m),
  \qquad
  V_{nn}=0,
  \label{eq:V_kernel_again}
\end{equation}
and we define the finite-size mixing ratio
\begin{equation}
  \chi_N(\alpha)
  \equiv
  \frac{\langle\psi^{(0)}_2|V|\psi^{(0)}_1\rangle}{\Delta E}.
  \label{eq:chiN_def_LRLR}
\end{equation}
The numerator is
\begin{align}
  \langle\psi^{(0)}_2|V|\psi^{(0)}_1\rangle
  &=
  \frac{4}{N}\sum_{n>m}
  \frac{
    \sin\!\left(\frac{2\pi n}{N}\right)\sin\!\left(\frac{\pi m}{N}\right)
    -
    \sin\!\left(\frac{2\pi m}{N}\right)\sin\!\left(\frac{\pi n}{N}\right)
  }{(n-m)^\alpha}.
  \label{eq:V21_explicit_again}
\end{align}

The large-$N$ behavior of this matrix element changes at $\alpha=2$.
Writing $r=n-m$, the antisymmetric sine combination in
Eq.~\eqref{eq:V21_explicit_again} is of order $r/N$ for fixed $r\ll N$.
After summing over $m$, this gives
\begin{equation}
  \langle\psi^{(0)}_2|V|\psi^{(0)}_1\rangle
  \sim
  \frac{1}{N}\sum_{r=1}^{N}r^{1-\alpha}
  \sim
  \begin{cases}
    N^{1-\alpha}, & 0<\alpha<2,\\[1mm]
    N^{-1}\ln N, & \alpha=2,\\[1mm]
    N^{-1}, & \alpha>2.
  \end{cases}
  \label{eq:V21_piecewise_scaling}
\end{equation}
For the same top-of-band pair, the single-chain gap obeys
\begin{equation}
  \Delta E
  \simeq
  C(\alpha)N^{1-\alpha},
  \qquad 0<\alpha<3,
  \label{appeq:DeltaE_LR_scaling}
\end{equation}
while at $\alpha=3$ it crosses to
$\Delta E\sim N^{-2}\ln N$.  Combining
Eqs.~\eqref{eq:V21_piecewise_scaling} and
\eqref{appeq:DeltaE_LR_scaling} gives
\begin{equation}
  \chi_N(\alpha)
  \sim
  \begin{cases}
    \chi(\alpha), & 0<\alpha<2,\\[1mm]
    \bar\chi_2\ln N, & \alpha=2,\\[1mm]
    \bar\chi(\alpha)N^{\alpha-2}, & 2<\alpha<3,
  \end{cases}
  \label{eq:chiN_piecewise_scaling}
\end{equation}
where the barred quantities are $N$-independent prefactors.  Thus the
size-independent mixing ratio assumed in the scale-covariant derivation is
controlled only for $0<\alpha<2$.  At $\alpha=2$ it acquires a marginal
logarithm, while for $2<\alpha<3$ it grows algebraically with $N$.

To first order in the wavefunction mixing, the projected rung elements are
\begin{equation}
  u_{11}
  =
  \frac{2}{ND^\alpha}\cot^2\frac{\pi}{2N},
  \qquad
  u_{12}\simeq\gamma\chi_N(\alpha)u_{11},
  \qquad
  u_{22}\simeq-\gamma^2|\chi_N(\alpha)|^2u_{11},
  \qquad
  v_{21}\simeq-u_{12}.
  \label{eq:deltas_with_Ialpha}
\end{equation}
The relevant branch of Eq.~\eqref{appeq:critical_condition_LRLR} is therefore
equivalent to the window
\begin{equation}
  u_{11}\left(1-\gamma|\chi_N(\alpha)|\right)^2
  <
  \Delta E
  <
  u_{11}\left(1+\gamma|\chi_N(\alpha)|\right)^2.
  \label{eq:DE_window_Ialpha}
\end{equation}
Using
\begin{equation}
  u_{11}\simeq\frac{8N}{\pi^2D^\alpha},
  \label{eq:u11_largeN_Ialpha}
\end{equation}
the first onset is obtained from the upper edge,
\begin{equation}
  C(\alpha)\left[N_c^{\mathrm{III}}\right]^{1-\alpha}
  \simeq
  \frac{8N_c^{\mathrm{III}}}{\pi^2D^\alpha}
  \left[1+\gamma|\chi_{N_c^{\mathrm{III}}}(\alpha)|\right]^2.
  \label{eq:Nc_EP_general_LRLR}
\end{equation}
Equation~\eqref{eq:Nc_EP_general_LRLR} is the common starting point for the
three regimes below.

\paragraph{$0<\alpha<2$.}
This is the regime used for the scale-covariant main-text result and for the Model-III curves in Fig.~\ref{fig:scalings}(d,e). The purpose of the following reduction is to show why the two independent lengths enter only through $N_c/D$, rather than to assume linearity from the outset.

In this regime $\chi_N(\alpha)\to\chi(\alpha)$ is asymptotically independent
of $N$.  Equation~\eqref{eq:Nc_EP_general_LRLR} gives
\begin{equation}
  \frac{N_c^{\mathrm{III}}}{D}
  =
  \left[\frac{\pi^2C(\alpha)}{8}\right]^{1/\alpha}
  \left(1+\gamma|\chi(\alpha)|\right)^{-2/\alpha}.
  \label{eq:Nc_EP_final_Ialpha}
\end{equation}
This is the scale-covariant branch: the onset fixes the ratio
$N_c^{\mathrm{III}}/D$.  For
$\gamma|\chi(\alpha)|\ll1$,
\begin{equation}
  \frac{N_c^{\mathrm{III}}}{D}
  \approx
  \left[\frac{\pi^2C(\alpha)}{8}\right]^{1/\alpha}
  \left[1-\frac{2\gamma}{\alpha}|\chi(\alpha)|\right]
  \equiv A(\alpha)-\gamma B(\alpha).
  \label{eq:Nc_EP_AB_compact}
\end{equation}
Therefore the $A(\alpha)-\gamma B(\alpha)$ form is the weak-nonreciprocity
expansion of the $0<\alpha<2$ scale-covariant result; it is not a uniform
formula for $\alpha\ge2$.  Any polynomial parametrization of $A$ and $B$ is
accordingly fitted only to the unsaturated $(1,2)$ data with $\alpha<2$. 

\begin{figure}
  \subfigure[$N_c$ vs $D$. ]{\includegraphics[width=0.98\linewidth]{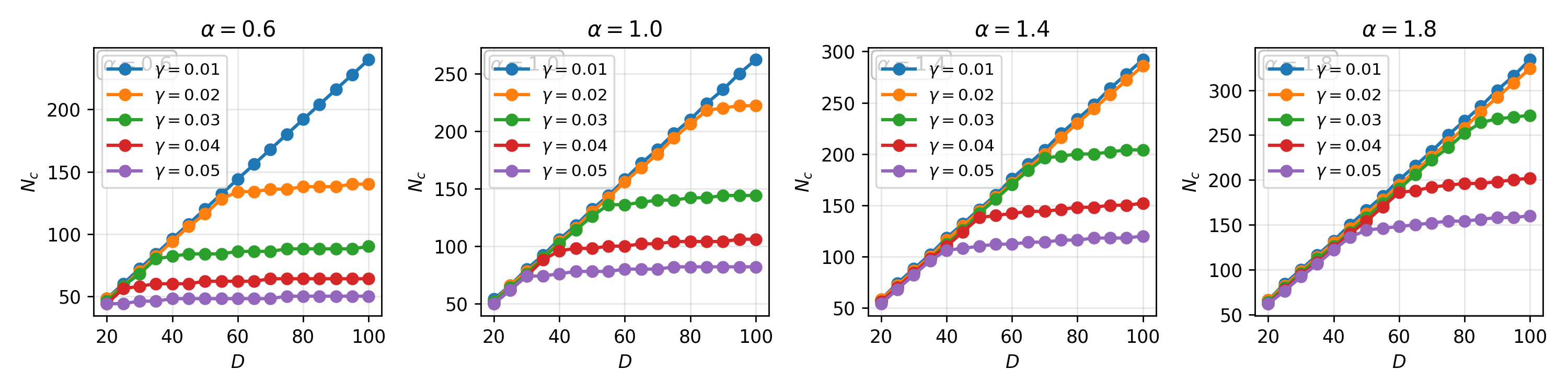}}\\
  \subfigure[$N_c/D$ vs $\gamma$. ]{\includegraphics[width=0.98\linewidth]{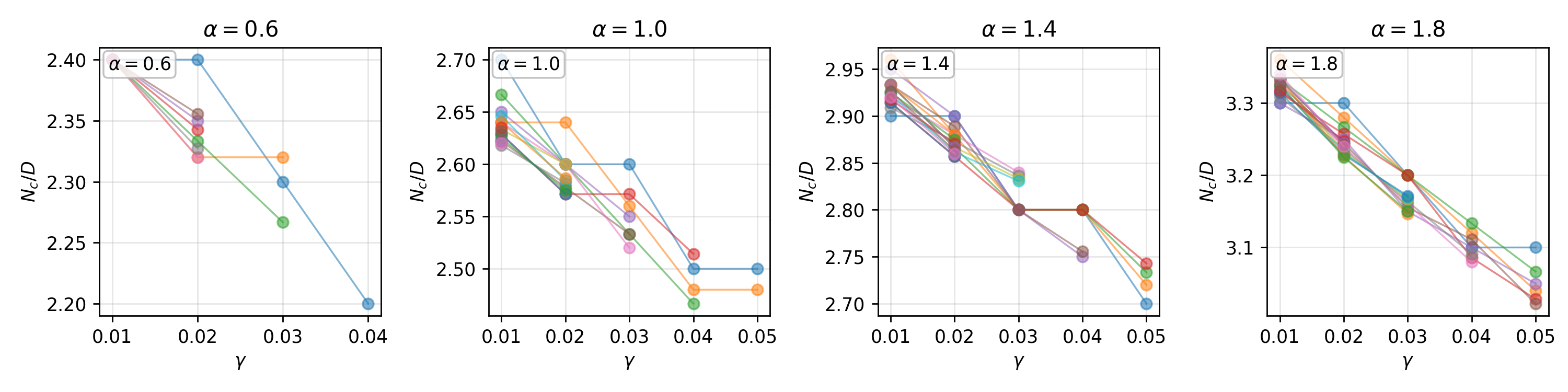}}\\
  \subfigure[$N_c/D$ vs $\alpha$. ]{\includegraphics[width=0.98\linewidth]{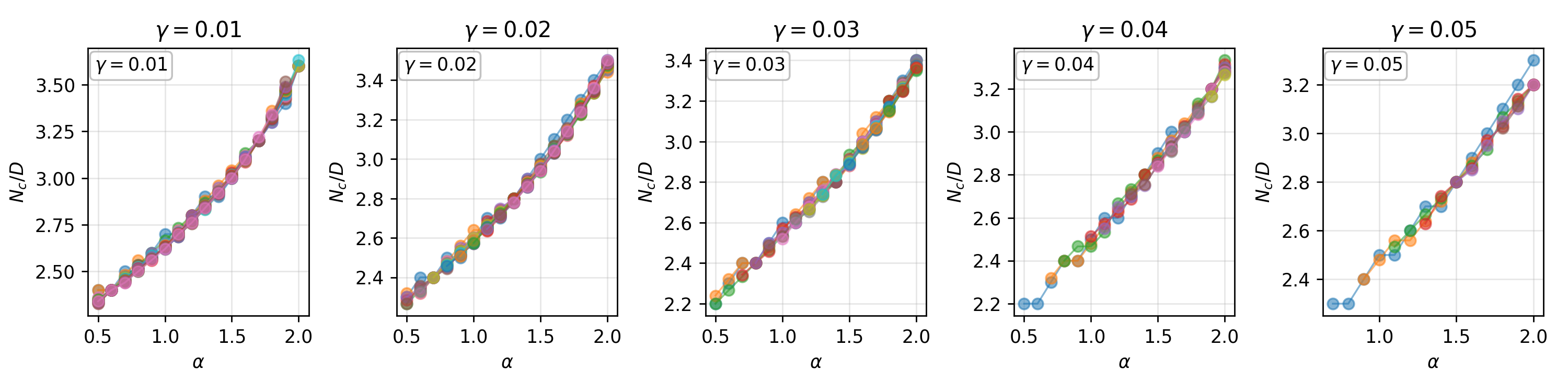}}
  \caption{Scaling of the transition length $N_c$ in the fully nonlocal LR--LR ladder (Model III) on the unsaturated $(1,2)$ branch with $\alpha<2$. (a) $N_c$ is approximately proportional to $D$. (b) The ratio $N_c/D$ is approximately linear in $\gamma$ in the weak-mixing window. (c) After rescaling by $D$, the remaining dependence is controlled by $\alpha$ and $\gamma$.}
  \label{fig:modelIII_Nc_vs_D_gamma}
\end{figure}

\begin{figure}
  \subfigure[]{\includegraphics[width=0.35\linewidth]{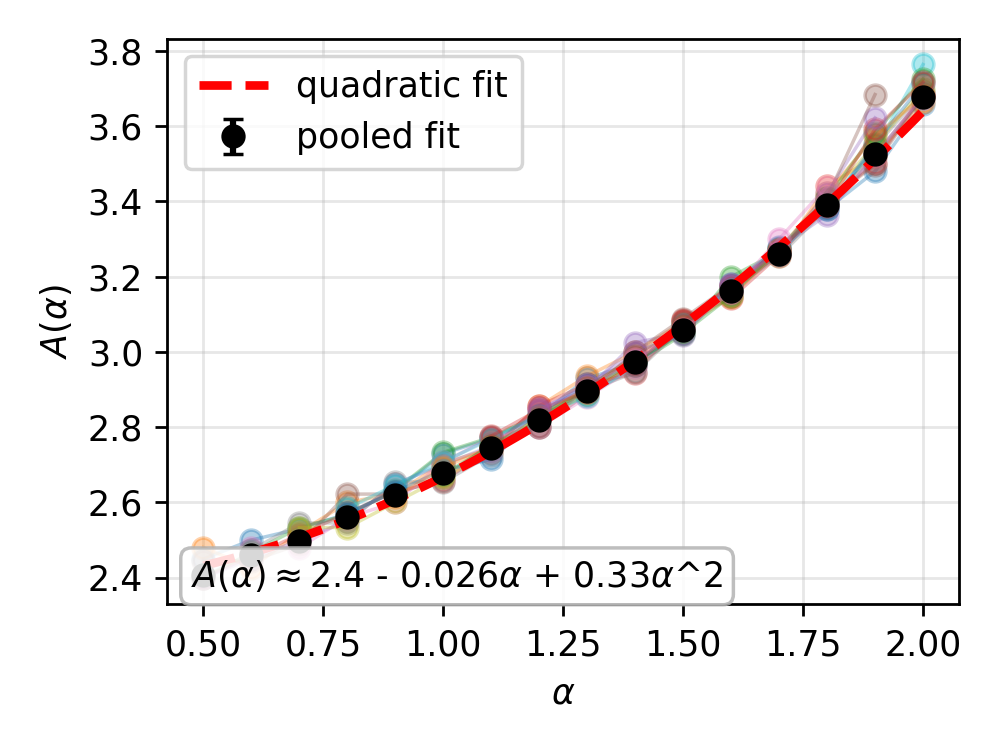}}
  \subfigure[]{\includegraphics[width=0.35\linewidth]{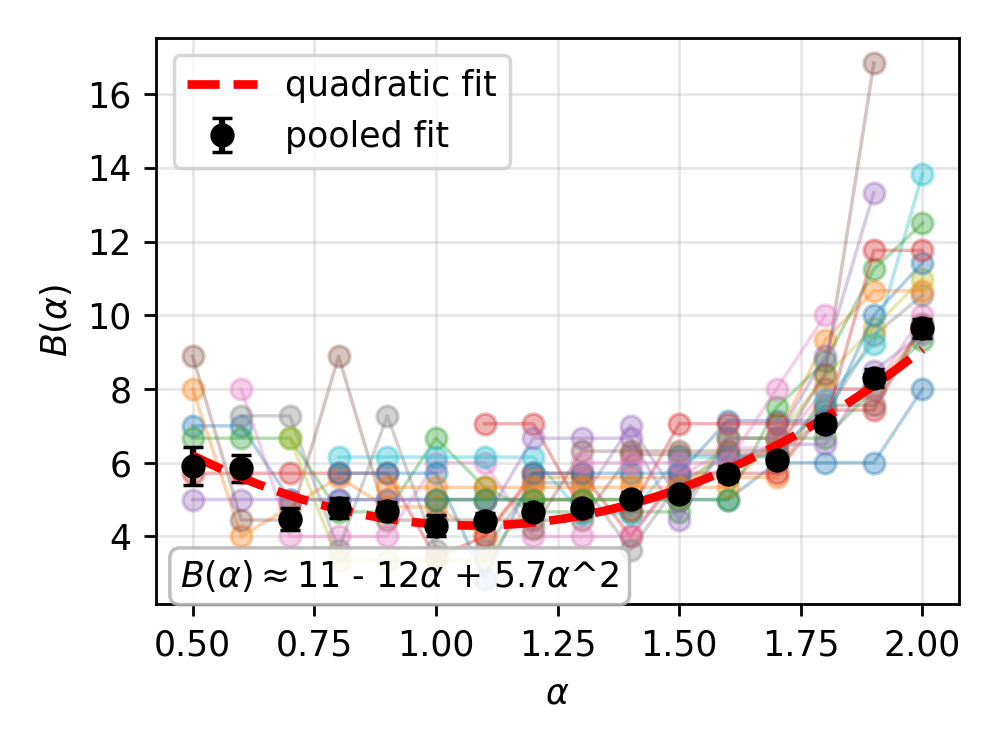}}
  \caption{Finite-window quadratic parametrization of $A(\alpha)$ and $B(\alpha)$ for the unsaturated fully nonlocal LR--LR branch. Only data with $\alpha<2$ and first-colliding pair $(p,q)=(1,2)$ enter this fit. The parametrization is used as a compact representation of Eq.~\eqref{eq:Nc_EP_AB_compact}, not as an extension to $\alpha\ge2$.}
  \label{fig:modelIII_fitAB}
\end{figure}

\paragraph{$\alpha=2$.}
This marginal derivation gives main-text Eqs.~\eqref{eq:LRLR_alpha2_main} and \eqref{eq:LRLR_alpha2_weak_main}. Although the representative panels of Fig.~\ref{fig:scalings} use $\alpha=0.5$, $1$, and $2.4$ rather than exactly $2$, this result is needed to connect the two neighboring regimes and shows that the apparent linear law acquires an intrinsic logarithmic finite-size correction at the boundary.

At the marginal exponent,
\begin{equation}
  \chi_N(2)\simeq\bar\chi_2\ln N,
\end{equation}
and Eq.~\eqref{eq:Nc_EP_general_LRLR} becomes
\begin{equation}
  \left[N_c^{\mathrm{III}}\right]^2
  \left(1+\gamma|\bar\chi_2|\ln N_c^{\mathrm{III}}\right)^2
  \simeq
  \frac{\pi^2C(2)}{8}D^2.
  \label{eq:Nc_EP_alpha2}
\end{equation}
The leading $N_c\propto D$ behavior therefore acquires a logarithmic
finite-size correction.  In the weak-mixing regime,
$\gamma|\bar\chi_2|\ln N_c\ll1$,
\begin{equation}
  \frac{N_c^{\mathrm{III}}}{D}
  \simeq
  \left[\frac{\pi^2C(2)}{8}\right]^{1/2}
  \left[
    1-\gamma|\bar\chi_2|
    \ln\!\left(
      \left[\frac{\pi^2C(2)}{8}\right]^{1/2}D
    \right)
  \right].
  \label{eq:Nc_EP_alpha2_weak}
\end{equation}
Hence the effective linear-in-$\gamma$ coefficient depends logarithmically on
$D$, and cannot be represented by a function $B(\alpha)$ alone.

\paragraph{$2<\alpha<3$.}
This is the branch used for the $\alpha=2.4$ Model-III curve in main-text Fig.~\ref{fig:scalings}(f) and for the unsaturated Model-I--Model-III hierarchy boundary in Fig.~\ref{fig:phase_diagram}. Its significance is that the mixing ratio now grows with $N$, so the aspect ratio is no longer fixed by dilation and the threshold must be solved implicitly.

In this interval,
\begin{equation}
  \chi_N(\alpha)
  \simeq
  \bar\chi(\alpha)N^{\alpha-2},
\end{equation}
and Eq.~\eqref{eq:Nc_EP_general_LRLR} becomes
\begin{equation}
  \left[N_c^{\mathrm{III}}\right]^\alpha
  \left[
    1+\gamma|\bar\chi(\alpha)|
    \left[N_c^{\mathrm{III}}\right]^{\alpha-2}
  \right]^2
  \simeq
  \frac{\pi^2C(\alpha)}{8}D^\alpha.
  \label{eq:Nc_EP_alpha23}
\end{equation}
It is useful to introduce two coefficients that play the same role as
$A(\alpha)$ and $B(\alpha)$ in the $0<\alpha<2$ regime:
\begin{equation}
  A_{>}(\alpha)
  \equiv
  \left[\frac{\pi^2C(\alpha)}{8}\right]^{1/\alpha},
  \qquad
  B_{>}(\alpha)
  \equiv
  |\bar\chi(\alpha)|.
  \label{eq:ABgt_alpha23_def}
\end{equation}
The onset equation can then be written compactly as
\begin{equation}
  \frac{N_c^{\mathrm{III}}}{D}
  =
  A_{>}(\alpha)
  \left[
    1+\gamma B_{>}(\alpha)
    \left[N_c^{\mathrm{III}}\right]^{\alpha-2}
  \right]^{-2/\alpha},
  \qquad
  2<\alpha<3.
  \label{eq:Nc_EP_alpha23_ABgt}
\end{equation}
Unlike Eq.~\eqref{eq:Nc_EP_AB_compact}, this relation is implicit because
$N_c^{\mathrm{III}}$ also appears inside the mixing correction.

To extract $A_{>}(\alpha)$ and $B_{>}(\alpha)$ from the numerical thresholds,
we first fit each value of $\alpha$ independently.  Equation
\eqref{eq:Nc_EP_alpha23_ABgt} can be rearranged into the linear form
\begin{equation}
  Y_\alpha
  =
  c_0(\alpha)+c_1(\alpha)X_\alpha,
  \label{eq:alpha23_linear_fit_form}
\end{equation}
where
\begin{equation}
  X_\alpha
  \equiv
  \gamma N_c^{\alpha-2},
  \qquad
  Y_\alpha
  \equiv
  \left(\frac{D}{N_c}\right)^{\alpha/2}.
  \label{eq:alpha23_XY_def}
\end{equation}
Comparing Eqs.~\eqref{eq:Nc_EP_alpha23_ABgt} and
\eqref{eq:alpha23_linear_fit_form} gives
\begin{equation}
  c_0(\alpha)=A_{>}(\alpha)^{-\alpha/2},
  \qquad
  c_1(\alpha)=c_0(\alpha)B_{>}(\alpha),
\end{equation}
and therefore
\begin{equation}
  A_{>}(\alpha)
  =
  c_0(\alpha)^{-2/\alpha},
  \qquad
  B_{>}(\alpha)
  =
  \frac{c_1(\alpha)}{c_0(\alpha)}.
  \label{eq:alpha23_AB_from_linear_fit}
\end{equation}
Only unsaturated points with first-colliding pair $(p,q)=(1,2)$ are retained
in this extraction.

For a compact finite-window representation across the interval, we then write
$x=\alpha-2$ and parametrize
\begin{align}
  A_{>}(\alpha)
  &=a_0+a_1x+a_2x^2,
  \label{eq:Agt_alpha23_quadratic}\\
  B_{>}(\alpha)
  &=b_0+b_1x+b_2x^2.
  \label{eq:Bgt_alpha23_quadratic}
\end{align}
The solid curves are obtained from a simultaneous nonlinear fit of
Eqs.~\eqref{eq:Nc_EP_alpha23_ABgt}--\eqref{eq:Bgt_alpha23_quadratic} to all
$209$ retained $(D,\gamma,\alpha)$ points.  For each numerical point, the
positive solution $N_{c,\mathrm{fit}}^{\mathrm{III}}$ of
Eq.~\eqref{eq:Nc_EP_alpha23_ABgt} is evaluated, and the six coefficients are
determined by minimizing
\begin{equation}
  \mathcal L
  =
  \sum_i
  \left[
    \ln
    \left(
      \frac{N_{c,\mathrm{fit}}^{\mathrm{III}}
      (D_i,\gamma_i,\alpha_i)}
      {N_{c,i}^{\mathrm{num}}}
    \right)
  \right]^2.
  \label{eq:alpha23_global_fit_loss}
\end{equation}

The resulting coefficients are
\begin{equation}
  A_{>}(\alpha)
  =
  3.62867
  +2.17682(\alpha-2)
  -0.097608(\alpha-2)^2
  \label{eq:Agt_alpha23_empirical}
\end{equation}
and
\begin{equation}
  B_{>}(\alpha)
  =
  2.63458
  -5.02404(\alpha-2)
  +2.47832(\alpha-2)^2.
  \label{eq:Bgt_alpha23_empirical}
\end{equation}
Thus the practical unsaturated threshold equation in this interval is
\begin{align}
  \frac{N_c^{\mathrm{III}}}{D}
  &=
  \left[
    3.62867
    +2.17682(\alpha-2)
    -0.097608(\alpha-2)^2
  \right]
  \nonumber\\
  &\quad\times
  \left\{
    1+\gamma
    \left[
      2.63458
      -5.02404(\alpha-2)
      +2.47832(\alpha-2)^2
    \right]
    \left[N_c^{\mathrm{III}}\right]^{\alpha-2}
  \right\}^{-2/\alpha},
  \qquad
  2<\alpha<3.
  \label{eq:Nc_EP_alpha23_empirical}
\end{align}

In the weak-mixing regime,
$\gamma B_{>}(\alpha)\left[N_c^{\mathrm{III}}\right]^{\alpha-2}\ll1$, define
\begin{equation}
  N_0=A_{>}(\alpha)D.
\end{equation}
Expanding Eq.~\eqref{eq:Nc_EP_alpha23_ABgt} gives
\begin{equation}
  \frac{N_c^{\mathrm{III}}}{D}
  \simeq
  A_{>}(\alpha)
  \left[
    1-\frac{2\gamma}{\alpha}
    B_{>}(\alpha)N_0^{\alpha-2}
  \right].
  \label{eq:Nc_EP_alpha23_weak}
\end{equation}
The correction is proportional to $\gamma D^{\alpha-2}$, so the
$0<\alpha<2$ form $A(\alpha)-\gamma B(\alpha)$ cannot be continued into this
interval.  The fitted functions in
Eqs.~\eqref{eq:Agt_alpha23_empirical} and
\eqref{eq:Bgt_alpha23_empirical} are finite-window parametrizations of the
unsaturated $(1,2)$ branch.  Close to $\alpha=3$, additional finite-size
corrections associated with the marginal band-edge gap become increasingly
important.

The change at $\alpha=2$ has a simple origin.  For $\alpha<2$, the long-distance
part of the antisymmetric hopping kernel controls both the mixing matrix element
and the top-band gap, so their ratio is size independent.  For $\alpha>2$, the
mixing matrix element is instead dominated by short hopping distances and scales
as $N^{-1}$, whereas the gap continues to scale as $N^{1-\alpha}$ up to
$\alpha=3$.  Their ratio therefore grows as $N^{\alpha-2}$.

\paragraph{Finite-window cutoff.}
This cutoff produces the horizontal $N_{\rm sat}$ line and the eventual flattening of the Model-III curves in main-text Fig.~\ref{fig:scalings}. It is also essential for the large-$D$ branches of the hierarchy diagram: once the inter-chain EP estimate exceeds this intrinsic single-chain scale, the physical threshold is no longer set by the inter-chain balance alone.

The inter-chain onset derived above competes with the intrinsic crossover of an
isolated long-range leg.  Removing the nearest-neighbor non-reciprocity by the
imaginary gauge transformation gives the longest enhanced hopping, relative to
the gauged nearest-neighbor scale,
\begin{equation}
  \left(\frac{1+\gamma}{1-\gamma}\right)^{(N-2)/2}(N-1)^{-\alpha}.
  \label{eq:longest_hopping_ratio}
\end{equation}
The single-chain crossover length is therefore defined by
\begin{equation}
  \left(\frac{1+\gamma}{1-\gamma}\right)^{(N_{\mathrm{sat}}-2)/2}
  =
  (N_{\mathrm{sat}}-1)^\alpha,
  \label{eq:Nsat_single_chain_condition}
\end{equation}
or, equivalently,
\begin{equation}
  e^{\kappa(N_{\mathrm{sat}}-2)}
  =
  (N_{\mathrm{sat}}-1)^\alpha,
  \qquad
  \kappa=\frac12\ln\frac{1+\gamma}{1-\gamma}.
  \label{eq:LRLR_Lsat_supp}
\end{equation}
The physical Model-III threshold is estimated as
\begin{equation}
  N_{c,\mathrm{full}}^{\mathrm{III}}
  \simeq
  \min\left[N_c^{\mathrm{III}},N_{\mathrm{sat}}\right],
  \label{eq:NcIII_min_NcEP_Nsat}
\end{equation}
where the candidate threshold $N_c^{\mathrm{III}}$ is obtained from
Eq.~\eqref{eq:Nc_EP_final_Ialpha} for $0<\alpha<2$,
Eq.~\eqref{eq:Nc_EP_alpha2} at $\alpha=2$, and
Eq.~\eqref{eq:Nc_EP_alpha23} for $2<\alpha<3$.  This cutoff does not restore
the $A(\alpha)-\gamma B(\alpha)$ form outside $\alpha<2$; it only limits the
largest observable threshold.  If another pair $(p,q)$ reaches the EP
condition first, the corresponding pair-resolved threshold replaces the
$(1,2)$ estimate in Eq.~\eqref{eq:NcIII_min_NcEP_Nsat}.

\subsection{LR--NN nonlocal-legs control geometry (Model IV)}
\label{sec:app:Nc_modelIV}

The LR--NN ladder is excluded from the three-model main-text hierarchy, and this subsection explains why. As a control, it isolates the effect of long-range legs without collective long-range rungs; the result shows that the first collision can move to the band bottom and become substantially more sensitive to nearby levels and finite-size corrections.

We retain this geometry only to test what changes when long-range hopping is placed on the legs but the inter-chain rungs remain local.

\begin{figure}[!htp]
  \begin{minipage}{0.48\linewidth}
    (a)\\[1mm]
    \includegraphics[width=\linewidth]{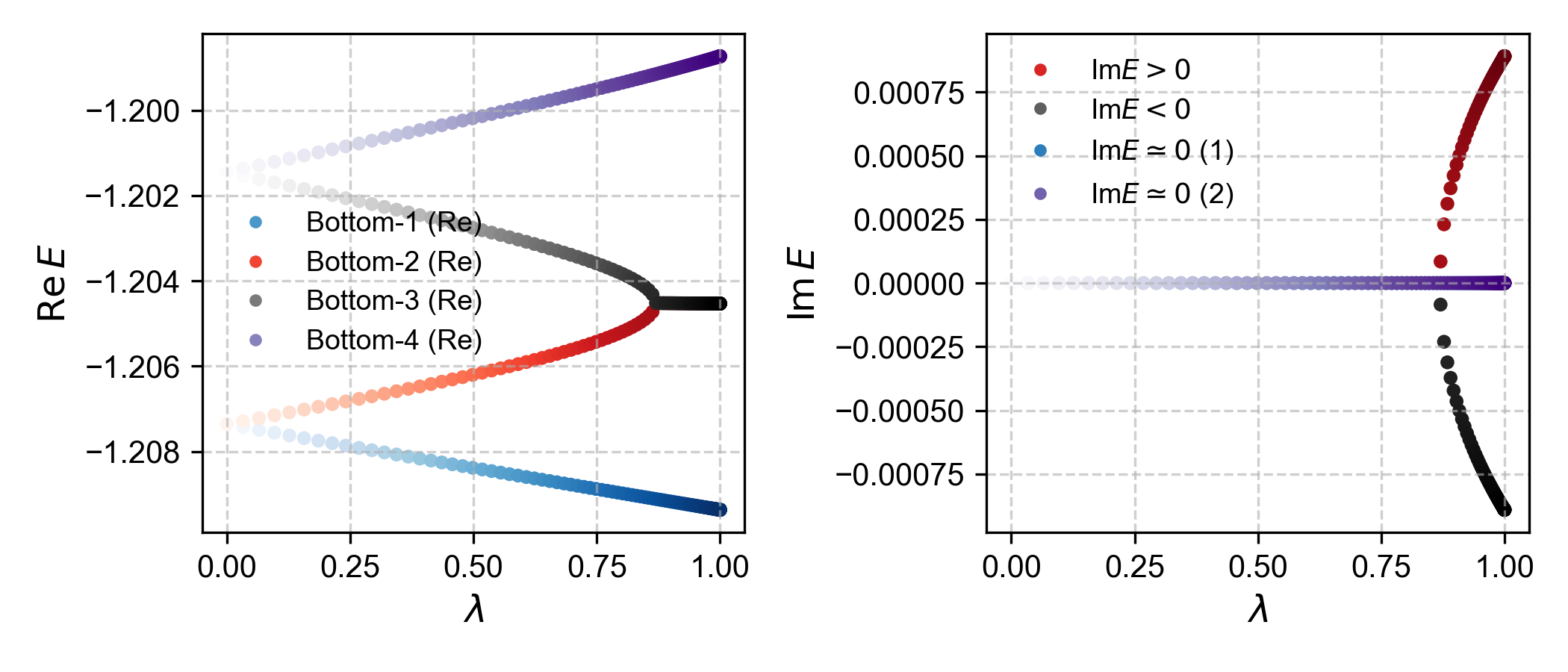}
  \end{minipage}
  \hfill
  \begin{minipage}{0.48\linewidth}
    (b)\\[1mm]
    \includegraphics[width=\linewidth]{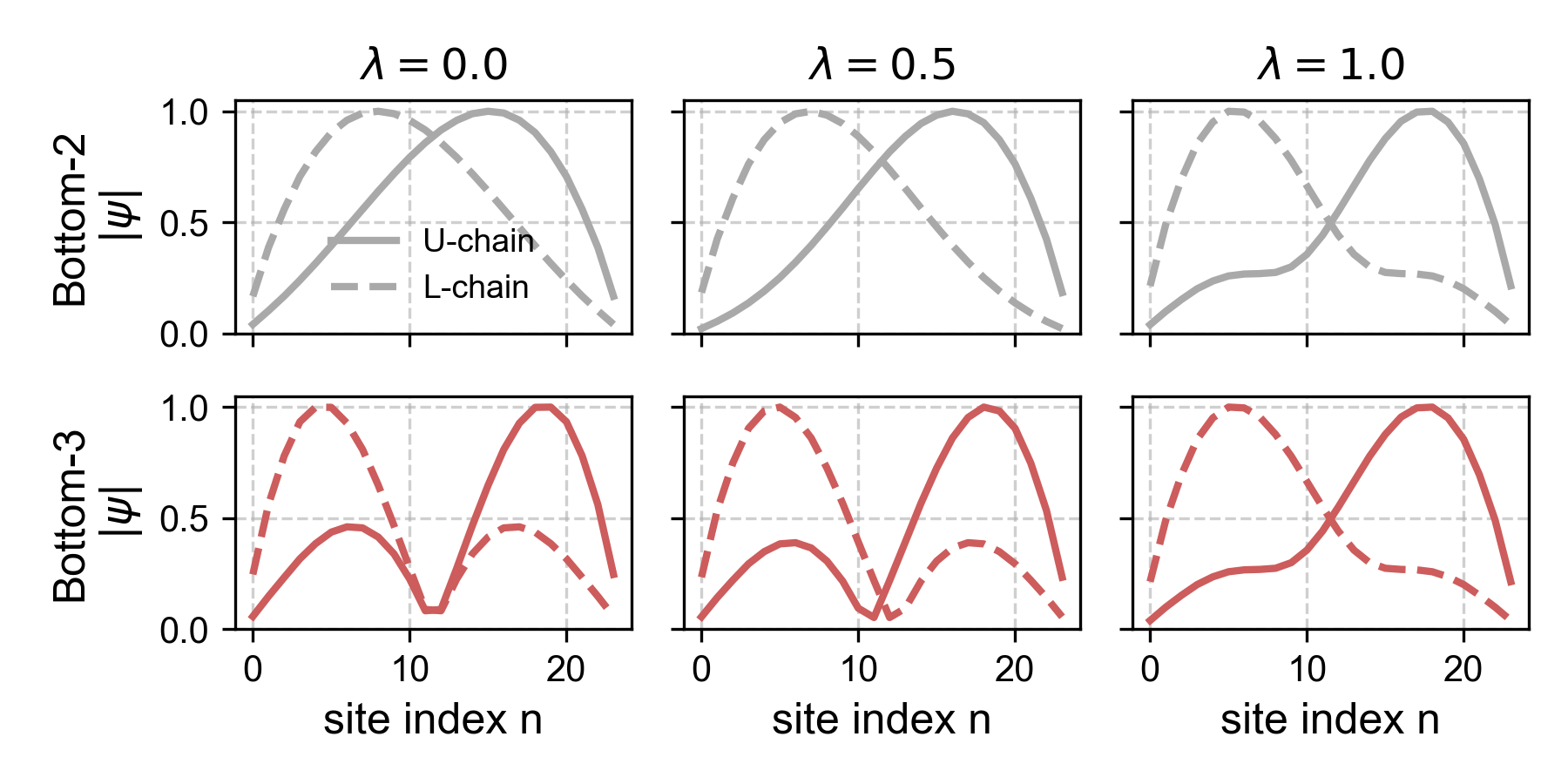}
  \end{minipage}
  \caption{LR--NN nonlocal-legs control geometry (Model IV).
    (a) Complex energy spectrum of the four states closest to the band bottom, plotted as a function of the inter-chain coupling ratio $\lambda$. One bottom-edge first-colliding pair approaches and becomes the first to complexify as $\lambda$ increases.
  (b) Spatial distributions $|\psi_n|$ of the corresponding participating bottom states at $\lambda=0.0$ (left), $0.5$ (middle), and $1.0$ (right). Solid and dashed curves denote the $(+)$ and $(-)$ legs, respectively. The system parameters are $N=44$, $\gamma=0.05$, $\alpha=2$, and $D=30$.}
  \label{fig:modelIV_bottom_pair_evolution}
\end{figure}

For this ladder, the real-to-complex transition is governed by the \emph{band-bottom} first-colliding pair rather than the topmost states. The relevant single-chain ingredients are therefore the band-bottom gap $\Delta E'$ and the corresponding perturbed wavefunctions derived in Sec.~\ref{sec:single_chain_properties}. Here we only summarize how they enter the coupled-chain problem and why they lead to a more delicate onset condition than in the other three cases.

In the LR--NN control geometry (Model IV), the ladder employs \emph{local} inter-chain coupling, so the two legs are coupled only site by site:
\begin{equation}
  H_{\perp}^{\rm NN}
  =
  D^{-\alpha}\sum_{n=1}^{N}
  c_{+,n}^\dagger c_{-,n},
\end{equation}
so that the projected upper-block matrix elements are
\begin{equation}
  u_{ij}
  = \langle L_{i,+} | H_{\perp}^{\rm NN} | R_{j,-} \rangle
  = \frac{1}{D^{\alpha}} \sum_{n=1}^{N}
  \psi_{+,i}^*(n)\psi_{-,j}(n).
  \label{eq:deltaij_local}
\end{equation}

We specialize to the band-bottom first-colliding pair
\begin{equation}
  p=N-1,\qquad q=N-2,
\end{equation}
for which the counter-pumped long-range legs acquire opposite $\mathcal O(\gamma)$ distortions on the two legs. Within the same two-state approximation introduced in Sec.~\ref{subsec:bottom_wavefunctions_LR}, this is captured by
\begin{align}
  \psi_{+,p}(n) &\simeq \psi^{(0)}_{p}(n)+\gamma J\,\psi^{(0)}_{q}(n),\qquad
  \psi_{-,p}(n) \simeq \psi^{(0)}_{p}(n)-\gamma J\,\psi^{(0)}_{q}(n), \\
  \psi_{+,q}(n) &\simeq \psi^{(0)}_{q}(n)-\gamma J\,\psi^{(0)}_{p}(n),\qquad
  \psi_{-,q}(n) \simeq \psi^{(0)}_{q}(n)+\gamma J\,\psi^{(0)}_{p}(n),
\end{align}
where $J$ denotes the first-order mixing amplitude induced by the antisymmetric long-range kernel,
\begin{equation}
  J \equiv \frac{1}{\Delta E'}\sum_{n\neq m}\frac{\mathrm{sgn}(n-m)}{|n-m|^\alpha}\,
  \psi^{(0)}_{q}(m)\psi^{(0)}_{p}(n).
  \label{eq:J_bottom_def}
\end{equation}

Substituting these expressions into Eq.~\eqref{eq:deltaij_local}, and using the orthonormality of the unperturbed standing waves, we obtain to leading order
\begin{align}
  u_{pp} &\simeq \frac{1}{D^\alpha}\left[1+\mathcal{O}(\gamma^2)\right], \\
  u_{qq} &\simeq \frac{1}{D^\alpha}\left[1+\mathcal{O}(\gamma^2)\right], \\
  u_{pq} &\simeq \frac{2\gamma J}{D^\alpha}, \\
  v_{qp} &\simeq -\frac{2\gamma J}{D^\alpha}.
\end{align}
Hence the off-diagonal hybridizations have opposite signs, so that
\begin{equation}
  u_{pq}v_{qp}\simeq -\frac{4\gamma^2 J^2}{D^{2\alpha}}<0.
\end{equation}
This sign structure is the essential mechanism behind the onset of complexification in the LR--NN control geometry (Model IV).

Since $u_{pp}-u_{qq}=\mathcal{O}(\gamma^2)$ at this order, the projected-sector criterion in Eq.~\eqref{eq:central_condition} reduces to
\begin{equation}
  (\Delta E')^2 + 4u_{pq}v_{qp}<0
  \quad \Longleftrightarrow \quad
  (\Delta E')^2 - \frac{16\gamma^2 J^2}{D^{2\alpha}}<0.
  \label{eq:central_condition_modelIV_bottom}
\end{equation}
Accordingly, the onset is controlled parametrically by the competition between the band-bottom spacing $\Delta E'$ and the off-diagonal mixing scale $4\gamma|J|/D^\alpha$, namely
\begin{equation}
  \Delta E' \sim \frac{4\gamma |J|}{D^\alpha}.
  \label{eq:Nc_condition_modelIV}
\end{equation}

At this level, Eq.~\eqref{eq:Nc_condition_modelIV} captures the qualitative mechanism: increasing $N$ reduces the band-bottom gap, while the local rung coupling probes the $\gamma$-induced overlap between the two distorted bottom states. Their balance determines when the first complex pair emerges.

We do not push this case to a high-precision closed-form prefactor for $N_c$ here. The reason is that, unlike the other three ladders, the present case is quantitatively more sensitive to subleading details: the first-colliding pair is identified only approximately by the bottom doublet, the two-state truncation neglects nearby modes, the estimate of $J$ is more sensitive to finite-size corrections, and $\mathcal O(\gamma^2)$ contributions to the diagonal terms can become non-negligible near onset. For this reason, we regard Eq.~\eqref{eq:Nc_condition_modelIV} as the core onset criterion, while any more explicit scaling estimate for this control would require additional asymptotic information about $J$.

This control geometry is more sensitive to finite-size corrections, neighboring-level admixture, and higher-order terms, and is therefore not used as one of the three main-text scaling branches.

Taken together, Appendix~\ref{sec:app_Nc_derivations} supplies the analytical curves and boundary inputs used in both main-text figures. The same EP criterion is used for the models considered here, while the placement of nonlocality changes the gap, the collective projection, or both; this is the precise origin of the hierarchy rather than a model-dependent adjustment of prefactors.

\section{Experimental extraction of the real-to-complex threshold}
\label{sec:app_experimental_extraction}

This Appendix supports the experimental claims in the main-text Discussion by converting the mathematical condition $\operatorname{Im}E\neq0$ into measurable spectral and time-domain criteria. It also explains how the scans underlying main-text Fig.~\ref{fig:scalings} and Fig.~\ref{fig:phase_diagram} could be reconstructed in programmable platforms. The central experimental point is that one must detect a relative imaginary splitting after subtracting uniform background loss or gain, rather than identify instability from an absolute increase of signal alone.

In this section we describe how the real-to-complex threshold discussed in the main text can be extracted experimentally. The goal is not to prescribe a single platform, but to state an operational procedure for measuring the critical length $N_c(D,\alpha,\gamma)$ and for reconstructing the scaling curves and hierarchy diagram shown in the main text.

The central observable is the appearance of a nonzero imaginary splitting in the spectrum, after subtracting any uniform background loss or gain. In an idealized Hamiltonian description, the transition is the point where a pair of real eigenvalues coalesces at an exceptional point and then becomes a complex-conjugate pair. In an experiment, all modes may have a common decay rate or a common frequency offset. This common shift is not part of the real-to-complex transition. Therefore one should track the relative imaginary spread
\begin{equation}
    \Delta_{\rm Im}(N,D,\alpha,\gamma)
    =
    \max_j
    \left|
    \Im E_j
    -
    \overline{\Im E}
    \right|,
    \qquad
    \overline{\Im E}
    =
    \frac{1}{2N}\sum_{j=1}^{2N}\Im E_j .
    \label{eq:exp_imaginary_spread}
\end{equation}
The experimentally extracted threshold is then
\begin{equation}
    N_c^{\rm exp}(D,\alpha,\gamma)
    =
    \min_N
    \left\{
    N:
    \Delta_{\rm Im}(N,D,\alpha,\gamma)>\Delta_{\rm res}
    \right\},
    \label{eq:exp_Nc_spectral}
\end{equation}
where $\Delta_{\rm res}$ is the spectral resolution floor. Equivalently, in a time-domain measurement with amplification, one may define
\begin{equation}
    \Gamma_{\max}(N,D,\alpha,\gamma)
    =
    \max_j \Im\left(E_j-\overline{E}\right),
    \label{eq:exp_Gamma_max}
\end{equation}
and identify the threshold from
\begin{equation}
    N_c^{\rm exp}
    =
    \min_N
    \left\{
    N:
    \Gamma_{\max}(N,D,\alpha,\gamma)>\Gamma_{\rm res}
    \right\}.
    \label{eq:exp_Nc_growth}
\end{equation}
Here $\Gamma_{\rm res}$ is set by the finite observation time, noise level, and calibration accuracy. Equations~\eqref{eq:exp_Nc_spectral} and \eqref{eq:exp_Nc_growth} are two equivalent operational versions of the same real-to-complex threshold.

\subsection{Topoelectrical circuit implementation}
\label{sec:app_circuit_implementation}

This subsection provides the concrete circuit protocol referred to in the main-text Discussion. It shows how the abstract parameters $(D,\alpha,\gamma)$ are encoded in an admittance matrix and how $N_c$ can be extracted either from the reconstructed complex spectrum or from calibrated modal growth.

A direct implementation is possible in a non-reciprocal topoelectrical circuit. The circuit Laplacian $J(\omega)$ plays the role of the finite open-boundary ladder Hamiltonian, up to an overall frequency-dependent and platform-specific affine transformation. The two legs of the ladder are represented by two rows of circuit nodes. The intra-chain and inter-chain admittances are chosen to realize, respectively,
\begin{align}
    J_{\pm,n;\pm,m}
    &\propto
    \frac{1\pm\gamma\,{\rm sgn}(n-m)}
    {|n-m|^\alpha},
    \qquad n\neq m ,
    \label{eq:exp_circuit_leg}\\
    J_{+,n;-,m}
    &\propto
    \frac{1}
    {[D^2+(n-m)^2]^{\alpha/2}},
    \label{eq:exp_circuit_rung}
\end{align}
for the LR--LR geometry, with the corresponding truncations used for the NN--NN and NN--LR geometries. The transverse separation $D$ is therefore not necessarily a literal physical distance between circuit boards; it is an effective geometric parameter encoded in the programmed or wired inter-chain admittance profile.

A frequency-domain measurement proceeds as follows. First, inject an AC current at node $\mu$ and measure the complex voltage response at all nodes $\nu$. Repeating this for all injection nodes reconstructs the response matrix
\begin{equation}
    G_{\nu\mu}(\omega)=\frac{V_\nu(\omega)}{I_\mu(\omega)} .
\end{equation}
Since $G(\omega)=J^{-1}(\omega)$, this gives the circuit Laplacian after calibration. Diagonalizing the reconstructed Laplacian yields the complex eigenvalues of the effective finite ladder. The real-to-complex threshold is then obtained by scanning $N$ and applying Eq.~\eqref{eq:exp_Nc_spectral}.

A time-domain measurement gives a complementary route. After a short pulse or a controlled initial voltage profile, the node voltages can be expanded in normal modes,
\begin{equation}
    \mathbf V(t)
    =
    \sum_j a_j\,\mathbf R_j\,e^{-iE_j t}.
    \label{eq:exp_modal_expansion}
\end{equation}
Below threshold, all modes have the same calibrated imaginary background, so no relative modal growth appears after subtracting this background. Above threshold, the first complex pair produces a finite $\Gamma_{\max}$, and the envelope of the most unstable component grows as
\begin{equation}
    \|\mathbf V(t)\|_{\rm unstable}
    \propto
    e^{\Gamma_{\max} t}.
    \label{eq:exp_voltage_growth}
\end{equation}
Thus the measurable signal is not merely that voltages become larger, but that a fitted modal growth rate becomes positive relative to the calibrated background and exceeds the resolution floor in Eq.~\eqref{eq:exp_Nc_growth}.

\subsection{Mapping the main-text scaling curves}
\label{sec:app_mapping_main_figures}

The protocol below is written specifically to reproduce the two principal main-text outputs: the three $N_c(D)$ curves of Fig.~\ref{fig:scalings} and the ordering map of Fig.~\ref{fig:phase_diagram}. This makes explicit which quantities must be held fixed and why the hierarchy boundaries are obtained from pairwise threshold crossings rather than from a conventional thermodynamic order parameter.

To reproduce the $N_c$ versus $D$ curves in the main text, one may use the following protocol.

\begin{enumerate}
    \item Choose one ladder architecture: NN--NN, NN--LR, or LR--LR.
    \item Fix the decay exponent $\alpha$ and non-reciprocity $\gamma$.
    \item Choose a transverse separation parameter $D$ by setting the inter-chain coupling profile.
    \item Build or program a sequence of finite open-boundary ladders with increasing chain length $N$.
    \item For each $N$, reconstruct the complex spectrum or measure the largest modal growth rate.
    \item Define $N_c(D,\alpha,\gamma)$ as the first $N$ satisfying Eq.~\eqref{eq:exp_Nc_spectral} or Eq.~\eqref{eq:exp_Nc_growth}.
    \item Repeat the scan over $D$.
\end{enumerate}

This produces the experimental analogue of the main-text curves $N_c(D)$ for each ladder. In practice, $N$ is an integer, so the uncertainty in the extracted threshold is at least $\pm1$ site, with an additional uncertainty set by $\Delta_{\rm res}$ or $\Gamma_{\rm res}$. The logarithmic, algebraic, and scale-covariant trends should therefore be extracted from the overall scaling over a range of $D$, rather than from a single threshold point.

To reconstruct the hierarchy diagram in the $(D,\alpha)$ plane, the same procedure is repeated for all three architectures at the same $(D,\alpha,\gamma)$. This gives three measured thresholds,
\begin{equation}
    N_c^{\rm I}(D,\alpha,\gamma),\qquad
    N_c^{\rm II}(D,\alpha,\gamma),\qquad
    N_c^{\rm III}(D,\alpha,\gamma).
\end{equation}
The experimental color assigned to that point is then determined by the ordering of these three values. The model with the smallest $N_c$ is the one that complexifies first. Pairwise crossings of the measured thresholds give the boundaries between the hierarchy regions. These boundaries should be interpreted as finite-size threshold crossings, not as thermodynamic phase transitions.

\subsection{Photonic, mechanical, and simulator implementations}
\label{sec:app_other_platforms}

This subsection explains the platform-independent significance of the operational definition above. Although the measured response differs between photonic, mechanical, and quantum-simulator settings, each platform can identify the same relative linewidth, decay-rate, or growth-rate splitting that defines the main-text threshold.

The same operational criterion applies beyond circuits. In photonic resonator or waveguide arrays, the complex eigenvalues can be extracted from transmission or reflection spectra. Below threshold, two resonances approach one another while their linewidths remain equal after subtracting common loss. Above threshold, the exceptional-point pair develops a measurable linewidth splitting. In a propagation experiment, the same transition appears as the emergence of a component whose intensity grows relative to the calibrated background during propagation. The extracted linewidth splitting or propagation growth rate plays the role of $\Delta_{\rm Im}$ or $\Gamma_{\max}$ in Eqs.~\eqref{eq:exp_imaginary_spread}--\eqref{eq:exp_Gamma_max}.

In mechanical or acoustic networks, the response matrix can be measured by driving one site at a time and recording the displacement or pressure response at all sites. The complex normal-mode frequencies are then obtained by fitting the response peaks or by reconstructing the dynamical matrix. The threshold is again the first chain length at which a relative imaginary splitting appears.

Programmable quantum simulators can test the same finite-size spectral transition by implementing non-unitary evolution or an effective non-Hermitian generator. In this setting, one does not usually observe a classical voltage growth. Instead, the relevant quantities are decay-rate splittings, postselected mode weights, or reconstructed eigenvalues of the effective generator. The same definition of $N_c$ applies once the complex spectrum or the dominant decay/growth rate has been extracted.

\subsection{Practical considerations}
\label{sec:app_experimental_practical}

These considerations specify which experimental imperfections change only the resolution of $N_c$ and which can distort the intended coupling matrix. They are included so that the flattening, crossings, and scaling exponents emphasized in the main text are not confused with background loss, finite-time resolution, or nonlinear saturation.

Several experimental details are important for identifying the threshold reliably.

First, uniform background loss or gain must be calibrated and subtracted. A global imaginary shift of all eigenvalues does not indicate the real-to-complex transition. The transition is instead signaled by a relative imaginary splitting of an eigenvalue pair.

Second, the finite observation time limits the smallest resolvable growth or decay-rate difference. If the observation window is $T_{\rm obs}$, then the resolution floor is typically of order $1/T_{\rm obs}$, up to noise and fitting uncertainties. Very close to threshold, where $\Gamma_{\max}$ is small, this sets the main experimental uncertainty in $N_c$.

Third, nonlinear saturation should be avoided when extracting the linear spectrum. The thresholds discussed here are linear spectral thresholds. In a gain-assisted setup, the measurement should therefore be performed either from small initial amplitudes or by fitting the early-time response before nonlinear gain compression or component saturation becomes important.

Fourth, imperfections in the long-range coupling profile can shift the thresholds. This is not necessarily fatal: the scaling laws are extracted by comparing how $N_c$ changes with $D$ and $\alpha$, not by relying on a single absolute value. A useful calibration procedure is to first measure the isolated single-chain spectrum and verify the expected band-edge spacing, then measure the rung-coupled ladder and extract the first complexifying pair.

Finally, the finite-window saturation of the fully nonlocal LR--LR ladder has a direct experimental signature. In the unsaturated regime, the measured threshold grows approximately linearly with $D$. Once the predicted linear branch exceeds the single-chain saturation scale, increasing $D$ no longer increases the observed $N_c$ substantially. Experimentally, this appears as a flattening of the LR--LR threshold curve, corresponding to the horizontal saturation line in the main-text scaling plots.

These protocols show how the main-text thresholds can be tested without relying on platform-specific notions of absolute amplification. Across all implementations, the experimentally relevant event is the first resolvable relative imaginary splitting, and repeating that measurement across $N$, $D$, and the three architectures directly reconstructs the scaling curves and hierarchy ordering.

\section{Numerical and approximation checks}
\label{sec:app_checks}

This Appendix benchmarks the principal analytical simplification used for the long-range rungs in Models II and III. The main-text numerical points always use the full distance-dependent kernel, while the compact derivations factorize it as $D^{-\alpha}$. The comparison below tests whether the reported hierarchy is a property of the physical model or an artifact of that factorization.

\subsection{Distance-dependent versus constant long-range rung kernels}
\label{sec:app_kernel_benchmark}

This benchmark is used to justify the constant-kernel steps in Appendices~\ref{sec:app:Nc_NNLR} and \ref{sec:app:Nc_LRLR}. Its significance is that the simplified projection preserves the observed scaling laws and closely tracks the quantitative thresholds over the range explicitly benchmarked below, while the numerical figures themselves remain based on the full distance-dependent kernel.

In the main text,
we use the physically motivated, distance-dependent inter-chain coupling block
\begin{equation}
  H_{\perp}^{\rm LR} = \sum_{n,m} \frac{c_{+,n}^\dagger c_{-,m}}{[D^2+(x_n-x_m)^2]^{\alpha/2}}\label{eq:interchain_distance}
\end{equation}
which explicitly accounts for the longitudinal displacement $\Delta x=x_n-x_m$ between sites $n$ and $m$ on different legs. For analytical tractability in the derivations above, we instead adopt the compact approximation where the coupling is treated as a constant dependent only on the vertical separation $D$,
\begin{equation}
  K_{nm}=[D^2+(x_n-x_m)^2]^{-\alpha/2}\simeq D^{-\alpha},
  \qquad
  H_{\perp}^{\rm LR} \approx D^{-\alpha}\sum_{n,m}c_{+,n}^\dagger c_{-,m}.
  \label{eq:interchain_simplified}
\end{equation}
The constant-kernel approximation is used for analytic estimates of long-range rungs in Models II and III. It is not needed for Model I, whose rung is local by definition.

\begin{figure}
  \centering
  \begin{minipage}{0.32\linewidth}
    (a)\\[1mm]
    \includegraphics[width=\linewidth]{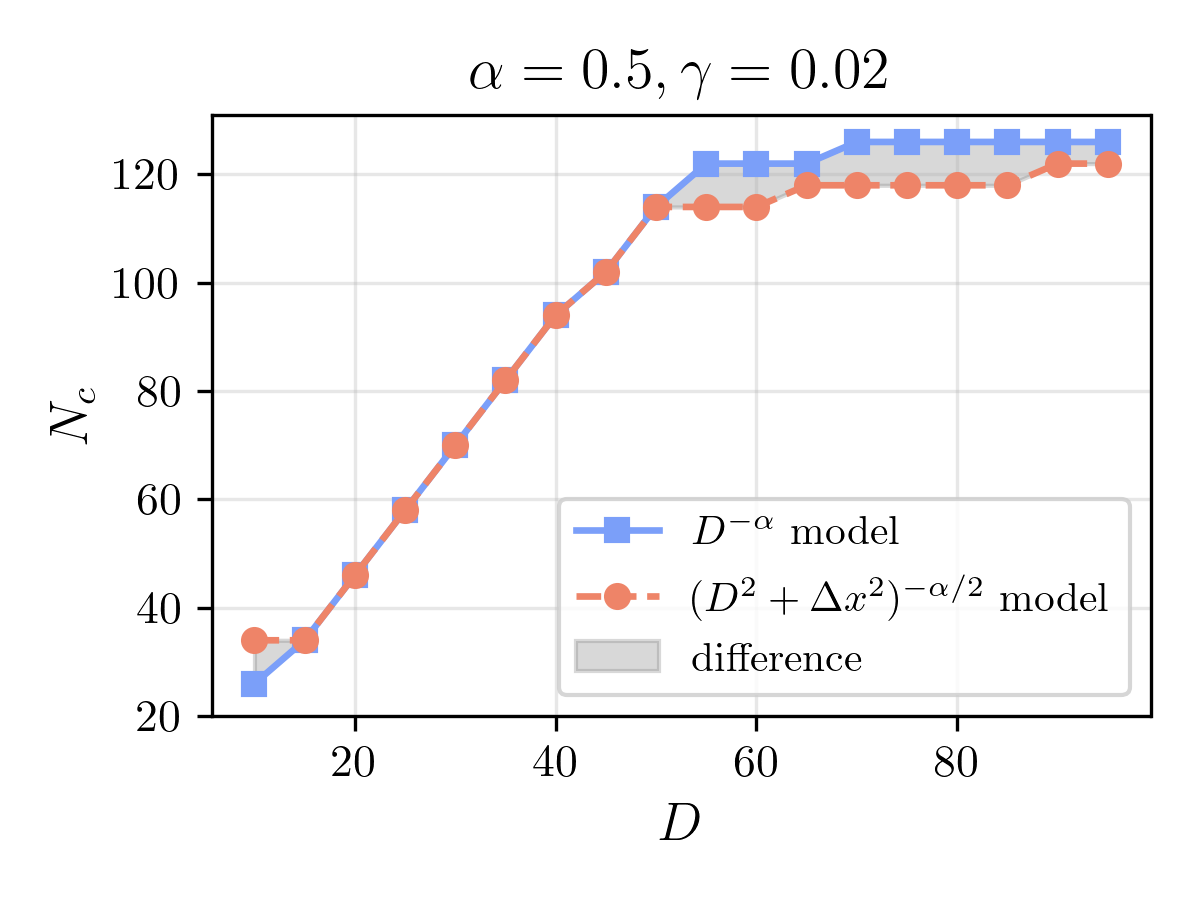}
  \end{minipage}
  \hfill
  \begin{minipage}{0.32\linewidth}
    (b)\\[1mm]
    \includegraphics[width=\linewidth]{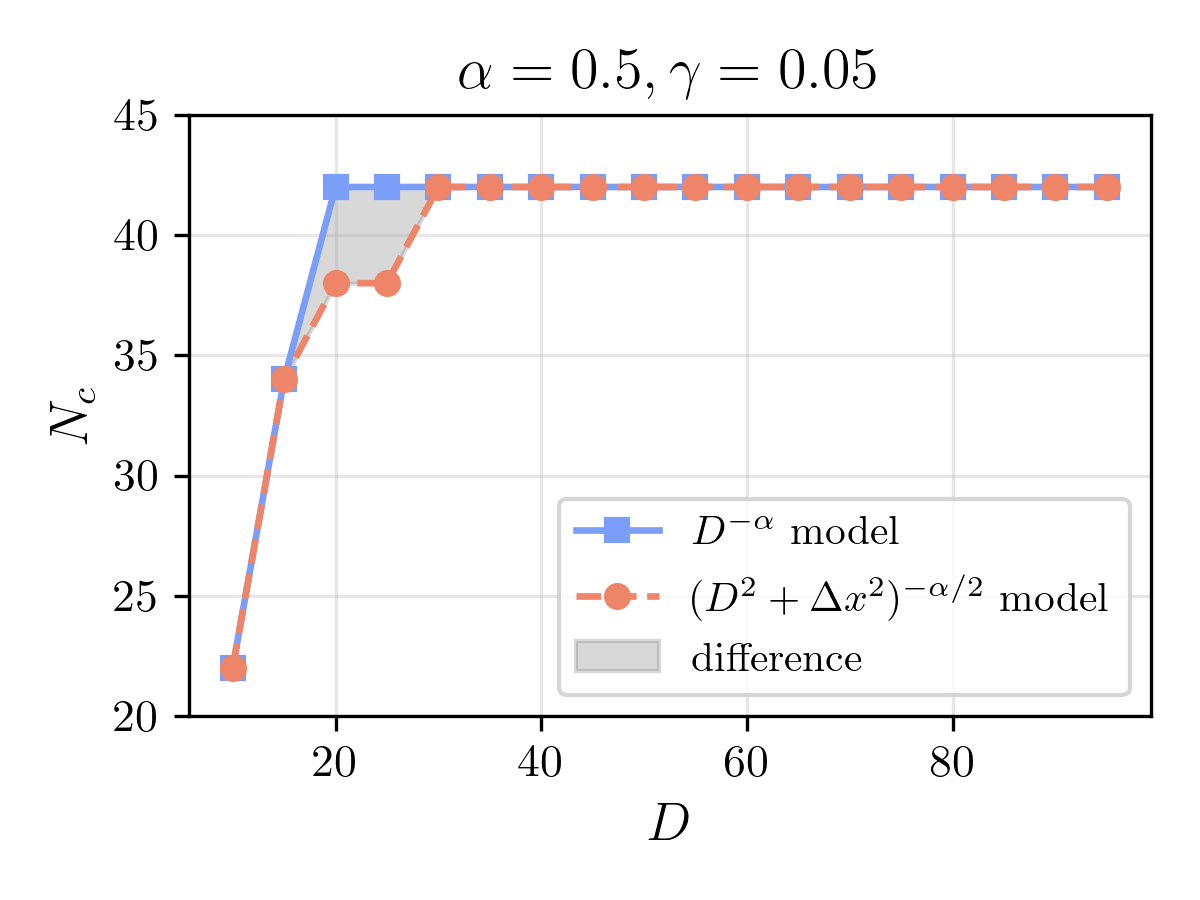}
  \end{minipage}
  \hfill
  \begin{minipage}{0.32\linewidth}
    (c)\\[1mm]
    \includegraphics[width=\linewidth]{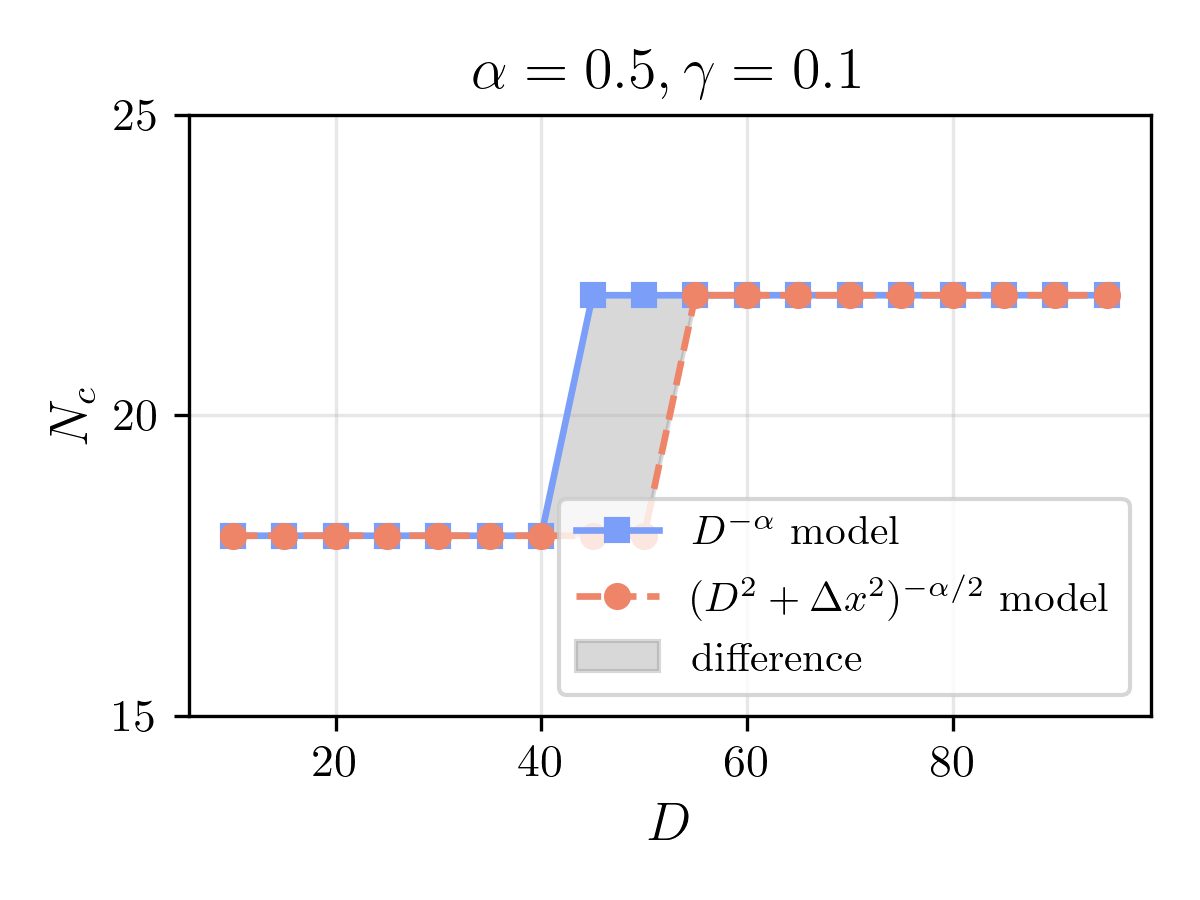}
  \end{minipage}
  \vspace{2mm}
  \begin{minipage}{0.32\linewidth}
    (d)\\[1mm]
    \includegraphics[width=\linewidth]{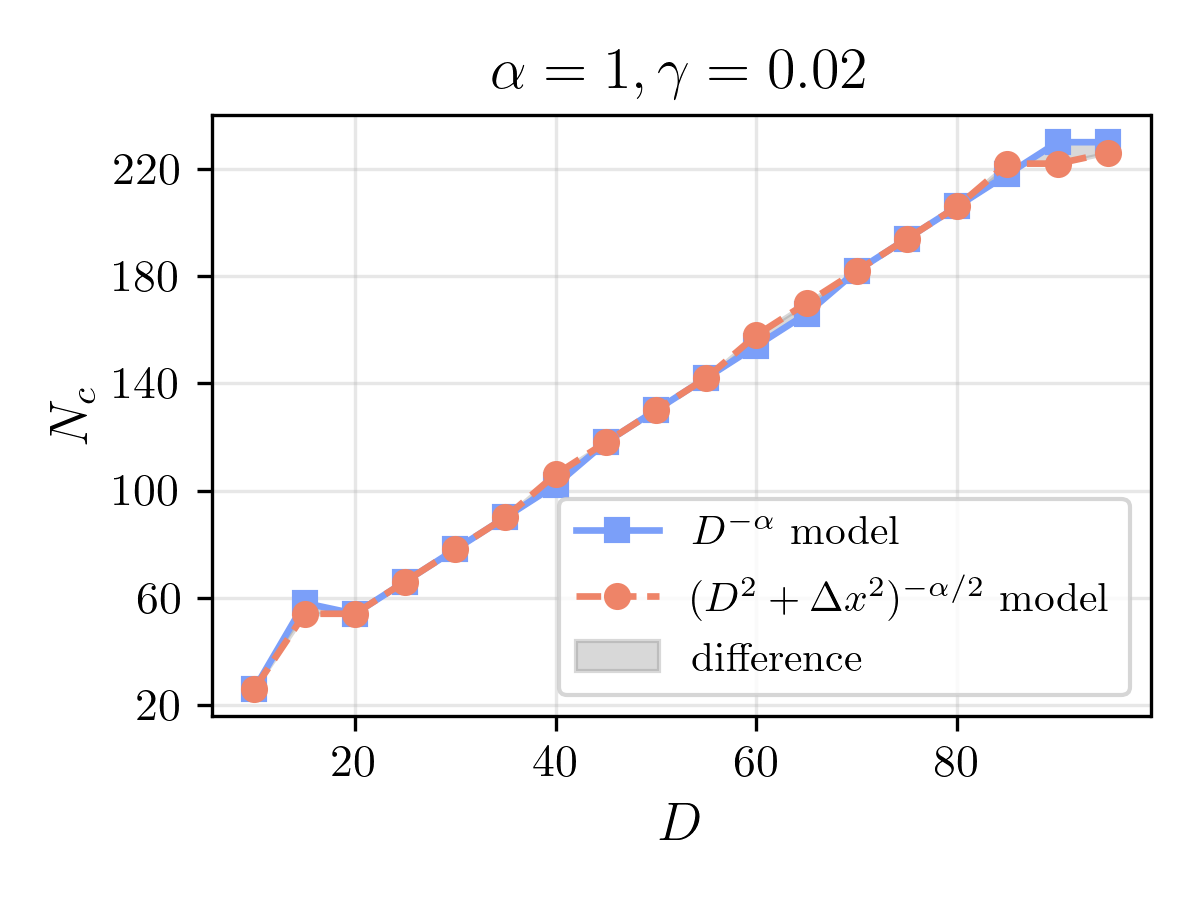}
  \end{minipage}
  \hfill
  \begin{minipage}{0.32\linewidth}
    (e)\\[1mm]
    \includegraphics[width=\linewidth]{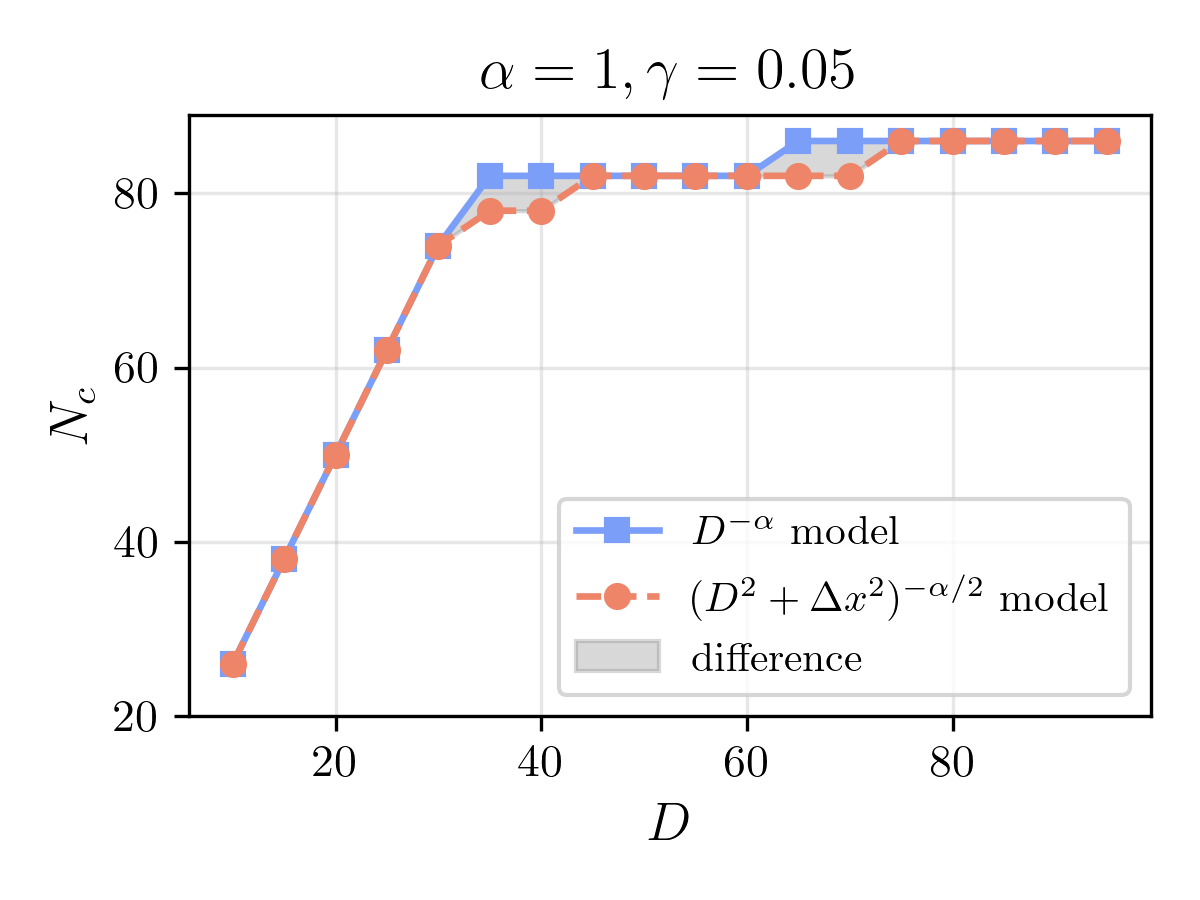}
  \end{minipage}
  \hfill
  \begin{minipage}{0.32\linewidth}
    (f)\\[1mm]
    \includegraphics[width=\linewidth]{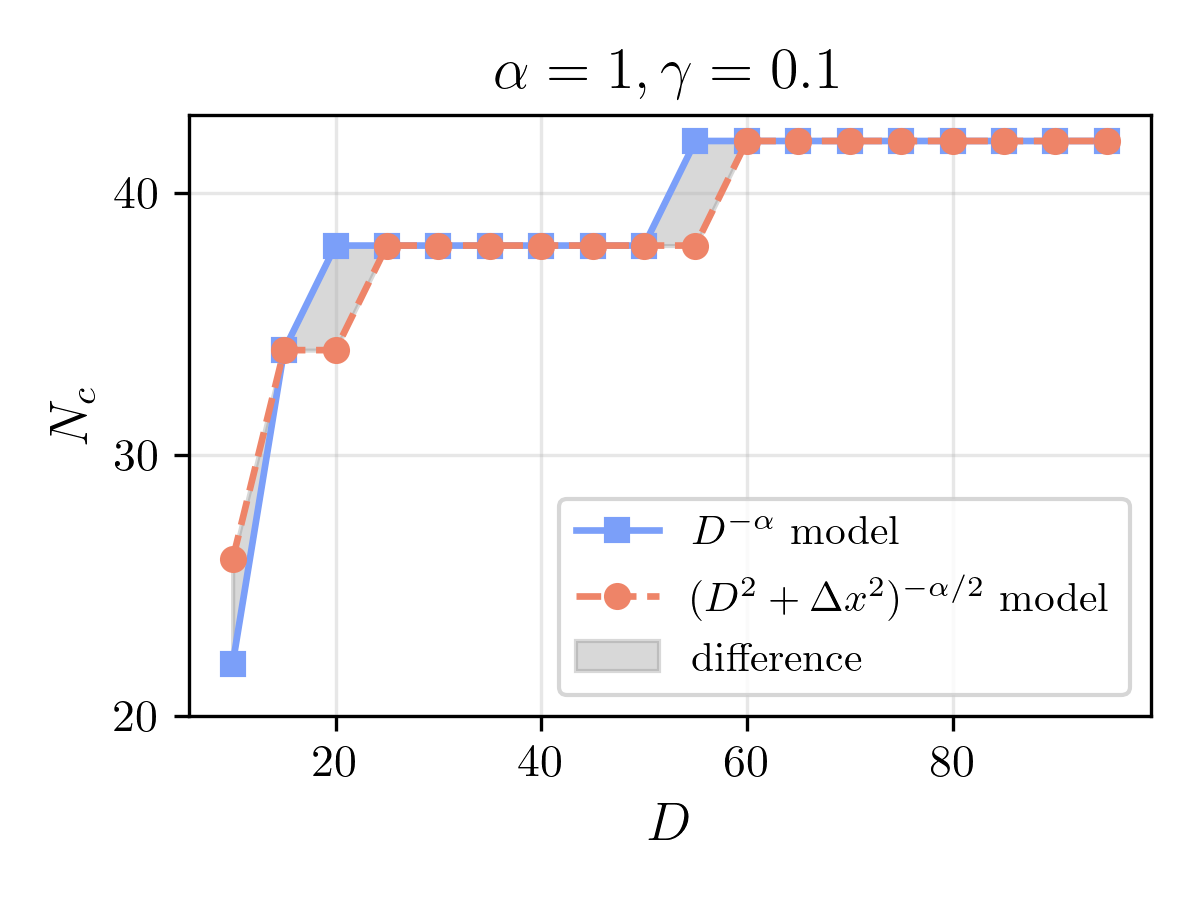}
  \end{minipage}
  \vspace{2mm}
  \begin{minipage}{0.32\linewidth}
    (g)\\[1mm]
    \includegraphics[width=\linewidth]{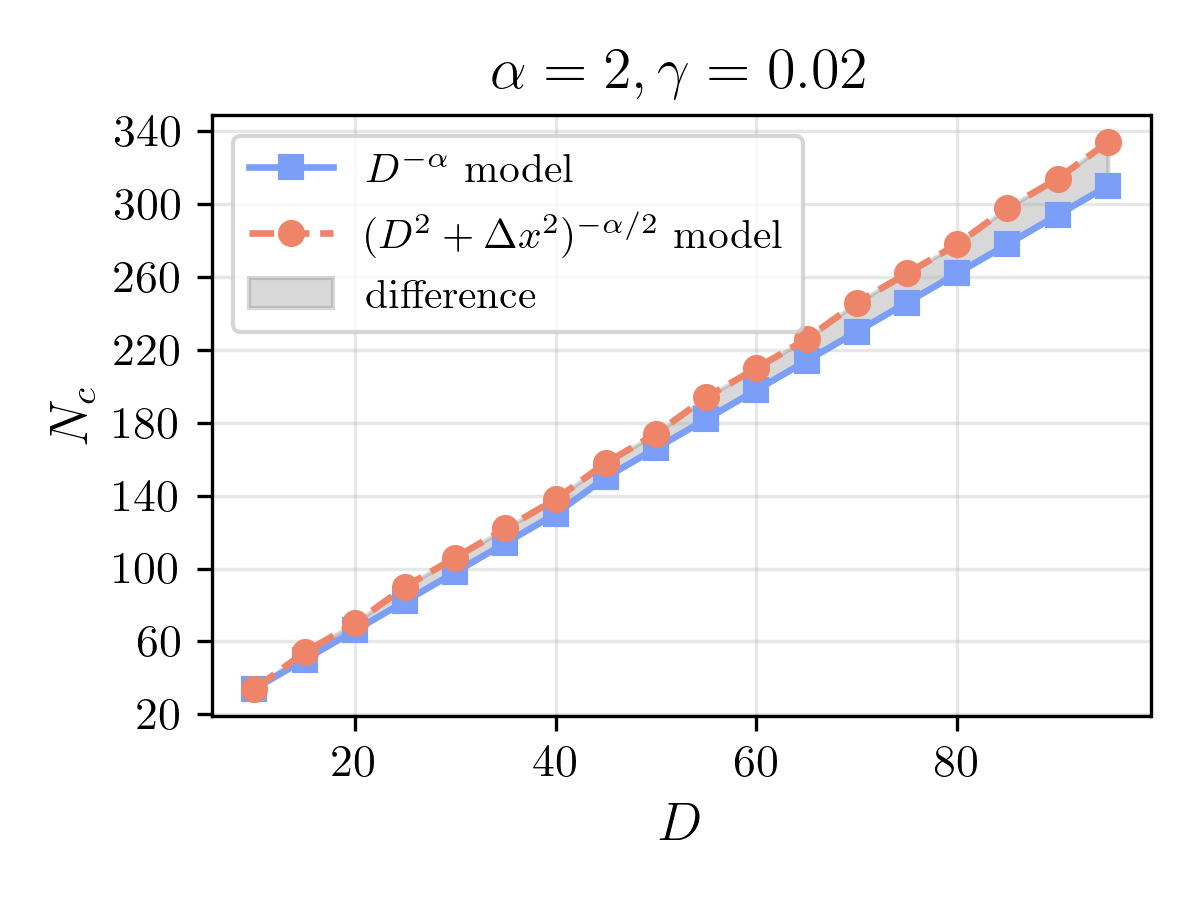}
  \end{minipage}
  \hfill
  \begin{minipage}{0.32\linewidth}
    (h)\\[1mm]
    \includegraphics[width=\linewidth]{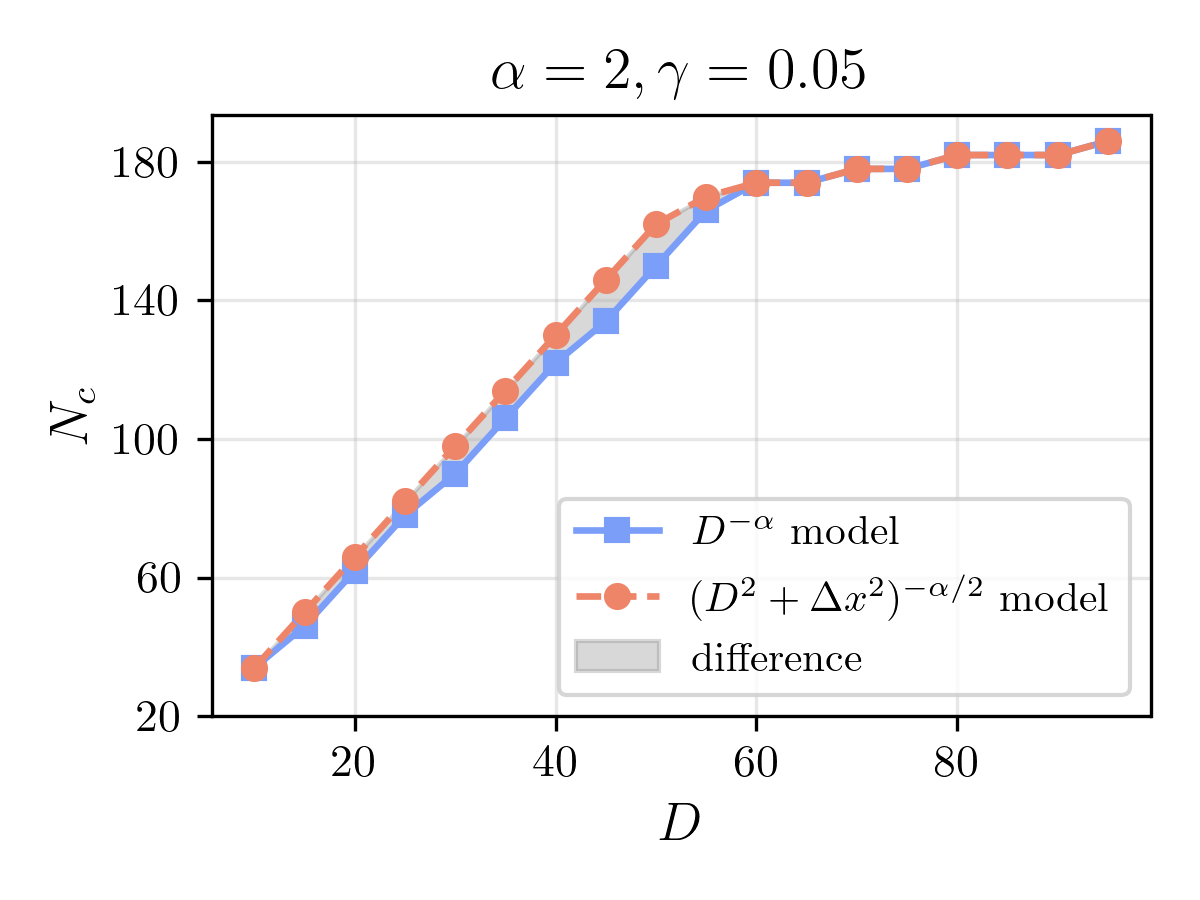}
  \end{minipage}
  \hfill
  \begin{minipage}{0.32\linewidth}
    (i)\\[1mm]
    \includegraphics[width=\linewidth]{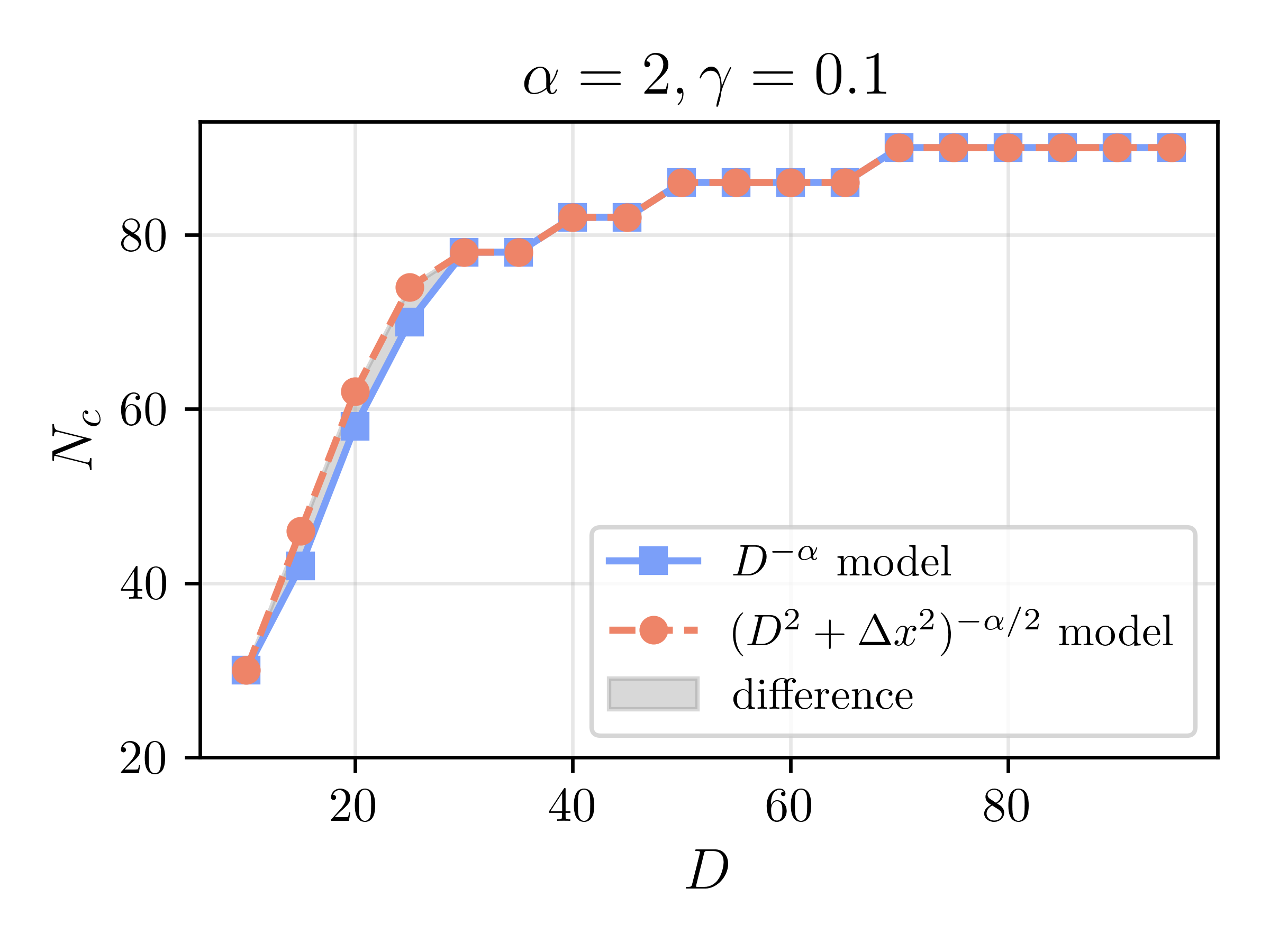}
  \end{minipage}
  \caption{Comparison between the distance-dependent inter-chain kernel $(D^{2}+\Delta x^{2})^{-\alpha/2}$ (red circles; used in the main text numerics) and the simplified constant-kernel $D^{-\alpha}$ (blue squares; used for analytic estimates in this Appendix). Plotted is the critical length $N_c(D)$ under OBCs for several $\alpha$ and $\gamma$. The two prescriptions agree quantitatively for $\alpha\ge 1$, and differ only slightly for $\alpha=0.5$ at the largest $D$ and $\gamma$ considered.}
  \label{app_fig:dist_depend_compare}
\end{figure}

Fig.~\ref{app_fig:dist_depend_compare} benchmarks these two kernel prescriptions by comparing the numerically extracted critical length $N_c(D)$ under OBCs. The numerical scans are performed for $\alpha\in\{0.5,1,2\}$ and $\gamma\in\{0.02,0.05,0.1\}$.

The comparison shows excellent quantitative agreement for $\alpha\ge 1$. For the smallest exponent $\alpha=0.5$, minor deviations appear only at the largest separations $D$ and the largest $\gamma$ considered. Therefore, the simplified kernel in Eq.~\eqref{eq:interchain_simplified} is justified for the analytic work, as it captures the same scaling laws and provides accurate estimates of $N_c$ in the parameter regimes of interest. The main numerical results in the paper remain based on the more physically distance-dependent coupling in Eq.~\eqref{eq:interchain_distance}.

Taken together, this benchmark shows that the simplified inter-chain kernel reproduces the same scaling behavior as the full distance-dependent form within the explored parameter range. Consequently, the logarithmic--algebraic--scale-covariant hierarchy and the main-text analytical--numerical agreement do not originate from replacing the physical rung distance by a constant; that replacement only makes the collective projection analytically transparent.


\begin{thebibliography}{154}%
\makeatletter
\providecommand \@ifxundefined [1]{%
 \@ifx{#1\undefined}
}%
\providecommand \@ifnum [1]{%
 \ifnum #1\expandafter \@firstoftwo
 \else \expandafter \@secondoftwo
 \fi
}%
\providecommand \@ifx [1]{%
 \ifx #1\expandafter \@firstoftwo
 \else \expandafter \@secondoftwo
 \fi
}%
\providecommand \natexlab [1]{#1}%
\providecommand \enquote  [1]{``#1''}%
\providecommand \bibnamefont  [1]{#1}%
\providecommand \bibfnamefont [1]{#1}%
\providecommand \citenamefont [1]{#1}%
\providecommand \href@noop [0]{\@secondoftwo}%
\providecommand \href [0]{\begingroup \@sanitize@url \@href}%
\providecommand \@href[1]{\@@startlink{#1}\@@href}%
\providecommand \@@href[1]{\endgroup#1\@@endlink}%
\providecommand \@sanitize@url [0]{\catcode `\\12\catcode `\$12\catcode `\&12\catcode `\#12\catcode `\^12\catcode `\_12\catcode `\%12\relax}%
\providecommand \@@startlink[1]{}%
\providecommand \@@endlink[0]{}%
\providecommand \url  [0]{\begingroup\@sanitize@url \@url }%
\providecommand \@url [1]{\endgroup\@href {#1}{\urlprefix }}%
\providecommand \urlprefix  [0]{URL }%
\providecommand \Eprint [0]{\href }%
\providecommand \doibase [0]{https://doi.org/}%
\providecommand \selectlanguage [0]{\@gobble}%
\providecommand \bibinfo  [0]{\@secondoftwo}%
\providecommand \bibfield  [0]{\@secondoftwo}%
\providecommand \translation [1]{[#1]}%
\providecommand \BibitemOpen [0]{}%
\providecommand \bibitemStop [0]{}%
\providecommand \bibitemNoStop [0]{.\EOS\space}%
\providecommand \EOS [0]{\spacefactor3000\relax}%
\providecommand \BibitemShut  [1]{\csname bibitem#1\endcsname}%
\let\auto@bib@innerbib\@empty
\bibitem [{\citenamefont {Lee}(2016)}]{lee2016anomalous}%
  \BibitemOpen
  \bibfield  {author} {\bibinfo {author} {\bibfnamefont {T.~E.}\ \bibnamefont {Lee}},\ }\bibfield  {title} {\bibinfo {title} {Anomalous edge state in a non-{H}ermitian lattice},\ }\href {https://doi.org/10.1103/PhysRevLett.116.133903} {\bibfield  {journal} {\bibinfo  {journal} {Phys. Rev. Lett.}\ }\textbf {\bibinfo {volume} {116}},\ \bibinfo {pages} {133903} (\bibinfo {year} {2016})}\BibitemShut {NoStop}%
\bibitem [{\citenamefont {Alvarez}\ \emph {et~al.}(2018)\citenamefont {Alvarez}, \citenamefont {Vargas},\ and\ \citenamefont {Torres}}]{alvarez2018non}%
  \BibitemOpen
  \bibfield  {author} {\bibinfo {author} {\bibfnamefont {V.~M.}\ \bibnamefont {Alvarez}}, \bibinfo {author} {\bibfnamefont {J.~B.}\ \bibnamefont {Vargas}},\ and\ \bibinfo {author} {\bibfnamefont {L.~F.}\ \bibnamefont {Torres}},\ }\bibfield  {title} {\bibinfo {title} {Non-{H}ermitian robust edge states in one dimension: Anomalous localization and eigenspace condensation at exceptional points},\ }\href {https://doi.org/10.1103/PhysRevB.97.121401} {\bibfield  {journal} {\bibinfo  {journal} {Phys. Rev. B}\ }\textbf {\bibinfo {volume} {97}},\ \bibinfo {pages} {121401} (\bibinfo {year} {2018})}\BibitemShut {NoStop}%
\bibitem [{\citenamefont {Yao}\ and\ \citenamefont {Wang}(2018)}]{yao2018edge}%
  \BibitemOpen
  \bibfield  {author} {\bibinfo {author} {\bibfnamefont {S.}~\bibnamefont {Yao}}\ and\ \bibinfo {author} {\bibfnamefont {Z.}~\bibnamefont {Wang}},\ }\bibfield  {title} {\bibinfo {title} {Edge states and topological invariants of non-{H}ermitian systems},\ }\href {https://doi.org/10.1103/PhysRevLett.121.086803} {\bibfield  {journal} {\bibinfo  {journal} {Phys. Rev. Lett.}\ }\textbf {\bibinfo {volume} {121}},\ \bibinfo {pages} {086803} (\bibinfo {year} {2018})}\BibitemShut {NoStop}%
\bibitem [{\citenamefont {Kunst}\ \emph {et~al.}(2018)\citenamefont {Kunst}, \citenamefont {Edvardsson}, \citenamefont {Budich},\ and\ \citenamefont {Bergholtz}}]{kunst2018biorthogonal}%
  \BibitemOpen
  \bibfield  {author} {\bibinfo {author} {\bibfnamefont {F.~K.}\ \bibnamefont {Kunst}}, \bibinfo {author} {\bibfnamefont {E.}~\bibnamefont {Edvardsson}}, \bibinfo {author} {\bibfnamefont {J.~C.}\ \bibnamefont {Budich}},\ and\ \bibinfo {author} {\bibfnamefont {E.~J.}\ \bibnamefont {Bergholtz}},\ }\bibfield  {title} {\bibinfo {title} {Biorthogonal bulk-boundary correspondence in non-{H}ermitian systems},\ }\href {https://doi.org/10.1103/PhysRevLett.121.026808} {\bibfield  {journal} {\bibinfo  {journal} {Phys. Rev. Lett.}\ }\textbf {\bibinfo {volume} {121}},\ \bibinfo {pages} {026808} (\bibinfo {year} {2018})}\BibitemShut {NoStop}%
\bibitem [{\citenamefont {Lee}\ and\ \citenamefont {Thomale}(2019)}]{lee2019anatomy}%
  \BibitemOpen
  \bibfield  {author} {\bibinfo {author} {\bibfnamefont {C.~H.}\ \bibnamefont {Lee}}\ and\ \bibinfo {author} {\bibfnamefont {R.}~\bibnamefont {Thomale}},\ }\bibfield  {title} {\bibinfo {title} {Anatomy of skin modes and topology in non-{H}ermitian systems},\ }\href {https://doi.org/10.1103/PhysRevB.99.201103} {\bibfield  {journal} {\bibinfo  {journal} {Phys. Rev. B}\ }\textbf {\bibinfo {volume} {99}},\ \bibinfo {pages} {201103} (\bibinfo {year} {2019})}\BibitemShut {NoStop}%
\bibitem [{\citenamefont {Okuma}\ \emph {et~al.}(2020)\citenamefont {Okuma}, \citenamefont {Kawabata}, \citenamefont {Shiozaki},\ and\ \citenamefont {Sato}}]{okuma2020topological}%
  \BibitemOpen
  \bibfield  {author} {\bibinfo {author} {\bibfnamefont {N.}~\bibnamefont {Okuma}}, \bibinfo {author} {\bibfnamefont {K.}~\bibnamefont {Kawabata}}, \bibinfo {author} {\bibfnamefont {K.}~\bibnamefont {Shiozaki}},\ and\ \bibinfo {author} {\bibfnamefont {M.}~\bibnamefont {Sato}},\ }\bibfield  {title} {\bibinfo {title} {Topological origin of non-{H}ermitian skin effects},\ }\href {https://doi.org/10.1103/PhysRevLett.124.086801} {\bibfield  {journal} {\bibinfo  {journal} {Phys. Rev. Lett.}\ }\textbf {\bibinfo {volume} {124}},\ \bibinfo {pages} {086801} (\bibinfo {year} {2020})}\BibitemShut {NoStop}%
\bibitem [{\citenamefont {Lin}\ \emph {et~al.}(2023)\citenamefont {Lin}, \citenamefont {Tai}, \citenamefont {Li},\ and\ \citenamefont {Lee}}]{lin2023topological}%
  \BibitemOpen
  \bibfield  {author} {\bibinfo {author} {\bibfnamefont {R.}~\bibnamefont {Lin}}, \bibinfo {author} {\bibfnamefont {T.}~\bibnamefont {Tai}}, \bibinfo {author} {\bibfnamefont {L.}~\bibnamefont {Li}},\ and\ \bibinfo {author} {\bibfnamefont {C.~H.}\ \bibnamefont {Lee}},\ }\bibfield  {title} {\bibinfo {title} {Topological non-{H}ermitian skin effect},\ }\href {https://doi.org/10.1007/s11467-023-1309-z} {\bibfield  {journal} {\bibinfo  {journal} {Front. Phys.}\ }\textbf {\bibinfo {volume} {18}},\ \bibinfo {pages} {53605} (\bibinfo {year} {2023})}\BibitemShut {NoStop}%
\bibitem [{\citenamefont {Okuma}\ and\ \citenamefont {Sato}(2023)}]{okuma2023non}%
  \BibitemOpen
  \bibfield  {author} {\bibinfo {author} {\bibfnamefont {N.}~\bibnamefont {Okuma}}\ and\ \bibinfo {author} {\bibfnamefont {M.}~\bibnamefont {Sato}},\ }\bibfield  {title} {\bibinfo {title} {Non-{H}ermitian topological phenomena: A review},\ }\href {https://doi.org/10.1146/annurev-conmatphys-040521-033133} {\bibfield  {journal} {\bibinfo  {journal} {Annu. Rev. Condens. Matter Phys.}\ }\textbf {\bibinfo {volume} {14}},\ \bibinfo {pages} {83} (\bibinfo {year} {2023})}\BibitemShut {NoStop}%
\bibitem [{\citenamefont {Longhi}(2019)}]{longhi2019topological}%
  \BibitemOpen
  \bibfield  {author} {\bibinfo {author} {\bibfnamefont {S.}~\bibnamefont {Longhi}},\ }\bibfield  {title} {\bibinfo {title} {Topological phase transition in non-{H}ermitian quasicrystals},\ }\href {https://doi.org/10.1103/PhysRevLett.122.237601} {\bibfield  {journal} {\bibinfo  {journal} {Phys. Rev. Lett.}\ }\textbf {\bibinfo {volume} {122}},\ \bibinfo {pages} {237601} (\bibinfo {year} {2019})}\BibitemShut {NoStop}%
\bibitem [{\citenamefont {Kawabata}\ \emph {et~al.}(2019)\citenamefont {Kawabata}, \citenamefont {Shiozaki}, \citenamefont {Ueda},\ and\ \citenamefont {Sato}}]{kawabata2019symmetry}%
  \BibitemOpen
  \bibfield  {author} {\bibinfo {author} {\bibfnamefont {K.}~\bibnamefont {Kawabata}}, \bibinfo {author} {\bibfnamefont {K.}~\bibnamefont {Shiozaki}}, \bibinfo {author} {\bibfnamefont {M.}~\bibnamefont {Ueda}},\ and\ \bibinfo {author} {\bibfnamefont {M.}~\bibnamefont {Sato}},\ }\bibfield  {title} {\bibinfo {title} {Symmetry and topology in non-{H}ermitian physics},\ }\href {https://doi.org/10.1103/PhysRevX.9.041015} {\bibfield  {journal} {\bibinfo  {journal} {Phys. Rev. X}\ }\textbf {\bibinfo {volume} {9}},\ \bibinfo {pages} {041015} (\bibinfo {year} {2019})}\BibitemShut {NoStop}%
\bibitem [{\citenamefont {Lee}\ \emph {et~al.}(2019)\citenamefont {Lee}, \citenamefont {Li},\ and\ \citenamefont {Gong}}]{lee2019hybrid}%
  \BibitemOpen
  \bibfield  {author} {\bibinfo {author} {\bibfnamefont {C.~H.}\ \bibnamefont {Lee}}, \bibinfo {author} {\bibfnamefont {L.}~\bibnamefont {Li}},\ and\ \bibinfo {author} {\bibfnamefont {J.}~\bibnamefont {Gong}},\ }\bibfield  {title} {\bibinfo {title} {Hybrid higher-order skin-topological modes in nonreciprocal systems},\ }\href {https://doi.org/10.1103/PhysRevLett.123.016805} {\bibfield  {journal} {\bibinfo  {journal} {Phys. Rev. Lett.}\ }\textbf {\bibinfo {volume} {123}},\ \bibinfo {pages} {016805} (\bibinfo {year} {2019})}\BibitemShut {NoStop}%
\bibitem [{\citenamefont {Song}\ \emph {et~al.}(2019)\citenamefont {Song}, \citenamefont {Yao},\ and\ \citenamefont {Wang}}]{song2019non}%
  \BibitemOpen
  \bibfield  {author} {\bibinfo {author} {\bibfnamefont {F.}~\bibnamefont {Song}}, \bibinfo {author} {\bibfnamefont {S.}~\bibnamefont {Yao}},\ and\ \bibinfo {author} {\bibfnamefont {Z.}~\bibnamefont {Wang}},\ }\bibfield  {title} {\bibinfo {title} {Non-{H}ermitian skin effect and chiral damping in open quantum systems},\ }\href {https://doi.org/10.1103/PhysRevLett.123.170401} {\bibfield  {journal} {\bibinfo  {journal} {Phys. Rev. Lett.}\ }\textbf {\bibinfo {volume} {123}},\ \bibinfo {pages} {170401} (\bibinfo {year} {2019})}\BibitemShut {NoStop}%
\bibitem [{\citenamefont {Borgnia}\ \emph {et~al.}(2020)\citenamefont {Borgnia}, \citenamefont {Kruchkov},\ and\ \citenamefont {Slager}}]{borgnia2020non}%
  \BibitemOpen
  \bibfield  {author} {\bibinfo {author} {\bibfnamefont {D.~S.}\ \bibnamefont {Borgnia}}, \bibinfo {author} {\bibfnamefont {A.~J.}\ \bibnamefont {Kruchkov}},\ and\ \bibinfo {author} {\bibfnamefont {R.-J.}\ \bibnamefont {Slager}},\ }\bibfield  {title} {\bibinfo {title} {Non-{H}ermitian boundary modes and topology},\ }\href {https://doi.org/10.1103/PhysRevLett.124.056802} {\bibfield  {journal} {\bibinfo  {journal} {Phys. Rev. Lett.}\ }\textbf {\bibinfo {volume} {124}},\ \bibinfo {pages} {056802} (\bibinfo {year} {2020})}\BibitemShut {NoStop}%
\bibitem [{\citenamefont {Helbig}\ \emph {et~al.}(2020)\citenamefont {Helbig}, \citenamefont {Hofmann}, \citenamefont {Imhof}, \citenamefont {Abdelghany}, \citenamefont {Kiessling}, \citenamefont {Molenkamp}, \citenamefont {Lee}, \citenamefont {Szameit}, \citenamefont {Greiter},\ and\ \citenamefont {Thomale}}]{helbig2020generalized}%
  \BibitemOpen
  \bibfield  {author} {\bibinfo {author} {\bibfnamefont {T.}~\bibnamefont {Helbig}}, \bibinfo {author} {\bibfnamefont {T.}~\bibnamefont {Hofmann}}, \bibinfo {author} {\bibfnamefont {S.}~\bibnamefont {Imhof}}, \bibinfo {author} {\bibfnamefont {M.}~\bibnamefont {Abdelghany}}, \bibinfo {author} {\bibfnamefont {T.}~\bibnamefont {Kiessling}}, \bibinfo {author} {\bibfnamefont {L.}~\bibnamefont {Molenkamp}}, \bibinfo {author} {\bibfnamefont {C.}~\bibnamefont {Lee}}, \bibinfo {author} {\bibfnamefont {A.}~\bibnamefont {Szameit}}, \bibinfo {author} {\bibfnamefont {M.}~\bibnamefont {Greiter}},\ and\ \bibinfo {author} {\bibfnamefont {R.}~\bibnamefont {Thomale}},\ }\bibfield  {title} {\bibinfo {title} {Generalized bulk--boundary correspondence in non-{H}ermitian topolectrical circuits},\ }\href {https://doi.org/10.1038/s41567-020-0922-9} {\bibfield  {journal} {\bibinfo  {journal} {Nat. Phys.}\ }\textbf {\bibinfo {volume} {16}},\ \bibinfo {pages} {747} (\bibinfo {year} {2020})}\BibitemShut {NoStop}%
\bibitem [{\citenamefont {Li}\ \emph {et~al.}(2020)\citenamefont {Li}, \citenamefont {Lee}, \citenamefont {Mu},\ and\ \citenamefont {Gong}}]{li2020critical}%
  \BibitemOpen
  \bibfield  {author} {\bibinfo {author} {\bibfnamefont {L.}~\bibnamefont {Li}}, \bibinfo {author} {\bibfnamefont {C.~H.}\ \bibnamefont {Lee}}, \bibinfo {author} {\bibfnamefont {S.}~\bibnamefont {Mu}},\ and\ \bibinfo {author} {\bibfnamefont {J.}~\bibnamefont {Gong}},\ }\bibfield  {title} {\bibinfo {title} {Critical non-{H}ermitian skin effect},\ }\href {https://doi.org/10.1038/s41467-020-18917-4} {\bibfield  {journal} {\bibinfo  {journal} {Nat. Commun.}\ }\textbf {\bibinfo {volume} {11}},\ \bibinfo {pages} {5491} (\bibinfo {year} {2020})}\BibitemShut {NoStop}%
\bibitem [{\citenamefont {Longhi}(2020)}]{longhi2020non}%
  \BibitemOpen
  \bibfield  {author} {\bibinfo {author} {\bibfnamefont {S.}~\bibnamefont {Longhi}},\ }\bibfield  {title} {\bibinfo {title} {Non-{B}loch-band collapse and chiral {Z}ener tunneling},\ }\href {https://doi.org/10.1103/PhysRevLett.124.066602} {\bibfield  {journal} {\bibinfo  {journal} {Phys. Rev. Lett.}\ }\textbf {\bibinfo {volume} {124}},\ \bibinfo {pages} {066602} (\bibinfo {year} {2020})}\BibitemShut {NoStop}%
\bibitem [{\citenamefont {Lee}\ \emph {et~al.}(2020)\citenamefont {Lee}, \citenamefont {Li}, \citenamefont {Thomale},\ and\ \citenamefont {Gong}}]{lee2020unraveling}%
  \BibitemOpen
  \bibfield  {author} {\bibinfo {author} {\bibfnamefont {C.~H.}\ \bibnamefont {Lee}}, \bibinfo {author} {\bibfnamefont {L.}~\bibnamefont {Li}}, \bibinfo {author} {\bibfnamefont {R.}~\bibnamefont {Thomale}},\ and\ \bibinfo {author} {\bibfnamefont {J.}~\bibnamefont {Gong}},\ }\bibfield  {title} {\bibinfo {title} {Unraveling non-{H}ermitian pumping: emergent spectral singularities and anomalous responses},\ }\href {https://doi.org/10.1103/PhysRevB.102.085151} {\bibfield  {journal} {\bibinfo  {journal} {Phys. Rev. B}\ }\textbf {\bibinfo {volume} {102}},\ \bibinfo {pages} {085151} (\bibinfo {year} {2020})}\BibitemShut {NoStop}%
\bibitem [{\citenamefont {Zou}\ \emph {et~al.}(2021)\citenamefont {Zou}, \citenamefont {Chen}, \citenamefont {He}, \citenamefont {Bao}, \citenamefont {Lee}, \citenamefont {Sun},\ and\ \citenamefont {Zhang}}]{zou2021observation}%
  \BibitemOpen
  \bibfield  {author} {\bibinfo {author} {\bibfnamefont {D.}~\bibnamefont {Zou}}, \bibinfo {author} {\bibfnamefont {T.}~\bibnamefont {Chen}}, \bibinfo {author} {\bibfnamefont {W.}~\bibnamefont {He}}, \bibinfo {author} {\bibfnamefont {J.}~\bibnamefont {Bao}}, \bibinfo {author} {\bibfnamefont {C.~H.}\ \bibnamefont {Lee}}, \bibinfo {author} {\bibfnamefont {H.}~\bibnamefont {Sun}},\ and\ \bibinfo {author} {\bibfnamefont {X.}~\bibnamefont {Zhang}},\ }\bibfield  {title} {\bibinfo {title} {Observation of hybrid higher-order skin-topological effect in non-{H}ermitian topolectrical circuits},\ }\href {https://doi.org/10.1038/s41467-021-26414-5} {\bibfield  {journal} {\bibinfo  {journal} {Nat. Commun.}\ }\textbf {\bibinfo {volume} {12}},\ \bibinfo {pages} {7201} (\bibinfo {year} {2021})}\BibitemShut {NoStop}%
\bibitem [{\citenamefont {Zhang}\ \emph {et~al.}(2021{\natexlab{a}})\citenamefont {Zhang}, \citenamefont {Tian}, \citenamefont {Jiang}, \citenamefont {Lu},\ and\ \citenamefont {Chen}}]{zhang2021observation}%
  \BibitemOpen
  \bibfield  {author} {\bibinfo {author} {\bibfnamefont {X.}~\bibnamefont {Zhang}}, \bibinfo {author} {\bibfnamefont {Y.}~\bibnamefont {Tian}}, \bibinfo {author} {\bibfnamefont {J.-H.}\ \bibnamefont {Jiang}}, \bibinfo {author} {\bibfnamefont {M.-H.}\ \bibnamefont {Lu}},\ and\ \bibinfo {author} {\bibfnamefont {Y.-F.}\ \bibnamefont {Chen}},\ }\bibfield  {title} {\bibinfo {title} {Observation of higher-order non-{H}ermitian skin effect},\ }\href {https://doi.org/10.1038/s41467-021-25716-y} {\bibfield  {journal} {\bibinfo  {journal} {Nat. Commun.}\ }\textbf {\bibinfo {volume} {12}},\ \bibinfo {pages} {5377} (\bibinfo {year} {2021}{\natexlab{a}})}\BibitemShut {NoStop}%
\bibitem [{\citenamefont {Yang}\ \emph {et~al.}(2022{\natexlab{a}})\citenamefont {Yang}, \citenamefont {Wang}, \citenamefont {Wu}, \citenamefont {Xiao}, \citenamefont {Yu}, \citenamefont {Yuan},\ and\ \citenamefont {Chen}}]{yang2022concentrated}%
  \BibitemOpen
  \bibfield  {author} {\bibinfo {author} {\bibfnamefont {M.}~\bibnamefont {Yang}}, \bibinfo {author} {\bibfnamefont {L.}~\bibnamefont {Wang}}, \bibinfo {author} {\bibfnamefont {X.}~\bibnamefont {Wu}}, \bibinfo {author} {\bibfnamefont {H.}~\bibnamefont {Xiao}}, \bibinfo {author} {\bibfnamefont {D.}~\bibnamefont {Yu}}, \bibinfo {author} {\bibfnamefont {L.}~\bibnamefont {Yuan}},\ and\ \bibinfo {author} {\bibfnamefont {X.}~\bibnamefont {Chen}},\ }\bibfield  {title} {\bibinfo {title} {Concentrated subradiant modes in a one-dimensional atomic array coupled with chiral waveguides},\ }\href {https://doi.org/10.1103/PhysRevA.106.043717} {\bibfield  {journal} {\bibinfo  {journal} {Phys. Rev. A}\ }\textbf {\bibinfo {volume} {106}},\ \bibinfo {pages} {043717} (\bibinfo {year} {2022}{\natexlab{a}})}\BibitemShut {NoStop}%
\bibitem [{\citenamefont {Zhang}\ \emph {et~al.}(2021{\natexlab{b}})\citenamefont {Zhang}, \citenamefont {Li}, \citenamefont {Liu}, \citenamefont {Tai}, \citenamefont {Thomale},\ and\ \citenamefont {Lee}}]{zhang2021tidal}%
  \BibitemOpen
  \bibfield  {author} {\bibinfo {author} {\bibfnamefont {X.}~\bibnamefont {Zhang}}, \bibinfo {author} {\bibfnamefont {G.}~\bibnamefont {Li}}, \bibinfo {author} {\bibfnamefont {Y.}~\bibnamefont {Liu}}, \bibinfo {author} {\bibfnamefont {T.}~\bibnamefont {Tai}}, \bibinfo {author} {\bibfnamefont {R.}~\bibnamefont {Thomale}},\ and\ \bibinfo {author} {\bibfnamefont {C.~H.}\ \bibnamefont {Lee}},\ }\bibfield  {title} {\bibinfo {title} {Tidal surface states as fingerprints of non-{H}ermitian nodal knot metals},\ }\href {https://doi.org/10.1038/s42005-021-00535-1} {\bibfield  {journal} {\bibinfo  {journal} {Commun. Phys.}\ }\textbf {\bibinfo {volume} {4}},\ \bibinfo {pages} {47} (\bibinfo {year} {2021}{\natexlab{b}})}\BibitemShut {NoStop}%
\bibitem [{\citenamefont {Li}\ and\ \citenamefont {Lee}(2022)}]{li2022non}%
  \BibitemOpen
  \bibfield  {author} {\bibinfo {author} {\bibfnamefont {L.}~\bibnamefont {Li}}\ and\ \bibinfo {author} {\bibfnamefont {C.~H.}\ \bibnamefont {Lee}},\ }\bibfield  {title} {\bibinfo {title} {Non-{H}ermitian pseudo-gaps},\ }\href {https://doi.org/10.1016/j.scib.2022.01.017} {\bibfield  {journal} {\bibinfo  {journal} {Sci. Bull.}\ }\textbf {\bibinfo {volume} {67}},\ \bibinfo {pages} {685} (\bibinfo {year} {2022})}\BibitemShut {NoStop}%
\bibitem [{\citenamefont {Yang}\ \emph {et~al.}(2022{\natexlab{b}})\citenamefont {Yang}, \citenamefont {Tan}, \citenamefont {Tai}, \citenamefont {Koh}, \citenamefont {Li}, \citenamefont {Longhi},\ and\ \citenamefont {Lee}}]{yang2022designing}%
  \BibitemOpen
  \bibfield  {author} {\bibinfo {author} {\bibfnamefont {R.}~\bibnamefont {Yang}}, \bibinfo {author} {\bibfnamefont {J.~W.}\ \bibnamefont {Tan}}, \bibinfo {author} {\bibfnamefont {T.}~\bibnamefont {Tai}}, \bibinfo {author} {\bibfnamefont {J.~M.}\ \bibnamefont {Koh}}, \bibinfo {author} {\bibfnamefont {L.}~\bibnamefont {Li}}, \bibinfo {author} {\bibfnamefont {S.}~\bibnamefont {Longhi}},\ and\ \bibinfo {author} {\bibfnamefont {C.~H.}\ \bibnamefont {Lee}},\ }\bibfield  {title} {\bibinfo {title} {Designing non-{H}ermitian real spectra through electrostatics},\ }\href {https://doi.org/10.1016/j.scib.2022.08.005} {\bibfield  {journal} {\bibinfo  {journal} {Sci. Bull.}\ }\textbf {\bibinfo {volume} {67}},\ \bibinfo {pages} {1865} (\bibinfo {year} {2022}{\natexlab{b}})}\BibitemShut {NoStop}%
\bibitem [{\citenamefont {Xue}\ \emph {et~al.}(2022)\citenamefont {Xue}, \citenamefont {Hu}, \citenamefont {Song},\ and\ \citenamefont {Wang}}]{xue2022non}%
  \BibitemOpen
  \bibfield  {author} {\bibinfo {author} {\bibfnamefont {W.-T.}\ \bibnamefont {Xue}}, \bibinfo {author} {\bibfnamefont {Y.-M.}\ \bibnamefont {Hu}}, \bibinfo {author} {\bibfnamefont {F.}~\bibnamefont {Song}},\ and\ \bibinfo {author} {\bibfnamefont {Z.}~\bibnamefont {Wang}},\ }\bibfield  {title} {\bibinfo {title} {Non-{H}ermitian edge burst},\ }\href {https://doi.org/10.1103/PhysRevLett.128.120401} {\bibfield  {journal} {\bibinfo  {journal} {Phys. Rev. Lett.}\ }\textbf {\bibinfo {volume} {128}},\ \bibinfo {pages} {120401} (\bibinfo {year} {2022})}\BibitemShut {NoStop}%
\bibitem [{\citenamefont {Longhi}(2022)}]{longhi2022self}%
  \BibitemOpen
  \bibfield  {author} {\bibinfo {author} {\bibfnamefont {S.}~\bibnamefont {Longhi}},\ }\bibfield  {title} {\bibinfo {title} {Self-healing of non-{H}ermitian topological skin modes},\ }\href {https://doi.org/10.1103/PhysRevLett.128.157601} {\bibfield  {journal} {\bibinfo  {journal} {Phys. Rev. Lett.}\ }\textbf {\bibinfo {volume} {128}},\ \bibinfo {pages} {157601} (\bibinfo {year} {2022})}\BibitemShut {NoStop}%
\bibitem [{\citenamefont {Gu}\ \emph {et~al.}(2022)\citenamefont {Gu}, \citenamefont {Gao}, \citenamefont {Xue}, \citenamefont {Li}, \citenamefont {Su},\ and\ \citenamefont {Zhu}}]{gu2022transient}%
  \BibitemOpen
  \bibfield  {author} {\bibinfo {author} {\bibfnamefont {Z.}~\bibnamefont {Gu}}, \bibinfo {author} {\bibfnamefont {H.}~\bibnamefont {Gao}}, \bibinfo {author} {\bibfnamefont {H.}~\bibnamefont {Xue}}, \bibinfo {author} {\bibfnamefont {J.}~\bibnamefont {Li}}, \bibinfo {author} {\bibfnamefont {Z.}~\bibnamefont {Su}},\ and\ \bibinfo {author} {\bibfnamefont {J.}~\bibnamefont {Zhu}},\ }\bibfield  {title} {\bibinfo {title} {Transient non-{H}ermitian skin effect},\ }\href {https://doi.org/10.1038/s41467-022-35448-2} {\bibfield  {journal} {\bibinfo  {journal} {Nat. Commun.}\ }\textbf {\bibinfo {volume} {13}},\ \bibinfo {pages} {7668} (\bibinfo {year} {2022})}\BibitemShut {NoStop}%
\bibitem [{\citenamefont {Jiang}\ and\ \citenamefont {Lee}(2023)}]{jiang2023dimensional}%
  \BibitemOpen
  \bibfield  {author} {\bibinfo {author} {\bibfnamefont {H.}~\bibnamefont {Jiang}}\ and\ \bibinfo {author} {\bibfnamefont {C.~H.}\ \bibnamefont {Lee}},\ }\bibfield  {title} {\bibinfo {title} {Dimensional transmutation from non-{H}ermiticity},\ }\href {https://doi.org/10.1103/PhysRevLett.131.076401} {\bibfield  {journal} {\bibinfo  {journal} {Phys. Rev. Lett.}\ }\textbf {\bibinfo {volume} {131}},\ \bibinfo {pages} {076401} (\bibinfo {year} {2023})}\BibitemShut {NoStop}%
\bibitem [{\citenamefont {Manna}\ and\ \citenamefont {Roy}(2023)}]{manna2023inner}%
  \BibitemOpen
  \bibfield  {author} {\bibinfo {author} {\bibfnamefont {S.}~\bibnamefont {Manna}}\ and\ \bibinfo {author} {\bibfnamefont {B.}~\bibnamefont {Roy}},\ }\bibfield  {title} {\bibinfo {title} {Inner skin effects on non-{H}ermitian topological fractals},\ }\href {https://doi.org/10.1038/s42005-023-01130-2} {\bibfield  {journal} {\bibinfo  {journal} {Commun. Phys.}\ }\textbf {\bibinfo {volume} {6}},\ \bibinfo {pages} {10} (\bibinfo {year} {2023})}\BibitemShut {NoStop}%
\bibitem [{\citenamefont {Xiong}\ \emph {et~al.}(2024)\citenamefont {Xiong}, \citenamefont {Xing},\ and\ \citenamefont {Hu}}]{xiong2024non}%
  \BibitemOpen
  \bibfield  {author} {\bibinfo {author} {\bibfnamefont {Y.}~\bibnamefont {Xiong}}, \bibinfo {author} {\bibfnamefont {Z.-Y.}\ \bibnamefont {Xing}},\ and\ \bibinfo {author} {\bibfnamefont {H.}~\bibnamefont {Hu}},\ }\bibfield  {title} {\bibinfo {title} {Non-{H}ermitian skin effect in arbitrary dimensions: non-{B}loch band theory and classification},\ }\bibfield  {journal} {\bibinfo  {journal} {arXiv preprint arXiv:2407.01296}\ }\href {https://doi.org/10.48550/arXiv.2407.01296} {10.48550/arXiv.2407.01296} (\bibinfo {year} {2024})\BibitemShut {NoStop}%
\bibitem [{\citenamefont {Xue}\ and\ \citenamefont {Lee}(2026)}]{xue2024topologically}%
  \BibitemOpen
  \bibfield  {author} {\bibinfo {author} {\bibfnamefont {W.-T.}\ \bibnamefont {Xue}}\ and\ \bibinfo {author} {\bibfnamefont {C.~H.}\ \bibnamefont {Lee}},\ }\bibfield  {title} {\bibinfo {title} {Topologically protected negative entanglement},\ }\href {https://doi.org/10.1002/advs.202513868} {\bibfield  {journal} {\bibinfo  {journal} {Adv. Sci.}\ }\textbf {\bibinfo {volume} {13}},\ \bibinfo {pages} {e13868} (\bibinfo {year} {2026})}\BibitemShut {NoStop}%
\bibitem [{\citenamefont {Tai}\ and\ \citenamefont {Lee}(2023)}]{tai2023zoology}%
  \BibitemOpen
  \bibfield  {author} {\bibinfo {author} {\bibfnamefont {T.}~\bibnamefont {Tai}}\ and\ \bibinfo {author} {\bibfnamefont {C.~H.}\ \bibnamefont {Lee}},\ }\bibfield  {title} {\bibinfo {title} {Zoology of non-{H}ermitian spectra and their graph topology},\ }\href {https://doi.org/10.1103/PhysRevB.107.L220301} {\bibfield  {journal} {\bibinfo  {journal} {Phys. Rev. B}\ }\textbf {\bibinfo {volume} {107}},\ \bibinfo {pages} {L220301} (\bibinfo {year} {2023})}\BibitemShut {NoStop}%
\bibitem [{\citenamefont {Lei}\ \emph {et~al.}(2024)\citenamefont {Lei}, \citenamefont {Lee},\ and\ \citenamefont {Li}}]{lei2024activating}%
  \BibitemOpen
  \bibfield  {author} {\bibinfo {author} {\bibfnamefont {Z.}~\bibnamefont {Lei}}, \bibinfo {author} {\bibfnamefont {C.~H.}\ \bibnamefont {Lee}},\ and\ \bibinfo {author} {\bibfnamefont {L.}~\bibnamefont {Li}},\ }\bibfield  {title} {\bibinfo {title} {Activating non-{H}ermitian skin modes by parity-time symmetry breaking},\ }\href {https://doi.org/10.1038/s42005-024-01591-z} {\bibfield  {journal} {\bibinfo  {journal} {Commun. Phys.}\ }\textbf {\bibinfo {volume} {7}},\ \bibinfo {pages} {100} (\bibinfo {year} {2024})}\BibitemShut {NoStop}%
\bibitem [{\citenamefont {Shimomura}\ and\ \citenamefont {Sato}(2024)}]{shimomura2024general}%
  \BibitemOpen
  \bibfield  {author} {\bibinfo {author} {\bibfnamefont {K.}~\bibnamefont {Shimomura}}\ and\ \bibinfo {author} {\bibfnamefont {M.}~\bibnamefont {Sato}},\ }\bibfield  {title} {\bibinfo {title} {General criterion for non-{H}ermitian skin effects and application: {F}ock space skin effects in many-body systems},\ }\href {https://doi.org/10.1103/PhysRevLett.133.136502} {\bibfield  {journal} {\bibinfo  {journal} {Phys. Rev. Lett.}\ }\textbf {\bibinfo {volume} {133}},\ \bibinfo {pages} {136502} (\bibinfo {year} {2024})}\BibitemShut {NoStop}%
\bibitem [{\citenamefont {Yang}\ and\ \citenamefont {Lee}(2024)}]{yang2024percolation}%
  \BibitemOpen
  \bibfield  {author} {\bibinfo {author} {\bibfnamefont {M.}~\bibnamefont {Yang}}\ and\ \bibinfo {author} {\bibfnamefont {C.~H.}\ \bibnamefont {Lee}},\ }\bibfield  {title} {\bibinfo {title} {Percolation-induced {PT} symmetry breaking},\ }\href {https://doi.org/10.1103/PhysRevLett.133.136602} {\bibfield  {journal} {\bibinfo  {journal} {Phys. Rev. Lett.}\ }\textbf {\bibinfo {volume} {133}},\ \bibinfo {pages} {136602} (\bibinfo {year} {2024})}\BibitemShut {NoStop}%
\bibitem [{\citenamefont {Gliozzi}\ \emph {et~al.}(2024)\citenamefont {Gliozzi}, \citenamefont {De~Tomasi},\ and\ \citenamefont {Hughes}}]{gliozzi2024many}%
  \BibitemOpen
  \bibfield  {author} {\bibinfo {author} {\bibfnamefont {J.}~\bibnamefont {Gliozzi}}, \bibinfo {author} {\bibfnamefont {G.}~\bibnamefont {De~Tomasi}},\ and\ \bibinfo {author} {\bibfnamefont {T.~L.}\ \bibnamefont {Hughes}},\ }\bibfield  {title} {\bibinfo {title} {Many-body non-{H}ermitian skin effect for multipoles},\ }\href {https://doi.org/10.1103/PhysRevLett.133.136503} {\bibfield  {journal} {\bibinfo  {journal} {Phys. Rev. Lett.}\ }\textbf {\bibinfo {volume} {133}},\ \bibinfo {pages} {136503} (\bibinfo {year} {2024})}\BibitemShut {NoStop}%
\bibitem [{\citenamefont {Yang}\ \emph {et~al.}(2025)\citenamefont {Yang}, \citenamefont {Yuan},\ and\ \citenamefont {Lee}}]{yang2025non}%
  \BibitemOpen
  \bibfield  {author} {\bibinfo {author} {\bibfnamefont {M.}~\bibnamefont {Yang}}, \bibinfo {author} {\bibfnamefont {L.}~\bibnamefont {Yuan}},\ and\ \bibinfo {author} {\bibfnamefont {C.~H.}\ \bibnamefont {Lee}},\ }\bibfield  {title} {\bibinfo {title} {Non-{H}ermitian strong bosonic clustering through interaction-induced caging},\ }\href {https://doi.org/10.1038/s42005-025-02274-z} {\bibfield  {journal} {\bibinfo  {journal} {Commun. Phys.}\ }\textbf {\bibinfo {volume} {8}},\ \bibinfo {pages} {388} (\bibinfo {year} {2025})}\BibitemShut {NoStop}%
\bibitem [{\citenamefont {Li}\ \emph {et~al.}(2025{\natexlab{a}})\citenamefont {Li}, \citenamefont {Jiang},\ and\ \citenamefont {Lee}}]{li2025phase}%
  \BibitemOpen
  \bibfield  {author} {\bibinfo {author} {\bibfnamefont {Q.}~\bibnamefont {Li}}, \bibinfo {author} {\bibfnamefont {H.}~\bibnamefont {Jiang}},\ and\ \bibinfo {author} {\bibfnamefont {C.~H.}\ \bibnamefont {Lee}},\ }\bibfield  {title} {\bibinfo {title} {Phase-space generalized {B}rillouin zone for spatially inhomogeneous non-{H}ermitian systems},\ }\bibfield  {journal} {\bibinfo  {journal} {arXiv preprint arXiv:2501.09785}\ }\href {https://doi.org/10.48550/arXiv.2501.09785} {10.48550/arXiv.2501.09785} (\bibinfo {year} {2025}{\natexlab{a}})\BibitemShut {NoStop}%
\bibitem [{\citenamefont {Yoshida}\ \emph {et~al.}(2024)\citenamefont {Yoshida}, \citenamefont {Zhang}, \citenamefont {Neupert},\ and\ \citenamefont {Kawakami}}]{yoshida2024non}%
  \BibitemOpen
  \bibfield  {author} {\bibinfo {author} {\bibfnamefont {T.}~\bibnamefont {Yoshida}}, \bibinfo {author} {\bibfnamefont {S.-B.}\ \bibnamefont {Zhang}}, \bibinfo {author} {\bibfnamefont {T.}~\bibnamefont {Neupert}},\ and\ \bibinfo {author} {\bibfnamefont {N.}~\bibnamefont {Kawakami}},\ }\bibfield  {title} {\bibinfo {title} {Non-{H}ermitian {M}ott skin effect},\ }\href {https://doi.org/10.1103/PhysRevLett.133.076502} {\bibfield  {journal} {\bibinfo  {journal} {Phys. Rev. Lett.}\ }\textbf {\bibinfo {volume} {133}},\ \bibinfo {pages} {076502} (\bibinfo {year} {2024})}\BibitemShut {NoStop}%
\bibitem [{\citenamefont {Hamanaka}\ and\ \citenamefont {Kawabata}(2025)}]{hamanaka2025multifractality}%
  \BibitemOpen
  \bibfield  {author} {\bibinfo {author} {\bibfnamefont {S.}~\bibnamefont {Hamanaka}}\ and\ \bibinfo {author} {\bibfnamefont {K.}~\bibnamefont {Kawabata}},\ }\bibfield  {title} {\bibinfo {title} {Multifractality of the many-body non-{H}ermitian skin effect},\ }\href {https://doi.org/10.1103/PhysRevB.111.035144} {\bibfield  {journal} {\bibinfo  {journal} {Phys. Rev. B}\ }\textbf {\bibinfo {volume} {111}},\ \bibinfo {pages} {035144} (\bibinfo {year} {2025})}\BibitemShut {NoStop}%
\bibitem [{\citenamefont {Yan}\ \emph {et~al.}(2024)\citenamefont {Yan}, \citenamefont {Li}, \citenamefont {Sun},\ and\ \citenamefont {Xie}}]{yan2024transport}%
  \BibitemOpen
  \bibfield  {author} {\bibinfo {author} {\bibfnamefont {Q.}~\bibnamefont {Yan}}, \bibinfo {author} {\bibfnamefont {H.}~\bibnamefont {Li}}, \bibinfo {author} {\bibfnamefont {Q.-F.}\ \bibnamefont {Sun}},\ and\ \bibinfo {author} {\bibfnamefont {X.}~\bibnamefont {Xie}},\ }\bibfield  {title} {\bibinfo {title} {Transport theory in non-{H}ermitian systems},\ }\href {https://doi.org/10.1103/PhysRevB.110.045138} {\bibfield  {journal} {\bibinfo  {journal} {Phys. Rev. B}\ }\textbf {\bibinfo {volume} {110}},\ \bibinfo {pages} {045138} (\bibinfo {year} {2024})}\BibitemShut {NoStop}%
\bibitem [{\citenamefont {Das}\ and\ \citenamefont {Roy}(2024)}]{das2024quantized}%
  \BibitemOpen
  \bibfield  {author} {\bibinfo {author} {\bibfnamefont {S.~K.}\ \bibnamefont {Das}}\ and\ \bibinfo {author} {\bibfnamefont {B.}~\bibnamefont {Roy}},\ }\bibfield  {title} {\bibinfo {title} {Quantized electrical, thermal, and spin transports of non-{H}ermitian clean and dirty two-dimensional topological insulators and superconductors},\ }\bibfield  {journal} {\bibinfo  {journal} {arXiv preprint arXiv:2408.00763}\ }\href {https://doi.org/10.48550/arXiv.2408.00763} {10.48550/arXiv.2408.00763} (\bibinfo {year} {2024})\BibitemShut {NoStop}%
\bibitem [{\citenamefont {Yang}\ and\ \citenamefont {Lee}(2025)}]{yang2025beyond}%
  \BibitemOpen
  \bibfield  {author} {\bibinfo {author} {\bibfnamefont {M.}~\bibnamefont {Yang}}\ and\ \bibinfo {author} {\bibfnamefont {C.~H.}\ \bibnamefont {Lee}},\ }\bibfield  {title} {\bibinfo {title} {Beyond symmetry protection: Robust feedback-enforced edge states in non-{H}ermitian stacked quantum spin {H}all systems},\ }\bibfield  {journal} {\bibinfo  {journal} {arXiv preprint arXiv:2507.17295}\ }\href {https://doi.org/10.48550/arXiv.2507.17295} {10.48550/arXiv.2507.17295} (\bibinfo {year} {2025})\BibitemShut {NoStop}%
\bibitem [{\citenamefont {Cheng}\ \emph {et~al.}(2026)\citenamefont {Cheng}, \citenamefont {Jiang}, \citenamefont {Chen}, \citenamefont {Zhang}, \citenamefont {Ang},\ and\ \citenamefont {Lee}}]{cheng2026non}%
  \BibitemOpen
  \bibfield  {author} {\bibinfo {author} {\bibfnamefont {X.}~\bibnamefont {Cheng}}, \bibinfo {author} {\bibfnamefont {H.}~\bibnamefont {Jiang}}, \bibinfo {author} {\bibfnamefont {J.}~\bibnamefont {Chen}}, \bibinfo {author} {\bibfnamefont {L.}~\bibnamefont {Zhang}}, \bibinfo {author} {\bibfnamefont {Y.~S.}\ \bibnamefont {Ang}},\ and\ \bibinfo {author} {\bibfnamefont {C.~H.}\ \bibnamefont {Lee}},\ }\bibfield  {title} {\bibinfo {title} {Non-{H}ermitian effective {PT}-symmetry restoration from structural disorder},\ }\href {https://doi.org/10.15302/frontphys.2026.035201} {\bibfield  {journal} {\bibinfo  {journal} {Front. Phys.}\ }\textbf {\bibinfo {volume} {21}},\ \bibinfo {pages} {035201} (\bibinfo {year} {2026})}\BibitemShut {NoStop}%
\bibitem [{\citenamefont {Shi}\ and\ \citenamefont {Poddubny}(2025)}]{shi2025chiral}%
  \BibitemOpen
  \bibfield  {author} {\bibinfo {author} {\bibfnamefont {J.}~\bibnamefont {Shi}}\ and\ \bibinfo {author} {\bibfnamefont {A.~N.}\ \bibnamefont {Poddubny}},\ }\bibfield  {title} {\bibinfo {title} {Chiral dissociation of bound photon pairs for a non-{H}ermitian skin effect},\ }\href {https://doi.org/10.1103/PhysRevLett.134.233602} {\bibfield  {journal} {\bibinfo  {journal} {Phys. Rev. Lett.}\ }\textbf {\bibinfo {volume} {134}},\ \bibinfo {pages} {233602} (\bibinfo {year} {2025})}\BibitemShut {NoStop}%
\bibitem [{\citenamefont {Yang}\ and\ \citenamefont {Lee}(2026)}]{yang2026reversing}%
  \BibitemOpen
  \bibfield  {author} {\bibinfo {author} {\bibfnamefont {M.}~\bibnamefont {Yang}}\ and\ \bibinfo {author} {\bibfnamefont {C.~H.}\ \bibnamefont {Lee}},\ }\bibfield  {title} {\bibinfo {title} {Reversing non-hermitian skin accumulation with a non-local transverse switch},\ }\href {https://doi.org/10.1038/s42005-026-02608-5} {\bibfield  {journal} {\bibinfo  {journal} {Communications Physics}\ }\textbf {\bibinfo {volume} {9}},\ \bibinfo {pages} {210} (\bibinfo {year} {2026})}\BibitemShut {NoStop}%
\bibitem [{\citenamefont {Yang}\ \emph {et~al.}(2020)\citenamefont {Yang}, \citenamefont {Zhang}, \citenamefont {Fang},\ and\ \citenamefont {Hu}}]{yang2020non}%
  \BibitemOpen
  \bibfield  {author} {\bibinfo {author} {\bibfnamefont {Z.}~\bibnamefont {Yang}}, \bibinfo {author} {\bibfnamefont {K.}~\bibnamefont {Zhang}}, \bibinfo {author} {\bibfnamefont {C.}~\bibnamefont {Fang}},\ and\ \bibinfo {author} {\bibfnamefont {J.}~\bibnamefont {Hu}},\ }\bibfield  {title} {\bibinfo {title} {Non-{H}ermitian bulk-boundary correspondence and auxiliary generalized {B}rillouin zone theory},\ }\href {https://doi.org/10.1103/PhysRevLett.125.226402} {\bibfield  {journal} {\bibinfo  {journal} {Phys. Rev. Lett.}\ }\textbf {\bibinfo {volume} {125}},\ \bibinfo {pages} {226402} (\bibinfo {year} {2020})}\BibitemShut {NoStop}%
\bibitem [{\citenamefont {Meng}\ \emph {et~al.}(2025)\citenamefont {Meng}, \citenamefont {Ang},\ and\ \citenamefont {Lee}}]{meng2025generalized}%
  \BibitemOpen
  \bibfield  {author} {\bibinfo {author} {\bibfnamefont {H.}~\bibnamefont {Meng}}, \bibinfo {author} {\bibfnamefont {Y.~S.}\ \bibnamefont {Ang}},\ and\ \bibinfo {author} {\bibfnamefont {C.~H.}\ \bibnamefont {Lee}},\ }\bibfield  {title} {\bibinfo {title} {Generalized {B}rillouin zone fragmentation},\ }\bibfield  {journal} {\bibinfo  {journal} {arXiv preprint arXiv:2508.13275}\ }\href {https://doi.org/10.48550/arXiv.2508.13275} {10.48550/arXiv.2508.13275} (\bibinfo {year} {2025})\BibitemShut {NoStop}%
\bibitem [{\citenamefont {Yuan}\ \emph {et~al.}(2026)\citenamefont {Yuan}, \citenamefont {Xue},\ and\ \citenamefont {Lee}}]{yuan2026non}%
  \BibitemOpen
  \bibfield  {author} {\bibinfo {author} {\bibfnamefont {S.-Y.}\ \bibnamefont {Yuan}}, \bibinfo {author} {\bibfnamefont {W.-T.}\ \bibnamefont {Xue}},\ and\ \bibinfo {author} {\bibfnamefont {C.~H.}\ \bibnamefont {Lee}},\ }\bibfield  {title} {\bibinfo {title} {Non-{H}ermitian edge state endocytosis},\ }\bibfield  {journal} {\bibinfo  {journal} {arXiv preprint arXiv:2607.07703}\ }\href {https://doi.org/10.48550/arXiv.2607.07703} {10.48550/arXiv.2607.07703} (\bibinfo {year} {2026})\BibitemShut {NoStop}%
\bibitem [{\citenamefont {Kunst}\ and\ \citenamefont {Dwivedi}(2019)}]{kunst2019non}%
  \BibitemOpen
  \bibfield  {author} {\bibinfo {author} {\bibfnamefont {F.~K.}\ \bibnamefont {Kunst}}\ and\ \bibinfo {author} {\bibfnamefont {V.}~\bibnamefont {Dwivedi}},\ }\bibfield  {title} {\bibinfo {title} {Non-{H}ermitian systems and topology: A transfer-matrix perspective},\ }\href {https://doi.org/10.1103/PhysRevB.99.245116} {\bibfield  {journal} {\bibinfo  {journal} {Phys. Rev. B}\ }\textbf {\bibinfo {volume} {99}},\ \bibinfo {pages} {245116} (\bibinfo {year} {2019})}\BibitemShut {NoStop}%
\bibitem [{\citenamefont {Luo}\ \emph {et~al.}(2021)\citenamefont {Luo}, \citenamefont {Ohtsuki},\ and\ \citenamefont {Shindou}}]{luo2021transfer}%
  \BibitemOpen
  \bibfield  {author} {\bibinfo {author} {\bibfnamefont {X.}~\bibnamefont {Luo}}, \bibinfo {author} {\bibfnamefont {T.}~\bibnamefont {Ohtsuki}},\ and\ \bibinfo {author} {\bibfnamefont {R.}~\bibnamefont {Shindou}},\ }\bibfield  {title} {\bibinfo {title} {Transfer matrix study of the {A}nderson transition in non-{H}ermitian systems},\ }\href {https://doi.org/10.1103/PhysRevB.104.104203} {\bibfield  {journal} {\bibinfo  {journal} {Phys. Rev. B}\ }\textbf {\bibinfo {volume} {104}},\ \bibinfo {pages} {104203} (\bibinfo {year} {2021})}\BibitemShut {NoStop}%
\bibitem [{\citenamefont {Li}\ \emph {et~al.}(2026)\citenamefont {Li}, \citenamefont {Wei}, \citenamefont {Ruan}, \citenamefont {Wu}, \citenamefont {Wang}, \citenamefont {Chen}, \citenamefont {Lin}, \citenamefont {Lee},\ and\ \citenamefont {Ni}}]{li2026non}%
  \BibitemOpen
  \bibfield  {author} {\bibinfo {author} {\bibfnamefont {L.}~\bibnamefont {Li}}, \bibinfo {author} {\bibfnamefont {Y.}~\bibnamefont {Wei}}, \bibinfo {author} {\bibfnamefont {Y.}~\bibnamefont {Ruan}}, \bibinfo {author} {\bibfnamefont {G.}~\bibnamefont {Wu}}, \bibinfo {author} {\bibfnamefont {J.}~\bibnamefont {Wang}}, \bibinfo {author} {\bibfnamefont {S.}~\bibnamefont {Chen}}, \bibinfo {author} {\bibfnamefont {T.}~\bibnamefont {Lin}}, \bibinfo {author} {\bibfnamefont {C.~H.}\ \bibnamefont {Lee}},\ and\ \bibinfo {author} {\bibfnamefont {Z.}~\bibnamefont {Ni}},\ }\bibfield  {title} {\bibinfo {title} {Non-{H}ermitian topology driven by an identity term: An exactly solvable paradigm},\ }\bibfield  {journal} {\bibinfo  {journal} {arXiv preprint arXiv:2607.08469}\ }\href {https://doi.org/10.48550/arXiv.2607.08469} {10.48550/arXiv.2607.08469} (\bibinfo {year} {2026})\BibitemShut {NoStop}%
\bibitem [{\citenamefont {Lepori}\ and\ \citenamefont {Dell'Anna}(2017)}]{lepori2017long}%
  \BibitemOpen
  \bibfield  {author} {\bibinfo {author} {\bibfnamefont {L.}~\bibnamefont {Lepori}}\ and\ \bibinfo {author} {\bibfnamefont {L.}~\bibnamefont {Dell'Anna}},\ }\bibfield  {title} {\bibinfo {title} {Long-range topological insulators and weakened bulk-boundary correspondence},\ }\href {https://doi.org/10.48550/arXiv.1612.08155} {\bibfield  {journal} {\bibinfo  {journal} {New J. Phys.}\ }\textbf {\bibinfo {volume} {19}},\ \bibinfo {pages} {103030} (\bibinfo {year} {2017})},\ \Eprint {https://arxiv.org/abs/1612.08155} {arXiv:1612.08155 [cond-mat.str-el]} \BibitemShut {NoStop}%
\bibitem [{\citenamefont {Gu}\ \emph {et~al.}(2016)\citenamefont {Gu}, \citenamefont {Lee}, \citenamefont {Wen}, \citenamefont {Cho}, \citenamefont {Ryu},\ and\ \citenamefont {Qi}}]{gu2016holographic}%
  \BibitemOpen
  \bibfield  {author} {\bibinfo {author} {\bibfnamefont {Y.}~\bibnamefont {Gu}}, \bibinfo {author} {\bibfnamefont {C.~H.}\ \bibnamefont {Lee}}, \bibinfo {author} {\bibfnamefont {X.}~\bibnamefont {Wen}}, \bibinfo {author} {\bibfnamefont {G.~Y.}\ \bibnamefont {Cho}}, \bibinfo {author} {\bibfnamefont {S.}~\bibnamefont {Ryu}},\ and\ \bibinfo {author} {\bibfnamefont {X.-L.}\ \bibnamefont {Qi}},\ }\bibfield  {title} {\bibinfo {title} {Holographic duality between (2+ 1)-dimensional quantum anomalous {H}all state and (3+ 1)-dimensional topological insulators},\ }\href {https://doi.org/10.1103/PhysRevB.94.125107} {\bibfield  {journal} {\bibinfo  {journal} {Phys. Rev. B}\ }\textbf {\bibinfo {volume} {94}},\ \bibinfo {pages} {125107} (\bibinfo {year} {2016})}\BibitemShut {NoStop}%
\bibitem [{\citenamefont {Jones}\ \emph {et~al.}(2023)\citenamefont {Jones}, \citenamefont {Thorngren},\ and\ \citenamefont {Verresen}}]{jones2023bulk}%
  \BibitemOpen
  \bibfield  {author} {\bibinfo {author} {\bibfnamefont {N.~G.}\ \bibnamefont {Jones}}, \bibinfo {author} {\bibfnamefont {R.}~\bibnamefont {Thorngren}},\ and\ \bibinfo {author} {\bibfnamefont {R.}~\bibnamefont {Verresen}},\ }\bibfield  {title} {\bibinfo {title} {Bulk-boundary correspondence and singularity-filling in long-range free-fermion chains},\ }\href {https://doi.org/10.1103/PhysRevLett.130.246601} {\bibfield  {journal} {\bibinfo  {journal} {Phys. Rev. Lett.}\ }\textbf {\bibinfo {volume} {130}},\ \bibinfo {pages} {246601} (\bibinfo {year} {2023})},\ \Eprint {https://arxiv.org/abs/2211.15690} {arXiv:2211.15690 [cond-mat.str-el]} \BibitemShut {NoStop}%
\bibitem [{\citenamefont {P{\'e}rez-Gonz{\'a}lez}\ \emph {et~al.}(2019)\citenamefont {P{\'e}rez-Gonz{\'a}lez}, \citenamefont {Bello}, \citenamefont {G{\'o}mez-Le{\'o}n},\ and\ \citenamefont {Platero}}]{perezgonzalez2019ssh}%
  \BibitemOpen
  \bibfield  {author} {\bibinfo {author} {\bibfnamefont {B.}~\bibnamefont {P{\'e}rez-Gonz{\'a}lez}}, \bibinfo {author} {\bibfnamefont {M.}~\bibnamefont {Bello}}, \bibinfo {author} {\bibfnamefont {{\'A}.}~\bibnamefont {G{\'o}mez-Le{\'o}n}},\ and\ \bibinfo {author} {\bibfnamefont {G.}~\bibnamefont {Platero}},\ }\bibfield  {title} {\bibinfo {title} {Interplay between long-range hopping and disorder in topological systems},\ }\href {https://doi.org/10.1103/PhysRevB.99.035146} {\bibfield  {journal} {\bibinfo  {journal} {Phys. Rev. B}\ }\textbf {\bibinfo {volume} {99}},\ \bibinfo {pages} {035146} (\bibinfo {year} {2019})},\ \Eprint {https://arxiv.org/abs/1810.02779} {arXiv:1810.02779 [cond-mat.mes-hall]} \BibitemShut {NoStop}%
\bibitem [{\citenamefont {Vodola}\ \emph {et~al.}(2014)\citenamefont {Vodola}, \citenamefont {Lepori}, \citenamefont {Ercolessi}, \citenamefont {Gorshkov},\ and\ \citenamefont {Pupillo}}]{vodola2014kitaev}%
  \BibitemOpen
  \bibfield  {author} {\bibinfo {author} {\bibfnamefont {D.}~\bibnamefont {Vodola}}, \bibinfo {author} {\bibfnamefont {L.}~\bibnamefont {Lepori}}, \bibinfo {author} {\bibfnamefont {E.}~\bibnamefont {Ercolessi}}, \bibinfo {author} {\bibfnamefont {A.~V.}\ \bibnamefont {Gorshkov}},\ and\ \bibinfo {author} {\bibfnamefont {G.}~\bibnamefont {Pupillo}},\ }\bibfield  {title} {\bibinfo {title} {{Kitaev} chains with long-range pairing},\ }\href {https://doi.org/10.1103/PhysRevLett.113.156402} {\bibfield  {journal} {\bibinfo  {journal} {Physical Review Letters}\ }\textbf {\bibinfo {volume} {113}},\ \bibinfo {pages} {156402} (\bibinfo {year} {2014})}\BibitemShut {NoStop}%
\bibitem [{\citenamefont {Viyuela}\ \emph {et~al.}(2016)\citenamefont {Viyuela}, \citenamefont {Vodola}, \citenamefont {Pupillo},\ and\ \citenamefont {Mart{\'i}n-Delgado}}]{viyuela2016topological}%
  \BibitemOpen
  \bibfield  {author} {\bibinfo {author} {\bibfnamefont {O.}~\bibnamefont {Viyuela}}, \bibinfo {author} {\bibfnamefont {D.}~\bibnamefont {Vodola}}, \bibinfo {author} {\bibfnamefont {G.}~\bibnamefont {Pupillo}},\ and\ \bibinfo {author} {\bibfnamefont {M.~A.}\ \bibnamefont {Mart{\'i}n-Delgado}},\ }\bibfield  {title} {\bibinfo {title} {Topological massive {Dirac} edge modes and long-range superconducting hamiltonians},\ }\href {https://doi.org/10.1103/PhysRevB.94.125121} {\bibfield  {journal} {\bibinfo  {journal} {Physical Review B}\ }\textbf {\bibinfo {volume} {94}},\ \bibinfo {pages} {125121} (\bibinfo {year} {2016})}\BibitemShut {NoStop}%
\bibitem [{\citenamefont {Alecce}\ and\ \citenamefont {Dell'Anna}(2017)}]{alecce2017extended}%
  \BibitemOpen
  \bibfield  {author} {\bibinfo {author} {\bibfnamefont {A.}~\bibnamefont {Alecce}}\ and\ \bibinfo {author} {\bibfnamefont {L.}~\bibnamefont {Dell'Anna}},\ }\bibfield  {title} {\bibinfo {title} {Extended {Kitaev} chain with longer-range hopping and pairing},\ }\href {https://doi.org/10.1103/PhysRevB.95.195160} {\bibfield  {journal} {\bibinfo  {journal} {Physical Review B}\ }\textbf {\bibinfo {volume} {95}},\ \bibinfo {pages} {195160} (\bibinfo {year} {2017})}\BibitemShut {NoStop}%
\bibitem [{\citenamefont {J{\"a}ger}\ \emph {et~al.}(2020)\citenamefont {J{\"a}ger}, \citenamefont {Dell'Anna},\ and\ \citenamefont {Morigi}}]{jager2020edge}%
  \BibitemOpen
  \bibfield  {author} {\bibinfo {author} {\bibfnamefont {S.~B.}\ \bibnamefont {J{\"a}ger}}, \bibinfo {author} {\bibfnamefont {L.}~\bibnamefont {Dell'Anna}},\ and\ \bibinfo {author} {\bibfnamefont {G.}~\bibnamefont {Morigi}},\ }\bibfield  {title} {\bibinfo {title} {Edge states of the long-range {Kitaev} chain: An analytical study},\ }\href {https://doi.org/10.1103/PhysRevB.102.035152} {\bibfield  {journal} {\bibinfo  {journal} {Physical Review B}\ }\textbf {\bibinfo {volume} {102}},\ \bibinfo {pages} {035152} (\bibinfo {year} {2020})}\BibitemShut {NoStop}%
\bibitem [{\citenamefont {Patrick}\ \emph {et~al.}(2017)\citenamefont {Patrick}, \citenamefont {Neupert},\ and\ \citenamefont {Pachos}}]{patrick2017topological}%
  \BibitemOpen
  \bibfield  {author} {\bibinfo {author} {\bibfnamefont {K.}~\bibnamefont {Patrick}}, \bibinfo {author} {\bibfnamefont {T.}~\bibnamefont {Neupert}},\ and\ \bibinfo {author} {\bibfnamefont {J.~K.}\ \bibnamefont {Pachos}},\ }\bibfield  {title} {\bibinfo {title} {Topological quantum liquids with long-range couplings},\ }\href {https://doi.org/10.1103/PhysRevLett.118.267002} {\bibfield  {journal} {\bibinfo  {journal} {Physical Review Letters}\ }\textbf {\bibinfo {volume} {118}},\ \bibinfo {pages} {267002} (\bibinfo {year} {2017})}\BibitemShut {NoStop}%
\bibitem [{\citenamefont {Qi}\ \emph {et~al.}(2021)\citenamefont {Qi}, \citenamefont {Yan}, \citenamefont {Xing}, \citenamefont {Zhao}, \citenamefont {Liu}, \citenamefont {Cui}, \citenamefont {Han}, \citenamefont {Zhang},\ and\ \citenamefont {Wang}}]{qi2021topological}%
  \BibitemOpen
  \bibfield  {author} {\bibinfo {author} {\bibfnamefont {L.}~\bibnamefont {Qi}}, \bibinfo {author} {\bibfnamefont {Y.}~\bibnamefont {Yan}}, \bibinfo {author} {\bibfnamefont {Y.}~\bibnamefont {Xing}}, \bibinfo {author} {\bibfnamefont {X.-D.}\ \bibnamefont {Zhao}}, \bibinfo {author} {\bibfnamefont {S.}~\bibnamefont {Liu}}, \bibinfo {author} {\bibfnamefont {W.-X.}\ \bibnamefont {Cui}}, \bibinfo {author} {\bibfnamefont {X.}~\bibnamefont {Han}}, \bibinfo {author} {\bibfnamefont {S.}~\bibnamefont {Zhang}},\ and\ \bibinfo {author} {\bibfnamefont {H.-F.}\ \bibnamefont {Wang}},\ }\bibfield  {title} {\bibinfo {title} {Topological router induced via long-range hopping in a {Su--Schrieffer--Heeger} chain},\ }\href {https://doi.org/10.1103/PhysRevResearch.3.023037} {\bibfield  {journal} {\bibinfo  {journal} {Physical Review Research}\ }\textbf {\bibinfo {volume} {3}},\ \bibinfo {pages} {023037} (\bibinfo {year} {2021})}\BibitemShut {NoStop}%
\bibitem [{\citenamefont {Ch{\'a}vez}\ \emph {et~al.}(2021)\citenamefont {Ch{\'a}vez}, \citenamefont {Mattiotti}, \citenamefont {M{\'e}ndez-Berm{\'u}dez}, \citenamefont {Borgonovi},\ and\ \citenamefont {Celardo}}]{chavez2021disorder}%
  \BibitemOpen
  \bibfield  {author} {\bibinfo {author} {\bibfnamefont {N.~C.}\ \bibnamefont {Ch{\'a}vez}}, \bibinfo {author} {\bibfnamefont {F.}~\bibnamefont {Mattiotti}}, \bibinfo {author} {\bibfnamefont {J.~A.}\ \bibnamefont {M{\'e}ndez-Berm{\'u}dez}}, \bibinfo {author} {\bibfnamefont {F.}~\bibnamefont {Borgonovi}},\ and\ \bibinfo {author} {\bibfnamefont {G.~L.}\ \bibnamefont {Celardo}},\ }\bibfield  {title} {\bibinfo {title} {Disorder-enhanced and disorder-independent transport with long-range hopping: Application to molecular chains in optical cavities},\ }\href {https://doi.org/10.1103/PhysRevLett.126.153201} {\bibfield  {journal} {\bibinfo  {journal} {Physical Review Letters}\ }\textbf {\bibinfo {volume} {126}},\ \bibinfo {pages} {153201} (\bibinfo {year} {2021})}\BibitemShut {NoStop}%
\bibitem [{\citenamefont {Singh}\ \emph {et~al.}(2017)\citenamefont {Singh}, \citenamefont {Moessner},\ and\ \citenamefont {Roy}}]{singh2017effect}%
  \BibitemOpen
  \bibfield  {author} {\bibinfo {author} {\bibfnamefont {R.}~\bibnamefont {Singh}}, \bibinfo {author} {\bibfnamefont {R.}~\bibnamefont {Moessner}},\ and\ \bibinfo {author} {\bibfnamefont {D.}~\bibnamefont {Roy}},\ }\bibfield  {title} {\bibinfo {title} {Effect of long-range hopping and interactions on entanglement dynamics and many-body localization},\ }\href {https://doi.org/10.1103/PhysRevB.95.094205} {\bibfield  {journal} {\bibinfo  {journal} {Physical Review B}\ }\textbf {\bibinfo {volume} {95}},\ \bibinfo {pages} {094205} (\bibinfo {year} {2017})}\BibitemShut {NoStop}%
\bibitem [{\citenamefont {Gong}\ \emph {et~al.}(2016)\citenamefont {Gong}, \citenamefont {Maghrebi}, \citenamefont {Hu}, \citenamefont {Wall}, \citenamefont {Foss-Feig},\ and\ \citenamefont {Gorshkov}}]{gong2016topological}%
  \BibitemOpen
  \bibfield  {author} {\bibinfo {author} {\bibfnamefont {Z.-X.}\ \bibnamefont {Gong}}, \bibinfo {author} {\bibfnamefont {M.~F.}\ \bibnamefont {Maghrebi}}, \bibinfo {author} {\bibfnamefont {A.}~\bibnamefont {Hu}}, \bibinfo {author} {\bibfnamefont {M.~L.}\ \bibnamefont {Wall}}, \bibinfo {author} {\bibfnamefont {M.}~\bibnamefont {Foss-Feig}},\ and\ \bibinfo {author} {\bibfnamefont {A.~V.}\ \bibnamefont {Gorshkov}},\ }\bibfield  {title} {\bibinfo {title} {Topological phases with long-range interactions},\ }\href {https://doi.org/10.1103/PhysRevB.93.041102} {\bibfield  {journal} {\bibinfo  {journal} {Physical Review B}\ }\textbf {\bibinfo {volume} {93}},\ \bibinfo {pages} {041102} (\bibinfo {year} {2016})}\BibitemShut {NoStop}%
\bibitem [{\citenamefont {Kim}\ \emph {et~al.}(2024)\citenamefont {Kim}, \citenamefont {Suh}, \citenamefont {Lee}, \citenamefont {Park},\ and\ \citenamefont {Yu}}]{kim2024longrange}%
  \BibitemOpen
  \bibfield  {author} {\bibinfo {author} {\bibfnamefont {G.}~\bibnamefont {Kim}}, \bibinfo {author} {\bibfnamefont {J.}~\bibnamefont {Suh}}, \bibinfo {author} {\bibfnamefont {D.}~\bibnamefont {Lee}}, \bibinfo {author} {\bibfnamefont {N.}~\bibnamefont {Park}},\ and\ \bibinfo {author} {\bibfnamefont {S.}~\bibnamefont {Yu}},\ }\bibfield  {title} {\bibinfo {title} {Long-range-interacting topological photonic lattices breaking channel-bandwidth limit},\ }\href {https://doi.org/10.1038/s41377-024-01557-4} {\bibfield  {journal} {\bibinfo  {journal} {Light: Science \& Applications}\ }\textbf {\bibinfo {volume} {13}},\ \bibinfo {pages} {189} (\bibinfo {year} {2024})}\BibitemShut {NoStop}%
\bibitem [{\citenamefont {Wang}\ \emph {et~al.}(2023)\citenamefont {Wang}, \citenamefont {Jen},\ and\ \citenamefont {You}}]{wang2023scaling}%
  \BibitemOpen
  \bibfield  {author} {\bibinfo {author} {\bibfnamefont {Y.-C.}\ \bibnamefont {Wang}}, \bibinfo {author} {\bibfnamefont {H.}~\bibnamefont {Jen}},\ and\ \bibinfo {author} {\bibfnamefont {J.-S.}\ \bibnamefont {You}},\ }\bibfield  {title} {\bibinfo {title} {Scaling laws for non-{H}ermitian skin effect with long-range couplings},\ }\href {https://doi.org/10.1103/PhysRevB.108.085418} {\bibfield  {journal} {\bibinfo  {journal} {Phys. Rev. B}\ }\textbf {\bibinfo {volume} {108}},\ \bibinfo {pages} {085418} (\bibinfo {year} {2023})}\BibitemShut {NoStop}%
\bibitem [{\citenamefont {Xu}\ \emph {et~al.}(2021)\citenamefont {Xu}, \citenamefont {Xia},\ and\ \citenamefont {Chen}}]{xu2021non}%
  \BibitemOpen
  \bibfield  {author} {\bibinfo {author} {\bibfnamefont {Z.}~\bibnamefont {Xu}}, \bibinfo {author} {\bibfnamefont {X.}~\bibnamefont {Xia}},\ and\ \bibinfo {author} {\bibfnamefont {S.}~\bibnamefont {Chen}},\ }\bibfield  {title} {\bibinfo {title} {Non-{H}ermitian {A}ubry-andr{\'e} model with power-law hopping},\ }\href {https://doi.org/10.1103/PhysRevB.104.224204} {\bibfield  {journal} {\bibinfo  {journal} {Phys. Rev. B}\ }\textbf {\bibinfo {volume} {104}},\ \bibinfo {pages} {224204} (\bibinfo {year} {2021})},\ \Eprint {https://arxiv.org/abs/2109.02072} {arXiv:2109.02072 [cond-mat.dis-nn]} \BibitemShut {NoStop}%
\bibitem [{\citenamefont {Guo}\ \emph {et~al.}(2024)\citenamefont {Guo}, \citenamefont {Su}, \citenamefont {Wang}, \citenamefont {Li}, \citenamefont {Wang}, \citenamefont {Ruan}, \citenamefont {Du}, \citenamefont {Zheng}, \citenamefont {Chen},\ and\ \citenamefont {Hu}}]{guo2024scale}%
  \BibitemOpen
  \bibfield  {author} {\bibinfo {author} {\bibfnamefont {C.-X.}\ \bibnamefont {Guo}}, \bibinfo {author} {\bibfnamefont {L.}~\bibnamefont {Su}}, \bibinfo {author} {\bibfnamefont {Y.}~\bibnamefont {Wang}}, \bibinfo {author} {\bibfnamefont {L.}~\bibnamefont {Li}}, \bibinfo {author} {\bibfnamefont {J.}~\bibnamefont {Wang}}, \bibinfo {author} {\bibfnamefont {X.}~\bibnamefont {Ruan}}, \bibinfo {author} {\bibfnamefont {Y.}~\bibnamefont {Du}}, \bibinfo {author} {\bibfnamefont {D.}~\bibnamefont {Zheng}}, \bibinfo {author} {\bibfnamefont {S.}~\bibnamefont {Chen}},\ and\ \bibinfo {author} {\bibfnamefont {H.}~\bibnamefont {Hu}},\ }\bibfield  {title} {\bibinfo {title} {Scale-tailored localization and its observation in non-{H}ermitian electrical circuits},\ }\href {https://doi.org/10.1038/s41467-024-53434-8} {\bibfield  {journal} {\bibinfo  {journal} {Nat. Commun.}\ }\textbf {\bibinfo {volume} {15}},\ \bibinfo {pages} {9120} (\bibinfo {year} {2024})}\BibitemShut {NoStop}%
\bibitem [{\citenamefont {Liu}\ \emph {et~al.}(2025)\citenamefont {Liu}, \citenamefont {Jiang}, \citenamefont {Xue}, \citenamefont {Li}, \citenamefont {Gong}, \citenamefont {Liu},\ and\ \citenamefont {Lee}}]{liu2025non}%
  \BibitemOpen
  \bibfield  {author} {\bibinfo {author} {\bibfnamefont {S.}~\bibnamefont {Liu}}, \bibinfo {author} {\bibfnamefont {H.}~\bibnamefont {Jiang}}, \bibinfo {author} {\bibfnamefont {W.-T.}\ \bibnamefont {Xue}}, \bibinfo {author} {\bibfnamefont {Q.}~\bibnamefont {Li}}, \bibinfo {author} {\bibfnamefont {J.}~\bibnamefont {Gong}}, \bibinfo {author} {\bibfnamefont {X.}~\bibnamefont {Liu}},\ and\ \bibinfo {author} {\bibfnamefont {C.~H.}\ \bibnamefont {Lee}},\ }\bibfield  {title} {\bibinfo {title} {Non-{H}ermitian entanglement dip from scaling-induced exceptional criticality},\ }\href {https://doi.org/10.1016/j.scib.2025.07.011} {\bibfield  {journal} {\bibinfo  {journal} {Sci. Bull.}\ }\textbf {\bibinfo {volume} {70}},\ \bibinfo {pages} {2929} (\bibinfo {year} {2025})}\BibitemShut {NoStop}%
\bibitem [{\citenamefont {Qin}\ \emph {et~al.}(2026)\citenamefont {Qin}, \citenamefont {Ang}, \citenamefont {Li},\ and\ \citenamefont {Lee}}]{qin2026anyon}%
  \BibitemOpen
  \bibfield  {author} {\bibinfo {author} {\bibfnamefont {Y.}~\bibnamefont {Qin}}, \bibinfo {author} {\bibfnamefont {Y.~S.}\ \bibnamefont {Ang}}, \bibinfo {author} {\bibfnamefont {L.}~\bibnamefont {Li}},\ and\ \bibinfo {author} {\bibfnamefont {C.~H.}\ \bibnamefont {Lee}},\ }\bibfield  {title} {\bibinfo {title} {Anyon-induced criticality and dynamical stability in non-{H}ermitian many-body systems},\ }\bibfield  {journal} {\bibinfo  {journal} {arXiv preprint arXiv:2603.17494}\ }\href {https://doi.org/10.48550/arXiv.2603.17494} {10.48550/arXiv.2603.17494} (\bibinfo {year} {2026})\BibitemShut {NoStop}%
\bibitem [{\citenamefont {Britton}\ \emph {et~al.}(2012)\citenamefont {Britton}, \citenamefont {Sawyer}, \citenamefont {Keith}, \citenamefont {Wang}, \citenamefont {Freericks}, \citenamefont {Uys}, \citenamefont {Biercuk},\ and\ \citenamefont {Bollinger}}]{britton2012engineered}%
  \BibitemOpen
  \bibfield  {author} {\bibinfo {author} {\bibfnamefont {J.~W.}\ \bibnamefont {Britton}}, \bibinfo {author} {\bibfnamefont {B.~C.}\ \bibnamefont {Sawyer}}, \bibinfo {author} {\bibfnamefont {A.~C.}\ \bibnamefont {Keith}}, \bibinfo {author} {\bibfnamefont {C.-C.~J.}\ \bibnamefont {Wang}}, \bibinfo {author} {\bibfnamefont {J.~K.}\ \bibnamefont {Freericks}}, \bibinfo {author} {\bibfnamefont {H.}~\bibnamefont {Uys}}, \bibinfo {author} {\bibfnamefont {M.~J.}\ \bibnamefont {Biercuk}},\ and\ \bibinfo {author} {\bibfnamefont {J.~J.}\ \bibnamefont {Bollinger}},\ }\bibfield  {title} {\bibinfo {title} {Engineered two-dimensional {Ising} interactions in a trapped-ion quantum simulator with hundreds of spins},\ }\href {https://doi.org/10.1038/nature10981} {\bibfield  {journal} {\bibinfo  {journal} {Nature}\ }\textbf {\bibinfo {volume} {484}},\ \bibinfo {pages} {489} (\bibinfo {year} {2012})}\BibitemShut {NoStop}%
\bibitem [{\citenamefont {Landig}\ \emph {et~al.}(2016)\citenamefont {Landig}, \citenamefont {Hruby}, \citenamefont {Dogra}, \citenamefont {Landini}, \citenamefont {Mottl}, \citenamefont {Donner},\ and\ \citenamefont {Esslinger}}]{landig2016quantum}%
  \BibitemOpen
  \bibfield  {author} {\bibinfo {author} {\bibfnamefont {R.}~\bibnamefont {Landig}}, \bibinfo {author} {\bibfnamefont {L.}~\bibnamefont {Hruby}}, \bibinfo {author} {\bibfnamefont {N.}~\bibnamefont {Dogra}}, \bibinfo {author} {\bibfnamefont {M.}~\bibnamefont {Landini}}, \bibinfo {author} {\bibfnamefont {R.}~\bibnamefont {Mottl}}, \bibinfo {author} {\bibfnamefont {T.}~\bibnamefont {Donner}},\ and\ \bibinfo {author} {\bibfnamefont {T.}~\bibnamefont {Esslinger}},\ }\bibfield  {title} {\bibinfo {title} {Quantum phases from competing short- and long-range interactions in an optical lattice},\ }\href {https://doi.org/10.1038/nature17409} {\bibfield  {journal} {\bibinfo  {journal} {Nature}\ }\textbf {\bibinfo {volume} {532}},\ \bibinfo {pages} {476} (\bibinfo {year} {2016})}\BibitemShut {NoStop}%
\bibitem [{\citenamefont {Evans}\ \emph {et~al.}(2018)\citenamefont {Evans}, \citenamefont {Bhaskar}, \citenamefont {Sukachev}, \citenamefont {Nguyen}, \citenamefont {Sipahigil}, \citenamefont {Burek}, \citenamefont {Machielse}, \citenamefont {Zhang}, \citenamefont {Zibrov}, \citenamefont {Bielejec}, \citenamefont {Park}, \citenamefont {Lon{\v c}ar},\ and\ \citenamefont {Lukin}}]{evans2018photon}%
  \BibitemOpen
  \bibfield  {author} {\bibinfo {author} {\bibfnamefont {R.~E.}\ \bibnamefont {Evans}}, \bibinfo {author} {\bibfnamefont {M.~K.}\ \bibnamefont {Bhaskar}}, \bibinfo {author} {\bibfnamefont {D.~D.}\ \bibnamefont {Sukachev}}, \bibinfo {author} {\bibfnamefont {C.~T.}\ \bibnamefont {Nguyen}}, \bibinfo {author} {\bibfnamefont {A.}~\bibnamefont {Sipahigil}}, \bibinfo {author} {\bibfnamefont {M.~J.}\ \bibnamefont {Burek}}, \bibinfo {author} {\bibfnamefont {B.}~\bibnamefont {Machielse}}, \bibinfo {author} {\bibfnamefont {G.~H.}\ \bibnamefont {Zhang}}, \bibinfo {author} {\bibfnamefont {A.~S.}\ \bibnamefont {Zibrov}}, \bibinfo {author} {\bibfnamefont {E.}~\bibnamefont {Bielejec}}, \bibinfo {author} {\bibfnamefont {H.}~\bibnamefont {Park}}, \bibinfo {author} {\bibfnamefont {M.}~\bibnamefont {Lon{\v c}ar}},\ and\ \bibinfo {author} {\bibfnamefont {M.~D.}\ \bibnamefont {Lukin}},\ }\bibfield  {title} {\bibinfo {title} {Photon-mediated interactions between quantum emitters in a diamond nanocavity},\ }\href
  {https://doi.org/10.1126/science.aau4691} {\bibfield  {journal} {\bibinfo  {journal} {Science}\ }\textbf {\bibinfo {volume} {362}},\ \bibinfo {pages} {662} (\bibinfo {year} {2018})}\BibitemShut {NoStop}%
\bibitem [{\citenamefont {Zhang}\ \emph {et~al.}(2023{\natexlab{a}})\citenamefont {Zhang}, \citenamefont {Kim}, \citenamefont {Mark}, \citenamefont {Choi},\ and\ \citenamefont {Painter}}]{zhang2023superconducting}%
  \BibitemOpen
  \bibfield  {author} {\bibinfo {author} {\bibfnamefont {X.}~\bibnamefont {Zhang}}, \bibinfo {author} {\bibfnamefont {E.}~\bibnamefont {Kim}}, \bibinfo {author} {\bibfnamefont {D.~K.}\ \bibnamefont {Mark}}, \bibinfo {author} {\bibfnamefont {S.}~\bibnamefont {Choi}},\ and\ \bibinfo {author} {\bibfnamefont {O.}~\bibnamefont {Painter}},\ }\bibfield  {title} {\bibinfo {title} {A superconducting quantum simulator based on a photonic-bandgap metamaterial},\ }\href {https://doi.org/10.1126/science.ade7651} {\bibfield  {journal} {\bibinfo  {journal} {Science}\ }\textbf {\bibinfo {volume} {379}},\ \bibinfo {pages} {278} (\bibinfo {year} {2023}{\natexlab{a}})}\BibitemShut {NoStop}%
\bibitem [{\citenamefont {Pellerin}\ \emph {et~al.}(2024)\citenamefont {Pellerin}, \citenamefont {Houvenaghel}, \citenamefont {Coish}, \citenamefont {Carusotto},\ and\ \citenamefont {St-Jean}}]{pellerin2024wavefunction}%
  \BibitemOpen
  \bibfield  {author} {\bibinfo {author} {\bibfnamefont {F.}~\bibnamefont {Pellerin}}, \bibinfo {author} {\bibfnamefont {R.}~\bibnamefont {Houvenaghel}}, \bibinfo {author} {\bibfnamefont {W.~A.}\ \bibnamefont {Coish}}, \bibinfo {author} {\bibfnamefont {I.}~\bibnamefont {Carusotto}},\ and\ \bibinfo {author} {\bibfnamefont {P.}~\bibnamefont {St-Jean}},\ }\bibfield  {title} {\bibinfo {title} {Wave-function tomography of topological dimer chains with long-range couplings},\ }\href {https://doi.org/10.1103/PhysRevLett.132.183802} {\bibfield  {journal} {\bibinfo  {journal} {Physical Review Letters}\ }\textbf {\bibinfo {volume} {132}},\ \bibinfo {pages} {183802} (\bibinfo {year} {2024})}\BibitemShut {NoStop}%
\bibitem [{\citenamefont {Hatano}\ and\ \citenamefont {Nelson}(1996)}]{hatano1996localization}%
  \BibitemOpen
  \bibfield  {author} {\bibinfo {author} {\bibfnamefont {N.}~\bibnamefont {Hatano}}\ and\ \bibinfo {author} {\bibfnamefont {D.~R.}\ \bibnamefont {Nelson}},\ }\bibfield  {title} {\bibinfo {title} {Localization transitions in non-{H}ermitian quantum mechanics},\ }\href {https://doi.org/10.1103/PhysRevLett.77.570} {\bibfield  {journal} {\bibinfo  {journal} {Phys. Rev. Lett.}\ }\textbf {\bibinfo {volume} {77}},\ \bibinfo {pages} {570} (\bibinfo {year} {1996})},\ \Eprint {https://arxiv.org/abs/cond-mat/9603165} {arXiv:cond-mat/9603165 [cond-mat]} \BibitemShut {NoStop}%
\bibitem [{\citenamefont {Hatano}\ and\ \citenamefont {Nelson}(1997)}]{hatano1997vortex}%
  \BibitemOpen
  \bibfield  {author} {\bibinfo {author} {\bibfnamefont {N.}~\bibnamefont {Hatano}}\ and\ \bibinfo {author} {\bibfnamefont {D.~R.}\ \bibnamefont {Nelson}},\ }\bibfield  {title} {\bibinfo {title} {Vortex pinning and non-{H}ermitian quantum mechanics},\ }\href {https://doi.org/10.1103/PhysRevB.56.8651} {\bibfield  {journal} {\bibinfo  {journal} {Phys. Rev. B}\ }\textbf {\bibinfo {volume} {56}},\ \bibinfo {pages} {8651} (\bibinfo {year} {1997})}\BibitemShut {NoStop}%
\bibitem [{\citenamefont {Li}\ \emph {et~al.}(2021)\citenamefont {Li}, \citenamefont {Liu}, \citenamefont {Li},\ and\ \citenamefont {Liu}}]{li2021extended}%
  \BibitemOpen
  \bibfield  {author} {\bibinfo {author} {\bibfnamefont {S.}~\bibnamefont {Li}}, \bibinfo {author} {\bibfnamefont {M.}~\bibnamefont {Liu}}, \bibinfo {author} {\bibfnamefont {F.}~\bibnamefont {Li}},\ and\ \bibinfo {author} {\bibfnamefont {B.}~\bibnamefont {Liu}},\ }\bibfield  {title} {\bibinfo {title} {Topological phase transition of the extended non-hermitian {Su--Schrieffer--Heeger} model},\ }\href {https://doi.org/10.1088/1402-4896/abc580} {\bibfield  {journal} {\bibinfo  {journal} {Physica Scripta}\ }\textbf {\bibinfo {volume} {96}},\ \bibinfo {pages} {015402} (\bibinfo {year} {2021})}\BibitemShut {NoStop}%
\bibitem [{\citenamefont {Wu}\ \emph {et~al.}(2022)\citenamefont {Wu}, \citenamefont {Liu}, \citenamefont {Chen},\ and\ \citenamefont {Jia}}]{wu2022nonhermiticity}%
  \BibitemOpen
  \bibfield  {author} {\bibinfo {author} {\bibfnamefont {C.}~\bibnamefont {Wu}}, \bibinfo {author} {\bibfnamefont {N.}~\bibnamefont {Liu}}, \bibinfo {author} {\bibfnamefont {G.}~\bibnamefont {Chen}},\ and\ \bibinfo {author} {\bibfnamefont {S.}~\bibnamefont {Jia}},\ }\bibfield  {title} {\bibinfo {title} {Non-hermiticity-induced topological transitions in long-range {Su--Schrieffer--Heeger} models},\ }\href {https://doi.org/10.1103/PhysRevA.106.012211} {\bibfield  {journal} {\bibinfo  {journal} {Physical Review A}\ }\textbf {\bibinfo {volume} {106}},\ \bibinfo {pages} {012211} (\bibinfo {year} {2022})}\BibitemShut {NoStop}%
\bibitem [{\citenamefont {Shi}\ \emph {et~al.}(2024)\citenamefont {Shi}, \citenamefont {Dong}, \citenamefont {Bao},\ and\ \citenamefont {Guo}}]{shi2024entanglement}%
  \BibitemOpen
  \bibfield  {author} {\bibinfo {author} {\bibfnamefont {S.}~\bibnamefont {Shi}}, \bibinfo {author} {\bibfnamefont {L.}~\bibnamefont {Dong}}, \bibinfo {author} {\bibfnamefont {J.}~\bibnamefont {Bao}},\ and\ \bibinfo {author} {\bibfnamefont {B.}~\bibnamefont {Guo}},\ }\bibfield  {title} {\bibinfo {title} {Entanglement entropy and topological properties in a long-range non-hermitian {Su--Schrieffer--Heeger} model},\ }\href {https://doi.org/10.1016/j.physb.2023.415601} {\bibfield  {journal} {\bibinfo  {journal} {Physica B: Condensed Matter}\ }\textbf {\bibinfo {volume} {674}},\ \bibinfo {pages} {415601} (\bibinfo {year} {2024})}\BibitemShut {NoStop}%
\bibitem [{\citenamefont {Rafi-Ul-Islam}\ \emph {et~al.}(2024)\citenamefont {Rafi-Ul-Islam}, \citenamefont {Siu}, \citenamefont {Razo}, \citenamefont {Sahin},\ and\ \citenamefont {Jalil}}]{rafiulislam2024knots}%
  \BibitemOpen
  \bibfield  {author} {\bibinfo {author} {\bibfnamefont {S.~M.}\ \bibnamefont {Rafi-Ul-Islam}}, \bibinfo {author} {\bibfnamefont {Z.~B.}\ \bibnamefont {Siu}}, \bibinfo {author} {\bibfnamefont {M.~S.~H.}\ \bibnamefont {Razo}}, \bibinfo {author} {\bibfnamefont {H.}~\bibnamefont {Sahin}},\ and\ \bibinfo {author} {\bibfnamefont {M.~B.~A.}\ \bibnamefont {Jalil}},\ }\bibfield  {title} {\bibinfo {title} {From knots to exceptional points: Emergence of topological features in non-hermitian systems with long-range coupling},\ }\href {https://doi.org/10.1103/PhysRevB.110.045444} {\bibfield  {journal} {\bibinfo  {journal} {Physical Review B}\ }\textbf {\bibinfo {volume} {110}},\ \bibinfo {pages} {045444} (\bibinfo {year} {2024})}\BibitemShut {NoStop}%
\bibitem [{\citenamefont {Lai}\ \emph {et~al.}(2025)\citenamefont {Lai}, \citenamefont {Fang}, \citenamefont {Su}, \citenamefont {Li},\ and\ \citenamefont {Wu}}]{lai2025skin}%
  \BibitemOpen
  \bibfield  {author} {\bibinfo {author} {\bibfnamefont {Y.-P.}\ \bibnamefont {Lai}}, \bibinfo {author} {\bibfnamefont {Y.-X.}\ \bibnamefont {Fang}}, \bibinfo {author} {\bibfnamefont {C.-Q.}\ \bibnamefont {Su}}, \bibinfo {author} {\bibfnamefont {Y.}~\bibnamefont {Li}},\ and\ \bibinfo {author} {\bibfnamefont {S.-Q.}\ \bibnamefont {Wu}},\ }\bibfield  {title} {\bibinfo {title} {Non-hermitian skin effect in an extended {Su--Schrieffer--Heeger} model with balanced gain and loss and nonreciprocal next-nearest-neighbor hopping phase},\ }\href {https://doi.org/10.1103/PhysRevB.111.085102} {\bibfield  {journal} {\bibinfo  {journal} {Physical Review B}\ }\textbf {\bibinfo {volume} {111}},\ \bibinfo {pages} {085102} (\bibinfo {year} {2025})}\BibitemShut {NoStop}%
\bibitem [{\citenamefont {Zhao}\ and\ \citenamefont {Hu}(2025)}]{zhao2025tentacle}%
  \BibitemOpen
  \bibfield  {author} {\bibinfo {author} {\bibfnamefont {X.}~\bibnamefont {Zhao}}\ and\ \bibinfo {author} {\bibfnamefont {H.}~\bibnamefont {Hu}},\ }\bibfield  {title} {\bibinfo {title} {Tentaclelike spectra and bound states in the {Hatano--Nelson} chain with long-range impurity coupling},\ }\href {https://doi.org/10.1103/PhysRevB.111.054204} {\bibfield  {journal} {\bibinfo  {journal} {Physical Review B}\ }\textbf {\bibinfo {volume} {111}},\ \bibinfo {pages} {054204} (\bibinfo {year} {2025})}\BibitemShut {NoStop}%
\bibitem [{Note1()}]{Note1}%
  \BibitemOpen
  \bibinfo {note} {In the $\alpha \rightarrow \infty $ local limit, $E_{\protect \rm LR}^{(N)}(k)$ is just a simple ellipse, corresponding to the one and only amplification channel around the PBC loop}\BibitemShut {NoStop}%
\bibitem [{Note2()}]{Note2}%
  \BibitemOpen
  \bibinfo {note} {This purely anti-symmetrical NHSE configuration can be easily gauge-transformed into a pair with any combination of NHSE strengths. For a pair of chains with inverse skin depths $\pm \kappa =\pm \ln \protect \sqrt \protect \frac {1+\gamma }{1-\gamma }$, threading an imaginary flux $h$ transforms their inverse skin depths viz. $\kappa \rightarrow \kappa +h$ and $-\kappa \rightarrow -\kappa +h$, which can take on any desired pair of values with suitable $\kappa $ and $h$.}\BibitemShut {Stop}%
\bibitem [{Note3()}]{Note3}%
  \BibitemOpen
  \bibinfo {note} {There exists another possible configuration LR--NN, but it is mathematically more subtle without illustrating additional new physics, and is discussed separately in Appendices~\ref {sec:app_model_definitions} and \ref {sec:app:Nc_modelIV}.}\BibitemShut {Stop}%
\bibitem [{Note4()}]{Note4}%
  \BibitemOpen
  \bibinfo {note} {This is generally valid very close to the real-to-complex transition, where this pair is much closer to each other than any other eigenmode.}\BibitemShut {Stop}%
\bibitem [{Note5()}]{Note5}%
  \BibitemOpen
  \bibinfo {note} {This is because oppositely localized skin modes overlap less with increasing system length. Being a single-chain property, it is \protect \emph {not} a consequence of the critical NHSE.}\BibitemShut {Stop}%
\bibitem [{\citenamefont {Song}(2008)}]{Song2008DoubleWell}%
  \BibitemOpen
  \bibfield  {author} {\bibinfo {author} {\bibfnamefont {D.-Y.}\ \bibnamefont {Song}},\ }\bibfield  {title} {\bibinfo {title} {Tunneling and energy splitting in an asymmetric double-well potential},\ }\href {https://doi.org/10.1016/j.aop.2008.09.004} {\bibfield  {journal} {\bibinfo  {journal} {Ann. Phys.}\ }\textbf {\bibinfo {volume} {323}},\ \bibinfo {pages} {2991} (\bibinfo {year} {2008})}\BibitemShut {NoStop}%
\bibitem [{\citenamefont {Mott}(1968)}]{Mott1968Localized}%
  \BibitemOpen
  \bibfield  {author} {\bibinfo {author} {\bibfnamefont {N.~F.}\ \bibnamefont {Mott}},\ }\bibfield  {title} {\bibinfo {title} {Conduction in non-crystalline systems. i. localized electronic states in disordered systems},\ }\href {https://doi.org/10.1080/14786436808223200} {\bibfield  {journal} {\bibinfo  {journal} {Philos. Mag.}\ }\textbf {\bibinfo {volume} {17}},\ \bibinfo {pages} {1259} (\bibinfo {year} {1968})}\BibitemShut {NoStop}%
\bibitem [{\citenamefont {Falco}\ \emph {et~al.}(2017)\citenamefont {Falco}, \citenamefont {Fedorenko},\ and\ \citenamefont {Gruzberg}}]{Falco2017MottBerezinskii}%
  \BibitemOpen
  \bibfield  {author} {\bibinfo {author} {\bibfnamefont {G.~M.}\ \bibnamefont {Falco}}, \bibinfo {author} {\bibfnamefont {A.~A.}\ \bibnamefont {Fedorenko}},\ and\ \bibinfo {author} {\bibfnamefont {I.~A.}\ \bibnamefont {Gruzberg}},\ }\bibfield  {title} {\bibinfo {title} {Wave function correlations and the {AC} conductivity of disordered wires beyond the {Mott--Berezinskii} law},\ }\href {https://doi.org/10.1209/0295-5075/120/37004} {\bibfield  {journal} {\bibinfo  {journal} {EPL}\ }\textbf {\bibinfo {volume} {120}},\ \bibinfo {pages} {37004} (\bibinfo {year} {2017})}\BibitemShut {NoStop}%
\bibitem [{\citenamefont {Tavis}\ and\ \citenamefont {Cummings}(1968)}]{Tavis1968Exact}%
  \BibitemOpen
  \bibfield  {author} {\bibinfo {author} {\bibfnamefont {M.}~\bibnamefont {Tavis}}\ and\ \bibinfo {author} {\bibfnamefont {F.~W.}\ \bibnamefont {Cummings}},\ }\bibfield  {title} {\bibinfo {title} {Exact solution for an {$N$}-molecule---radiation-field hamiltonian},\ }\href {https://doi.org/10.1103/PhysRev.170.379} {\bibfield  {journal} {\bibinfo  {journal} {Phys. Rev.}\ }\textbf {\bibinfo {volume} {170}},\ \bibinfo {pages} {379} (\bibinfo {year} {1968})}\BibitemShut {NoStop}%
\bibitem [{\citenamefont {Breeze}\ \emph {et~al.}(2017)\citenamefont {Breeze}, \citenamefont {Salvadori}, \citenamefont {Sathian}, \citenamefont {Alford},\ and\ \citenamefont {Kay}}]{Breeze2017Dicke}%
  \BibitemOpen
  \bibfield  {author} {\bibinfo {author} {\bibfnamefont {J.~D.}\ \bibnamefont {Breeze}}, \bibinfo {author} {\bibfnamefont {E.}~\bibnamefont {Salvadori}}, \bibinfo {author} {\bibfnamefont {J.}~\bibnamefont {Sathian}}, \bibinfo {author} {\bibfnamefont {N.~M.}\ \bibnamefont {Alford}},\ and\ \bibinfo {author} {\bibfnamefont {C.~W.~M.}\ \bibnamefont {Kay}},\ }\bibfield  {title} {\bibinfo {title} {Room-temperature cavity quantum electrodynamics with strongly coupled {Dicke} states},\ }\href {https://doi.org/10.1038/s41534-017-0041-3} {\bibfield  {journal} {\bibinfo  {journal} {npj Quantum Information}\ }\textbf {\bibinfo {volume} {3}},\ \bibinfo {pages} {40} (\bibinfo {year} {2017})}\BibitemShut {NoStop}%
\bibitem [{Note6()}]{Note6}%
  \BibitemOpen
  \bibinfo {note} {The physical threshold is always determined by the minimum condition in Eq.~\protect \eqref {eq:min_threshold_main}; the band-edge formulas below apply to the parameter window in which $(1,2)$ is the first-colliding pair.}\BibitemShut {Stop}%
\bibitem [{Note7()}]{Note7}%
  \BibitemOpen
  \bibinfo {note} {In this case, this is because both $\Delta E_{\protect \rm LR}$ and $u_{11}$ acquire the same factor $\lambda ^{1-\alpha }$, but that does not hold for $\alpha \geq 2$.}\BibitemShut {Stop}%
\bibitem [{\citenamefont {Pan}\ \emph {et~al.}(2018)\citenamefont {Pan}, \citenamefont {Zhao}, \citenamefont {Miao}, \citenamefont {Longhi},\ and\ \citenamefont {Feng}}]{pan2018photonic}%
  \BibitemOpen
  \bibfield  {author} {\bibinfo {author} {\bibfnamefont {M.}~\bibnamefont {Pan}}, \bibinfo {author} {\bibfnamefont {H.}~\bibnamefont {Zhao}}, \bibinfo {author} {\bibfnamefont {P.}~\bibnamefont {Miao}}, \bibinfo {author} {\bibfnamefont {S.}~\bibnamefont {Longhi}},\ and\ \bibinfo {author} {\bibfnamefont {L.}~\bibnamefont {Feng}},\ }\bibfield  {title} {\bibinfo {title} {Photonic zero mode in a non-{H}ermitian photonic lattice},\ }\href {https://doi.org/10.1038/s41467-018-03822-8} {\bibfield  {journal} {\bibinfo  {journal} {Nat. Commun.}\ }\textbf {\bibinfo {volume} {9}},\ \bibinfo {pages} {1308} (\bibinfo {year} {2018})}\BibitemShut {NoStop}%
\bibitem [{\citenamefont {Xiao}\ \emph {et~al.}(2020)\citenamefont {Xiao}, \citenamefont {Deng}, \citenamefont {Wang}, \citenamefont {Zhu}, \citenamefont {Wang}, \citenamefont {Yi},\ and\ \citenamefont {Xue}}]{xiao2020non}%
  \BibitemOpen
  \bibfield  {author} {\bibinfo {author} {\bibfnamefont {L.}~\bibnamefont {Xiao}}, \bibinfo {author} {\bibfnamefont {T.}~\bibnamefont {Deng}}, \bibinfo {author} {\bibfnamefont {K.}~\bibnamefont {Wang}}, \bibinfo {author} {\bibfnamefont {G.}~\bibnamefont {Zhu}}, \bibinfo {author} {\bibfnamefont {Z.}~\bibnamefont {Wang}}, \bibinfo {author} {\bibfnamefont {W.}~\bibnamefont {Yi}},\ and\ \bibinfo {author} {\bibfnamefont {P.}~\bibnamefont {Xue}},\ }\bibfield  {title} {\bibinfo {title} {Non-{H}ermitian bulk--boundary correspondence in quantum dynamics},\ }\href {https://doi.org/10.1038/s41567-020-0836-6} {\bibfield  {journal} {\bibinfo  {journal} {Nat. Phys.}\ }\textbf {\bibinfo {volume} {16}},\ \bibinfo {pages} {761} (\bibinfo {year} {2020})}\BibitemShut {NoStop}%
\bibitem [{\citenamefont {Zhu}\ \emph {et~al.}(2020)\citenamefont {Zhu}, \citenamefont {Wang}, \citenamefont {Gupta}, \citenamefont {Zhang}, \citenamefont {Xie}, \citenamefont {Lu},\ and\ \citenamefont {Chen}}]{zhu2020photonic}%
  \BibitemOpen
  \bibfield  {author} {\bibinfo {author} {\bibfnamefont {X.}~\bibnamefont {Zhu}}, \bibinfo {author} {\bibfnamefont {H.}~\bibnamefont {Wang}}, \bibinfo {author} {\bibfnamefont {S.~K.}\ \bibnamefont {Gupta}}, \bibinfo {author} {\bibfnamefont {H.}~\bibnamefont {Zhang}}, \bibinfo {author} {\bibfnamefont {B.}~\bibnamefont {Xie}}, \bibinfo {author} {\bibfnamefont {M.}~\bibnamefont {Lu}},\ and\ \bibinfo {author} {\bibfnamefont {Y.}~\bibnamefont {Chen}},\ }\bibfield  {title} {\bibinfo {title} {Photonic non-{H}ermitian skin effect and non-{B}loch bulk-boundary correspondence},\ }\href {https://doi.org/10.1103/PhysRevResearch.2.013280} {\bibfield  {journal} {\bibinfo  {journal} {Phys. Rev. Research}\ }\textbf {\bibinfo {volume} {2}},\ \bibinfo {pages} {013280} (\bibinfo {year} {2020})}\BibitemShut {NoStop}%
\bibitem [{\citenamefont {Song}\ \emph {et~al.}(2020)\citenamefont {Song}, \citenamefont {Liu}, \citenamefont {Zheng}, \citenamefont {Zhang}, \citenamefont {Wang},\ and\ \citenamefont {Lu}}]{song2020two}%
  \BibitemOpen
  \bibfield  {author} {\bibinfo {author} {\bibfnamefont {Y.}~\bibnamefont {Song}}, \bibinfo {author} {\bibfnamefont {W.}~\bibnamefont {Liu}}, \bibinfo {author} {\bibfnamefont {L.}~\bibnamefont {Zheng}}, \bibinfo {author} {\bibfnamefont {Y.}~\bibnamefont {Zhang}}, \bibinfo {author} {\bibfnamefont {B.}~\bibnamefont {Wang}},\ and\ \bibinfo {author} {\bibfnamefont {P.}~\bibnamefont {Lu}},\ }\bibfield  {title} {\bibinfo {title} {Two-dimensional non-{H}ermitian skin effect in a synthetic photonic lattice},\ }\href {https://doi.org/10.1103/PhysRevApplied.14.064076} {\bibfield  {journal} {\bibinfo  {journal} {Phys. Rev. Applied}\ }\textbf {\bibinfo {volume} {14}},\ \bibinfo {pages} {064076} (\bibinfo {year} {2020})}\BibitemShut {NoStop}%
\bibitem [{\citenamefont {Ao}\ \emph {et~al.}(2020)\citenamefont {Ao}, \citenamefont {Hu}, \citenamefont {You}, \citenamefont {Lu}, \citenamefont {Fu}, \citenamefont {Wang},\ and\ \citenamefont {Gong}}]{ao2020topological}%
  \BibitemOpen
  \bibfield  {author} {\bibinfo {author} {\bibfnamefont {Y.}~\bibnamefont {Ao}}, \bibinfo {author} {\bibfnamefont {X.}~\bibnamefont {Hu}}, \bibinfo {author} {\bibfnamefont {Y.}~\bibnamefont {You}}, \bibinfo {author} {\bibfnamefont {C.}~\bibnamefont {Lu}}, \bibinfo {author} {\bibfnamefont {Y.}~\bibnamefont {Fu}}, \bibinfo {author} {\bibfnamefont {X.}~\bibnamefont {Wang}},\ and\ \bibinfo {author} {\bibfnamefont {Q.}~\bibnamefont {Gong}},\ }\bibfield  {title} {\bibinfo {title} {Topological phase transition in the non-{H}ermitian coupled resonator array},\ }\href {https://doi.org/10.1103/PhysRevLett.125.013902} {\bibfield  {journal} {\bibinfo  {journal} {Phys. Rev. Lett.}\ }\textbf {\bibinfo {volume} {125}},\ \bibinfo {pages} {013902} (\bibinfo {year} {2020})}\BibitemShut {NoStop}%
\bibitem [{\citenamefont {Lin}\ \emph {et~al.}(2024)\citenamefont {Lin}, \citenamefont {Song}, \citenamefont {Wang}, \citenamefont {Xin}, \citenamefont {Sun}, \citenamefont {Wu}, \citenamefont {Huang}, \citenamefont {Zhu}, \citenamefont {Jiang},\ and\ \citenamefont {Li}}]{lin2024observation}%
  \BibitemOpen
  \bibfield  {author} {\bibinfo {author} {\bibfnamefont {Z.}~\bibnamefont {Lin}}, \bibinfo {author} {\bibfnamefont {W.}~\bibnamefont {Song}}, \bibinfo {author} {\bibfnamefont {L.-W.}\ \bibnamefont {Wang}}, \bibinfo {author} {\bibfnamefont {H.}~\bibnamefont {Xin}}, \bibinfo {author} {\bibfnamefont {J.}~\bibnamefont {Sun}}, \bibinfo {author} {\bibfnamefont {S.}~\bibnamefont {Wu}}, \bibinfo {author} {\bibfnamefont {C.}~\bibnamefont {Huang}}, \bibinfo {author} {\bibfnamefont {S.}~\bibnamefont {Zhu}}, \bibinfo {author} {\bibfnamefont {J.-H.}\ \bibnamefont {Jiang}},\ and\ \bibinfo {author} {\bibfnamefont {T.}~\bibnamefont {Li}},\ }\bibfield  {title} {\bibinfo {title} {Observation of topological transition in {F}loquet non-{H}ermitian skin effects in silicon photonics},\ }\href {https://doi.org/10.1103/PhysRevLett.133.073803} {\bibfield  {journal} {\bibinfo  {journal} {Phys. Rev. Lett.}\ }\textbf {\bibinfo {volume} {133}},\ \bibinfo {pages} {073803} (\bibinfo {year} {2024})}\BibitemShut {NoStop}%
\bibitem [{\citenamefont {Liu}\ \emph {et~al.}(2024{\natexlab{a}})\citenamefont {Liu}, \citenamefont {Cao}, \citenamefont {Qi}, \citenamefont {Huang}, \citenamefont {Gao}, \citenamefont {Peng}, \citenamefont {Li},\ and\ \citenamefont {Zhu}}]{liu2024observation}%
  \BibitemOpen
  \bibfield  {author} {\bibinfo {author} {\bibfnamefont {Y.-K.}\ \bibnamefont {Liu}}, \bibinfo {author} {\bibfnamefont {P.-C.}\ \bibnamefont {Cao}}, \bibinfo {author} {\bibfnamefont {M.}~\bibnamefont {Qi}}, \bibinfo {author} {\bibfnamefont {Q.-K.-L.}\ \bibnamefont {Huang}}, \bibinfo {author} {\bibfnamefont {F.}~\bibnamefont {Gao}}, \bibinfo {author} {\bibfnamefont {Y.-G.}\ \bibnamefont {Peng}}, \bibinfo {author} {\bibfnamefont {Y.}~\bibnamefont {Li}},\ and\ \bibinfo {author} {\bibfnamefont {X.-F.}\ \bibnamefont {Zhu}},\ }\bibfield  {title} {\bibinfo {title} {Observation of non-{H}ermitian skin effect in thermal diffusion},\ }\href {https://doi.org/10.1016/j.scib.2024.02.040} {\bibfield  {journal} {\bibinfo  {journal} {Sci. Bull.}\ }\textbf {\bibinfo {volume} {69}},\ \bibinfo {pages} {1228} (\bibinfo {year} {2024}{\natexlab{a}})}\BibitemShut {NoStop}%
\bibitem [{\citenamefont {Brandenbourger}\ \emph {et~al.}(2019)\citenamefont {Brandenbourger}, \citenamefont {Locsin}, \citenamefont {Lerner},\ and\ \citenamefont {Coulais}}]{brandenbourger2019non}%
  \BibitemOpen
  \bibfield  {author} {\bibinfo {author} {\bibfnamefont {M.}~\bibnamefont {Brandenbourger}}, \bibinfo {author} {\bibfnamefont {X.}~\bibnamefont {Locsin}}, \bibinfo {author} {\bibfnamefont {E.}~\bibnamefont {Lerner}},\ and\ \bibinfo {author} {\bibfnamefont {C.}~\bibnamefont {Coulais}},\ }\bibfield  {title} {\bibinfo {title} {Non-reciprocal robotic metamaterials},\ }\href {https://doi.org/10.1038/s41467-019-12599-3} {\bibfield  {journal} {\bibinfo  {journal} {Nat. Commun.}\ }\textbf {\bibinfo {volume} {10}},\ \bibinfo {pages} {4608} (\bibinfo {year} {2019})}\BibitemShut {NoStop}%
\bibitem [{\citenamefont {Ghatak}\ \emph {et~al.}(2020)\citenamefont {Ghatak}, \citenamefont {Brandenbourger}, \citenamefont {Van~Wezel},\ and\ \citenamefont {Coulais}}]{ghatak2020observation}%
  \BibitemOpen
  \bibfield  {author} {\bibinfo {author} {\bibfnamefont {A.}~\bibnamefont {Ghatak}}, \bibinfo {author} {\bibfnamefont {M.}~\bibnamefont {Brandenbourger}}, \bibinfo {author} {\bibfnamefont {J.}~\bibnamefont {Van~Wezel}},\ and\ \bibinfo {author} {\bibfnamefont {C.}~\bibnamefont {Coulais}},\ }\bibfield  {title} {\bibinfo {title} {Observation of non-{H}ermitian topology and its bulk--edge correspondence in an active mechanical metamaterial},\ }\href {https://doi.org/10.1073/pnas.2010580117} {\bibfield  {journal} {\bibinfo  {journal} {Proc. Natl. Acad. Sci. U.S.A.}\ }\textbf {\bibinfo {volume} {117}},\ \bibinfo {pages} {29561} (\bibinfo {year} {2020})}\BibitemShut {NoStop}%
\bibitem [{\citenamefont {Wen}\ \emph {et~al.}(2022)\citenamefont {Wen}, \citenamefont {Zhu}, \citenamefont {Fan}, \citenamefont {Tam}, \citenamefont {Zhu}, \citenamefont {Wu}, \citenamefont {Lemoult}, \citenamefont {Fink},\ and\ \citenamefont {Li}}]{wen2022unidirectional}%
  \BibitemOpen
  \bibfield  {author} {\bibinfo {author} {\bibfnamefont {X.}~\bibnamefont {Wen}}, \bibinfo {author} {\bibfnamefont {X.}~\bibnamefont {Zhu}}, \bibinfo {author} {\bibfnamefont {A.}~\bibnamefont {Fan}}, \bibinfo {author} {\bibfnamefont {W.~Y.}\ \bibnamefont {Tam}}, \bibinfo {author} {\bibfnamefont {J.}~\bibnamefont {Zhu}}, \bibinfo {author} {\bibfnamefont {H.~W.}\ \bibnamefont {Wu}}, \bibinfo {author} {\bibfnamefont {F.}~\bibnamefont {Lemoult}}, \bibinfo {author} {\bibfnamefont {M.}~\bibnamefont {Fink}},\ and\ \bibinfo {author} {\bibfnamefont {J.}~\bibnamefont {Li}},\ }\bibfield  {title} {\bibinfo {title} {Unidirectional amplification with acoustic non-{H}ermitian space- time varying metamaterial},\ }\href {https://doi.org/10.1038/s42005-021-00790-2} {\bibfield  {journal} {\bibinfo  {journal} {Commun. Phys.}\ }\textbf {\bibinfo {volume} {5}},\ \bibinfo {pages} {18} (\bibinfo {year} {2022})}\BibitemShut {NoStop}%
\bibitem [{\citenamefont {Xiu}\ \emph {et~al.}(2023)\citenamefont {Xiu}, \citenamefont {Frankel}, \citenamefont {Liu}, \citenamefont {Qian}, \citenamefont {Sarkar}, \citenamefont {MacNider}, \citenamefont {Chen}, \citenamefont {Boechler},\ and\ \citenamefont {Mao}}]{xiu2023synthetically}%
  \BibitemOpen
  \bibfield  {author} {\bibinfo {author} {\bibfnamefont {H.}~\bibnamefont {Xiu}}, \bibinfo {author} {\bibfnamefont {I.}~\bibnamefont {Frankel}}, \bibinfo {author} {\bibfnamefont {H.}~\bibnamefont {Liu}}, \bibinfo {author} {\bibfnamefont {K.}~\bibnamefont {Qian}}, \bibinfo {author} {\bibfnamefont {S.}~\bibnamefont {Sarkar}}, \bibinfo {author} {\bibfnamefont {B.}~\bibnamefont {MacNider}}, \bibinfo {author} {\bibfnamefont {Z.}~\bibnamefont {Chen}}, \bibinfo {author} {\bibfnamefont {N.}~\bibnamefont {Boechler}},\ and\ \bibinfo {author} {\bibfnamefont {X.}~\bibnamefont {Mao}},\ }\bibfield  {title} {\bibinfo {title} {Synthetically non-{H}ermitian nonlinear wave-like behavior in a topological mechanical metamaterial},\ }\href {https://doi.org/10.1073/pnas.2217928120} {\bibfield  {journal} {\bibinfo  {journal} {Proc. Natl. Acad. Sci. U.S.A.}\ }\textbf {\bibinfo {volume} {120}},\ \bibinfo {pages} {e2217928120} (\bibinfo {year} {2023})}\BibitemShut {NoStop}%
\bibitem [{\citenamefont {Li}\ \emph {et~al.}(2024)\citenamefont {Li}, \citenamefont {Wang}, \citenamefont {Wang}, \citenamefont {Lin}, \citenamefont {Ma},\ and\ \citenamefont {Jiang}}]{li2024observation}%
  \BibitemOpen
  \bibfield  {author} {\bibinfo {author} {\bibfnamefont {Z.}~\bibnamefont {Li}}, \bibinfo {author} {\bibfnamefont {L.-W.}\ \bibnamefont {Wang}}, \bibinfo {author} {\bibfnamefont {X.}~\bibnamefont {Wang}}, \bibinfo {author} {\bibfnamefont {Z.-K.}\ \bibnamefont {Lin}}, \bibinfo {author} {\bibfnamefont {G.}~\bibnamefont {Ma}},\ and\ \bibinfo {author} {\bibfnamefont {J.-H.}\ \bibnamefont {Jiang}},\ }\bibfield  {title} {\bibinfo {title} {Observation of dynamic non-{H}ermitian skin effects},\ }\href {https://doi.org/10.1038/s41467-024-50776-1} {\bibfield  {journal} {\bibinfo  {journal} {Nat. Commun.}\ }\textbf {\bibinfo {volume} {15}},\ \bibinfo {pages} {6544} (\bibinfo {year} {2024})}\BibitemShut {NoStop}%
\bibitem [{\citenamefont {Sahin}\ \emph {et~al.}(2025{\natexlab{a}})\citenamefont {Sahin}, \citenamefont {Jalil},\ and\ \citenamefont {Lee}}]{sahin2025topolectrical}%
  \BibitemOpen
  \bibfield  {author} {\bibinfo {author} {\bibfnamefont {H.}~\bibnamefont {Sahin}}, \bibinfo {author} {\bibfnamefont {M.}~\bibnamefont {Jalil}},\ and\ \bibinfo {author} {\bibfnamefont {C.~H.}\ \bibnamefont {Lee}},\ }\bibfield  {title} {\bibinfo {title} {Topolectrical circuits—recent experimental advances and developments},\ }\bibfield  {journal} {\bibinfo  {journal} {APL Electronic Devices}\ }\textbf {\bibinfo {volume} {1}},\ \href {https://doi.org/10.1063/5.0265293} {10.1063/5.0265293} (\bibinfo {year} {2025}{\natexlab{a}})\BibitemShut {NoStop}%
\bibitem [{\citenamefont {Hofmann}\ \emph {et~al.}(2019)\citenamefont {Hofmann}, \citenamefont {Helbig}, \citenamefont {Lee}, \citenamefont {Greiter},\ and\ \citenamefont {Thomale}}]{hofmann2019chiral}%
  \BibitemOpen
  \bibfield  {author} {\bibinfo {author} {\bibfnamefont {T.}~\bibnamefont {Hofmann}}, \bibinfo {author} {\bibfnamefont {T.}~\bibnamefont {Helbig}}, \bibinfo {author} {\bibfnamefont {C.~H.}\ \bibnamefont {Lee}}, \bibinfo {author} {\bibfnamefont {M.}~\bibnamefont {Greiter}},\ and\ \bibinfo {author} {\bibfnamefont {R.}~\bibnamefont {Thomale}},\ }\bibfield  {title} {\bibinfo {title} {Chiral voltage propagation and calibration in a topolectrical {C}hern circuit},\ }\href {https://doi.org/10.1103/PhysRevLett.122.247702} {\bibfield  {journal} {\bibinfo  {journal} {Phys. Rev. Lett.}\ }\textbf {\bibinfo {volume} {122}},\ \bibinfo {pages} {247702} (\bibinfo {year} {2019})}\BibitemShut {NoStop}%
\bibitem [{\citenamefont {Ezawa}(2019)}]{ezawa2019electric}%
  \BibitemOpen
  \bibfield  {author} {\bibinfo {author} {\bibfnamefont {M.}~\bibnamefont {Ezawa}},\ }\bibfield  {title} {\bibinfo {title} {Electric circuits for non-{H}ermitian {C}hern insulators},\ }\href {https://doi.org/10.1103/PhysRevB.100.081401} {\bibfield  {journal} {\bibinfo  {journal} {Phys. Rev. B}\ }\textbf {\bibinfo {volume} {100}},\ \bibinfo {pages} {081401} (\bibinfo {year} {2019})}\BibitemShut {NoStop}%
\bibitem [{\citenamefont {Hofmann}\ \emph {et~al.}(2020)\citenamefont {Hofmann}, \citenamefont {Helbig}, \citenamefont {Schindler}, \citenamefont {Salgo}, \citenamefont {Brzezi{\'n}ska}, \citenamefont {Greiter}, \citenamefont {Kiessling}, \citenamefont {Wolf}, \citenamefont {Vollhardt}, \citenamefont {Kaba{\v{s}}i} \emph {et~al.}}]{hofmann2020reciprocal}%
  \BibitemOpen
  \bibfield  {author} {\bibinfo {author} {\bibfnamefont {T.}~\bibnamefont {Hofmann}}, \bibinfo {author} {\bibfnamefont {T.}~\bibnamefont {Helbig}}, \bibinfo {author} {\bibfnamefont {F.}~\bibnamefont {Schindler}}, \bibinfo {author} {\bibfnamefont {N.}~\bibnamefont {Salgo}}, \bibinfo {author} {\bibfnamefont {M.}~\bibnamefont {Brzezi{\'n}ska}}, \bibinfo {author} {\bibfnamefont {M.}~\bibnamefont {Greiter}}, \bibinfo {author} {\bibfnamefont {T.}~\bibnamefont {Kiessling}}, \bibinfo {author} {\bibfnamefont {D.}~\bibnamefont {Wolf}}, \bibinfo {author} {\bibfnamefont {A.}~\bibnamefont {Vollhardt}}, \bibinfo {author} {\bibfnamefont {A.}~\bibnamefont {Kaba{\v{s}}i}}, \emph {et~al.},\ }\bibfield  {title} {\bibinfo {title} {Reciprocal skin effect and its realization in a topolectrical circuit},\ }\href {https://doi.org/10.1103/PhysRevResearch.2.023265} {\bibfield  {journal} {\bibinfo  {journal} {Phys. Rev. Research}\ }\textbf {\bibinfo {volume} {2}},\ \bibinfo {pages} {023265} (\bibinfo {year} {2020})}\BibitemShut
  {NoStop}%
\bibitem [{\citenamefont {Liu}\ \emph {et~al.}(2020)\citenamefont {Liu}, \citenamefont {Ma}, \citenamefont {Yang}, \citenamefont {Zhang}, \citenamefont {Gao}, \citenamefont {Xiang}, \citenamefont {Cui},\ and\ \citenamefont {Zhang}}]{liu2020gain}%
  \BibitemOpen
  \bibfield  {author} {\bibinfo {author} {\bibfnamefont {S.}~\bibnamefont {Liu}}, \bibinfo {author} {\bibfnamefont {S.}~\bibnamefont {Ma}}, \bibinfo {author} {\bibfnamefont {C.}~\bibnamefont {Yang}}, \bibinfo {author} {\bibfnamefont {L.}~\bibnamefont {Zhang}}, \bibinfo {author} {\bibfnamefont {W.}~\bibnamefont {Gao}}, \bibinfo {author} {\bibfnamefont {Y.~J.}\ \bibnamefont {Xiang}}, \bibinfo {author} {\bibfnamefont {T.~J.}\ \bibnamefont {Cui}},\ and\ \bibinfo {author} {\bibfnamefont {S.}~\bibnamefont {Zhang}},\ }\bibfield  {title} {\bibinfo {title} {Gain-and loss-induced topological insulating phase in a non-{H}ermitian electrical circuit},\ }\href {https://doi.org/10.1103/PhysRevApplied.13.014047} {\bibfield  {journal} {\bibinfo  {journal} {Phys. Rev. Applied}\ }\textbf {\bibinfo {volume} {13}},\ \bibinfo {pages} {014047} (\bibinfo {year} {2020})}\BibitemShut {NoStop}%
\bibitem [{\citenamefont {Liu}\ \emph {et~al.}(2021)\citenamefont {Liu}, \citenamefont {Shao}, \citenamefont {Ma}, \citenamefont {Zhang}, \citenamefont {You}, \citenamefont {Wu}, \citenamefont {Xiang}, \citenamefont {Cui},\ and\ \citenamefont {Zhang}}]{liu2021non}%
  \BibitemOpen
  \bibfield  {author} {\bibinfo {author} {\bibfnamefont {S.}~\bibnamefont {Liu}}, \bibinfo {author} {\bibfnamefont {R.}~\bibnamefont {Shao}}, \bibinfo {author} {\bibfnamefont {S.}~\bibnamefont {Ma}}, \bibinfo {author} {\bibfnamefont {L.}~\bibnamefont {Zhang}}, \bibinfo {author} {\bibfnamefont {O.}~\bibnamefont {You}}, \bibinfo {author} {\bibfnamefont {H.}~\bibnamefont {Wu}}, \bibinfo {author} {\bibfnamefont {Y.~J.}\ \bibnamefont {Xiang}}, \bibinfo {author} {\bibfnamefont {T.~J.}\ \bibnamefont {Cui}},\ and\ \bibinfo {author} {\bibfnamefont {S.}~\bibnamefont {Zhang}},\ }\bibfield  {title} {\bibinfo {title} {Non-{H}ermitian skin effect in a non-{H}ermitian electrical circuit},\ }\bibfield  {journal} {\bibinfo  {journal} {Research}\ }\href {https://doi.org/10.34133/2021/5608038} {10.34133/2021/5608038} (\bibinfo {year} {2021})\BibitemShut {NoStop}%
\bibitem [{\citenamefont {Stegmaier}\ \emph {et~al.}(2021)\citenamefont {Stegmaier}, \citenamefont {Imhof}, \citenamefont {Helbig}, \citenamefont {Hofmann}, \citenamefont {Lee}, \citenamefont {Kremer}, \citenamefont {Fritzsche}, \citenamefont {Feichtner}, \citenamefont {Klembt}, \citenamefont {H{\"o}fling} \emph {et~al.}}]{stegmaier2021topological}%
  \BibitemOpen
  \bibfield  {author} {\bibinfo {author} {\bibfnamefont {A.}~\bibnamefont {Stegmaier}}, \bibinfo {author} {\bibfnamefont {S.}~\bibnamefont {Imhof}}, \bibinfo {author} {\bibfnamefont {T.}~\bibnamefont {Helbig}}, \bibinfo {author} {\bibfnamefont {T.}~\bibnamefont {Hofmann}}, \bibinfo {author} {\bibfnamefont {C.~H.}\ \bibnamefont {Lee}}, \bibinfo {author} {\bibfnamefont {M.}~\bibnamefont {Kremer}}, \bibinfo {author} {\bibfnamefont {A.}~\bibnamefont {Fritzsche}}, \bibinfo {author} {\bibfnamefont {T.}~\bibnamefont {Feichtner}}, \bibinfo {author} {\bibfnamefont {S.}~\bibnamefont {Klembt}}, \bibinfo {author} {\bibfnamefont {S.}~\bibnamefont {H{\"o}fling}}, \emph {et~al.},\ }\bibfield  {title} {\bibinfo {title} {Topological defect engineering and {PT} symmetry in non-{H}ermitian electrical circuits},\ }\href {https://doi.org/10.1103/PhysRevLett.126.215302} {\bibfield  {journal} {\bibinfo  {journal} {Phys. Rev. Lett.}\ }\textbf {\bibinfo {volume} {126}},\ \bibinfo {pages} {215302} (\bibinfo {year} {2021})}\BibitemShut
  {NoStop}%
\bibitem [{\citenamefont {Zhang}\ and\ \citenamefont {Franz}(2020)}]{zhang2020non}%
  \BibitemOpen
  \bibfield  {author} {\bibinfo {author} {\bibfnamefont {X.-X.}\ \bibnamefont {Zhang}}\ and\ \bibinfo {author} {\bibfnamefont {M.}~\bibnamefont {Franz}},\ }\bibfield  {title} {\bibinfo {title} {Non-{H}ermitian exceptional {L}andau quantization in electric circuits},\ }\href {https://doi.org/10.1103/PhysRevLett.124.046401} {\bibfield  {journal} {\bibinfo  {journal} {Phys. Rev. Lett.}\ }\textbf {\bibinfo {volume} {124}},\ \bibinfo {pages} {046401} (\bibinfo {year} {2020})}\BibitemShut {NoStop}%
\bibitem [{\citenamefont {Zhang}\ \emph {et~al.}(2024{\natexlab{a}})\citenamefont {Zhang}, \citenamefont {Zhang}, \citenamefont {Zhao},\ and\ \citenamefont {Lee}}]{zhang2022observation}%
  \BibitemOpen
  \bibfield  {author} {\bibinfo {author} {\bibfnamefont {X.}~\bibnamefont {Zhang}}, \bibinfo {author} {\bibfnamefont {B.}~\bibnamefont {Zhang}}, \bibinfo {author} {\bibfnamefont {W.}~\bibnamefont {Zhao}},\ and\ \bibinfo {author} {\bibfnamefont {C.~H.}\ \bibnamefont {Lee}},\ }\bibfield  {title} {\bibinfo {title} {Observation of non-local impedance response in a passive electrical circuit},\ }\href {https://doi.org/10.21468/SciPostPhys.16.1.002} {\bibfield  {journal} {\bibinfo  {journal} {SciPost Phys.}\ }\textbf {\bibinfo {volume} {16}},\ \bibinfo {pages} {002} (\bibinfo {year} {2024}{\natexlab{a}})}\BibitemShut {NoStop}%
\bibitem [{\citenamefont {Yuan}\ \emph {et~al.}(2023)\citenamefont {Yuan}, \citenamefont {Zhang}, \citenamefont {Zhou}, \citenamefont {Wang}, \citenamefont {Pan}, \citenamefont {Feng}, \citenamefont {Sun},\ and\ \citenamefont {Zhang}}]{yuan2023non}%
  \BibitemOpen
  \bibfield  {author} {\bibinfo {author} {\bibfnamefont {H.}~\bibnamefont {Yuan}}, \bibinfo {author} {\bibfnamefont {W.}~\bibnamefont {Zhang}}, \bibinfo {author} {\bibfnamefont {Z.}~\bibnamefont {Zhou}}, \bibinfo {author} {\bibfnamefont {W.}~\bibnamefont {Wang}}, \bibinfo {author} {\bibfnamefont {N.}~\bibnamefont {Pan}}, \bibinfo {author} {\bibfnamefont {Y.}~\bibnamefont {Feng}}, \bibinfo {author} {\bibfnamefont {H.}~\bibnamefont {Sun}},\ and\ \bibinfo {author} {\bibfnamefont {X.}~\bibnamefont {Zhang}},\ }\bibfield  {title} {\bibinfo {title} {Non-{H}ermitian topolectrical circuit sensor with high sensitivity},\ }\href {https://doi.org/10.1002/advs.202301128} {\bibfield  {journal} {\bibinfo  {journal} {Adv. Sci.}\ ,\ \bibinfo {pages} {2301128}} (\bibinfo {year} {2023})}\BibitemShut {NoStop}%
\bibitem [{\citenamefont {Zhu}\ \emph {et~al.}(2023)\citenamefont {Zhu}, \citenamefont {Sun}, \citenamefont {Hughes},\ and\ \citenamefont {Bahl}}]{zhu2023higher}%
  \BibitemOpen
  \bibfield  {author} {\bibinfo {author} {\bibfnamefont {P.}~\bibnamefont {Zhu}}, \bibinfo {author} {\bibfnamefont {X.-Q.}\ \bibnamefont {Sun}}, \bibinfo {author} {\bibfnamefont {T.~L.}\ \bibnamefont {Hughes}},\ and\ \bibinfo {author} {\bibfnamefont {G.}~\bibnamefont {Bahl}},\ }\bibfield  {title} {\bibinfo {title} {Higher rank chirality and non-{H}ermitian skin effect in a topolectrical circuit},\ }\href {https://doi.org/10.1038/s41467-023-36130-x} {\bibfield  {journal} {\bibinfo  {journal} {Nat. Commun.}\ }\textbf {\bibinfo {volume} {14}},\ \bibinfo {pages} {720} (\bibinfo {year} {2023})}\BibitemShut {NoStop}%
\bibitem [{\citenamefont {Hohmann}\ \emph {et~al.}(2023)\citenamefont {Hohmann}, \citenamefont {Hofmann}, \citenamefont {Helbig}, \citenamefont {Imhof}, \citenamefont {Brand}, \citenamefont {Upreti}, \citenamefont {Stegmaier}, \citenamefont {Fritzsche}, \citenamefont {M{\"u}ller}, \citenamefont {Schwingenschl{\"o}gl} \emph {et~al.}}]{hohmann2023observation}%
  \BibitemOpen
  \bibfield  {author} {\bibinfo {author} {\bibfnamefont {H.}~\bibnamefont {Hohmann}}, \bibinfo {author} {\bibfnamefont {T.}~\bibnamefont {Hofmann}}, \bibinfo {author} {\bibfnamefont {T.}~\bibnamefont {Helbig}}, \bibinfo {author} {\bibfnamefont {S.}~\bibnamefont {Imhof}}, \bibinfo {author} {\bibfnamefont {H.}~\bibnamefont {Brand}}, \bibinfo {author} {\bibfnamefont {L.~K.}\ \bibnamefont {Upreti}}, \bibinfo {author} {\bibfnamefont {A.}~\bibnamefont {Stegmaier}}, \bibinfo {author} {\bibfnamefont {A.}~\bibnamefont {Fritzsche}}, \bibinfo {author} {\bibfnamefont {T.}~\bibnamefont {M{\"u}ller}}, \bibinfo {author} {\bibfnamefont {U.}~\bibnamefont {Schwingenschl{\"o}gl}}, \emph {et~al.},\ }\bibfield  {title} {\bibinfo {title} {Observation of cnoidal wave localization in nonlinear topolectric circuits},\ }\href {https://doi.org/10.1103/PhysRevResearch.5.L012041} {\bibfield  {journal} {\bibinfo  {journal} {Phys. Rev. Research}\ }\textbf {\bibinfo {volume} {5}},\ \bibinfo {pages} {L012041} (\bibinfo {year}
  {2023})}\BibitemShut {NoStop}%
\bibitem [{\citenamefont {Zhang}\ \emph {et~al.}(2024{\natexlab{b}})\citenamefont {Zhang}, \citenamefont {Zhang}, \citenamefont {Zhao},\ and\ \citenamefont {Lee}}]{zhang2024observation}%
  \BibitemOpen
  \bibfield  {author} {\bibinfo {author} {\bibfnamefont {X.}~\bibnamefont {Zhang}}, \bibinfo {author} {\bibfnamefont {B.}~\bibnamefont {Zhang}}, \bibinfo {author} {\bibfnamefont {W.}~\bibnamefont {Zhao}},\ and\ \bibinfo {author} {\bibfnamefont {C.~H.}\ \bibnamefont {Lee}},\ }\bibfield  {title} {\bibinfo {title} {Observation of non-local impedance response in a passive electrical circuit},\ }\href {https://doi.org/10.21468/SciPostPhys.16.1.002} {\bibfield  {journal} {\bibinfo  {journal} {SciPost Phys.}\ }\textbf {\bibinfo {volume} {16}},\ \bibinfo {pages} {002} (\bibinfo {year} {2024}{\natexlab{b}})}\BibitemShut {NoStop}%
\bibitem [{\citenamefont {Zou}\ \emph {et~al.}(2024)\citenamefont {Zou}, \citenamefont {Chen}, \citenamefont {Meng}, \citenamefont {Ang}, \citenamefont {Zhang},\ and\ \citenamefont {Lee}}]{zou2024experimental}%
  \BibitemOpen
  \bibfield  {author} {\bibinfo {author} {\bibfnamefont {D.}~\bibnamefont {Zou}}, \bibinfo {author} {\bibfnamefont {T.}~\bibnamefont {Chen}}, \bibinfo {author} {\bibfnamefont {H.}~\bibnamefont {Meng}}, \bibinfo {author} {\bibfnamefont {Y.~S.}\ \bibnamefont {Ang}}, \bibinfo {author} {\bibfnamefont {X.}~\bibnamefont {Zhang}},\ and\ \bibinfo {author} {\bibfnamefont {C.~H.}\ \bibnamefont {Lee}},\ }\bibfield  {title} {\bibinfo {title} {Experimental observation of exceptional bound states in a classical circuit network},\ }\bibfield  {journal} {\bibinfo  {journal} {Sci. Bull.}\ }\href {https://doi.org/10.1016/j.scib.2024.05.036} {10.1016/j.scib.2024.05.036} (\bibinfo {year} {2024})\BibitemShut {NoStop}%
\bibitem [{\citenamefont {Zhang}\ \emph {et~al.}(2023{\natexlab{b}})\citenamefont {Zhang}, \citenamefont {Chen}, \citenamefont {Li}, \citenamefont {Lee},\ and\ \citenamefont {Zhang}}]{zhang2023electrical}%
  \BibitemOpen
  \bibfield  {author} {\bibinfo {author} {\bibfnamefont {H.}~\bibnamefont {Zhang}}, \bibinfo {author} {\bibfnamefont {T.}~\bibnamefont {Chen}}, \bibinfo {author} {\bibfnamefont {L.}~\bibnamefont {Li}}, \bibinfo {author} {\bibfnamefont {C.~H.}\ \bibnamefont {Lee}},\ and\ \bibinfo {author} {\bibfnamefont {X.}~\bibnamefont {Zhang}},\ }\bibfield  {title} {\bibinfo {title} {Electrical circuit realization of topological switching for the non-{H}ermitian skin effect},\ }\href {https://doi.org/10.1103/PhysRevB.107.085426} {\bibfield  {journal} {\bibinfo  {journal} {Phys. Rev. B}\ }\textbf {\bibinfo {volume} {107}},\ \bibinfo {pages} {085426} (\bibinfo {year} {2023}{\natexlab{b}})}\BibitemShut {NoStop}%
\bibitem [{\citenamefont {Stegmaier}\ \emph {et~al.}(2024{\natexlab{a}})\citenamefont {Stegmaier}, \citenamefont {Brand}, \citenamefont {Imhof}, \citenamefont {Fritzsche}, \citenamefont {Helbig}, \citenamefont {Hofmann}, \citenamefont {Boettcher}, \citenamefont {Greiter}, \citenamefont {Lee}, \citenamefont {Bahl} \emph {et~al.}}]{stegmaier2024realizing}%
  \BibitemOpen
  \bibfield  {author} {\bibinfo {author} {\bibfnamefont {A.}~\bibnamefont {Stegmaier}}, \bibinfo {author} {\bibfnamefont {H.}~\bibnamefont {Brand}}, \bibinfo {author} {\bibfnamefont {S.}~\bibnamefont {Imhof}}, \bibinfo {author} {\bibfnamefont {A.}~\bibnamefont {Fritzsche}}, \bibinfo {author} {\bibfnamefont {T.}~\bibnamefont {Helbig}}, \bibinfo {author} {\bibfnamefont {T.}~\bibnamefont {Hofmann}}, \bibinfo {author} {\bibfnamefont {I.}~\bibnamefont {Boettcher}}, \bibinfo {author} {\bibfnamefont {M.}~\bibnamefont {Greiter}}, \bibinfo {author} {\bibfnamefont {C.~H.}\ \bibnamefont {Lee}}, \bibinfo {author} {\bibfnamefont {G.}~\bibnamefont {Bahl}}, \emph {et~al.},\ }\bibfield  {title} {\bibinfo {title} {Realizing efficient topological temporal pumping in electrical circuits},\ }\href {https://doi.org/10.1103/PhysRevResearch.6.023010} {\bibfield  {journal} {\bibinfo  {journal} {Phys. Rev. Research}\ }\textbf {\bibinfo {volume} {6}},\ \bibinfo {pages} {023010} (\bibinfo {year} {2024}{\natexlab{a}})}\BibitemShut
  {NoStop}%
\bibitem [{\citenamefont {Stegmaier}\ \emph {et~al.}(2024{\natexlab{b}})\citenamefont {Stegmaier}, \citenamefont {Fritzsche}, \citenamefont {Sorbello}, \citenamefont {Greiter}, \citenamefont {Brand}, \citenamefont {Barko}, \citenamefont {Hofer}, \citenamefont {Schwingenschl{\"o}gl}, \citenamefont {Moessner}, \citenamefont {Lee} \emph {et~al.}}]{stegmaier2024topological}%
  \BibitemOpen
  \bibfield  {author} {\bibinfo {author} {\bibfnamefont {A.}~\bibnamefont {Stegmaier}}, \bibinfo {author} {\bibfnamefont {A.}~\bibnamefont {Fritzsche}}, \bibinfo {author} {\bibfnamefont {R.}~\bibnamefont {Sorbello}}, \bibinfo {author} {\bibfnamefont {M.}~\bibnamefont {Greiter}}, \bibinfo {author} {\bibfnamefont {H.}~\bibnamefont {Brand}}, \bibinfo {author} {\bibfnamefont {C.}~\bibnamefont {Barko}}, \bibinfo {author} {\bibfnamefont {M.}~\bibnamefont {Hofer}}, \bibinfo {author} {\bibfnamefont {U.}~\bibnamefont {Schwingenschl{\"o}gl}}, \bibinfo {author} {\bibfnamefont {R.}~\bibnamefont {Moessner}}, \bibinfo {author} {\bibfnamefont {C.~H.}\ \bibnamefont {Lee}}, \emph {et~al.},\ }\bibfield  {title} {\bibinfo {title} {Topological edge state nucleation in frequency space and its realization with {F}loquet electrical circuits},\ }\bibfield  {journal} {\bibinfo  {journal} {arXiv preprint arXiv:2407.10191}\ }\href {https://doi.org/10.48550/arXiv.2407.10191} {10.48550/arXiv.2407.10191} (\bibinfo {year}
  {2024}{\natexlab{b}})\BibitemShut {NoStop}%
\bibitem [{\citenamefont {Shang}\ \emph {et~al.}(2024)\citenamefont {Shang}, \citenamefont {Liu}, \citenamefont {Jiang}, \citenamefont {Shao}, \citenamefont {Zang}, \citenamefont {Lee}, \citenamefont {Thomale}, \citenamefont {Manchon}, \citenamefont {Cui},\ and\ \citenamefont {Schwingenschl{\"o}gl}}]{shang2024observation}%
  \BibitemOpen
  \bibfield  {author} {\bibinfo {author} {\bibfnamefont {C.}~\bibnamefont {Shang}}, \bibinfo {author} {\bibfnamefont {S.}~\bibnamefont {Liu}}, \bibinfo {author} {\bibfnamefont {C.}~\bibnamefont {Jiang}}, \bibinfo {author} {\bibfnamefont {R.}~\bibnamefont {Shao}}, \bibinfo {author} {\bibfnamefont {X.}~\bibnamefont {Zang}}, \bibinfo {author} {\bibfnamefont {C.~H.}\ \bibnamefont {Lee}}, \bibinfo {author} {\bibfnamefont {R.}~\bibnamefont {Thomale}}, \bibinfo {author} {\bibfnamefont {A.}~\bibnamefont {Manchon}}, \bibinfo {author} {\bibfnamefont {T.~J.}\ \bibnamefont {Cui}},\ and\ \bibinfo {author} {\bibfnamefont {U.}~\bibnamefont {Schwingenschl{\"o}gl}},\ }\bibfield  {title} {\bibinfo {title} {Observation of a higher-order end topological insulator in a real projective lattice},\ }\href {https://doi.org/10.1002/advs.202303222} {\bibfield  {journal} {\bibinfo  {journal} {Adv. Sci.}\ }\textbf {\bibinfo {volume} {11}},\ \bibinfo {pages} {2303222} (\bibinfo {year} {2024})}\BibitemShut {NoStop}%
\bibitem [{\citenamefont {Zhang}\ \emph {et~al.}(2024{\natexlab{c}})\citenamefont {Zhang}, \citenamefont {Wu}, \citenamefont {Yan}, \citenamefont {Liu}, \citenamefont {Wang},\ and\ \citenamefont {Chen}}]{zhang2024observation2}%
  \BibitemOpen
  \bibfield  {author} {\bibinfo {author} {\bibfnamefont {X.}~\bibnamefont {Zhang}}, \bibinfo {author} {\bibfnamefont {C.}~\bibnamefont {Wu}}, \bibinfo {author} {\bibfnamefont {M.}~\bibnamefont {Yan}}, \bibinfo {author} {\bibfnamefont {N.}~\bibnamefont {Liu}}, \bibinfo {author} {\bibfnamefont {Z.}~\bibnamefont {Wang}},\ and\ \bibinfo {author} {\bibfnamefont {G.}~\bibnamefont {Chen}},\ }\bibfield  {title} {\bibinfo {title} {Observation of continuum {L}andau modes in non-{H}ermitian electric circuits},\ }\href {https://doi.org/10.1038/s41467-024-46122-0} {\bibfield  {journal} {\bibinfo  {journal} {Nat. Commun.}\ }\textbf {\bibinfo {volume} {15}},\ \bibinfo {pages} {1798} (\bibinfo {year} {2024}{\natexlab{c}})}\BibitemShut {NoStop}%
\bibitem [{\citenamefont {Halder}\ \emph {et~al.}(2024)\citenamefont {Halder}, \citenamefont {Thomale},\ and\ \citenamefont {Basu}}]{halder2024circuit}%
  \BibitemOpen
  \bibfield  {author} {\bibinfo {author} {\bibfnamefont {D.}~\bibnamefont {Halder}}, \bibinfo {author} {\bibfnamefont {R.}~\bibnamefont {Thomale}},\ and\ \bibinfo {author} {\bibfnamefont {S.}~\bibnamefont {Basu}},\ }\bibfield  {title} {\bibinfo {title} {Circuit realization of a two-orbital non-{H}ermitian tight-binding chain},\ }\href {https://doi.org/10.1103/PhysRevB.109.115407} {\bibfield  {journal} {\bibinfo  {journal} {Phys. Rev. B}\ }\textbf {\bibinfo {volume} {109}},\ \bibinfo {pages} {115407} (\bibinfo {year} {2024})}\BibitemShut {NoStop}%
\bibitem [{\citenamefont {Sahin}\ \emph {et~al.}(2025{\natexlab{b}})\citenamefont {Sahin}, \citenamefont {Akg{\"u}n}, \citenamefont {Siu}, \citenamefont {Rafi-Ul-Islam}, \citenamefont {Kong}, \citenamefont {Jalil},\ and\ \citenamefont {Lee}}]{sahin2025protected}%
  \BibitemOpen
  \bibfield  {author} {\bibinfo {author} {\bibfnamefont {H.}~\bibnamefont {Sahin}}, \bibinfo {author} {\bibfnamefont {H.}~\bibnamefont {Akg{\"u}n}}, \bibinfo {author} {\bibfnamefont {Z.~B.}\ \bibnamefont {Siu}}, \bibinfo {author} {\bibfnamefont {S.}~\bibnamefont {Rafi-Ul-Islam}}, \bibinfo {author} {\bibfnamefont {J.~F.}\ \bibnamefont {Kong}}, \bibinfo {author} {\bibfnamefont {M.~B.}\ \bibnamefont {Jalil}},\ and\ \bibinfo {author} {\bibfnamefont {C.~H.}\ \bibnamefont {Lee}},\ }\bibfield  {title} {\bibinfo {title} {Protected chaos in a topological lattice},\ }\href {https://doi.org/10.1002/advs.202503216} {\bibfield  {journal} {\bibinfo  {journal} {Adv. Sci.}\ }\textbf {\bibinfo {volume} {12}},\ \bibinfo {pages} {e03216} (\bibinfo {year} {2025}{\natexlab{b}})}\BibitemShut {NoStop}%
\bibitem [{\citenamefont {Li}\ \emph {et~al.}(2025{\natexlab{b}})\citenamefont {Li}, \citenamefont {Wang}, \citenamefont {Kong}, \citenamefont {Lv}, \citenamefont {Jia}, \citenamefont {Tao}, \citenamefont {Li},\ and\ \citenamefont {Liu}}]{li2025realization}%
  \BibitemOpen
  \bibfield  {author} {\bibinfo {author} {\bibfnamefont {R.}~\bibnamefont {Li}}, \bibinfo {author} {\bibfnamefont {W.}~\bibnamefont {Wang}}, \bibinfo {author} {\bibfnamefont {X.}~\bibnamefont {Kong}}, \bibinfo {author} {\bibfnamefont {B.}~\bibnamefont {Lv}}, \bibinfo {author} {\bibfnamefont {Y.}~\bibnamefont {Jia}}, \bibinfo {author} {\bibfnamefont {H.}~\bibnamefont {Tao}}, \bibinfo {author} {\bibfnamefont {P.}~\bibnamefont {Li}},\ and\ \bibinfo {author} {\bibfnamefont {Y.}~\bibnamefont {Liu}},\ }\bibfield  {title} {\bibinfo {title} {Realization of a non-{H}ermitian {H}aldane model in circuits},\ }\bibfield  {journal} {\bibinfo  {journal} {arXiv preprint arXiv:2503.23737}\ }\href {https://doi.org/10.48550/arXiv.2503.23737} {10.48550/arXiv.2503.23737} (\bibinfo {year} {2025}{\natexlab{b}})\BibitemShut {NoStop}%
\bibitem [{\citenamefont {Shen}\ \emph {et~al.}(2026)\citenamefont {Shen}, \citenamefont {Chen}, \citenamefont {Tai}, \citenamefont {Koh}, \citenamefont {Ghaemi},\ and\ \citenamefont {Lee}}]{shen2026simulating}%
  \BibitemOpen
  \bibfield  {author} {\bibinfo {author} {\bibfnamefont {R.}~\bibnamefont {Shen}}, \bibinfo {author} {\bibfnamefont {T.}~\bibnamefont {Chen}}, \bibinfo {author} {\bibfnamefont {T.}~\bibnamefont {Tai}}, \bibinfo {author} {\bibfnamefont {J.~M.}\ \bibnamefont {Koh}}, \bibinfo {author} {\bibfnamefont {P.}~\bibnamefont {Ghaemi}},\ and\ \bibinfo {author} {\bibfnamefont {C.~H.}\ \bibnamefont {Lee}},\ }\bibfield  {title} {\bibinfo {title} {Simulating condensed matter physics on quantum hardware},\ }\bibfield  {journal} {\bibinfo  {journal} {arXiv preprint arXiv:2606.02721}\ }\href {https://doi.org/10.48550/arXiv.2606.02721} {10.48550/arXiv.2606.02721} (\bibinfo {year} {2026})\BibitemShut {NoStop}%
\bibitem [{\citenamefont {Smith}\ \emph {et~al.}(2019)\citenamefont {Smith}, \citenamefont {Kim}, \citenamefont {Pollmann},\ and\ \citenamefont {Knolle}}]{smith2019simulating}%
  \BibitemOpen
  \bibfield  {author} {\bibinfo {author} {\bibfnamefont {A.}~\bibnamefont {Smith}}, \bibinfo {author} {\bibfnamefont {M.}~\bibnamefont {Kim}}, \bibinfo {author} {\bibfnamefont {F.}~\bibnamefont {Pollmann}},\ and\ \bibinfo {author} {\bibfnamefont {J.}~\bibnamefont {Knolle}},\ }\bibfield  {title} {\bibinfo {title} {Simulating quantum many-body dynamics on a current digital quantum computer},\ }\href {https://doi.org/10.1038/s41534-019-0217-0} {\bibfield  {journal} {\bibinfo  {journal} {npj Quantum Information}\ }\textbf {\bibinfo {volume} {5}},\ \bibinfo {pages} {106} (\bibinfo {year} {2019})}\BibitemShut {NoStop}%
\bibitem [{\citenamefont {Gou}\ \emph {et~al.}(2020)\citenamefont {Gou}, \citenamefont {Chen}, \citenamefont {Xie}, \citenamefont {Xiao}, \citenamefont {Deng}, \citenamefont {Gadway}, \citenamefont {Yi},\ and\ \citenamefont {Yan}}]{gou2020tunable}%
  \BibitemOpen
  \bibfield  {author} {\bibinfo {author} {\bibfnamefont {W.}~\bibnamefont {Gou}}, \bibinfo {author} {\bibfnamefont {T.}~\bibnamefont {Chen}}, \bibinfo {author} {\bibfnamefont {D.}~\bibnamefont {Xie}}, \bibinfo {author} {\bibfnamefont {T.}~\bibnamefont {Xiao}}, \bibinfo {author} {\bibfnamefont {T.-S.}\ \bibnamefont {Deng}}, \bibinfo {author} {\bibfnamefont {B.}~\bibnamefont {Gadway}}, \bibinfo {author} {\bibfnamefont {W.}~\bibnamefont {Yi}},\ and\ \bibinfo {author} {\bibfnamefont {B.}~\bibnamefont {Yan}},\ }\bibfield  {title} {\bibinfo {title} {Tunable nonreciprocal quantum transport through a dissipative {A}haronov-{B}ohm ring in ultracold atoms},\ }\href {https://doi.org/10.1103/PhysRevLett.124.070402} {\bibfield  {journal} {\bibinfo  {journal} {Phys. Rev. Lett.}\ }\textbf {\bibinfo {volume} {124}},\ \bibinfo {pages} {070402} (\bibinfo {year} {2020})}\BibitemShut {NoStop}%
\bibitem [{\citenamefont {Koh}\ \emph {et~al.}(2022)\citenamefont {Koh}, \citenamefont {Tai},\ and\ \citenamefont {Lee}}]{koh2022simulation}%
  \BibitemOpen
  \bibfield  {author} {\bibinfo {author} {\bibfnamefont {J.~M.}\ \bibnamefont {Koh}}, \bibinfo {author} {\bibfnamefont {T.}~\bibnamefont {Tai}},\ and\ \bibinfo {author} {\bibfnamefont {C.~H.}\ \bibnamefont {Lee}},\ }\bibfield  {title} {\bibinfo {title} {Simulation of interaction-induced chiral topological dynamics on a digital quantum computer},\ }\href {https://doi.org/10.1103/PhysRevLett.129.140502} {\bibfield  {journal} {\bibinfo  {journal} {Phys. Rev. Lett.}\ }\textbf {\bibinfo {volume} {129}},\ \bibinfo {pages} {140502} (\bibinfo {year} {2022})}\BibitemShut {NoStop}%
\bibitem [{\citenamefont {Kirmani}\ \emph {et~al.}(2022)\citenamefont {Kirmani}, \citenamefont {Bull}, \citenamefont {Hou}, \citenamefont {Saravanan}, \citenamefont {Saeed}, \citenamefont {Papi{\'c}}, \citenamefont {Rahmani},\ and\ \citenamefont {Ghaemi}}]{kirmani2022probing}%
  \BibitemOpen
  \bibfield  {author} {\bibinfo {author} {\bibfnamefont {A.}~\bibnamefont {Kirmani}}, \bibinfo {author} {\bibfnamefont {K.}~\bibnamefont {Bull}}, \bibinfo {author} {\bibfnamefont {C.-Y.}\ \bibnamefont {Hou}}, \bibinfo {author} {\bibfnamefont {V.}~\bibnamefont {Saravanan}}, \bibinfo {author} {\bibfnamefont {S.~M.}\ \bibnamefont {Saeed}}, \bibinfo {author} {\bibfnamefont {Z.}~\bibnamefont {Papi{\'c}}}, \bibinfo {author} {\bibfnamefont {A.}~\bibnamefont {Rahmani}},\ and\ \bibinfo {author} {\bibfnamefont {P.}~\bibnamefont {Ghaemi}},\ }\bibfield  {title} {\bibinfo {title} {Probing geometric excitations of fractional quantum {H}all states on quantum computers},\ }\href {https://doi.org/10.1103/PhysRevLett.129.056801} {\bibfield  {journal} {\bibinfo  {journal} {Phys. Rev. Lett.}\ }\textbf {\bibinfo {volume} {129}},\ \bibinfo {pages} {056801} (\bibinfo {year} {2022})}\BibitemShut {NoStop}%
\bibitem [{\citenamefont {Frey}\ and\ \citenamefont {Rachel}(2022)}]{frey2022realization}%
  \BibitemOpen
  \bibfield  {author} {\bibinfo {author} {\bibfnamefont {P.}~\bibnamefont {Frey}}\ and\ \bibinfo {author} {\bibfnamefont {S.}~\bibnamefont {Rachel}},\ }\bibfield  {title} {\bibinfo {title} {Realization of a discrete time crystal on 57 qubits of a quantum computer},\ }\href {https://doi.org/10.1126/sciadv.abm7652} {\bibfield  {journal} {\bibinfo  {journal} {Sci. Adv.}\ }\textbf {\bibinfo {volume} {8}},\ \bibinfo {pages} {eabm7652} (\bibinfo {year} {2022})}\BibitemShut {NoStop}%
\bibitem [{\citenamefont {Chertkov}\ \emph {et~al.}(2023)\citenamefont {Chertkov}, \citenamefont {Cheng}, \citenamefont {Potter}, \citenamefont {Gopalakrishnan}, \citenamefont {Gatterman}, \citenamefont {Gerber}, \citenamefont {Gilmore}, \citenamefont {Gresh}, \citenamefont {Hall}, \citenamefont {Hankin} \emph {et~al.}}]{chertkov2023characterizing}%
  \BibitemOpen
  \bibfield  {author} {\bibinfo {author} {\bibfnamefont {E.}~\bibnamefont {Chertkov}}, \bibinfo {author} {\bibfnamefont {Z.}~\bibnamefont {Cheng}}, \bibinfo {author} {\bibfnamefont {A.~C.}\ \bibnamefont {Potter}}, \bibinfo {author} {\bibfnamefont {S.}~\bibnamefont {Gopalakrishnan}}, \bibinfo {author} {\bibfnamefont {T.~M.}\ \bibnamefont {Gatterman}}, \bibinfo {author} {\bibfnamefont {J.~A.}\ \bibnamefont {Gerber}}, \bibinfo {author} {\bibfnamefont {K.}~\bibnamefont {Gilmore}}, \bibinfo {author} {\bibfnamefont {D.}~\bibnamefont {Gresh}}, \bibinfo {author} {\bibfnamefont {A.}~\bibnamefont {Hall}}, \bibinfo {author} {\bibfnamefont {A.}~\bibnamefont {Hankin}}, \emph {et~al.},\ }\bibfield  {title} {\bibinfo {title} {Characterizing a non-equilibrium phase transition on a quantum computer},\ }\href {https://doi.org/10.1038/s41567-023-02199-w} {\bibfield  {journal} {\bibinfo  {journal} {Nat. Phys.}\ ,\ \bibinfo {pages} {1}} (\bibinfo {year} {2023})}\BibitemShut {NoStop}%
\bibitem [{\citenamefont {Chen}\ \emph {et~al.}(2023)\citenamefont {Chen}, \citenamefont {Shen}, \citenamefont {Lee},\ and\ \citenamefont {Yang}}]{chen2023high}%
  \BibitemOpen
  \bibfield  {author} {\bibinfo {author} {\bibfnamefont {T.}~\bibnamefont {Chen}}, \bibinfo {author} {\bibfnamefont {R.}~\bibnamefont {Shen}}, \bibinfo {author} {\bibfnamefont {C.~H.}\ \bibnamefont {Lee}},\ and\ \bibinfo {author} {\bibfnamefont {B.}~\bibnamefont {Yang}},\ }\bibfield  {title} {\bibinfo {title} {High-fidelity realization of the {AKLT} state on a {NISQ}-era quantum processor},\ }\href {https://doi.org/10.21468/SciPostPhys.15.4.170} {\bibfield  {journal} {\bibinfo  {journal} {SciPost Phys.}\ }\textbf {\bibinfo {volume} {15}},\ \bibinfo {pages} {170} (\bibinfo {year} {2023})}\BibitemShut {NoStop}%
\bibitem [{\citenamefont {Liu}\ \emph {et~al.}(2024{\natexlab{b}})\citenamefont {Liu}, \citenamefont {Shtengel},\ and\ \citenamefont {Pollmann}}]{liu2024simulating}%
  \BibitemOpen
  \bibfield  {author} {\bibinfo {author} {\bibfnamefont {Y.-J.}\ \bibnamefont {Liu}}, \bibinfo {author} {\bibfnamefont {K.}~\bibnamefont {Shtengel}},\ and\ \bibinfo {author} {\bibfnamefont {F.}~\bibnamefont {Pollmann}},\ }\bibfield  {title} {\bibinfo {title} {Simulating two-dimensional topological quantum phase transitions on a digital quantum computer},\ }\href {https://doi.org/10.1103/PhysRevResearch.6.043256} {\bibfield  {journal} {\bibinfo  {journal} {Phys. Rev. Research}\ }\textbf {\bibinfo {volume} {6}},\ \bibinfo {pages} {043256} (\bibinfo {year} {2024}{\natexlab{b}})}\BibitemShut {NoStop}%
\bibitem [{\citenamefont {Yang}\ \emph {et~al.}(2023)\citenamefont {Yang}, \citenamefont {Christianen}, \citenamefont {Coll-Vinent}, \citenamefont {Smelyanskiy}, \citenamefont {Ba{\~n}uls}, \citenamefont {O'Brien}, \citenamefont {Wild},\ and\ \citenamefont {Cirac}}]{yang2023simulating}%
  \BibitemOpen
  \bibfield  {author} {\bibinfo {author} {\bibfnamefont {Y.}~\bibnamefont {Yang}}, \bibinfo {author} {\bibfnamefont {A.}~\bibnamefont {Christianen}}, \bibinfo {author} {\bibfnamefont {S.}~\bibnamefont {Coll-Vinent}}, \bibinfo {author} {\bibfnamefont {V.}~\bibnamefont {Smelyanskiy}}, \bibinfo {author} {\bibfnamefont {M.~C.}\ \bibnamefont {Ba{\~n}uls}}, \bibinfo {author} {\bibfnamefont {T.~E.}\ \bibnamefont {O'Brien}}, \bibinfo {author} {\bibfnamefont {D.~S.}\ \bibnamefont {Wild}},\ and\ \bibinfo {author} {\bibfnamefont {J.~I.}\ \bibnamefont {Cirac}},\ }\bibfield  {title} {\bibinfo {title} {Simulating prethermalization using near-term quantum computers},\ }\href {https://doi.org/10.1103/PRXQuantum.4.030320} {\bibfield  {journal} {\bibinfo  {journal} {PRX Quantum}\ }\textbf {\bibinfo {volume} {4}},\ \bibinfo {pages} {030320} (\bibinfo {year} {2023})}\BibitemShut {NoStop}%
\bibitem [{\citenamefont {Iqbal}\ \emph {et~al.}(2023)\citenamefont {Iqbal}, \citenamefont {Tantivasadakarn}, \citenamefont {Verresen}, \citenamefont {Campbell}, \citenamefont {Dreiling}, \citenamefont {Figgatt}, \citenamefont {Gaebler}, \citenamefont {Johansen}, \citenamefont {Mills}, \citenamefont {Moses} \emph {et~al.}}]{iqbal2023creation}%
  \BibitemOpen
  \bibfield  {author} {\bibinfo {author} {\bibfnamefont {M.}~\bibnamefont {Iqbal}}, \bibinfo {author} {\bibfnamefont {N.}~\bibnamefont {Tantivasadakarn}}, \bibinfo {author} {\bibfnamefont {R.}~\bibnamefont {Verresen}}, \bibinfo {author} {\bibfnamefont {S.~L.}\ \bibnamefont {Campbell}}, \bibinfo {author} {\bibfnamefont {J.~M.}\ \bibnamefont {Dreiling}}, \bibinfo {author} {\bibfnamefont {C.}~\bibnamefont {Figgatt}}, \bibinfo {author} {\bibfnamefont {J.~P.}\ \bibnamefont {Gaebler}}, \bibinfo {author} {\bibfnamefont {J.}~\bibnamefont {Johansen}}, \bibinfo {author} {\bibfnamefont {M.}~\bibnamefont {Mills}}, \bibinfo {author} {\bibfnamefont {S.~A.}\ \bibnamefont {Moses}}, \emph {et~al.},\ }\bibfield  {title} {\bibinfo {title} {Creation of non-{A}belian topological order and anyons on a trapped-ion processor},\ }\bibfield  {journal} {\bibinfo  {journal} {arXiv preprint arXiv:2305.03766}\ }\href {https://doi.org/10.48550/arXiv.2305.03766} {10.48550/arXiv.2305.03766} (\bibinfo {year} {2023})\BibitemShut {NoStop}%
\bibitem [{\citenamefont {Shen}\ \emph {et~al.}(2025{\natexlab{a}})\citenamefont {Shen}, \citenamefont {Chen}, \citenamefont {Yang},\ and\ \citenamefont {Lee}}]{shen2025observation}%
  \BibitemOpen
  \bibfield  {author} {\bibinfo {author} {\bibfnamefont {R.}~\bibnamefont {Shen}}, \bibinfo {author} {\bibfnamefont {T.}~\bibnamefont {Chen}}, \bibinfo {author} {\bibfnamefont {B.}~\bibnamefont {Yang}},\ and\ \bibinfo {author} {\bibfnamefont {C.~H.}\ \bibnamefont {Lee}},\ }\bibfield  {title} {\bibinfo {title} {Observation of the non-{H}ermitian skin effect and {F}ermi skin on a digital quantum computer},\ }\href {https://doi.org/10.1038/s41467-025-55953-4} {\bibfield  {journal} {\bibinfo  {journal} {Nat. Commun.}\ }\textbf {\bibinfo {volume} {16}},\ \bibinfo {pages} {1340} (\bibinfo {year} {2025}{\natexlab{a}})}\BibitemShut {NoStop}%
\bibitem [{\citenamefont {Koukoutsis}\ \emph {et~al.}(2024)\citenamefont {Koukoutsis}, \citenamefont {Papagiannis}, \citenamefont {Hizanidis}, \citenamefont {Ram}, \citenamefont {Vahala}, \citenamefont {Amaro}, \citenamefont {Gamiz},\ and\ \citenamefont {Vallis}}]{koukoutsis2024quantum}%
  \BibitemOpen
  \bibfield  {author} {\bibinfo {author} {\bibfnamefont {E.}~\bibnamefont {Koukoutsis}}, \bibinfo {author} {\bibfnamefont {P.}~\bibnamefont {Papagiannis}}, \bibinfo {author} {\bibfnamefont {K.}~\bibnamefont {Hizanidis}}, \bibinfo {author} {\bibfnamefont {A.~K.}\ \bibnamefont {Ram}}, \bibinfo {author} {\bibfnamefont {G.}~\bibnamefont {Vahala}}, \bibinfo {author} {\bibfnamefont {O.}~\bibnamefont {Amaro}}, \bibinfo {author} {\bibfnamefont {L.~I.~I.}\ \bibnamefont {Gamiz}},\ and\ \bibinfo {author} {\bibfnamefont {D.}~\bibnamefont {Vallis}},\ }\bibfield  {title} {\bibinfo {title} {Quantum implementation of non-unitary operations with biorthogonal representations},\ }\bibfield  {journal} {\bibinfo  {journal} {arXiv preprint arXiv:2410.22505}\ }\href {https://doi.org/10.48550/arXiv.2410.22505} {10.48550/arXiv.2410.22505} (\bibinfo {year} {2024})\BibitemShut {NoStop}%
\bibitem [{\citenamefont {Koh}\ \emph {et~al.}(2024)\citenamefont {Koh}, \citenamefont {Tai},\ and\ \citenamefont {Lee}}]{koh2024realization}%
  \BibitemOpen
  \bibfield  {author} {\bibinfo {author} {\bibfnamefont {J.~M.}\ \bibnamefont {Koh}}, \bibinfo {author} {\bibfnamefont {T.}~\bibnamefont {Tai}},\ and\ \bibinfo {author} {\bibfnamefont {C.~H.}\ \bibnamefont {Lee}},\ }\bibfield  {title} {\bibinfo {title} {Realization of higher-order topological lattices on a quantum computer},\ }\href {https://doi.org/10.1038/s41467-024-49648-5} {\bibfield  {journal} {\bibinfo  {journal} {Nat. Commun.}\ }\textbf {\bibinfo {volume} {15}},\ \bibinfo {pages} {5807} (\bibinfo {year} {2024})}\BibitemShut {NoStop}%
\bibitem [{\citenamefont {Koh}\ \emph {et~al.}(2025)\citenamefont {Koh}, \citenamefont {Xue}, \citenamefont {Tai}, \citenamefont {Koh},\ and\ \citenamefont {Lee}}]{koh2025interacting}%
  \BibitemOpen
  \bibfield  {author} {\bibinfo {author} {\bibfnamefont {J.~M.}\ \bibnamefont {Koh}}, \bibinfo {author} {\bibfnamefont {W.-T.}\ \bibnamefont {Xue}}, \bibinfo {author} {\bibfnamefont {T.}~\bibnamefont {Tai}}, \bibinfo {author} {\bibfnamefont {D.~E.}\ \bibnamefont {Koh}},\ and\ \bibinfo {author} {\bibfnamefont {C.~H.}\ \bibnamefont {Lee}},\ }\bibfield  {title} {\bibinfo {title} {Interacting non-{H}ermitian edge and cluster bursts on a digital quantum processor},\ }\bibfield  {journal} {\bibinfo  {journal} {arXiv preprint arXiv:2503.14595}\ }\href {https://doi.org/10.48550/arXiv.2503.14595} {10.48550/arXiv.2503.14595} (\bibinfo {year} {2025})\BibitemShut {NoStop}%
\bibitem [{\citenamefont {Shen}\ \emph {et~al.}(2025{\natexlab{b}})\citenamefont {Shen}, \citenamefont {Chen}, \citenamefont {Yang}, \citenamefont {Zhong},\ and\ \citenamefont {Lee}}]{shen2025robust}%
  \BibitemOpen
  \bibfield  {author} {\bibinfo {author} {\bibfnamefont {R.}~\bibnamefont {Shen}}, \bibinfo {author} {\bibfnamefont {T.}~\bibnamefont {Chen}}, \bibinfo {author} {\bibfnamefont {B.}~\bibnamefont {Yang}}, \bibinfo {author} {\bibfnamefont {Y.}~\bibnamefont {Zhong}},\ and\ \bibinfo {author} {\bibfnamefont {C.~H.}\ \bibnamefont {Lee}},\ }\bibfield  {title} {\bibinfo {title} {Robust simulations of many-body symmetry-protected topological phase transitions on a quantum processor},\ }\href {https://doi.org/10.1038/s41534-025-01122-w} {\bibfield  {journal} {\bibinfo  {journal} {npj Quantum Information}\ }\textbf {\bibinfo {volume} {11}},\ \bibinfo {pages} {179} (\bibinfo {year} {2025}{\natexlab{b}})}\BibitemShut {NoStop}%
\bibitem [{\citenamefont {Zhang}\ \emph {et~al.}(2025)\citenamefont {Zhang}, \citenamefont {Carrasquilla},\ and\ \citenamefont {Kim}}]{zhang2025observation}%
  \BibitemOpen
  \bibfield  {author} {\bibinfo {author} {\bibfnamefont {Y.}~\bibnamefont {Zhang}}, \bibinfo {author} {\bibfnamefont {J.}~\bibnamefont {Carrasquilla}},\ and\ \bibinfo {author} {\bibfnamefont {Y.~B.}\ \bibnamefont {Kim}},\ }\bibfield  {title} {\bibinfo {title} {Observation of a non-{H}ermitian supersonic mode on a trapped-ion quantum computer},\ }\href {https://doi.org/10.1038/s41467-025-57930-3} {\bibfield  {journal} {\bibinfo  {journal} {Nat. Commun.}\ }\textbf {\bibinfo {volume} {16}},\ \bibinfo {pages} {3286} (\bibinfo {year} {2025})}\BibitemShut {NoStop}%
\bibitem [{\citenamefont {Chen}\ \emph {et~al.}(2017)\citenamefont {Chen}, \citenamefont {Kaya~{\"O}zdemir}, \citenamefont {Zhao}, \citenamefont {Wiersig},\ and\ \citenamefont {Yang}}]{chen2017exceptional}%
  \BibitemOpen
  \bibfield  {author} {\bibinfo {author} {\bibfnamefont {W.}~\bibnamefont {Chen}}, \bibinfo {author} {\bibfnamefont {{\c{S}}.}~\bibnamefont {Kaya~{\"O}zdemir}}, \bibinfo {author} {\bibfnamefont {G.}~\bibnamefont {Zhao}}, \bibinfo {author} {\bibfnamefont {J.}~\bibnamefont {Wiersig}},\ and\ \bibinfo {author} {\bibfnamefont {L.}~\bibnamefont {Yang}},\ }\bibfield  {title} {\bibinfo {title} {Exceptional points enhance sensing in an optical microcavity},\ }\href {https://doi.org/10.1038/nature23281} {\bibfield  {journal} {\bibinfo  {journal} {Nature}\ }\textbf {\bibinfo {volume} {548}},\ \bibinfo {pages} {192} (\bibinfo {year} {2017})}\BibitemShut {NoStop}%
\bibitem [{\citenamefont {Hodaei}\ \emph {et~al.}(2017)\citenamefont {Hodaei}, \citenamefont {Hassan}, \citenamefont {Wittek}, \citenamefont {Garcia-Gracia}, \citenamefont {El-Ganainy}, \citenamefont {Christodoulides},\ and\ \citenamefont {Khajavikhan}}]{hodaei2017enhanced}%
  \BibitemOpen
  \bibfield  {author} {\bibinfo {author} {\bibfnamefont {H.}~\bibnamefont {Hodaei}}, \bibinfo {author} {\bibfnamefont {A.~U.}\ \bibnamefont {Hassan}}, \bibinfo {author} {\bibfnamefont {S.}~\bibnamefont {Wittek}}, \bibinfo {author} {\bibfnamefont {H.}~\bibnamefont {Garcia-Gracia}}, \bibinfo {author} {\bibfnamefont {R.}~\bibnamefont {El-Ganainy}}, \bibinfo {author} {\bibfnamefont {D.~N.}\ \bibnamefont {Christodoulides}},\ and\ \bibinfo {author} {\bibfnamefont {M.}~\bibnamefont {Khajavikhan}},\ }\bibfield  {title} {\bibinfo {title} {Enhanced sensitivity at higher-order exceptional points},\ }\href {https://doi.org/10.1038/nature23280} {\bibfield  {journal} {\bibinfo  {journal} {Nature}\ }\textbf {\bibinfo {volume} {548}},\ \bibinfo {pages} {187} (\bibinfo {year} {2017})}\BibitemShut {NoStop}%
\bibitem [{\citenamefont {Ashida}\ \emph {et~al.}(2020)\citenamefont {Ashida}, \citenamefont {Gong},\ and\ \citenamefont {Ueda}}]{ashida2020non}%
  \BibitemOpen
  \bibfield  {author} {\bibinfo {author} {\bibfnamefont {Y.}~\bibnamefont {Ashida}}, \bibinfo {author} {\bibfnamefont {Z.}~\bibnamefont {Gong}},\ and\ \bibinfo {author} {\bibfnamefont {M.}~\bibnamefont {Ueda}},\ }\bibfield  {title} {\bibinfo {title} {Non-{H}ermitian physics},\ }\href {https://doi.org/10.1080/00018732.2021.1876991} {\bibfield  {journal} {\bibinfo  {journal} {Adv. Phys.}\ }\textbf {\bibinfo {volume} {69}},\ \bibinfo {pages} {249} (\bibinfo {year} {2020})}\BibitemShut {NoStop}%
\bibitem [{\citenamefont {Bergholtz}\ \emph {et~al.}(2021)\citenamefont {Bergholtz}, \citenamefont {Budich},\ and\ \citenamefont {Kunst}}]{bergholtz2021exceptional}%
  \BibitemOpen
  \bibfield  {author} {\bibinfo {author} {\bibfnamefont {E.~J.}\ \bibnamefont {Bergholtz}}, \bibinfo {author} {\bibfnamefont {J.~C.}\ \bibnamefont {Budich}},\ and\ \bibinfo {author} {\bibfnamefont {F.~K.}\ \bibnamefont {Kunst}},\ }\bibfield  {title} {\bibinfo {title} {Exceptional topology of non-{H}ermitian systems},\ }\href {https://doi.org/10.1103/RevModPhys.93.015005} {\bibfield  {journal} {\bibinfo  {journal} {Rev. Mod. Phys.}\ }\textbf {\bibinfo {volume} {93}},\ \bibinfo {pages} {015005} (\bibinfo {year} {2021})}\BibitemShut {NoStop}%
\bibitem [{\citenamefont {Hsu}\ and\ \citenamefont {Chen}(2020)}]{hsu2020anderson}%
  \BibitemOpen
  \bibfield  {author} {\bibinfo {author} {\bibfnamefont {H.-C.}\ \bibnamefont {Hsu}}\ and\ \bibinfo {author} {\bibfnamefont {T.-W.}\ \bibnamefont {Chen}},\ }\bibfield  {title} {\bibinfo {title} {Topological anderson insulating phases in the long-range {Su--Schrieffer--Heeger} model},\ }\href {https://doi.org/10.1103/PhysRevB.102.205425} {\bibfield  {journal} {\bibinfo  {journal} {Physical Review B}\ }\textbf {\bibinfo {volume} {102}},\ \bibinfo {pages} {205425} (\bibinfo {year} {2020})}\BibitemShut {NoStop}%
\bibitem [{\citenamefont {Dias}\ and\ \citenamefont {Marques}(2022)}]{dias2022indexing}%
  \BibitemOpen
  \bibfield  {author} {\bibinfo {author} {\bibfnamefont {R.~G.}\ \bibnamefont {Dias}}\ and\ \bibinfo {author} {\bibfnamefont {A.~M.}\ \bibnamefont {Marques}},\ }\bibfield  {title} {\bibinfo {title} {Long-range hopping and indexing assumption in one-dimensional topological insulators},\ }\href {https://doi.org/10.1103/PhysRevB.105.035102} {\bibfield  {journal} {\bibinfo  {journal} {Physical Review B}\ }\textbf {\bibinfo {volume} {105}},\ \bibinfo {pages} {035102} (\bibinfo {year} {2022})}\BibitemShut {NoStop}%
\bibitem [{\citenamefont {Ghosh}\ \emph {et~al.}(2023)\citenamefont {Ghosh}, \citenamefont {Martin},\ and\ \citenamefont {Majumder}}]{ghosh2023quench}%
  \BibitemOpen
  \bibfield  {author} {\bibinfo {author} {\bibfnamefont {A.}~\bibnamefont {Ghosh}}, \bibinfo {author} {\bibfnamefont {A.~M.}\ \bibnamefont {Martin}},\ and\ \bibinfo {author} {\bibfnamefont {S.}~\bibnamefont {Majumder}},\ }\bibfield  {title} {\bibinfo {title} {Quench dynamics of edge states in a finite extended {Su--Schrieffer--Heeger} system},\ }\href {https://doi.org/10.1103/PhysRevE.108.034102} {\bibfield  {journal} {\bibinfo  {journal} {Physical Review E}\ }\textbf {\bibinfo {volume} {108}},\ \bibinfo {pages} {034102} (\bibinfo {year} {2023})}\BibitemShut {NoStop}%
\bibitem [{\citenamefont {Betancur-Ocampo}\ \emph {et~al.}(2024)\citenamefont {Betancur-Ocampo}, \citenamefont {Manjarrez-Monta{\~n}ez}, \citenamefont {Mart{\'i}nez-Arg{\"u}ello},\ and\ \citenamefont {M{\'e}ndez-S{\'a}nchez}}]{betancurocampo2024twofold}%
  \BibitemOpen
  \bibfield  {author} {\bibinfo {author} {\bibfnamefont {Y.}~\bibnamefont {Betancur-Ocampo}}, \bibinfo {author} {\bibfnamefont {B.}~\bibnamefont {Manjarrez-Monta{\~n}ez}}, \bibinfo {author} {\bibfnamefont {A.~M.}\ \bibnamefont {Mart{\'i}nez-Arg{\"u}ello}},\ and\ \bibinfo {author} {\bibfnamefont {R.~A.}\ \bibnamefont {M{\'e}ndez-S{\'a}nchez}},\ }\bibfield  {title} {\bibinfo {title} {Twofold topological phase transitions induced by third-nearest-neighbor hoppings in one-dimensional chains},\ }\href {https://doi.org/10.1103/PhysRevB.109.104111} {\bibfield  {journal} {\bibinfo  {journal} {Physical Review B}\ }\textbf {\bibinfo {volume} {109}},\ \bibinfo {pages} {104111} (\bibinfo {year} {2024})}\BibitemShut {NoStop}%
\end{thebibliography}
\end{document}